\documentclass[11pt]{article}
\pdfoutput=1
\usepackage{jheppub}
\usepackage{graphicx}
\usepackage{array,tikz-cd}
\usetikzlibrary{positioning, quotes, arrows}
\usepackage{color}
\usepackage{bm, dsfont, braket}

\graphicspath{{./vxfigures/}}

\newcommand{\pd}{{\partial}}
\newcommand{\ol}{\overline}
\newcommand{\wt}{\widetilde}
\newcommand{\mb}{\mathbf}
\newcommand{\ds}{\displaystyle}
\newcommand{\cp}{{\mathbb{CP}}}
\newcommand{\id}{{\mathds{1}}}

\renewcommand{\k}{\mathbf{k}}
\newcommand{\n}{\mathfrak{n}}

\newcommand{\Hom}{\text{Hom}}
\newcommand{\End}{\text{End}}

\newcommand{\rav}{\text{rav}}

\newcommand{\SQED}{\text{SQED}}

\newcommand{\SQM}{\text{SQM}}

\newcommand{\QB}{Q_B}

\newcommand{\Tr}{\text{Tr}}
\newcommand{\one}{\mathbf 1}

\newcommand{\cf}{\emph{cf.}}
\newcommand{\ie}{\emph{i.e.}}
\newcommand{\eg}{\emph{e.g.}}

\newcommand{\be}{\begin{equation}}
\newcommand{\ee}{\end{equation}}
\newcommand{\bp}{\begin{pmatrix}}
\newcommand{\ep}{\end{pmatrix}}
\newcommand{\bsp}{\left(\begin{smallmatrix}}
\newcommand{\esp}{\end{smallmatrix}\right)}

\newcommand{\R}{{\mathbb R}}
\renewcommand{\P}{{\mathbb P}}

\newcommand{\V}{{\mathbb V}}
\newcommand{\W}{{\mathbb W}}
\newcommand{\C}{{\mathbb C}}
\newcommand{\Z}{{\mathbb Z}}

\newcommand{\CA}{{\mathcal A}}
\newcommand{\CB}{{\mathcal B}}
\newcommand{\CC}{{\mathcal C}}

\newcommand{\CE}{{\mathcal E}}
\newcommand{\CF}{{\mathcal F}}
\newcommand{\CG}{{\mathcal G}}
\newcommand{\CH}{{\mathcal H}}
\newcommand{\CI}{{\mathcal I}}
\newcommand{\CK}{{\mathcal K}}
\newcommand{\CM}{{\mathcal M}}
\newcommand{\CN}{{\mathcal N}}
\newcommand{\CO}{{\mathcal O}}

\newcommand{\CS}{{\mathcal S}}
\newcommand{\CT}{{\mathcal T}}

\newcommand{\CV}{{\mathbb V}}  
\newcommand{\CX}{{\mathcal X}}

\newcommand{\CL}{{\mathcal L}}

\title{Mirror symmetry and line operators}
\author[1]{Tudor Dimofte,}
\author[1]{Niklas Garner,}
\author[1]{Michael Geracie,}
\author[2]{Justin Hilburn}

\affiliation[1]{Department of Mathematics and Center for Quantum Mathematics and Physics (QMAP), University of California, Davis, CA 95616, USA}
\affiliation[2]{Department of Mathematics, University of Pennsylvania, Philadelphia, PA 19104, USA}

\emailAdd{tudor@math.ucdavis.edu, nkgarner@ucdavis.edu, mgeracie@mail.tsinghua.edu.cn, jhilburn@math.upenn.edu}

\abstract{We study half-BPS line operators in 3d $\CN=4$ gauge theories, focusing in particular on the algebras of local operators at their junctions. It is known that there are two basic types of such line operators, distinguished by the SUSY subalgebras that they preserve; the two types can roughly be called ``Wilson lines'' and ``vortex lines,'' and are exchanged under 3d mirror symmetry. We describe a large class of vortex lines that can be characterized by basic algebraic data, and propose a mathematical scheme to compute the algebras of local operators at their junctions --- including monopole operators --- in terms of this data. The computation generalizes mathematical and physical definitions/analyses of the bulk Coulomb-branch chiral ring. We fully classify the junctions of half-BPS Wilson lines and of half-BPS vortex lines in abelian gauge theories with sufficient matter. We also test our computational scheme in a non-abelian quiver gauge theory, using a 3d-mirror-map of line operators from work of Assel and Gomis.
}

\begin{document}
\today
\maketitle


\newpage
\section{Introduction}

BPS line operators in supersymmetry gauge theories hold a wealth of algebraic and geometric structure. Such structure has been most extensively studied in four-dimensional supersymmetric gauge theories, where one encounters BPS Wilson lines \cite{Wilson, BlauThompson2, BLN-Wilson, Maldacena-Wilson, ReyYee-Wilson, Zarembo-Wilson}, BPS 't Hooft lines \cite{tHooft, GNO, Kapustin-Wt, KapustinWitten}, and hybrids thereof.
A few of the contexts in which these line operators have played a central role during the last decade and a half include 
the physics of geometric Langlands   \cite{KapustinWitten, GW-surface}, wall crossing phenomena  \cite{GMN-framed, ADJM-galaxies}, and the AGT correspondence \cite{AGT, AGGTV, DMO, DGOT}. It was also realized that the precise spectrum of line operators constitutes part of the very definition of a 4d gauge theory \cite{GMN-framed, AharonySeibergTachikawa}.

This paper focuses on half-BPS line operators in \emph{three} dimensions, specifically in 3d $\CN=4$ gauge theories.
Much as in 4d, line operators in 3d gauge theories come in two basic varieties: Wilson lines (`order' operators) and \emph{vortex} lines (disorder operators). 
Supersymmetric Wilson lines in pure 3d $\CN=4$ gauge theories were introduced by \cite{BlauThompson2}; and their analogues in sigma-models \cite{RW, KRS} played a central role in the construction of Rozansky-Witten invariants.
Supersymmetric vortex lines are codimension-two disorder operators, modeled on singular limits of the Nielsen-Olesen vortex \cite{NO} and its supersymmetric cousins, \eg\ \cite{BJSV, ENS-vortices, HananyTong-VIB}. They may also be understood as dimensional reductions of supersymmetric surface operators in 4d gauge theories: the basic Gukov-Witten surface operators \cite{GW-surface, GW-rigid, Witten-wild} and their 4d $\CN=2$ analogues \cite{Gukov-gaugeknot, KohYamaguchi,Tan-surface,AGGTV,Gaiotto-surface}, studied and generalized in many later works --- a small sampling includes \cite{DGH, GMN-2d4d, GGS,  KannoTachikawa,  GaiottoRastelliRazamat, GaddeGukov, FrenkelGukovTeschner, GorskyLeFloch, NekrasovIV} (see \cite{Gukov-surface} for a clear review). Compactifying further to two dimensions, vortex lines become twist fields, which played a fundamental role in T-duality/mirror symmetry \cite{HoriVafa-MS} and were recently reexamined by \cite{Okuda-2d, HLO-2d}.

Supersymmetric vortex lines in the 3d $\CN=6$ ABJM theory were constructed by \cite{DrukkerGomisYoung} (further studied in many works \emph{e.g.} \cite{Arai:2008kv,Kim:2009ny,Auzzi:2009es,Lee:2010hk}); then generalized and studied in 3d $\CN=2$ theories by \cite{DGG, KWY-vortex, DOP-vortex} using supersymmetric localization. Further physical aspects of vortex lines in abelian 3d $\CN=2$ and $\CN=4$ theories were developed in \cite{TongWong, HKT}. 
A systematic study of half-BPS vortex lines in 3d $\CN=4$ quiver gauge theories --- both abelian and nonabelian --- was initiated more recently by Assel and Gomis \cite{AsselGomis} using IIB brane constructions \cite{HananyWitten}, akin to the constructions of surface operators in \cite{HananyHori, DGH}. 
It was shown by \cite{AsselGomis} that 3d mirror symmetry \cite{IS, dBHOO, dBHOOY} swaps Wilson and vortex lines in quiver gauge theories. The rather nontrivial mirror map was verified with computations of supersymmetric partition functions, generalizing \cite{KWY-vortex, DOP-vortex, HoriKimYi, CordovaShao-QM}.

Our overarching goal in the current paper is to describe --- in both a theoretical and a computationally effective way --- the BPS local operators at \emph{junctions} of line operators, and their OPE.%
\footnote{Our use of the term ``line operator,'' as opposed to ``loop operator,'' is meant to emphasize the focus on such local structure.} %
We will expand on precisely what this means further below. For Wilson lines, achieving this goal is relatively straightforward.
Understanding junctions of vortex lines, however, requires some work. Indeed, to the best of our knowledge, even half-BPS vortex lines themselves have not been fully classified in general gauge theories. Part of this paper will be devoted to first characterizing a large class of vortex lines in terms of algebraic data, and then proposing a precise computational scheme (based on the algebraic data) to determine the spaces of local operators at junctions. We test this scheme in several nontrivial examples.

\subsection{Some general structure}
\label{sec:intro-struc}

The order/disorder distinction between Wilson and vortex lines is not truly intrinsic. For example, many vortex lines may equivalently be engineered by coupling a 3d theory to 1d degrees of freedom in a nonsingular way; and the order/disorder distinction is also not preserved across dualities. A better distinction comes from classifying the half-BPS SUSY subalgebras that a line operator can preserve. In this paper, we will always consider straight, parallel line operators,  preserving a 1d $\CN=4$ subalgebra of 3d $\CN=4$ that contains translations along the line. There are two inequivalent choices of 1d $\CN=4$ subalgebra, which we call $\SQM_A$ and $\SQM_B$.%
\footnote{These subalgebras were called $\SQM_V$ and $\SQM_W$, respectively, in \cite{AsselGomis}.} %
We refer to line operators that preserve $\SQM_A$ (resp. $\SQM_B$) as A-type (resp. B-type) lines; they include half-BPS vortex lines (resp. half-BPS Wilson lines). The subalgebras $\SQM_A$ and $\SQM_B$ --- and thus the entire collections of A-type and B-type line operators --- are exchanged by 3d mirror symmetry.

An important feature of $\SQM_A$ and $\SQM_B$ is that each subalgebra contains a topological supercharge, which we denote $Q_A$ and $Q_B$, respectively. These supercharges are nilpotent, and they make all translations (in fact, the entire stress-energy tensor) exact. The $Q_A$ supercharge is the 3d reduction of the supercharge used to define the Donaldson-Witten twist of 4d $\CN=2$ gauge theories \cite{Witten-Donaldson}; parts of its extended-TQFT structure in 3d gauge theories were discussed in \cite{KV, KSV}. The $Q_B$ supercharge defines a twist that was introduced in 3d gauge theories by \cite{BlauThompson2}; it is better known in 3d $\CN=4$ sigma-models as the Rozansky-Witten twist \cite{RW}, with extended-TQFT structure developed in \cite{KRS}.

\begin{figure}[htb]
\centering
\includegraphics[width=3.2in]{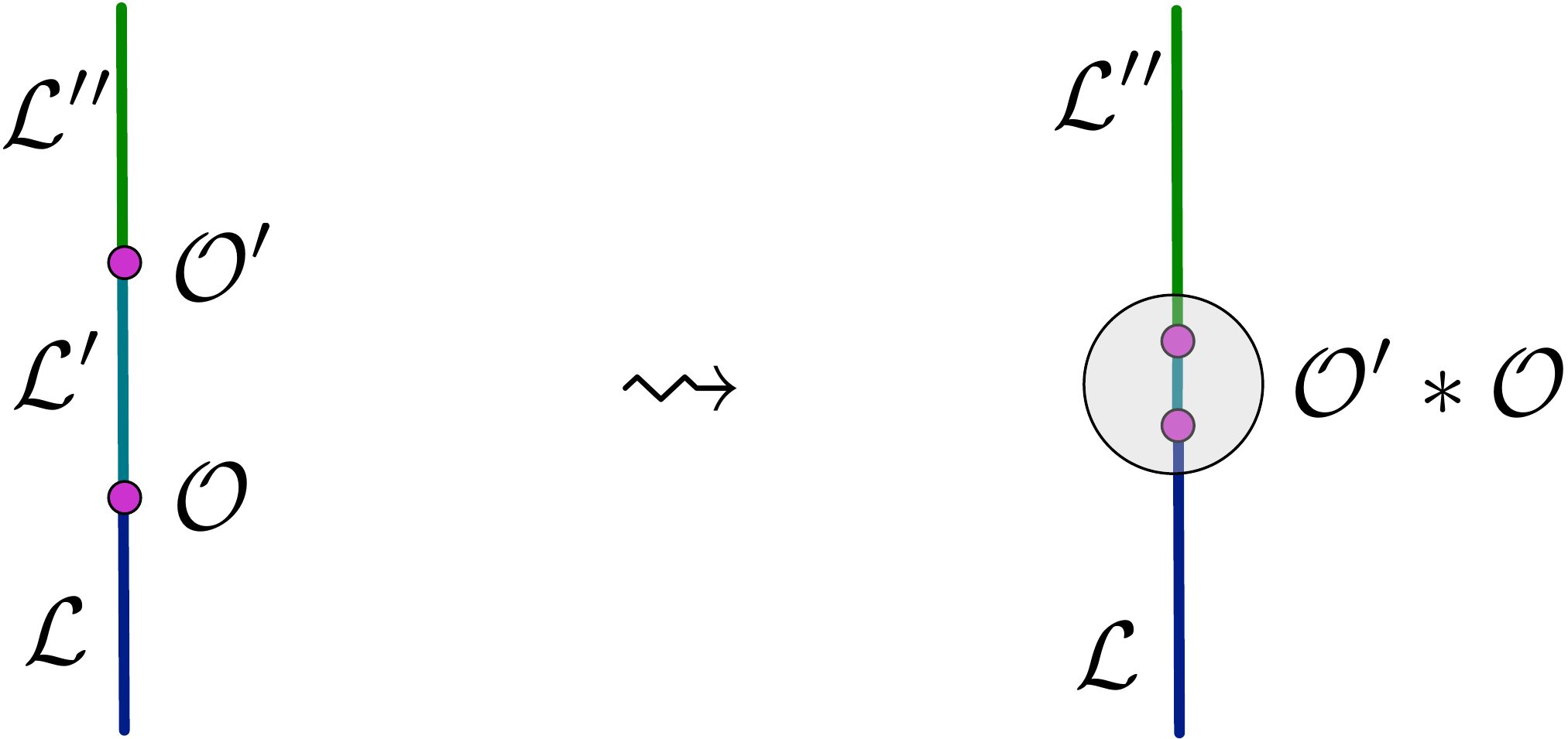}
\caption{Local operators at junctions of lines have products induced by collision.}
\label{fig:ops-intro}
\end{figure}

Now, any pair of line operators $\CL,\CL'$ defines a vector space $\text{Ops}(\CL,\CL')$ of local operators that can be inserted at the junction between $\CL$ and $\CL'$. (If no junction is possible, then $\text{Ops}(\CL,\CL')$ is simply declared to be the zero vector space.) If $\CL$ and $\CL'$ are both A-type or both B-type, then we can further consider the $Q_A$ or $Q_B$-cohomology of the space of local operators, denoted
\be \label{intro-HomAB}
 \text{Hom}_A(\CL,\CL') := H^\bullet(\text{Ops}(\CL,\CL'),Q_A)\qquad\text{or}\qquad \text{Hom}_B(\CL,\CL') := H^\bullet(\text{Ops}(\CL,\CL'),Q_B)\,.
 \ee
Moreover, since the cohomology of a topological supercharge is locally constant, there is an associative product on cohomology classes induced by collision of successive junctions, illustrated in Figure \ref{fig:ops-intro}. For example, if $\CL,\CL',\CL''$ are all A-type, then there is a product
\be \begin{array}{ccccc} \text{Hom}_A(\CL',\CL'')&\otimes& \text{Hom}_A(\CL,\CL')& \to& \text{Hom}_A(\CL,\CL'') \\
\CO'&&\CO & \mapsto & \CO' * \CO \end{array}\,, \label{intro-coll} \ee  
and similarly if $\CL,\CL',\CL''$ are all B-type.

Altogether, this endows the spaces $\text{Hom}_A$ or $\text{Hom}_B$ of local operators with an algebraic structure. 
It has some important specializations. For any single $\CL$, the space $\text{Hom}_*(\CL,\CL)$ of local operators bound to $\CL$ becomes a standard associative algebra with product \eqref{intro-coll}. Moreover, if we take $\CL=\id$ to be the trivial or ``empty'' line operator, then $\text{Ops}(\id,\id)$ simply consists of bulk local operators, and the $Q$-cohomologies
\be \label{intro-C} \text{Hom}_A(\id,\id) \supseteq \C[\CM_C]\,,\qquad \text{Hom}_B(\id,\id) \supseteq \C[\CM_H] \ee
turn out to contain the Coulomb-branch and Higgs-branch chiral rings. Thus, chiral rings can be recovered from knowing about junctions of line operators.

More broadly, one expects from general principles of extended TQFT (\cf\ \cite{Lurie, Kapustin-ICM}) that the set of all line operators preserved by a supercharge $Q_A$ (resp. $Q_B$) --- including the half-BPS line operators --- acquire the structure of an $E_2$-monoidal category $\CC_A$ (resp. $\CC_B$).%
\footnote{For a physical interpretation of the $E_2$-monoidal (in particular, braided monoidal) structure see \cite{descent}.} %
From this perspective, $\text{Hom}_*(\CL,\CL')$ is the cohomology of a morphism space between objects $\CL,\CL'$, and \eqref{intro-coll} is composition of morphisms. A geometric model of $\CC_B$ in 3d $\CN=4$ sigma-models was proposed by \cite{KRS}, and interesting examples of $\CC_B$ in gauge theories appeared in \cite{RO-homology, RO-Chern}.

Algebro-geometric models for both $\CC_A$ and $\CC_B$ in general 3d $\CN=4$ gauge theories with linear symplectic matter will be proposed in the upcoming \cite{lineops, HilburnYoo}. There are technical challenges to overcome in properly defining them, from both mathematical and physical perspectives, which are well outside the scope of the current paper. As a rough preview, given compact gauge group $G$ and hypermultiplet representation $R\oplus R^*$, we expect the category $\CC_A$ to be described as equivariant D-modules on the loop space of $R$, and $\CC_B$ to be described as quasi-coherent sheaves on the derived (homologically constant) loop space of the stack $R/G_\C$
\be \CC_A \approx \text{D-mod}_{LG_\C}(LR)\,,\qquad \CC_B  \approx \text{QCoh}\big(\text{Maps}(S^1_{\rm dR}, R/G_\C)\big)\,. \label{intro-CAB} \ee
If the Higgs branch $\CM_H$ happens to be smoothly resolved, then there is a functor from $\CC_B$ to quasi-coherent sheaves on $\CM_H$ itself, modulo a grading shift, which connects with the sigma-model analysis of \cite{KRS}. The $\CC_A$ category is also very closely related to the category of modules for boundary VOA's of 3d $\CN=4$ theories studied by \cite{CostelloGaiotto, CCG}. (The relation is analogous to that of line operators in Chern-Simons theory and modules for WZW \cite{Witten-Jones, EMSS}.) Some mathematical remarks on $\CC_A$, $\CC_B$, and 3d mirror symmetry appear in \cite{BF-notes}.

Our current purview is more pragmatic. We wish to describe half-BPS objects of $\CC_A$ and $\CC_B$ from a physical perspective, and to explain a way to compute their Hom's. The categorical perspective provides us with a powerful organizational framework, but we will use it only for organization and motivation --- we will not be doing any derived computations in the categories \eqref{intro-CAB} here.

\subsection{Wilson lines}
\label{sec:intro-Wilson}

In 3d $\CN=4$ gauge theories with linear hypermultiplet matter, half-BPS B-type Wilson lines are relatively simple. The outcome of a now-standard SUSY analysis \cite{Wilson, BlauThompson2, BLN-Wilson, Maldacena-Wilson, ReyYee-Wilson, Zarembo-Wilson}, reviewed in Section~\ref{sec:WL}, is that half-BPS Wilson lines are labeled by complex representations $V$ of the gauge group $G$.%
\footnote{There do exist half-BPS B-type line operators that are \emph{not} Wilson lines; they are disorder operators, defined as codimension-two defects in a flat complex $G_\C$ connection \cite{lineops}. We will not need them in this paper.} %
At a junction of Wilson lines $\W_V$, $\W_{V'}$ labeled by two such representations, one finds $G$-non-invariant local operators, transforming in the representation $V'\otimes V^*$. Moreover, as long as there is enough hypermultiplet matter (so that $G$ acts faithfully), the $Q_B$-cohomology of the space of local operators at a junction is entirely constructed from polynomials in the complex hypermultiplet scalars that transform in $V'\otimes V^*$. If the hypermultiplets come in a complex symplectic representation $R\oplus R^*$, then
\be \text{Hom}_B(\W_V,\W_{V'}) \simeq \big(\C[R\oplus R^*]\otimes V\otimes V'{}^*/(\mu)\big)^G\,, \ee
where the ideal $(\mu)$ sets the complex moment map to zero.

This is in close analogy with the Higgs-branch chiral ring $\C[\CM_H]$. Indeed, for $V=V'=\C$ the trivial representation, the Wilson line $\W_\C=\id$ is the trivial line operator, and
\be \text{Hom}_B(\id,\id) = H^\bullet(\text{bulk local ops},Q_B)= \big(\C[R\oplus R^*]/(\mu)\big)^G = \C[\CM_H] \ee
reproduces the familiar Higgs-branch chiral ring. Just as there are no quantum corrections to the Higgs-branch chiral ring (and for the same reason --- the gauge coupling does not sit in a hypermultiplet, \emph{cf.} \cite{AHISS, BDG}), we expect no quantum corrections to the spaces of local operators at junctions of Wilson lines, or their collision products.

An interesting deformation of the spaces $\text{Hom}_B(\W_V,\W_{V'})$ and their product comes from turning on an Omega background \cite{Nek-Omega} with parameter $\varepsilon$ in the plane transverse to line operators. It is well known that the Omega background induces a deformation quantization of the Higgs-branch chiral ring. This follows from dimensional reduction of similar quantization results in 4d \cite{NS-quantization} (and the 4d setups in \cite{GW-surface, GMN-framed}); it has also been verified by direct calculation in sigma-models \cite{Yagi-Omega}; and interpreted from a general TQFT perspective \cite{descent}. The quantization extends to junctions of Wilson lines, deforming
\be \text{Hom}_B(\W_V,\W_{V'}) \quad\leadsto\quad \text{Hom}_B^\varepsilon(\W_V,\W_{V'}) \ee
in a straightforward way that we formalize in Section \ref{sec:Wilson-Omega}, generalizing a prior analysis from~\cite{BDGH}. We carry out an explicit computation of such a deformed space, involving Wilson lines in nonabelian gauge theory, in Section \ref{sec:FundamentalWilsonLine} and Appendix \ref{app:Wilson}.

\subsection{A closer look at vortex lines}
\label{sec:intro-vortex}

As for A-type half-BPS line operators, which we simply refer to as ``vortex lines,'' there is more to say. Motivated by the Gukov-Witten construction of surface operators in 4d $\CN=4$ SYM \cite{GW-surface, GW-rigid, Witten-wild} and their 4d $\CN=2$ generalizations \cite{Gukov-gaugeknot, KohYamaguchi,Tan-surface,AGGTV,Gaiotto-surface} we expect vortex lines to admit two equivalent types of definitions:
\begin{itemize}
\item as disorder operators, modeled on a half-BPS codimension-two singularity in the gauge and matter fields; or
\item by adding 1d $\SQM_A$ degrees of freedom along the line, coupled to 3d bulk fields via gauging flavor symmetries and introducing 1d superpotentials.
\end{itemize}
The second description should be related to the first by integrating out the 1d fields.

We will consider 3d $\CN=4$ theories with gauge group $G$ and linear hypermultiplet matter in a symplectic representation of the form $R\oplus R^*$, where $R$ is a unitary representation. Then the relevant half-BPS equations in the plane transverse to a line operator are vortex equations, for a connection on a $G$ principal bundle and a section of an associated $R\oplus R^*$ bundle, supplemented by a complex moment-map constraint.

Vortex equations have a long history in physics and mathematics, initiated by the work of \cite{NO} and \cite{Taubes-vortex, JaffeTaubes} on abelian vortices and their moduli spaces. The vortex equations that we encounter here, involving arbitrary $G$ and $R$, were first studied mathematically in \cite{vortex-stab1, vortex-stab2, vortex-stab2II, vortex-mathrev}, and entered physics via SUSY field theory \cite{BJSV} and string theory \cite{HananyTong-VIB}. See \cite{Tong-TASI, Tong-review} for physically oriented reviews. The particular appearance of vortex equations in \cite{BJSV}, and later \cite{Kapustin-hol}, is directly related to our setup: there, and here, one considers half-BPS equations along a fixed complex plane in a gauge theory with eight supercharges. In the special case that $R=\text{adj}$ is the adjoint representation, the vortex equations include Hitchin's equations \cite{Hitchin}, entering the physics of gauge theories with sixteen supercharges \cite{BJSV, KapustinWitten}.

In order to describe A-type line operators, we must introduce singularities in the vortex equations, analogous to the treatment of surface operators in 4d $\CN=4$ and $\CN=2$ theories.
For trivial or adjoint $R$, the algebraic structure of singularities was developed in classic work of Mehta-Seshadri and Simpson \cite{MehtaSeshadri, Simpson}, generalized in \cite{Sabbah, BiquardBoalch}. More recent physical and mathematical analyses of singularities in abelian theories include  \cite{TongWong, HKT, BaptistaBiswas}.
However, we are not aware of any classification of such singularities that is sufficiently general for describing the full set of half-BPS A-type line operators in 3d $\CN=4$ theories with arbitrary $G$ and $R$ --- \emph{e.g.} a classification that would encompass a set of generators for the category $\CC_A$, or all vortex lines that are 3d-mirror to Wilson lines.

We make some progress toward such a classification in Section \ref{sec:VL}. We describe a large class of half-BPS vortex lines that are characterized by two pieces of algebraic data:
\begin{itemize}
\item[1)] A holomorphic Lagrangian subspace $\CL_0 \subset R(\CK)\oplus R^*(\CK)$.
\item[2)] An algebraic subgroup $\CG_0\subseteq G(\CO)$ that preserves $\CL_0$.
\end{itemize}
Here $\CK=\C(\!(z)\!)$ denotes the ring of formal Laurent series, \emph{a.k.a.} holomorphic functions on an infinitesimally small disc transverse to a line operator; and $\CL_0 \subset R(\CK)\oplus R^*(\CK)$ represents the allowed singularities in hypermultiplet scalars near the line. The space $R(\CK)\oplus R^*(\CK)$ is naturally endowed with a holomorphic symplectic structure, given by taking the residue $\text{Res}_{z=0}\Omega$ of the holomorphic symplectic form from $R\oplus R^*$; and $\CL_0$ is required to be Lagrangian for half-BPS singularities. We note that choices of $\CL_0$ may allow poles of arbitrarily high order in the hypermultiplet scalars, analogous to ``wild ramification'' for 4d $\CN=4$ surface operators \cite{Witten-wild}.
 Similarly, $G(\CO)$ denotes the algebraic group $G_\C$ defined over formal power series $\CO=\C[\![z]\!]$, \emph{i.e.} the group of holomorphic, complexified gauge transformations near the line. The subgroup $\CG_0 \subseteq G(\CO)$ defines a pattern of gauge-symmetry breaking. In these terms, the trivial line $\id$ is given by
\be \id\,:\quad \CL_0 = R(\CO)\oplus R^*(\CO)\,,\qquad \CG_0 = G(\CO)\,.\ee

We explain in Section \ref{sec:VL} how the algebraic data may be extracted from a singularity in the bulk physical fields \emph{or} from a coupling of the bulk fields to 1d $\SQM_A$ quantum mechanics. Conversely, given algebraic data, we explain how to construct a coupling to 1d $\SQM_A$ quantum mechanics that matches it.
There are additional real parameters associated with vortex lines as well, but their variations are $Q_A$-exact (they appear as K\"ahler parameters in the $\SQM_A$ quantum mechanics, where $Q_A$ is a de Rham differential), so they do not play an important role for us.

We \emph{expect} that the algebraic data $\CL_0,\CG_0$ is also sufficient to define moduli spaces of solutions to the half-BPS equations on a complex plane or a Riemann surface in the presence of the line operator, via a generalized Hitchin-Kobayashi correspondence. We do not, however, attempt to prove this rigorously.

Certain vortex-line operators with algebraic data of the above type already appeared in work on Coulomb branches of 3d $\CN=4$ theories and symplectic duality. In particular, \cite{BDGH} and \cite{BFN-lines} used ``flavor vortex lines'' --- where $\CL_0,\CG_0$ are associated to a singular flavor-symmetry transformation --- to describe resolutions of Coulomb branches. This perspective was also discussed in the lectures \cite{BZ-lectures}. Additionally, the mathematical work \cite{Web2016} implicitly used a large class of line operators with data as above%
\footnote{In fact, \cite{Web2016} allowed subgroups $\CG_0 \subset G(\CK)$ in addition to $\CG_0\subseteq G(\CO)$. This seems to be a reasonable generalization, compatible with the $\CC_A$ category in \eqref{intro-CAB}, though we will not need it in this paper.} %
 to provide a practical combinatorial construction of Coulomb-branch chiral rings. (This construction reproduces KLR algebras \cite{KhovanovLauda, Rouquier} in the case of linear quiver gauge theories.) All these works provided important motivation for us.
We finally note that the data $\CL_0,\CG_0$ immediately defines objects in the $\CC_A$ category of \eqref{intro-CAB} (with $LR$ and $LG_\C$ modeled as $R(\CK)$ and $G(\CK)$, respectively), giving another indication that our description is reasonable.

\subsection{Junctions of vortex lines}

In Section \ref{sec:comp}, we then propose a way to compute spaces of local operators $\text{Hom}_A$ at junctions of vortex lines. One should not expect this to be easy, since even for the trivial line $\text{Hom}_A(\id,\id)$ captures the Coulomb-branch chiral ring, which famously contains nonperturbative monopole operators \cite{SW-3d, IS, BKW-I, GW-Sduality}. On the other hand, recent years have seen spectacular progress in developing exact methods to compute the Coulomb-branch chiral ring \cite{CHZ-Hilbert, BDG, VV, DG-star, HananySperling-fans, BPR-defq, Pufu-Coulomb, Pufu-bubbling, CCG}, including fully mathematical definitions \cite{Teleman-ICM, Nak, BFNII}. (Closely related computations of the Coulomb branch of 4d $\CN=2$ theories on $S^1$, whose global functions are given by vevs of line operators, include \cite{Kapustin-hol,KapustinSaulina, GMN, GMN-Hitchin,  ItoOkudaTaki, BrennanDeyMooreI, BrennanDeyMooreII, CautisWilliams}.) Most of these approaches should generalize to include A-type line operators.

In this paper, we generalize the computational approach of \cite{VV}. We fix a massive vacuum $\nu$ at infinity in the plane transverse to line operators. Then, to every line operator $\CL$, this associates a vector space
\be \CH(\CL,\nu) = H^\bullet_{Q_A}(\text{Hilb}(\C;\CL,\nu))\,, 
 \label{intro-Hilb} \ee
the $Q_A$-cohomology of the Hilbert space on the plane punctured by $\CL$ at the origin, and with a vacuum boundary condition at infinity. Any $Q_A$-closed local operator $\CO \in \text{Hom}_A(\CL,\CL')$ at a junction of lines $\CL,\CL'$ gets represented as a linear map
\be  \CO : \, \CH(\CL,\nu) \to \CH(\CL',\nu)\,, \label{intro-Oaction} \ee
and we compute this representation in terms of a certain convolution algebra.

Concretely, the space $\CH(\CL,\nu)$ may be realized as the cohomology of a moduli space $\CM(\CL,\nu)$ of solutions to the half-BPS equations on the plane. For trivial line $\CL=\id$, this moduli space was given a rigorous algebraic construction in \cite{VW-vortices}, following physical examples in \emph{e.g.} \cite{HananyTong-VIB, Eto1, Eto2, DGH}, and the classic abelian constructions of \cite{Taubes-vortex, JaffeTaubes}. We propose an algebraic construction of $\CM(\CL,\nu)$ that incorporates the algebraic data $\CL_0,\CG_0$ of a given line operator. The construction is a straightforward but as-yet nonrigorous generalization of \cite{VW-vortices}. 

\begin{figure}[htb]
\centering
\includegraphics[width=2.7in]{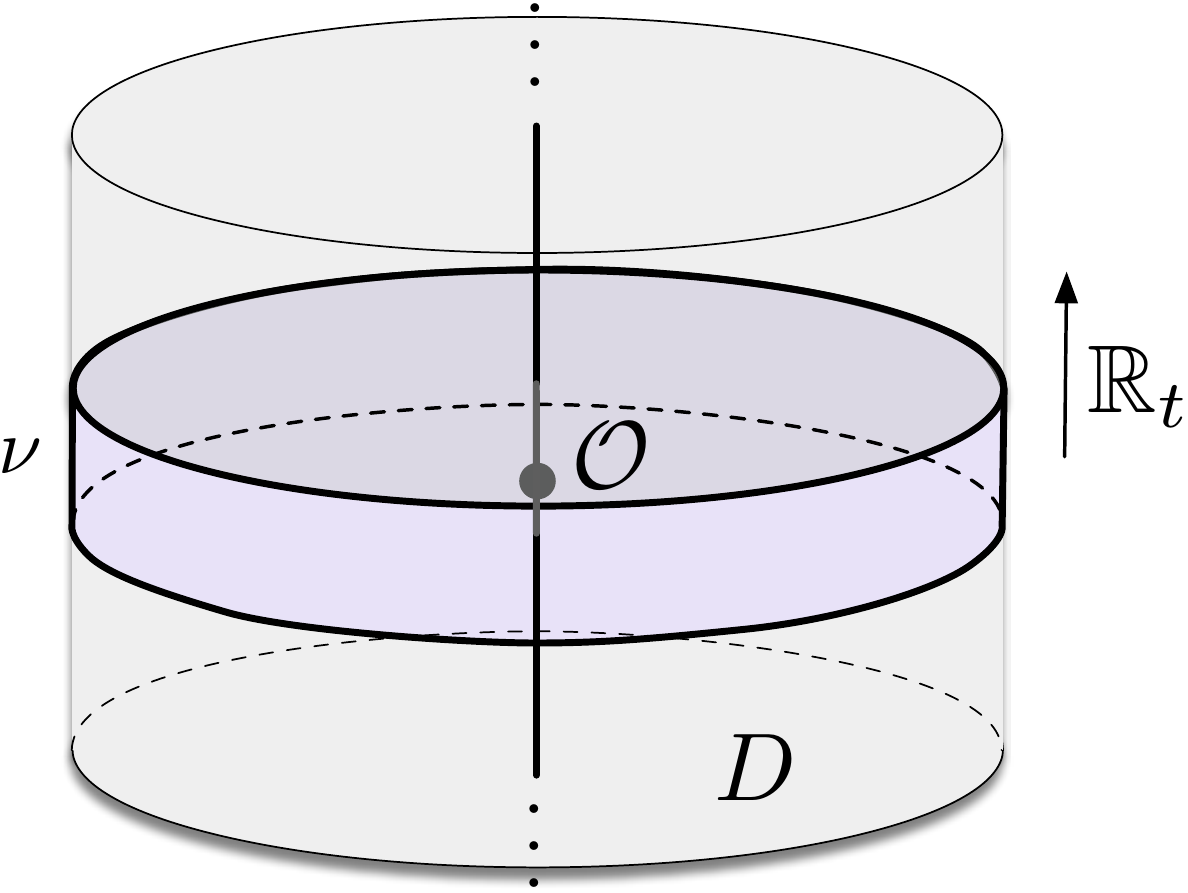}
\caption{The setup used to study local operators bound to vortex lines. We place the 3d theory in a solid cylinder $D\times \R_{t}$, with a vacuum boundary condition wrapping the outside, and a line operator along the axis. The space of local operators $\CO$ at a point $p$ may be computed as the cohomology of the moduli space of solutions to BPS equations on a ``Gaussian pillbox'' (or ``raviolo'') surrounding $p$.}
\label{fig:cyl-intro}
\end{figure}

The representation \eqref{intro-Oaction} is then obtained by interpreting each local operator $\CO$ as a cohomology class in a moduli space $\CM_{\rm rav}(\nu;\CL,\CL')$ of solutions to the half-BPS equations on a ``Gaussian pillbox'' surrounding the location of $\CO$, as in Figure \ref{fig:cyl-intro}. This is in essence a state-operator correspondence.
 Algebraically, the space $\CM_{\rm rav}(\nu;\CL,\CL')$ may also be thought of as solutions to the BPS equations on two planes $\C\cup_{\C^*}\C$, identified away from the origin, sometimes called a ``raviolo.'' It may also be thought of as a space of generalized Hecke modifications, analogous to that discussed in \cite[Sec. 10]{KapustinWitten}.
Altogether, we produce a map
\be \text{Hom}_A(\CL,\CL') \longrightarrow H^\bullet(\CM_{\rm rav}(\nu;\CL,\CL'))\,, \label{intro-rep} \ee
which is almost surjective (in a precise sense) and \emph{often} injective. Given a class $\CO \in H^\bullet(\CM_{\rm rav}(\nu;\CL,\CL'))$, the action 
\eqref{intro-Oaction} comes from a natural convolution in cohomology.

Computing the cohomology $H^\bullet(\CM_{\rm rav}(\nu;\CL,\CL'))$ in practice is quite subtle, because the spaces $\CM_{\rm rav}$ are typically singular and noncompact. To deal with this, we propose that $H^\bullet$ be interpreted as equivariant intersection cohomology. Some physical and mathematical justifications for this proposal are given in Section~\ref{sec:cohomologies}. (Mathematical justification includes the use of intersection cohomology in \cite{Braverman-W, BFFR}, which was interpreted in \cite{VV} as computing the algebra of bulk local operators $\CO \in H^\bullet(\CM_{\rm rav}(\nu;\id,\id))$ for linear quiver gauge theories.)
Most convincingly, we will \emph{test} this proposal with highly nontrivial examples in Sections~\ref{sec:IwahoriLine}--\ref{sec:conifold}.

In Appendix \ref{app:BFN}, we discuss an alternative algebraic approach to computing the spaces $\text{Hom}_A(\CL,\CL')$. Physically, it uses a ``Neumann'' boundary condition rather than a massive vacuum $\nu$ at infinity on the plane transverse to line operators. A major advantage is that the Neumann boundary condition is always available, even in theories that do not admit massive vacua $\nu$. Mathematically, this approach more directly generalizes the Braverman-Finkelberg-Nakajima definition of the Coulomb-branch chiral ring \cite{Nak,BFNII}, and matches morphism spaces in the $\CC_A$ category \eqref{intro-CAB}. Unfortunately, actual computations in this approach require the technology of Borel-Moore homology on infinite-dimensional stacks, which is beyond our current scope.

Just like B-type line operators, A-type line operators also admit an Omega deformation, which deforms all the spaces $\text{Hom}_A(\CL,\CL')\leadsto \text{Hom}_A^\varepsilon(\CL,\CL')$ and their products. On the RHS of \eqref{intro-rep}, the Omega background is easily realized by turning on equivariance with respect to ``loop rotation,'' \emph{i.e.} rotations in the plane $\C$. This deformation already appeared in the context of quantizing the Coulomb-branch chiral ring \cite{BDG, Nak, BFNII}, and is a dimensional reduction of angular momentum background that quantizes the product of line operators in 4d $\CN=2$ theories, as in \cite{NS-quantization, GMN-framed, DG-motivic, AGGTV, DGOT}.

In this paper, we only consider the local structure of line operators and their junctions, for which it suffices to analyze the BPS equations on $\C$ with a singularity/line operator puncturing the origin. It would be quite interesting to make the story more global --- construct, say, the cohomology of the Hilbert space of 3d $\CN=4$ gauge theories on compact Riemann surfaces $\Sigma\times \R_t$, with singularities/line operators at various points on $\Sigma$. This should make contact with many recent physical works on partition functions and Hilbert spaces of 3d SUSY gauge theories \cite{BeniniZaffaroni-twisted, BeniniZaffaroni-Riemann, GukovPei, ClossetKim, JockersMayr, Gaiotto-twisted, Bullimore-twisted, BFK}. A-type line operators should also combine nicely with a large body of mathematical work on quasi-maps and (K-theory lifts of) Gromov-Witten theory, as in the classic \cite{GiventalLee} and more recent \cite{GITquasimap, OkounkovSmirnov, Koroteev-Kthy, AganagicOkounkov, Kim-symplectic} (for example). We leave this for future investigation!

\subsection{Main examples}
\label{sec:intro-ex}

In Sections \ref{sec:abel}--\ref{sec:conifold}, we give several examples of line operators of increasing complexity. We compute their Hom spaces using the algebraic prescription of Section \ref{sec:comp}.

We begin in Section \ref{sec:abel} by considering abelian gauge theories, with enough matter to ensure that the gauge group acts faithfully. (This also ensures that they have abelian mirrors of the same type \cite{dBHOOY, KapustinStrassler}.) We give complete descriptions of the set of half-BPS Wilson lines, and of a basic set of half-BPS vortex lines that may be described as ``flavor vortices,'' generalizing discussions in \cite{HKT, TongWong, BDGH} (and \cite{Okuda-2d} for 2d theories). The two sets of line operators are precisely exchanged under 3d mirror symmetry. We fully describe the junctions of line operators within each set, and action of the mirror map on them. We also connect, in the case of quiver gauge theories, to the quiver/brane constructions of \cite{AsselGomis}. 

The simplest abelian example is instructive to mention here. Consider the mirror pair of 3d theories
\be \CT_{\rm hyper}:\,\text{free hypermultiplet}\quad\leftrightarrow\quad \text{SQED}_1:\,\text{$U(1)$\,+\,1 hypermultiplet}\,. \label{intro-hyper}\ee
The free hypermultiplet theory has a unique half-BPS Wilson line (the trivial line $\id$), but has a large collection of basic vortex lines $\V_k$ labeled by integers, with algebraic data
\be \CL_0 = z^k R(\CO)\oplus z^{-k}R^*(\CO) \ee
that allows one complex hypermultiplet scalar to have a pole of order $k$, while restricting the other to have a zero of order $k$. We compute that $\text{Hom}_B^\varepsilon(\id,\id) = \C_\varepsilon[\CM_H]$ is a Heisenberg algebra, generated by the hypermultiplet scalars, whereas $\text{Hom}_A^\varepsilon(\V_k,\V_{k'})=\C$ for all $k,k'$.

On the other hand, $\SQED_1$ has Wilson lines $\W_k$ labeled by the 1d representations of $U(1)$, whose junctions each contain a unique operator of the correct gauge charge, $\text{Hom}^\varepsilon_B(\W_k,\W_{k'}) = \C$. $\SQED_1$ also contains a unique flavor vortex line (the identity $\id$); all other basic vortex lines get screened and are equivalent to $\id$. The algebra $\text{Hom}_A^\varepsilon(\id,\id)=\C_\varepsilon[\CM_C]$ is an interesting realization of the Heisenberg algebra, generated by monopole operators. Altogether, 3d mirror symmetry swaps both the line operators and their Hom spaces in the pair \eqref{intro-hyper}.

In Sections \ref{sec:IwahoriLine} and \ref{sec:conifold}, we then explore two special examples of vortex lines in 3d $\CN=4$ SQCD, with gauge group $G=U(2)$ and four fundamental hypermultiplets. This theory is particularly convenient to work with, because it has enough matter to admit massive vacua (allowing the computational methods of Section \ref{sec:comp} to proceed); and it has a simple 3d mirror with gauge group $G^!=U(1)\times U(2)\times U(1)$ \cite{dBHOO, HananyWitten}.

We first consider a half-BPS vortex line $\V_\CI$ in $U(2)$ SQCD that breaks gauge symmetry to the torus $U(1)^2$, but does not affect the hypermultiplet fields. This is analogous to the simplest surface operators of \cite{GW-surface}. The algebraic data has $\CL_0 = R(\CO)\oplus R^*(\CO)$ and $\CG_0 = \CI$, known as the Iwahori subgroup of $G(\CO)$. We find by direct computation that $\text{Hom}_A^\varepsilon(\V_\CI,\V_\CI)$ is \emph{almost} trivial: it is a $2\times 2$ matrix algebra over the quantum Coulomb-branch algebra of SQCD. In other words, in $Q_A$ cohomology, $\V_\CI$ is isomorphic to two copies of the trivial line. The matrix algebra does arise in an interesting way, as the product of an abelianized Coulomb-branch algebra similar to that in \cite{BDG}, and the nil-Hecke algebra \cite{KostantKumar} for $SL(2)$. This structure is analogous to the affine Hecke algebra of \cite{GW-surface}, and has been studied extensively in the papers \cite{Weekes, Web2016}.

We explain in Section \ref{sec:Deligne} that the equivalence of $\V_\CI$ with a direct sum of trivial lines is no surprise. Indeed, in any gauge theory, any vortex line that breaks gauge symmetry but leaves the hypermultiplets untouched is expected to behave this way. Mathematically, this follows from a classic result of Deligne \cite{Deligne} and its generalization by Beilinson-Bernstein-Deligne-Gabber \cite{BBDG}.

In Section \ref{sec:conifold} we ``fix'' this triviality by introducing another vortex line operator $\V_{\rm con}$ that also has $\CG_0=\CI$, but introduces first-order poles in some of the hypermultiplet fields. The line operator can also be engineered by coupling the bulk SQCD to a 1d $\SQM_A$ sigma-model whose target is the resolved conifold, or to a 1d $\SQM_A$ quiver gauge theory that flows to the conifold. The quiver description coincides with a brane construction of \cite{AsselGomis}, which also predicts that the 3d mirror of $\V_{\rm con}$ will be a fundamental Wilson line $\W_{\mb 2}$ for the $U(2)$ factor of $G^!=U(1)\times U(2)\times U(1)$. We produce a detailed match of the algebra of local operators bound to $\V_{\rm con}$ in SQCD and the algebra of local operators bound to $\W_{\mb 2}$ in the mirror.

\subsection{Acknowledgements}

We thank David Ben-Zvi, Alexander Braverman, Mathew Bullimore, Kevin Costello, Davide Gaiotto, Alexei Oblomkov, Lev Rozansky, Ben Webster, and Philsang Yoo for many extended and enlightening discussions on the various aspects of this paper, as well as collaboration on several related projects. The work of T.D. was supported by NSF CAREER Grant DMS-1753077 and in part by ERC Starting Grant No. 335739. J.H. is part of the Simons Collaboration on Homological Mirror Symmetry supported by Simons Grant 390287.

Part of this work was carried out during the KITP program Quantum Knot Invariants and Supersymmetric Gauge Theories (fall 2018), supported by NSF Grant PHY-1748958. Part of this work was also conducted at the Perimeter Institute for Theoretical Physics; research at PI is supported by the Government of Canada through the Department of Innovation, Science and Economic Development, and by the Government of Ontario through the Ministry of Research, Innovation and Science.

\section{Conventions and SUSY algebras}
\label{sec:conventions}

In this section, we briefly review the form of the 3d $\CN=4$ SUSY algebras, and its half-BPS subalgebras that preserve various line operators and local operators.

We work in flat Euclidean spacetime $\R^3$. We will usually regroup the three real coordinates $x^1,x^2,x^3$ into a complex `spatial' coordinate $z=x^1+ix^2$ and a Euclidean time $t=x^3$, thinking of spacetime as $\R^3\simeq \C_z\times \R_t$. We consider line operators supported at $z=0$, extending along $\R_t$\,:
\be \raisebox{-.8in}{\includegraphics[width=1.8in]{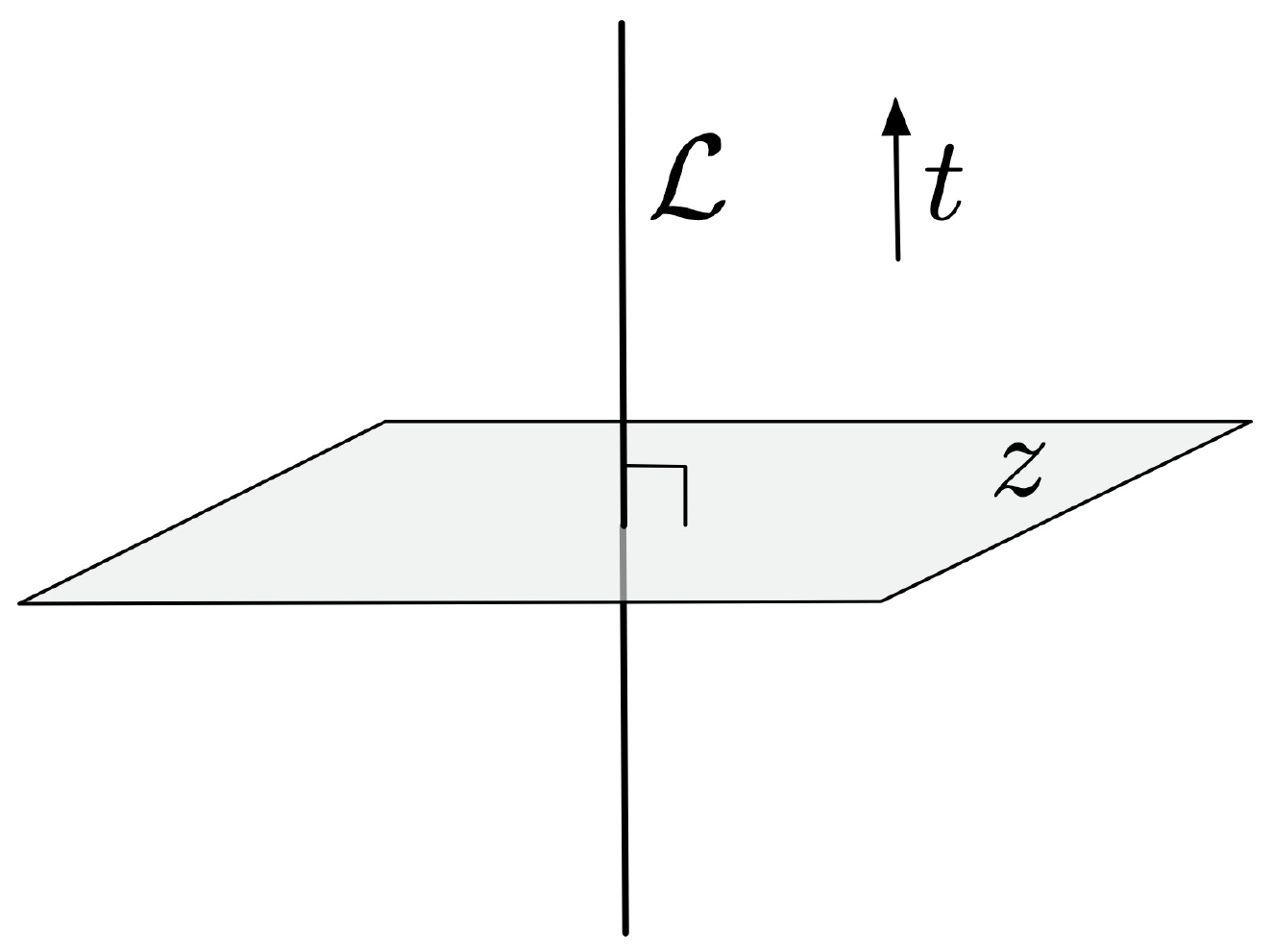}} \ee

The 3d $\CN=4$ algebra is generated by eight complex supercharges $Q_\alpha^{a\dot a}$, where $\alpha\in\{+,-\}$ is a spinor index for $\text{Spin}(3)_E\simeq SU(2)_E$, and $a,\dot a\in \{+,-\}$ are spinor indices for the $SU(2)_H\times SU(2)_C$ R-symmetry.%
\footnote{In Lorentzian signature, the supercharges would satisfy $(Q_\alpha^{a\dot a})^\dagger = \epsilon_{ab}\epsilon_{\dot a\dot b}Q_\alpha^{b\dot b}$.} %
The algebra with central charges takes the form
\be \{Q_\alpha^{a\dot a},Q_\beta^{b\dot b}\} = \epsilon^{ab}\epsilon^{\dot a\dot b} \sigma_{\alpha\beta}^\mu P_\mu -i \epsilon_{\alpha\beta}\big( \epsilon^{ab}m^{(\dot a\dot b)}+t^{(ab)}\epsilon^{\dot a\dot b}\big)\,, \ee
where $(\sigma^1)^\alpha{}_\beta = \bsp 0&1\\1&0\esp$, $(\sigma^2)^\alpha{}_\beta = \bsp 0&-i\\i&0 \esp$, $(\sigma^3)^\alpha{}_\beta = \bsp 1&0\\0&-1\esp$ are the Pauli matrices, and all $SU(2)$ indices are raised and lowered with $\epsilon^{\alpha\beta}$ (or $\epsilon^{ab},\epsilon^{\dot a\dot b}$, etc.), such that $\epsilon^{12}=\epsilon_{21}=1$.
The central charges
\be t^{(ab)} = \bp 2t_\C & -t_\R \\ -t_\R & -2\ol t_\C \ep\,,\qquad m^{(\dot a\dot b)} = \bp 2 m_\C & -m_\R \\ -m_\R & -2 \ol m_\C \ep\,. \ee
will be realized in gauge theory as hyperk\"ahler triplets of FI and mass parameters.
Splitting spacetime as $\R^3\simeq \C_z\times \R_t$, we may write the SUSY algebra more transparently as
\be \label{SUSYz-app} \begin{array}{c} \{Q_+^{a\dot a},Q_+^{b\dot b}\} = -2\epsilon^{ab}\epsilon^{\dot a\dot b} P_{\bar z}\,,\qquad \{Q_-^{a\dot a},Q_-^{b\dot b}\} = 2\epsilon^{ab}\epsilon^{\dot a\dot b} P_{ z}\,,\\[.2cm]
 \{Q_+^{a\dot a},Q_-^{b\dot b}\} = \{Q_-^{b\dot b},Q_+^{a\dot a}\} = \epsilon^{ab}\epsilon^{\dot a\dot b}P_t-i\big( \epsilon^{ab}m^{(\dot a\dot b)}+t^{(ab)}\epsilon^{\dot a\dot b}\big)\,. \end{array} \ee
A more extensive review of 3d $\CN=4$ SUSY algebra appears in Appendix \ref{app:SUSY}.

\subsection{1d subalgebras}

We are interested in half-BPS line operators supported along $\{z=\bar z=0\}\times \R_t$, which preserve a 1d $\CN=4$ SUSY subalgebra of the 3d $\CN=4$ algebra above. There are essentially two inequivalent choices of 1d $\CN=4$ subalgebras, which we will call $\text{SQM}_A$ and $\text{SQM}_B$. We will often refer to the half-BPS line operators preserved by $\text{SQM}_A$ and $\text{SQM}_B$, respectively, as A-type and B-type line operators.

The 1d $\CN=4$ algebra $\text{SQM}_A$ is generated by the four supercharges
\be Q_A^{\dot a} = \delta^\alpha{}_a Q_\alpha^{a\dot a}\,,\qquad \wt Q_A^{\dot a} = (\sigma^3)^\alpha{}_a Q_\alpha^{a\dot a}\,, \label{SQMA} \ee
which satisfy 
\be \{Q_A^{\dot a},\wt Q_A^{\dot b}\} = 2\epsilon^{\dot a\dot b}(P_t -i t_\R)\,,\qquad   \{Q_A^{\dot a},Q_A^{\dot b}\} = \{\wt Q_A^{\dot a},\wt Q_A^{\dot b}\}  = 2im^{(\dot a\dot b)}\,. \label{SQMA-rel}  \ee
Clearly this 1d subalgebra preserves the full 3d $SU(2)_C$ R-symmetry, but breaks $SU(2)_H$ to a diagonal $U(1)_H$ subgroup. In \cite{AsselGomis}, the $\text{SQM}_A$ algebra was denoted $\text{SQM}_V$, because it turns out to be preserved by vortex-line operators. (In \cite{VV}, it was similarly shown that $\text{SQM}_A$ is the subalgebra preserved by dynamical half-BPS vortex excitations.)

For completeness, we note that there is actually a $\cp^1$ family of $\text{SQM}_A$ algebras, parameterized by the choices of unbroken $U(1)_H$'s inside $SU(2)_H$. The different choices lead to different combinations of $t_\R$ and $t_\C, \bar t_\C$ appearing in the $\{Q_A^{\dot a},\wt Q_A^{\dot b}\}$ commutation relation. Equivalently, in a 3d $\CN=4$ gauge theory, different choices of $\text{SQM}_A$ algebra correlate with different choices of complex structure on the Higgs branch. We will work with \eqref{SQMA}, and thus fix a choice of complex structure on the Higgs branch once and for all.

Similarly, the 1d $\CN=4$ algebra $\text{SQM}_B$ is generated by the four supercharges
\be Q_B^{ a} = \delta^\alpha{}_{\dot a} Q_\alpha^{a\dot a}\,,\qquad \wt Q_B^{ a} = (\sigma^3)^\alpha{}_{\dot a} Q_\alpha^{a\dot a}\,, \label{SQMB} \ee
which satisfy 
\be \{Q_B^{ a},\wt Q_B^{ b}\} = 2\epsilon^{ a b}(P_t -i m_\R)\,,\qquad   \{Q_B^{ a},Q_B^{ b}\} = \{\wt Q_B^{ a},\wt Q_B^{ b}\}  = 2it^{( a b)}\,. \label{SQMB-rel}  \ee
This subalgebra preserves an $SU(2)_H\times U(1)_C$ subgroup of the bulk R-symmetry. It is again part of a $\cp^1$ family, parameterized by different choices of $U(1)_C$ inside $SU(2)_C$, or different choices of complex structure on the Coulomb branch (we fix this choice once and for all). In \cite{AsselGomis}, the $\text{SQM}_B$ algebra was denoted $\text{SQM}_W$, because it turns out to be preserved by Wilson lines.

\subsection{Topological twists}

There are two distinct topological twists of 3d $\CN=4$ gauge theories, which we will refer to as the A and B twists. The supercharges that define these respective twists in flat space --- in the usual sense that topologically twisting the theory amounts to working in the cohomology of a particular supercharge ---
are
\be Q_A := Q_A^{\dot +}\,,\qquad Q_B := Q_B^+\,. \ee
Thus, these are elements of the $\text{SQM}_A$ and $\text{SQM}_B$ algebras above. It will occasionally be useful for us to think of line operators from the perspective of topological twists.

The A-twist is a dimensional reduction of the 4d Donaldson-Witten twist \cite{Witten-Donaldson}, and is involved in the definition of Seiberg-Witten invariants of 3-manifolds \cite{MengTaubes}.
Some families of A-twisted 3d sigma-models were studied in \cite{KV, KSV}.
 The B-twist is intrinsically three-dimensional. It was first identified by Blau and Thompson \cite{BlauThompson2} in pure 3d $\CN=4$ gauge theories, and then studied by Rozansky and Witten \cite{RW} in 3d $\CN=4$ sigma-models (which could be thought of as 3d $\CN=4$ gauge theories on their Higgs branches). The extended TQFT defined by the B-twist of a sigma-model was described by Kapustin-Rozansky-Saulina~\cite{KRS}. The fact that the A and B twists are the \emph{only} topological twists in 3d $\CN=4$ theories follows from a basic algebraic classification of nilpotent supercharges whose commutators contain all translation \cite{ElliottSafronov,ESW}.

Some important properties of the A and B twists can already be seen from the 3d and 1d SUSY algebras above. For example, in a gauge theory with flavor symmetries, the supercharge $Q_A$ is not quite nilpotent, satisfying
\be Q_A^2 \sim \varphi+ m_\C \,, \label{Q2m}\ee
with an infinitesimal gauge and flavor transformation on the RHS.
Eventually, we will reduce computations involving vortex lines to 1d quantum mechanics (as in Figure \ref{fig:cyl-intro}). Then $Q_A$ will act as an equivariant de Rham differential, with equivariant parameters $\varphi,m_\C$.

\subsection{Local operators}

We are not just interested in half-BPS line operators, but in the BPS local operators that are bound to them.

Given any two line operators $\CL,\CL'$, there is a vector space $\text{Ops}(\CL,\CL')$ of local operators supported at their junction. If both $\CL$ and $\CL'$ are (say) A-type, 
the topological supercharge $Q_A$ will act on $\text{Ops}(\CL,\CL')$. We can then ask for eighth-BPS local operators preserved by $Q_A$. It is convenient to arrange them into cohomology classes, thinking of $Q_A$-closed operators that differ by a $Q_A$-exact operator as equivalent. This equivalence relation is automatically imposed in any correlation functions that only involve other $Q_A$-closed operators --- since in such correlation functions $Q_A$-exact operators will evaluate to zero. We denote the cohomology as
\be \text{Hom}_A(\CL,\CL') := H^\bullet(\text{Ops}(\CL,\CL'),Q_A)\,. \label{HomA} \ee 
Similarly, if both $\CL$ and $\CL'$ are B-type, we can consider
\be \text{Hom}_B(\CL,\CL') := H^\bullet(\text{Ops}(\CL,\CL'),Q_B)\,. \label{HomB} \ee 

The use of the notation ``Hom'' here is motivated by the structure of extended TQFT. In a topological twist, the line operators that preserve the twist have the structure of a braided tensor category, \emph{cf.} \cite{Lurie, Kapustin-ICM}. (This category was studied by \cite{RW, KRS} for the B-twist of 3d $\CN=4$ sigma-models.) The objects in the category are the line operators themselves, and the morphisms $\text{Hom}(\CL,\CL')$ are the cohomologies of spaces of local operators, as in \eqref{HomA}--\eqref{HomB}.

Since both $Q_A$ and $Q_B$ are topological supercharges, the OPE of eighth-BPS local operators supported at consecutive junctions defines a non-singular product in cohomology.
The careful way to describe this involves first enlarging the notion of local operators to include operators supported in the \emph{neighborhood} of a junction, \cf\ \cite{descent}. In cohomology, the actual size of the neighborhood does not matter. Moreover, the displacements of $Q$-closed local operators at junctions (including displacements of the junctions themselves) are $Q$-exact. Then we can bring two consecutive junctions, supporting (say) $\CO$ and $\CO'$, arbitrarily close together while keeping the cohomology class of the entire configuration constant. Eventually we find that $\CO$ and $\CO'$ are contained in the neighborhood of just a single junction, as on the RHS of \eqref{Jprod}, defining a new local operator $\CO'*\CO$,
\be \begin{array}{c} \hspace{.5in}\includegraphics[width=2.5in]{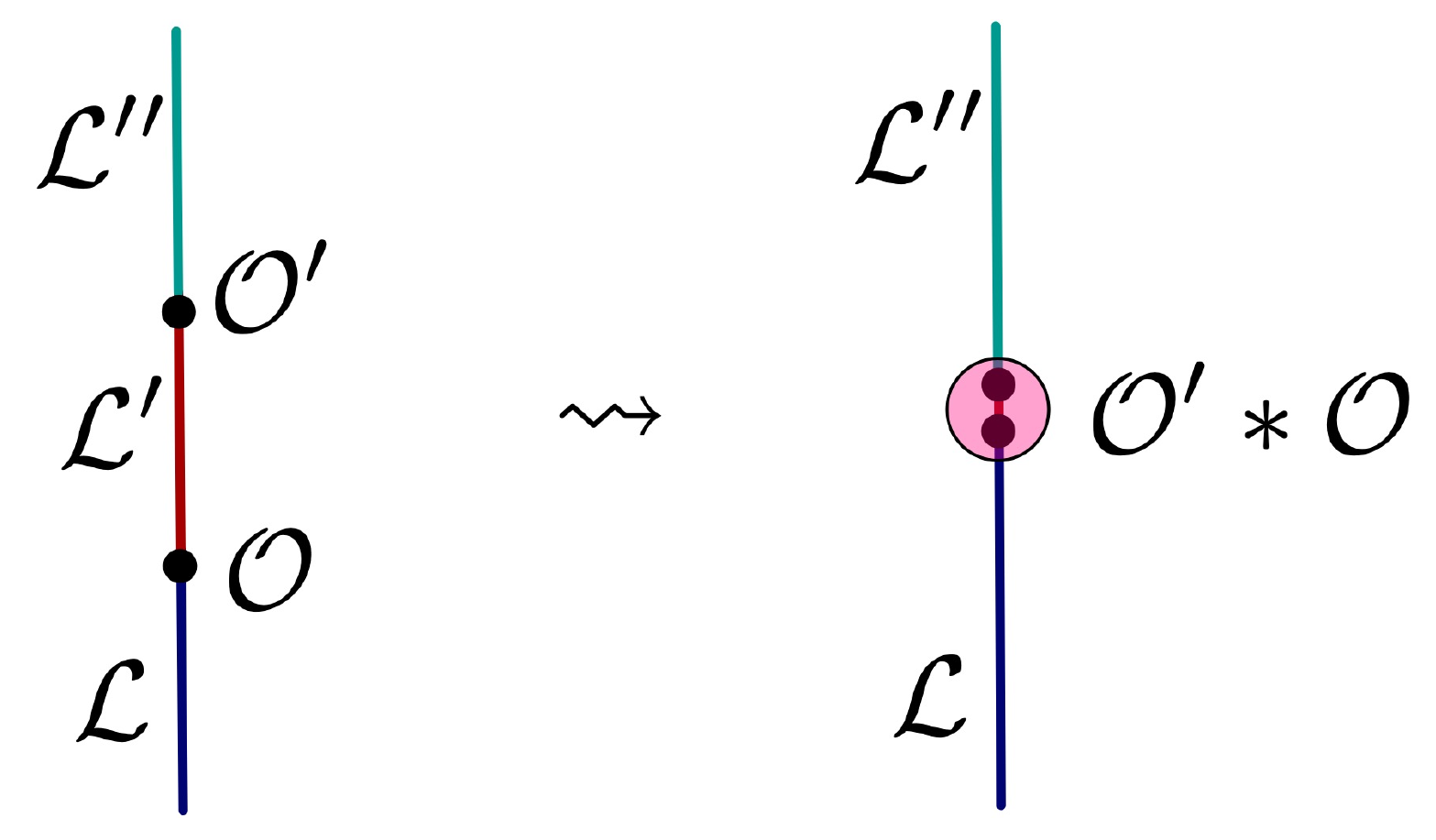} \\
\hspace{-.5cm}\begin{array}{r@{\quad}ccccc} & \text{Hom}_A(\CL',\CL'') & \otimes& \text{Hom}_A(\CL,\CL') &\to& \text{Hom}_A(\CL,\CL'') \\
\text{or}& \text{Hom}_B(\CL',\CL'') & \otimes& \text{Hom}_B(\CL,\CL') &\to& \text{Hom}_B(\CL,\CL'') \\
& \CO'&&\CO && \CO'*\CO\,. \end{array} 
 \end{array} \label{Jprod} \ee
We will usually write the product as simply $\CO'\CO$. The product is associative, because deforming one limit of consecutive collisions to another is a continuous, $Q$-exact operation.

In the special case that $\CL=\CL'=\CL''$ are all the same line operator (say of type A), $\text{End}_A(\CL):=\text{Hom}_A(\CL,\CL)$ simply denotes the cohomology of the space of local operators bound to $\CL$. The product \eqref{Jprod} then becomes
\be \text{End}_A(\CL) \otimes\text{End}_A(\CL) \to\text{End}_A(\CL)\,,  \label{HomLL} \ee
and endows the space $\text{End}_A(\CL)$ with the structure of an associative algebra. This structure should be extremely familiar from supersymmetric quantum mechanics \cite{Witten-Morse}. Indeed, if we did not have a bulk 3d theory, and were merely considering 1d SQM supported on a line, the algebra \eqref{HomLL} is the usual algebra of BPS local operators in SQM. Similarly, \eqref{Jprod} is analogous to a product of BPS interfaces between different SQM theories.

A line operator that exists in every 3d theory is the trivial, or empty line operator.  We'll denote this line operator as $\CL=\id$. It is the line-operator analogue of the identity local operator `1', and it plays a rather special role. It is simultaneously both A-type and B-type, so the spaces $\text{End}_A(\id)$ and $\text{End}_B(\id)$ both make sense. Indeed, they are simply the $Q_A$ and $Q_B$ cohomologies (respectively) of the space of \emph{bulk} local operators. Similarly, given any other half-BPS line operator $\CL$ (say, A-type), the spaces $\text{Hom}_A(\id,\CL)$  and $\text{Hom}_A(\CL,\id)$ denote the cohomology of the space of local operators at an endpoint of $\CL$. 

\subsubsection{Relation to chiral rings}

The local operators that we will actually compute in this paper (particularly by means of TQFT methods) are the eighth-BPS operators discussed above. However, they often turn out to be equivalent to other familiar classes of quarter-BPS and half-BPS local operators. 

For example, since our line operators preserve 1d $\CN=4$ SQM algebras, we could consider local operators that preserve a pair of mutually commuting supercharges, either
\be Q_A^{\dot +},\;Q_A^{\dot -}\qquad\text{or}\qquad Q_B^+,\;Q_B^-\,. \label{Qpair} \ee
(We should set $m_\R=0$ for the former to commute, and $t_\R=0$ for the latter.) In Section~\ref{sec:WL} we argue that, as long as the group $G$ acts faithfully on the hypermultiplet representation $R$, the local operators that preserve the pair $Q_B^+,\;Q_B^-$ are equivalent to the cohomology of just the single supercharge~$Q_B$. We expect the same to be true for A-type operators in a large class of gauge theories, due to 3d mirror symmetry.

In the special case of local operators bound to the trivial line operator $\CL=\id$ --- \emph{a.k.a.} ordinary bulk local operators --- we could ask for even more. Bulk local operators can be preserved by as many as four independent supercharges, either
\be \{Q_\alpha^{a\dot +}\}_{\alpha,a=\pm} \qquad\text{or}\qquad  \{Q_\alpha^{+\dot a}\}_{\alpha,\dot a=\pm}\,.  \label{Qchiralring}\ee
The corresponding spaces of half-BPS local operators are known as the Coulomb-branch and Higgs-branch chiral rings, respectively, \cf\ \cite{AHISS, BKW-II, BFHH-Hilbert,  CHZ-Hilbert, GW-Sduality, BDG}. We denote the chiral rings as $\C[\CM_C]$ and $\C[\CM_H]$, since they contain holomorphic functions on the Coulomb and Higgs branches of vacua.

Since $Q_A$ and $Q_B$ belong to the half-BPS algebras \eqref{Qchiralring}, it is clear that $\C[\CM_C]\subseteq \text{End}_A(\id)$ and that $\C[\CM_H]\subseteq \text{End}_B(\id)$.
We will argue in Section \ref{sec:WL} that in theories with sufficient matter content the $Q_B$-cohomology of bulk local operators is equivalent to the chiral ring
\be \text{End}_B(\id) \simeq \C[\CM_H]\,. \label{End=C} \ee
We similarly expect that $\text{End}_A(\id) \simeq \C[\CM_C]$. The expectation in this case is borne out by the Braverman-Finkelberg-Nakajima construction of the Coulomb-branch chiral ring \cite{Nak, BFNII}, which actually  computes $Q_A$-cohomology but nevertheless reproduces $\C[\CM_C]$ in all known examples.

\subsection{3d mirror symmetry}

At the level of the 3d $\CN=4$ SUSY algebra, 3d mirror symmetry \cite{IS, dBHOO, dBHOOY} is an involution that exchanges the roles of $SU(2)_H$ and $SU(2)_C$ R-symmetries. (This is directly analogous to the classic description of mirror symmetry in 2d $\CN=(2,2)$ theories \cite{HoriVafa-MS}, as exchanging the role of axial and vector R-symmetries.)
In sufficiently nice cases, 3d mirror symmetry also exchanges one gauge theory with linear matter for another. In this case, many of the structures discussed are swapped:
\be \begin{array}{c@{\quad\leftrightarrow\quad}c}
 \text{SQM}_A & \text{SQM}_B \\
  Q_A & Q_B \\
  \text{vortex lines} & \text{Wilson lines} \\
  \text{Hom}_A(\CL,\CL') & \text{Hom}_B(\CL^!, \CL'{}^!) \\
  \CM_C,\;\C[\CM_C] & \CM_H,\;\C[\CM_H]
\end{array} \ee
In particular, half-BPS vortex lines (and the BPS local operators bound to them) will be mapped to half-BPS Wilson lines (and the BPS local operators bound to them), and vice versa. In later sections, mirror symmetry will provide an important consistency check on our calculations.

\section{Wilson lines and their junctions}
\label{sec:WL}

We'll consider a 3d $\CN=4$ gauge theory with compact gauge group $G$ and hypermultiplet matter in representation $R\oplus R^*$, where $R\simeq \C^n$ is a finite-dimensional unitary representation of $G$ and $R^*$ its dual.%
\footnote{In general, linear hypermultiplet matter transforms in a symplectic representation of $G$, \emph{i.e.} with $G$ acting as a subgroup of $USp(n)$ for some $n$. The restriction that the representation is of the form $R\oplus R^*$ amounts to saying that the $G$ action factors through $U(n)\subset USp(n)$. For the purpose of analyzing Wilson lines, this restriction is purely a matter of convenience -- formulas below have obvious generalizations to general matter. In contrast, when describing vortex lines, the restriction will be essential for the methods herein to work.} %

The vectormultiplet contains a connection $A_\mu$ and an $SU(2)_C$ triplet of adjoint-valued scalars $\phi^{(\dot a \dot b)}\in \mathfrak g$, which we'll usually split into a real $\sigma\in \mathfrak g$ and a complex $\varphi\in \mathfrak g_\C$,
\be \phi^{(\dot a \dot b)} = \bp 2\varphi & \sigma \\ \sigma & -2\bar\varphi \ep\,.\ee
In addition, there are gauginos transforming as tri-spinors of $SU(2)_E\times SU(2)_H\times SU(2)_C$.
The SUSY transformations of the vectormultiplet fields are summarized in Appendix \ref{app:SUSY}.

One salient feature is that the complexified connection
\be \CA_\mu := A_\mu -\tfrac i2(\sigma_\mu)_{\dot a \dot b}\phi^{(\dot a\dot b)} \ee
is annihilated by both supercharges $Q_B^a = \delta^\alpha{}_{\dot a}Q_\alpha^{a\dot a}$. 
Its $\mu=3$ component, namely
\be \CA_t = A_t -i\sigma\,, \ee
is also annihilated by $\wt Q_B^a = (\sigma_3)^\alpha{}_{\dot a}Q_\alpha^{a\dot a}$. Thus $\CA_t$ is annihilated by the entire 1d $\CN=4$ algebra $\text{SQM}_B$ from \eqref{SQMB}.

This suggests a way to define half-BPS Wilson lines \cite{BlauThompson2, BLN-Wilson, Maldacena-Wilson, ReyYee-Wilson, Zarembo-Wilson, AsselGomis}.
Let $V=\C^k$ be another finite-dimensional unitary representation of $G$, or equivalently, a complex-linear representation of the complexified group $G_\C$. Let $\rho:\mathfrak g_\C \to \mathfrak{gl}(k)$ be the corresponding map of Lie algebras. 
Then a half-BPS Wilson line operator supported on $\ell = \{z=0\} \times \R_t$ is defined as
\be \W_V := \text{Hol}_\ell \big( \rho(\CA) \big) = P \exp \int_{\R_t} \rho(\CA_t) dt\,. \label{def-W}  \ee
If instead of the line $\ell$ we had considered a closed loop $\gamma$, we could take the trace of the holonomy to produce a gauge-invariant operator. Here, with a noncompact line $\ell$, gauge-invariance can be recovered with a suitable choice of boundary condition at $t\to\pm\infty$. This choice will not affect any of the local structure that we are interested in.

Alternatively, a half-BPS Wilson line may be defined by coupling the bulk 3d $\CN=4$ theory to 1d $\CN=4$ SQM degrees of freedom along the line $\ell$. 
The relevant quantum mechanics contains fermionic hypermultiplets, with a finite-dimensional Hilbert space, \emph{a.k.a.} a Chan-Paton bundle; and $\CA_t$ appears as a coupling in the 1d Hamiltonian, \emph{a.k.a.} a connection on the Chan-Paton bundle. This construction of Wilson lines was discussed in \cite{AsselGomis}, and is directly analogous to standard definitions of B-type boundary conditions for 2d $\CN=(2,2)$ theories \cite{HoriIqbalVafa, Douglas-categories} (reviewed in \cite{Dbranes}).

\subsection{Bulk local operators}
\label{sec:Wilson-bulk}

As a warmup to analyzing local operators bound to Wilson lines, we review some features of bulk local operators in $Q_B$-cohomology, and the Higgs-branch chiral ring. We wish to explain why the two are actually equivalent in theories with sufficient matter. (Readers who already believe this may safely move on.)
We work momentarily with mass and FI parameters turned off, then reintroduce them further below.

The Higgs branch of a 3d $\CN=4$ theory with gauge group $G$ and matter $R\oplus R^*\simeq \C^n\oplus \C^n$ is the hyperk\"ahler quotient $\CM_H = (R\oplus R^*)/\!/\!/G = \{\mu_\R = \mu_\C = 0\}/G$, where $\mu_\R$ and $\mu_\C$ are the real and complex moment maps for $G$. Recall that the moment maps are functions $\mu_\R:(R\oplus R^*)\to \mathfrak g^*$, $\mu_\C:(R\oplus R^*)\to \mathfrak g_\C^*$.
Let us denote the complex hypermultiplet scalars as $X = (X^1,...,X^n)^T \in R$ and $Y = (Y_1,...,Y_n)\in R^*$, denote the generators of $\mathfrak g$ as $\{T_k\}$, and denote their action on $R$ and $R^*$ as $\rho_R(T_k)$ and $\rho_{R^*}(T_k) = [\rho_R(T_k)]^\dagger=-\rho_R(T_k)$.
Then the components of the moment maps become (\emph{cf.} Appendix \ref{app:moment})
\be \langle \mu_\R,T_k\rangle = \text{Tr}\big[\rho_R(T_k) (X X^\dagger-Y^\dagger Y)^T\big]\,,\qquad \langle \mu_\C,T_k\rangle = \text{Tr}\big[\rho_R(T_k) (X Y)^T \big]\,. \ee

For analyzing B-type bulk local operators, it is sufficient to think of $\CM_H$ as a complex-symplectic manifold (in a fixed complex structure), rather than a hyperk\"ahler manifold. Then, trading the real moment-map constraint $\mu_\R=0$ for a complexification $G\to G_\C$ of the gauge group, we have
\be \CM_H \simeq (R\oplus R^*)/\!/G_\C = \{\mu_\C=0\}/G_\C \,.  \label{MH-cx} \ee
This is typically a singular cone. Its holomorphic symplectic form is induced from the canonical form $\Omega = \sum_i dX^i\wedge dY_i$ on $R\oplus R^*$.

The Higgs-branch chiral ring $\C[\CM_H]$ is usually identified with the ring of holomorphic (and polynomial) functions on the space \eqref{MH-cx}. These functions are easily constructed by starting with all polynomials in the $X^i$ and $Y^i$ fields, then imposing an equivalence relation that $\mu=0$, and finally restricting to $G$-invariants. In equations:
\be \C[\CM_H] = \big[\C[X,Y]/(\mu_\C) \big]^G\,,  \label{Higgsring} \ee
where $(\mu)$ denotes the double-sided ideal generated by the components of $\mu$.

More fundamentally, the ``Higgs-branch chiral ring'' should be defined as the subspace of bulk local operators annihilated by all four supercharges $Q_\alpha^{+\dot a}$ from \eqref{Qchiralring}, modulo operators of the form $\CO = Q_\alpha^{+\dot a}(...)$. A quick semiclassical analysis of the SUSY transformations of vectormultiplets and hypermultiplets (see Appendix \ref{app:SUSY}) reproduces the algebra \eqref{Higgsring}. In particular, the (zero modes of the) hypermultiplet scalars $X$ and $Y$ are the only operators annihilated by all the $Q_\alpha^{+\dot a}$, which are not themselves of the form $Q_\alpha^{+\dot a}(...)$. The complex moment map is set to zero because it appears in the image of $Q_\alpha^{+\dot a}$ acting on gauginos,
\be Q_\alpha^{+\dot a}(\lambda_\beta^{+\dot b}) = i\epsilon^{\dot a\dot b}\epsilon_{\alpha\beta}{\rm D}^{++} \sim i\epsilon^{\dot a\dot b}\epsilon_{\alpha\beta}\mu_\C \,. \label{Bbulklambda} \ee

In theories with insufficient matter for the gauge group to act faithfully, some components of $\mu$ might vanish automatically. For example, in pure gauge theory, $\mu \equiv 0$. In this case, one might think from \eqref{Bbulklambda} that corresponding components of the gauginos would appear in the chiral ring. This does not happen, because all of the $\lambda_\beta^{+\dot b}$'s themselves appear as $Q_\beta^{+\dot a}$ transformations of various components of $\phi^{\dot c\dot d}$; thus the gauginos are always set to zero.

Now let's compare the chiral ring \eqref{Higgsring} to the cohomology of just the single topological supercharge $Q_B = \delta^\alpha{}_{\dot a}Q_\alpha^{+\dot a}$, acting on the space of bulk local operators. The $Q_B$ transformations of the fields are most easily expressed if we regroup the fermions into scalars and one-forms with respect to the diagonal subgroup of $SU(2)_E\times SU(2)_C$. We rewrite the gauginos $\lambda_\alpha^{a\dot a}$ in terms of $\gamma = \delta^\alpha{}_{\dot a} \lambda_\alpha^{+\dot a}$, $\upsilon_\mu = (\sigma_\mu)^\alpha{}_{\dot a} \lambda_\alpha^{+\dot a}$, $\tilde \gamma = \delta^\alpha{}_{\dot a} \lambda_\alpha^{-\dot a}$, and  $\tilde \upsilon_\mu = (\sigma_\mu)^\alpha{}_{\dot a} \lambda_\alpha^{-\dot a}$. 
Then
\be \label{fullB-gauge}
\begin{array}{c}
 \QB\CA = 0\,,\\[.2cm]
  \QB\ol \CA \sim \upsilon\,,\qquad \QB\,\upsilon = 0\,, \qquad \QB \tilde\upsilon \sim *\CF\,,\\[.2cm]
   \QB \gamma \sim \mu_\C\,, \\[.2cm]
   \QB \tilde \gamma \sim * d_A * \phi +\mu_\R\,,
\end{array} \ee
with $\CF$ denoting the curvature of $\CA$ and with $\phi_\mu = \tfrac12 (\sigma_\mu)_{\dot a\dot b}\phi^{(\dot a\dot b)}$ in the last line. Similarly, the hypermultiplet fermions get regrouped into scalars $\eta_1^i,\eta_2^i$ and one-forms $\chi_{1,\mu}^i,\chi_{2,\mu}^i$. Then
\be \label{fullB-hyper} \begin{array}{c}\QB\,X^i = \QB\,Y_i = 0\,,\qquad \QB\,\ol X^i = \eta_1^i\,,\quad \QB\,\ol Y_i = \eta_2^i\,,\qquad \QB\,\eta_1^i=\QB\,\eta_2^i = 0\,, \\[.1cm]
\QB\,\chi_1^i \sim d_\CA X^i\,, \quad \QB\,\chi_2^i \sim\,d_\CA Y_i\,. \end{array}  \ee
(We write `$\sim$' to mean equal up to numerical factors.)

We find that (covariant) derivatives of all fields are exact, so we focus on operators constructed out of the zero-modes.
We further simplify the cohomology by removing the pairs $(\ol X^i,\eta_1^i)$, $(\ol Y^i,\eta_2^i)$, $(\ol\CA,\upsilon)$, and $(\tilde\upsilon,*\CF)$, each consisting of operators $(\CO_1,\CO_2)$ such that $Q_B(\CO_1)=\CO_2$. Local operators formed out of $\CA$ could contribute, but they are either not Lorentz-invariant or not gauge-invariant. We are left with a simple model for the $Q_B$ cohomology of local operators, which consists of polynomials $\C[X,Y,\gamma]$ in the scalars $X^i,Y_i$ and the components of the $\mathfrak g_\C$ valued gaugino $\gamma$, together with a differential that acts as 
\be Q_B(\gamma) = \mu_\C(X,Y)\,,\qquad Q_B(X) = Q_B(Y)=0\,. \label{QXYg} \ee
Then the $Q_B$-cohomology of the algebra of local operators, denoted $\text{End}_B(\id)$ as in \eqref{End=C}, becomes
\be \text{End}_B(\id) = H^\bullet(\C[X,Y,\gamma],Q_B)^G\,, \label{EndB1} \ee
\emph{i.e.} the $G$-invariant part of the cohomology of the algebra $\C[X,Y,\gamma]$.

As long as the matter representation $R$ is large enough so that $G$ acts faithfully, all components of the moment map $\mu_\C$ will be nontrivial functions of $X$ and $Y$. Then the cohomology $H^\bullet(\C[X,Y,\gamma],Q_B)$ is isomorphic to the ordinary quotient $\C[X,Y]/(\mu_\C)$, and
\be \text{End}_B(\id) \simeq \C[\CM_H]\,, \ee
as claimed in \eqref{End=C}.

If the representation $R$ is not faithful, then $\text{End}_B(\id)$ can be larger than $\C[\CM_H]$. In particular, it will contain additional gauginos. An extreme example is pure gauge theory ($R=0$), where $\text{End}_B(\id) = \C[\gamma]^G$ is nontrivial, and contains gauge-invariant polynomials in~$\gamma$. We will investigate the algebra $\text{End}_B(\id)$ in much greater detail in \cite{lineops}, and explain how the gauginos appears naturally in the context of derived algebraic geometry.

In principle, the semiclassical calculation of $\text{End}_B(\id)$ that we have just presented could acquire quantum corrections. An indirect way to check that quantum corrections do not enter is by appealing to a standard non-renormalization theorem for the Higgs-branch chiral ring in 3d $\CN=4$ theories.%
\footnote{In a 3d $\CN=4$ theory, neither the Higgs-branch nor Coulomb-branch chiral rings can be renormalized. 
This follows (\eg) from noting that quantum corrections are controlled by the gauge coupling; but the gauge coupling cannot enter an effective superpotential (in 3d $\CN=2$ terms) that encodes chiral-ring relations. This is a special case of non-renormalization in 3d $\CN=2$ theories \cite{AHISS}. In the case of the Coulomb branch, the argument has an extra subtlety: certain one-loop quantum corrections are effectively independent of the gauge coupling, and thus are allowed (see \cite{BDG}). For the Higgs branch, there are no such subtleties.\label{foot:renorm}} %
Thus, at least in the case that $R$ is faithful, so that $\text{End}_B(\id) \simeq \C[\CM_H]$, we expect the semi-classical computation of $Q_B$-cohomology of bulk local operators to be exact.

\subsubsection{FI and mass parameters}

We now briefly consider the effects of turning on mass and/or FI parameters.

FI parameters are associated with the abelian part of the gauge group $G$. Representation-theoretically, they are infinitesimal characters, \emph{i.e.} elements $t_\R\in \mathfrak g^*$, $t_\C\in \mathfrak g_\C^*$ that provide \emph{$G$-invariant} maps $t_\R :\mathfrak g\to \R$, $t_\C:\mathfrak g_\C\to \C$. They resolve or deform the Higgs branch by modifying the moment-map equations; as a hyperk\"ahler quotient one finds
\be \CM_H = \{\mu_\R + t_\R=0\,,\;\mu_\C+t_\C=0\} /G\,. \ee
Alternatively, as a complex-symplectic variety, the Higgs branch with FI parameters gets expressed as
\be \CM_H \simeq \{\mu_\C+t_\C = 0\}^{\text{stab}(t_\R)}/G_\C\,. \label{MHstab} \ee
Here the value of $t_\R$ goes into defining an appropriate stability condition.

As far as the chiral ring goes, the real FI parameter is invisible: the ring of holomorphic functions on \eqref{MHstab} is insensitive to the choice of stability condition. The complex FI parameter does deform the chiral ring, in an obvious way: we now have
\be \C[\CM_H] = \big[\C[X,Y]/(\mu_\C+t_\C)\big]^G\,.\ee
Similarly, in the $Q_B$ transformations \eqref{fullB-gauge}--\eqref{fullB-hyper}, $t_\C$ deforms $\mu$ while $t_\R$ deforms $\mu_\R$. The $Q_B$-cohomology of the algebra of local operators is still given by \eqref{EndB1}, with the modified differential
\be Q_B(\gamma) = \mu_\C + t_\C\,,\qquad Q_B(X)=Q_B(Y) = 0\,.\ee

Dually, mass parameters are associated with flavor symmetries that act as hyperk\"ahler isometries of $R\oplus R^*$.
If we let $F$ denote the flavor symmetry group, then the masses take values in its Lie algebra
\be m_\R \in \mathfrak f\,,\qquad m_\C \in \mathfrak f_\C \ee
(more precisely, they take values in a common Cartan subalgebra). Thus it makes sense to speak of the infinitesimal flavor symmetry generated by $m_\R$ and $m_\C$.

 Mass parameters restrict the Higgs branch to fixed points of the symmetry they generate. They reduce the chiral ring to the ring of polynomial functions on the fixed locus. In contrast, the effect of masses on $Q_B$-cohomology of local operators is much more subtle. In the SUSY transformations \eqref{fullB-gauge}--\eqref{fullB-hyper}, masses $m^{(\dot a\dot b)}$ enter the same way as the vectormultiplet scalars $\phi^{(\dot a\dot b)}$, \emph{i.e.} through the complexified connection $\CA$ and its covariant derivatives. The cohomology $\text{End}_B(\id)$ continues to be described by \eqref{EndB1}, so long as we interpret $X,Y$ (and~$\gamma$) as \emph{covariantly} constant modes of the corresponding fields. Gauge-invariant local operators $f(X,Y)$ become flat sections of a flat $F_\C$ bundle over spacetime, with connection determined by the masses.

\subsection{Local operators at junctions}
\label{sec:Wilson-junct}

We next generalize the above characterization of the $Q_B$-cohomology of bulk local operators to local operators bound to junctions of half-BPS Wilson lines. We will assume that $R$ is a faithful representation of $G$, so that gauginos do not enter the cohomology, and we can work entirely with polynomials in the complex hypermultiplet scalars.

\begin{figure}[htb]
\centering
\includegraphics[width=5.5in]{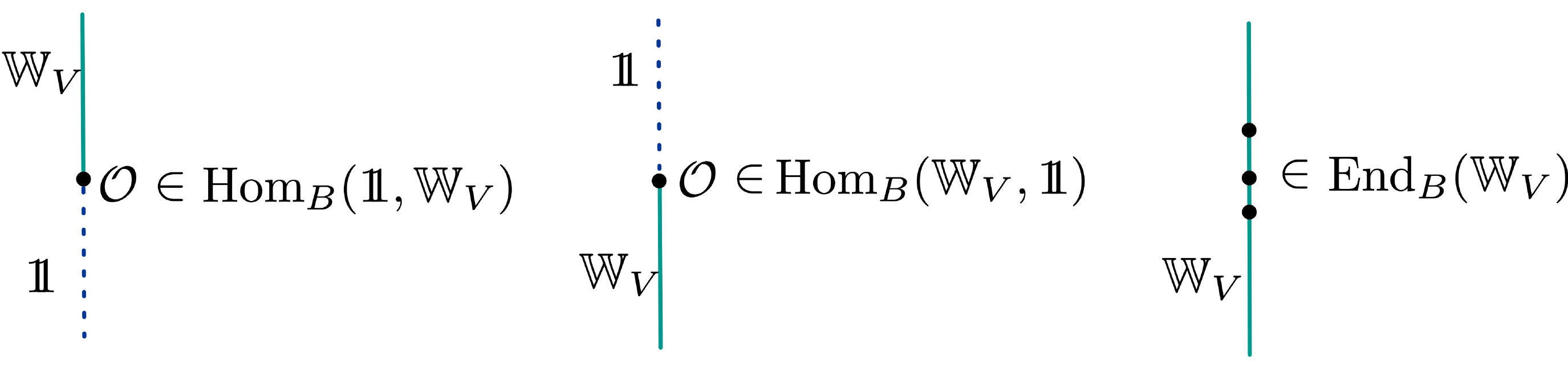}
\caption{Endpoints and endomorphisms of Wilson lines}
\label{fig:Wops}
\end{figure}

Suppose that a Wilson line in representation $V$ is supported on $\R_{t\geq 0}\times \{0\}$, as on the left of Figure \ref{fig:Wops}.
In order for this configuration to preserve gauge invariance, any local operator $\CO$ at the starting point of the Wilson line is required to transform in the representation $V$.
Letting $\text{Ops}(\id,\W_V)$ denote the space of local operators at the starting point of $\mathbb W_V$ in the full physical theory, a straightforward generalization of the computation of $Q_B$-cohomology in Section \ref{sec:Wilson-bulk} now leads to 
\begin{align} \text{Hom}_B(\id, \W_V) &= H^\bullet_{Q_B}\big( \text{Ops}(\id,\W_V)\big) \notag \\
 &= \big[(\C[X,Y]\otimes V^*)/(\mu_\C)\big]^G\,. \label{Hom1W} \end{align}
Here we have accounted algebraically for the fact that operators must transform in the representation $V$ by tensoring with the dual space $V^*$ before taking $G$-invariants. We find polynomials in the hypermultiplet scalars $X,Y$, restricted to transform in $V$, with $\mu_\C$ set to zero.

Similarly, at the opposite endpoint of the Wilson line, we expect local operators that transform in the representation $V^*$, with cohomology given by
\be \text{Hom}_B(\W_V,\id) = \big[(\C[X,Y]\otimes V)/(\mu_\C)\big]^G\,.   \label{HomW1}  \ee
On the Wilson line itself, we should have
\be \text{End}_B(\W_V) := \text{Hom}_B(\W_V,\W_V) = \big[(\C[X,Y]\otimes V^*\otimes V)/(\mu_\C)\big]^G\,, \label{EndW} \ee
which is now naturally an algebra, because the matrices $V^*\otimes V \simeq \text{End}(V^*)$ form an algebra. Finally, given a pair of distinct Wilson lines, we expect
\begin{align} \text{Hom}_B(\W_{V},\W_{V'}) &= \big[(\C[X,Y]\otimes V'{}^*\otimes V)/(\mu_\C)\big]^G \label{HomWW'} \\
 &\simeq   \big[(\C[X,Y]\otimes \text{Hom}(V^*,V'{}^*))/(\mu_\C)\big]^G  \notag\end{align}
Indeed, \eqref{Hom1W}--\eqref{EndW} are all special cases of \eqref{HomWW'} corresponding to $V=V'$, or $V=\C$, or $V'=\C$, where $\C$ is the trivial one-dimensional representation.

The validity of these expectations (and their generalization to theories where $R$ is not a faithful representation) is  justified from a TQFT perspective in \cite{lineops}. The idea is roughly as follows. In the topological B-twist, the category of line operators can be approximated as a category of $G$-equivariant modules for the differential graded algebra $\C[X,Y,\gamma]$, with differential as in \eqref{QXYg}. The ``approximation'' is sufficient to capture properties of Wilson lines, though it misses some other interesting B-type line operators.%
\footnote{In particular, the approximation misses vortex-like disorder operators that are defined by a monodromy defect for the flat connection $\CA$. They turn out to be half-BPS line operators, preserved by $\text{SQM}_B$ (not $\text{SQM}_A$!).  We will not discuss  them in the current paper.} %
The morphism space $\text{Hom}_B(\W_{V'},\W_{V})$ corresponding to a junction of Wilson lines is computed by elementary algebraic techniques in this module category, and reproduces \eqref{HomWW'} when $R$ is faithful.

\subsection{Sheaves on the Higgs branch}
\label{sec:Wilson-sheaves}

Suppose that we introduce FI parameters so that the Higgs branch $\CM_H$ of a 3d $\CN=4$ gauge theory becomes smooth (and the Coulomb branch is fully massive). Then, in the infrared, the gauge theory will flow to a sigma model on its Higgs branch. Any half-BPS line operators defined in the UV should similarly flow to half-BPS line operators in the sigma model. In particular, Wilson lines should flow to operators in a $\CM_H$ sigma model that are compatible with the B-twist. They are easy to describe geometrically. 

In \cite{RW, KRS}, line operators in the B-twist of a sigma model with target $\CX$ were identified as objects in the (derived) category of coherent sheaves, $D^b\text{Coh}(\CX)$. UV Wilson lines turn out to flow in the IR to the simplest types of coherent sheaves on the Higgs branch; namely, they flow to holomorphic vector bundles.

Given a gauge theory with group $G$ and matter $R\oplus R^*$, and a Wilson line $\W_V$ in representation $V$, we can construct a holomorphic vector bundle on the Higgs branch in the following (standard) way. First, let $E_V$ be the trivial vector bundle on the complex vector space $R\oplus R^*$, with complex fiber $V$. It is an equivariant bundle with respect to the complexified gauge group $G_\C$, which acts simultaneously on the base $R\oplus R^*$ and fiber $V$. The restriction of $E_V$ to the $G_\C$-invariant locus $\{\mu_\C + t_\C = 0\}^{\text{stab}(t_\R)}\subset R\oplus R^*$ inherits this equivariant structure. Therefore, $E_V$ descends to a holomorphic bundle $\CE_V$ on the quotient
\be  \CE_V \to \CM_H =  \{\mu_\C + t_\C=0\}^{\text{stab}(t_\R)}/G_\C\,. \ee
The sheaf $\CE_V$ is the IR image of $\mathbb W_V$.

As a (very) simple example, consider the trivial line operator $\id$, thought of as the Wilson line in the trivial one-dimensional representation $V=\C$. In this case, $E_V$ is the trivial line bundle on $R\oplus R^*$, with trivial equivariant structure. The sheaf $\CE_V$ on the Higgs branch to which $E_V$ descends is again a trivial line bundle, \emph{a.k.a.} the structure sheaf
\be \CE_\C \simeq \CO_{\CM_H}\,. \ee

The spaces of local operators at junctions of Wilson lines, discussed from a gauge-theory perspective in Section \ref{sec:Wilson-junct}, also have a nice geometric interpretation on the Higgs branch. Namely, the local operators at a junction of $\W_V$ and $\W_{V'}$ are realized in the IR as the space of morphisms of associated sheaves:
\be \text{Hom}_B(\W_V,\W_{V'}) \simeq \text{Hom}_{\text{Coh}(\CM_H)}(\CE_V,\CE_{V'})\,. \label{Ehom} \ee
Explicitly,  $\text{Hom}_{\text{Coh}(\CM_H)}(\CE_V,\CE_{V'})$ is the space of global holomorphic maps from the sections of $\CE_V$ to the sections of $\CE_{V'}$.%
\footnote{In making this statement, we have implicitly invoked a vanishing theorem. In general, the space of local operators at a junction of lines is a \emph{derived} morphism space in an appropriate category. Here we are dealing with locally free sheaves on the Higgs branch $\CM_H$, which is an affine variety. By a classic result in algebraic geometry, all higher derived morphism spaces vanish, \ie\ $\text{Ext}^{i>0}_{\text{Coh}(\CM_H)}(\CE_V,\CE_{V'})=0$.}

\subsection{Omega background}
\label{sec:Wilson-Omega}

\begin{figure}[htb]
\centering
\includegraphics[width=2.5in]{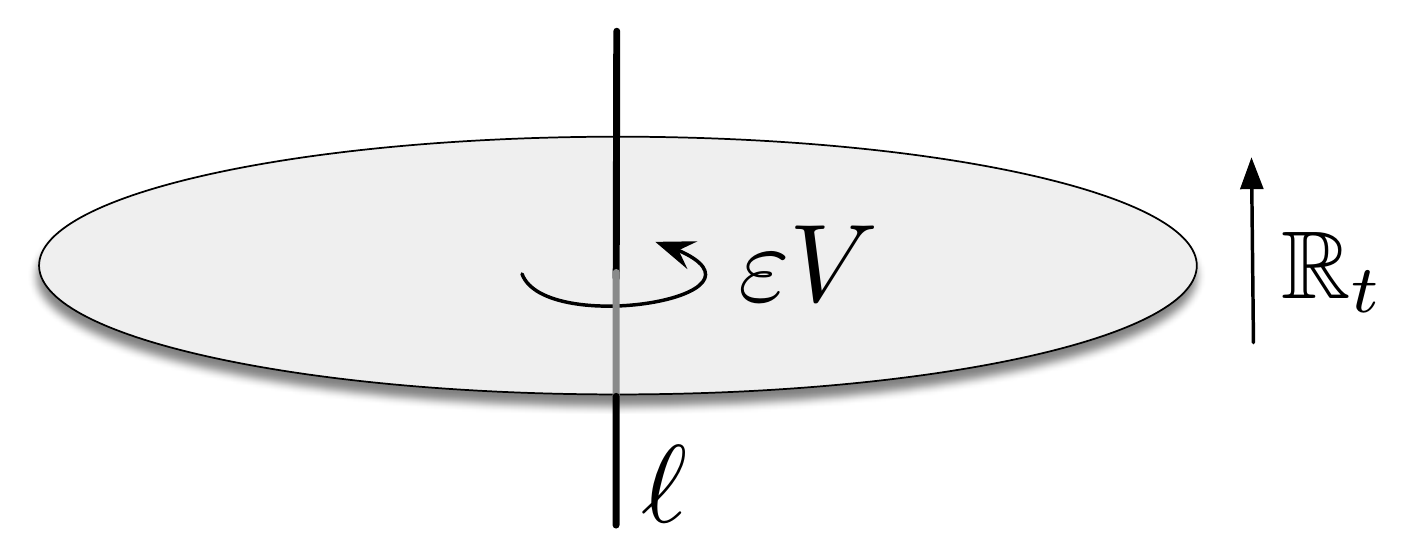}
\caption{Introducing an Omega background for rotations about $\ell$.}
\label{fig:omega}
\end{figure}

The A and B twists of 3d $\CN=4$ gauge theories are each compatible with (distinct) Omega deformations~\cite{Nek-Omega}.
An Omega deformation in three dimensions involves working equivariantly with respect to rotations about a fixed axis. The axis we choose is the usual line $\ell = {\{z=0\}}\times \R_t$, where putative line operators are supported (Fig. \ref{fig:omega}). 
Let $U(1)_E$ be the group of rotations about $\ell$, and let $V_A$ (resp. $V_B$) be the generator of the diagonal subgroup of $U(1)_E\times U(1)_H$ (resp. $U(1)_E\times U(1)_C$) that leaves $Q_A$ (resp. $Q_B$) invariant. Then the two Omega deformations deform the 3d $\CN=4$ theory in such a way that%
\footnote{Complex mass or FI parameters will also contribute to the RHS of \eqref{Omega-e}, as in \eqref{Q2m}. This just means that $Q_A$ or $Q_B$ act as equivariant differentials with respect to both flavor symmetries and spacetime rotations.}
\be Q_A^2 \sim \varepsilon V_A \qquad\text{or}\qquad Q_B^2 \sim \varepsilon V_B\,. \label{Omega-e} \ee
They each depend on a complex-valued equivariant parameter $\varepsilon$.

A useful way to think about these Omega backgrounds comes from dimensional reduction. (This perspective was discussed in \cite{VV}.) Namely, we can rewrite a 3d $\CN=4$ theory on $\C_z\times \R_t$ as 1d $\CN=4$ SQM on $\R_t$, with an infinite-dimensional gauge group and target space. In fact there are two ways to do this, using either the $\text{SQM}_A$ or $\text{SQM}_B$ subalgebras. From the perspective of the 1d SQM$_A$ (resp. SQM$_B$) theory, the diagonal of $U(1)_E\times U(1)_H$ (resp. $U(1)_E\times U(1)_C$) acts as an ordinary flavor symmetry --- indeed, these rotations are ordinary isometries of the infinite-dimensional target space. Then each Omega deformation is achieved by turning on a twisted mass $\varepsilon$ for the appropriate flavor symmetry. 

In (say) the B-type Omega deformation, both B-type line operators wrapping $\ell$ and B-type local operators at points on $\ell$ survive  --- in the sense that they remain in the cohomology of the $Q_B$ supercharge. However, the products of local operators induced by collisions of junctions may be deformed. We would like to spell out how this happens, explicitly and algebraically, in 3d $\CN=4$ gauge theories. We will restrict ourselves to the simple case that the representation $R$ is faithful, so that the local operators in question are just polynomials in $X^i$ and $Y_i$.

\subsubsection{Quantization of the bulk algebra}

In the case of bulk local operators, the effect of the B-type Omega deformation is well understood: it \emph{quantizes} the commutative algebra $\text{End}_B(\id)\simeq \C[\CM_H]$, in the sense of deformation quantization with respect to the holomorphic symplectic form. This quantization was derived for B-twisted sigma-models in \cite{Yagi-Omega}, and explained in a general TQFT context in \cite{descent}. (The idea that the Omega background is related to quantization goes back to work of Nekrasov and Shatashvili \cite{NS-quantization}.)

We can describe the quantization of $\C[\CM_H]$ explicitly, following \cite{BDG, BDGH}, as a quantum Hamiltonian reduction.%
\footnote{Quantum Hamiltonian reduction is a standard procedure in mathematics, often described in the language of D-modules, \emph{cf.} \cite{KR-quant, MN-quant} and references therein.} %
Namely, the polynomial algebra $\C[X,Y]$ of hypermultiplet scalars is first quantized to $n$ copies of the Heisenberg algebra, with commutation relations
\be [X^i,Y_j] = \varepsilon \delta^i{}_j \,.\ee
We'll call this algebra $\C_\varepsilon[X,Y]$. The complex moment map for the $G$ action is promoted to an operator $\mu_\C \in \C_\varepsilon[X,Y]\otimes \mathfrak g_\C^*$, with components given by the normal-ordered combinations
\be \langle\mu_\C,t_a\rangle = \text{Tr}\big[ \rho_R(t_a) :\!(XY)^T\!: \big]\,.\ee
(In practice, the normal-ordering is only important for abelian factors in $G$.) By construction, the commutators of components of the moment map in the Heisenberg algebra $\C_\varepsilon[X,Y]$ agree with the Lie bracket of generators of $\mathfrak g$, namely
\be [ \langle\mu_\C,t_a\rangle, \langle\mu_\C,t_a\rangle] = - \varepsilon  \langle\mu_\C,[t_a,t_b]\rangle\,. \ee
Moreover, the commutator of $\mu$ and any other element of $\C_\varepsilon[X,Y]$ generates an infinitesimal gauge transformation; schematically,  
\be [\langle \mu_\C,t_a\rangle,\CO] = - \varepsilon\, t_a\cdot \CO\,.\ee

We get from the Heisenberg algebra $\C_\varepsilon[X,Y]$ to the quantization of the chiral ring $\text{End}_B^\varepsilon(\id)=\C_\varepsilon[\CM_H]$ in two steps. First, we quotient by \emph{either} the left ideal $\C_\varepsilon[X,Y](\mu_\C)$ \emph{or} the right ideal $(\mu_\C)\C_\varepsilon[X,Y]$ generated by the components of $\mu_\C$. Notice that neither of these are two-sided ideals, since $\mu$ does not commute with general elements of $\C_\varepsilon[X,Y]$ (precisely because general elements are not gauge-invariant). Then we impose gauge-invariance, finding
\be \label{End1-q} \text{End}_B^\varepsilon(\id) = \big[ \C_\varepsilon[X,Y]/(\mu_\C)\big]^G \simeq \big[(\mu_\C) \backslash \C_\varepsilon[X,Y]\big]^G\,.\ee

The two quotients, by left and right ideals, are only equivalent after imposing $G$ invariance. The equivalence follows from the fact that any element $\CO\in \C_\varepsilon[X,Y]$ that is $G$-invariant commutes with $\mu_\C$. A consequence of the equivalence of the two quotients is that $\text{End}_B^\varepsilon(\id)$ is again an algebra, with a well-defined associative multiplication.

\subsubsection{Quantization of operators on Wilson lines}

The quantum Hamiltonian reduction above can be generalized to describe the Omega-deformed spaces of local operators bound to junctions of arbitrary Wilson lines.

Let us consider a single Wilson line $\W_V$, and the algebra of local operators bound to it. Prior to introducing the Omega background, this algebra \eqref{EndW} was computed as $\big[\C[X,Y]\otimes \text{End}(V^*)/(\mu_\C)\big]^G$. In the Omega background, we instead begin with the algebra $\C_\varepsilon[X,Y]\otimes \text{End}(V^*)$, consisting of $N\times N$ matrices ($N=\text{dim}(V)$) whose elements are entries of the Heisenberg algebra $\C_\varepsilon[X,Y]$. We would like to quotient by an appropriate left or right ideal, in order to set the moment map to zero, and then to restrict to gauge-invariant operators.

Identifying the correct ideals to quotient by requires some care. A naive guess would be to take a left or right ideal generated by components of $\mu_\C$ as in \eqref{End1-q}. However, this prescription fails to produce an algebra, because the left and right quotients do not agree. To see the problem, consider an arbitrary element $\CO\in \C_\varepsilon[X,Y]\otimes \text{End}(V^*)$, and suppose that $\CO$ is $G$-invariant with respect to the simultaneous action of $G$ on $X,Y$ and on $\text{End}(V^*)$. In other words, $\CO$ transforms as an element of $\text{End}(V)$. Then $\mu_\C$ does not commute with $\CO$. Rather, for any element $t_a\in \mathfrak g$, we have
\be [\langle \mu_\C,t_a\rangle \otimes \text{id}_V, \CO] = - \varepsilon [\rho_V(t_a),\CO]\,, \label{mu-rho} \ee
where $\rho_V:\mathfrak g\to \text{End}(V)$ is the representation of the Lie algebra generator on $V$. (To be clear, the LHS is a commutator in the quantum Heisenberg algebra, whereas the RHS is a commutator of matrices.)

The relation \eqref{mu-rho} tells us how to correct our naive guess. Namely, we consider ideals generated by the components of $\mu_\C \otimes \text{id}_V +\varepsilon\, \rho_V$, \emph{i.e.} the elements
\be \langle \mu_\C,t_a\rangle \otimes \text{id}_V + \varepsilon\,1\otimes \rho_V(t_a)  \ee
for all $t_a\in \mathfrak g$. It is also useful to rewrite $\mu_\C \otimes \text{id}_V +\varepsilon\, \rho_V \simeq \mu_\C\otimes \text{id}_{V^*} - \varepsilon\, \rho_{V^*}$, using unitarity of the representation $V$. Then the Omega-deformed algebra of operators bound to the line becomes
\begin{align}  \text{End}^\varepsilon_B(\W_{V}) &= \big[(\mu_\C \otimes \text{id}_{V^*} - \varepsilon \rho_{V^*}) \backslash \C_\varepsilon[X,Y]\otimes \text{End}(V^*)\big]^G \label{EndW-q}  \\ &\hspace{.5in} \simeq \big[ \C_\varepsilon[X,Y]\otimes \text{End}(V)/ (\mu_\C \otimes \text{id}_{V^*} - \varepsilon \rho_{V^*})\big]^G\,. \notag \end{align}
The equivalence of left and right quotients ensures that this will be an algebra, as expected.

Many examples of \eqref{EndW-q} appeared in \cite{BDGH}, for one-dimensional representations $V$ of $G$ (\emph{i.e.} for abelian representations).
In this case, $\rho_{V}$ itself acted as multiplication by a constant $q$, the quantized charge of the Wilson line. In the examples of \eqref{EndW-q}, related to symplectic duality \cite{SD-I, SD-II}, the algebras $\text{End}^\varepsilon_B(\W_{V})$ had familiar interpretations (\emph{e.g.} as quotients of enveloping algebras of semisimple Lie algebras), and the charge of the Wilson line specified the value of Casimir operators.

It is now easy to extend \eqref{EndW-q} to describe local operators at general junctions of Wilson lines. Given a pair of Wilson lines $\W_V, \W_{V'}$, we begin with the vector space $\C_\varepsilon[X,Y]\otimes \text{Hom}(V^*,V'{}^*)$, generalizing \eqref{HomWW'}. To set the moment map to zero, we quotient \emph{either} by the left ideal generated by components of $\mu \otimes \text{id}_{V^*} - \varepsilon \rho_{V^*}$ or the right ideal generated by components of $\mu \otimes \text{id}_{V'{}^*} - \varepsilon \rho_{V'{}^*}$. After restricting to $G$-invariant operators, the two quotients become equivalent, and we have
\begin{align} \label{HomWW'-q}
\text{Hom}_B^\varepsilon(\W_{V},\W_{V'}) &=
  \big[(\mu_\C \otimes \text{id}_{V'{}^*} - \varepsilon \rho_{V'{}^*}) \backslash \C_\varepsilon[X,Y]\otimes \text{Hom}(V^*,V'{}^*)\big]^G  \\ &\hspace{.5in} \simeq \big[ \C_\varepsilon[X,Y]\otimes \text{Hom}(V^*,V'{}^*)/ (\mu_\C \otimes \text{id}_{V^*} - \varepsilon \rho_{V^*})\big]^G\,. \notag \end{align}

Note that the space \eqref{HomWW'-q} is not an algebra unless $V=V'$. 
In general, $\text{Hom}_B^\varepsilon(\W_{V},\W_{V'})$ naturally has the structure of a bimodule for the two algebras $\text{End}_B^\varepsilon(\W_V)$ and $\text{End}_B^\varepsilon(\W_{V'})$.  Physically, collision of local operators bound to $\W_V$ with operators $\CO$ at the junction define a right action of $\text{End}_B^\varepsilon(\W_V)$ on $\text{Hom}_B^\varepsilon(\W_{V},\W_{V'})$; whereas collision of operators bound to $\W_{V'}$ with the junction define an independent, commuting left action of $\text{End}_B^\varepsilon(\W_{V'})$\,:
\be \raisebox{-.5in}{\includegraphics[width=2in]{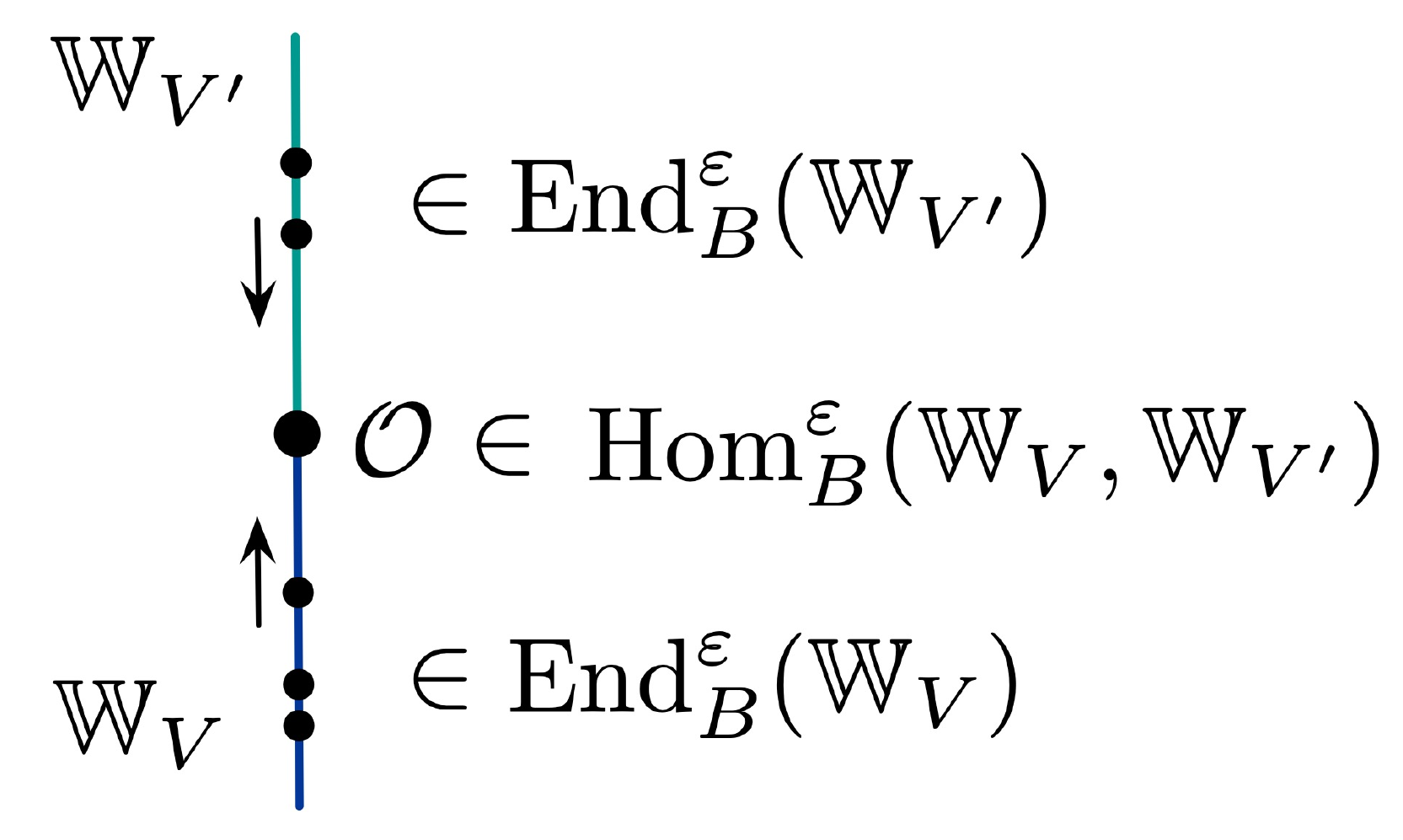}} \ee
(This bimodule structure exists with or without the Omega deformation.)
Similarly, given a triple of Wilson lines $\W_V,\W_{V'},\W_{V''}$, there is the usual composition of operators
\be  \begin{array}{ccc}\text{Hom}_B^\varepsilon(\W_{V'},\W_{V''}) \times \text{Hom}_B^\varepsilon(\W_{V},\W_{V'}) &\to& \text{Hom}_B^\varepsilon(\W_{V},\W_{V''})\,, \\[.1cm]
 \CO'\,, \hspace{.7in} \CO &\mapsto & \CO' \cdot \CO \end{array}  \ee
defined collision of junctions, \cf\ \eqref{Jprod}.

\subsubsection{FI parameters}

The presence of complex FI parameters $t_\C\in \mathfrak g^*$ deforms the quantum moment maps in all the expressions above, replacing $\mu_\C \leadsto \mu_\C +t_\C$. For example, the quantized algebra of operators bound to a Wilson line \eqref{EndW-q} involves quotients by elements of $\mu_\C\otimes \text{id}_{V^*}+t_\C\otimes \text{id}_{V^*}-\varepsilon \rho_{V*}$.

We note that when $V=\C$ is a one-dimensional (\emph{i.e.} abelian) representation of $G$, the terms $t_\C\otimes \text{id}_{V^*}$ and $\varepsilon \rho_{V*}$ can mix. Namely, if $q$ is the charge of $V$, we simply find that  $t_\C\otimes \text{id}_{V^*}-\varepsilon \rho_{V*} = t_\C-q\varepsilon$. From the perspective of operator algebras, only the single combination $t_\C-q\varepsilon$ can be detected. This is a reflection of a more fundamental physical phenomenon: in the Omega background, turning on a quantized FI parameter (quantized in units of $\varepsilon$) is equivalent to introducing an abelian Wilson line.  Roughly speaking, a quantized $t_\C$ induces a vortex for a topological $U(1)$ flavor symmetry, which is the same as an abelian Wilson line for the gauge group.
See \cite{BDGH} for some further discussion.

\section{Half-BPS vortex lines}
\label{sec:VL}

In this section we turn to A-type line operators, \ie\ half-BPS line operators that are preserved by the 1d $\CN=4$ algebra $\text{SQM}_A$.

As reviewed in the Introduction, many aspects of these extended operators have already been studied in the literature --- often under the guise of half-BPS surface operators in 4d $\CN=2$ gauge theories, which share much of the same structure. Moreover, surface operators in 4d $\CN=2$ gauge theories were themselves a generalization of the prototypical Gukov-Witten defects of 4d $\CN=4$ super-Yang-Mills theory, classified in \cite{GW-surface,Witten-wild}.

We know from the literature to expect several different --- but largely equivalent --- constructions of A-type line operators, as
\begin{itemize}
\setlength{\itemsep}{-.1cm}
\item[1)] disorder operators, modeled on singular solutions to the BPS equations for the $\SQM_A$ subalgebra of 3d $\CN=4$ \\
(the BPS equations are generalized vortex equations, whence we typically refer to A-type line operators as \emph{vortex lines});
\item[2)] coupled 3d-1d systems (coupling bulk 3d fields to 1d $\SQM_A$ quantum mechanics, by gauging 1d flavor symmetries and introducing superpotential interactions).
\end{itemize}
In addition, all A-type line operators should define objects in the category of line operators in the A-twist, so we may also hope for a description as
\begin{itemize}
\item[3)] objects of a (dg/A$_\infty$) braided tensor category, with some mathematical definition.
\end{itemize}

In this paper, we will largely focus on constructions (1) and (2). We consider a class of line operators characterized by
\begin{itemize}
\item A meromorphic singularity in the hypermultiplet scalars at $z=0$ in the $\C_z$ plane transverse to a line operator. In description (2), these singularities can be engineered by coupling 3d hypers to 1d chiral matter via a superpotential.
\item A breaking of gauge symmetry near $z=0$, compatible with the singular profile of hypermultiplets. In description (2), this breaking can be engineered by gauging flavor symmetries of a 1d sigma model (essentially a coset model) with the 3d gauge group.
\end{itemize}
It is essential for us to allow higher-order singularities in the matter fields, and breaking of gauge symmetry to higher order around $z=0$; correspondingly, when coupling to 1d quantum mechanics, we allow higher-order derivative couplings.
 In the context of geometric Langlands, such singularities were referred to as ``wild ramification,'' and studied from a physical perspective in \cite{Witten-wild}. In 3d $\CN=4$ gauge theories, A-type line operators defined by higher-order singularities turn out to be the 3d mirrors of ordinary B-type Wilson lines with higher (non-minuscule) charge.

Many standard brane constructions of surface operators in 4d $\CN=2$ theories and line operators in 3d $\CN=4$ (\emph{e.g.} \cite{HananyHori, HananyTong-VIB, DGH, AsselGomis} actually lead naturally to higher-order singularities.
In quiver quantum-mechanics descriptions of these operators, there are higher-derivative couplings present. These have often been overlooked in the literature; we will discuss a simple example in Section \ref{sec:quiver}.

We will also take some preliminary steps toward identifying the category (3) of A-type line operators, as in \eqref{intro-CAB} of the Introduction. Examining mathematical and physical properties of this category is a major objective of \cite{lineops}.

We begin in Section \ref{sec:BPSA} by reviewing the BPS equations for $\SQM_A$, their relation to 1d quantum mechanics, and their associated holomorphic data. Then in Section \ref{sec:VL-matter} we discuss in detail the structure of vortex lines in a theory of free hypermultiplets. Perhaps surprisingly, this turns out to be interesting and nontrivial, and gives us a concrete realization of all three constructions (1)-(3) above! Geometrically, we will associate vortex lines in a theory of $n$ free hypermultiplets with holomorphic Lagrangians in the loop space $L(T^*\C^n)$.

We then consider A-type lines in gauge theories in Section \ref{sec:VL-gauge}. Roughly speaking, this requires combining meromorphic singularities in hypermultiplet fields with compatible patterns of gauge-symmetry breaking in the neighborhood of a line. We give many examples, and define a general class of A-type line operators in gauge theories whose junctions we will study in the remainder of the paper.

\subsection{BPS equations for $\SQM_A$}
\label{sec:BPSA}

As preparation for studying half-BPS line operators preserved by the $\SQM_A$ subalgebra, we review the BPS equations for $\SQM_A$ and their moduli space of solutions. The analysis is very similar to that in \cite{BJSV, Kapustin-hol} for 4d $\CN=2$, and \cite{VV, Gaiotto-twisted, BFK} for 3d $\CN=4$.

As usual, we split spacetime as $\R^3\simeq \R_t \times \C_z$, and we consider a 3d $\CN=4$ gauge theory whose vectormultiplet contains scalars $\varphi\in \mathfrak g_\C$, $\sigma\in \mathfrak g$ and whose hypermultiplets contain pairs of complex scalars $X\in R$, $Y\in R^*$. The full SUSY transformations of these fields are summarized in Appendix \ref{app:SUSY}. By setting to zero the variations of gauginos and hypermultiplet fermions under the four supercharges $Q_A^{\dot a} = \delta^\alpha{}_a Q_\alpha^{a\dot a}$, $\wt Q_A^{\dot a}=(\sigma^3)^\alpha{}_a Q_\alpha^{a\dot a}$ that generate $\SQM_A$, we find bosonic BPS equations
\begin{subequations} \label{A-eqs}
\be [D_t,D_z]=[D_t,D_{\bar z}]=0\,,\qquad D_tX=D_tY = D_t\varphi=D_t\sigma=0\,, \label{A-time} \ee
\be \begin{array}{c} [\sigma,\varphi]=[\sigma,\varphi^\dagger]=[\varphi,\varphi^\dagger]=0\,,\\[.2cm] \sigma\cdot X=\sigma \cdot Y=0\,,\quad \varphi\cdot X=\varphi \cdot Y=0\,,\qquad \varphi^\dagger\cdot X=\varphi^\dagger\cdot Y = 0\,, \end{array}  \label{A-phi} \ee
\be
D_z\sigma=D_z\varphi=D_z\varphi^\dagger = 0\,,\qquad D_{\bar z}\sigma=D_{\bar z}\varphi=D_{\bar z}\varphi^\dagger = 0 \label{A-zphi} \ee
\be \boxed{F_{z\bar z} = \mu_\R\,,\qquad D_{\bar z}X=D_{\bar z}Y = 0\,,\qquad \mu_\C= 0}\,. \label{A-vortex}\ee
\end{subequations}
Here $D_t,D_z,D_{\bar z}$ are covariant derivatives with respect to the $G$-connection $A$; and in \eqref{A-phi} the schematic expressions $\sigma\cdot X,\varphi\cdot X, \sigma \cdot Y$, etc. denote the infinitesimal action of the $\sigma\in \mathfrak g$, $\varphi\in \mathfrak g_\C$ on the representation $X\in R$ and $Y\in R^*$. (More explicitly, one could write $\rho_R(\sigma) X=0$, $\rho_{R^*}(\sigma)Y=0$, etc.)

Observe that the first set of equations \eqref{A-time} guarantees that all fields are covariantly constant in time, as we would expect for BPS equations in quantum mechanics. This allows us to restrict our analysis of solutions to the plane $\C_z$ transverse to a line operator, knowing that solutions can then be extended along $\R_t$ in a unique way.

The equations \eqref{A-phi} restrict $\sigma,\varphi$ to lie in a common Cartan subalgebra, and say that $X$ and $Y$ should be fixed under the infinitesimal action of these fields. The third set \eqref{A-zphi} requires $\sigma,\varphi$ to be covariantly constant in the $\C_z$ plane as well.
So far, these are standard BPS vacuum equations for 3d $\CN=4$ SUSY.

 The final boxed set of equations \eqref{A-vortex} are the interesting ones: they are generalized vortex equations in the $\C_z$ plane \cite{vortex-stab1, vortex-stab2, vortex-stab2II, BJSV, HananyTong-VIB}, requiring $X$ and $Y$ to be covariantly holomorphic (not constant), and sourcing the magnetic flux $F_{z\bar z}$ with the real moment map. 

Mass and FI parameters can be included in the BPS equations in a standard way. Namely, the FI parameters deform the moment maps $(\mu_\R,\mu_\C)\leadsto (\mu_\R+t_\R,\mu_\C+t_\C)$ while masses $m_\R\in \mathfrak f$, $m_\C\in \mathfrak f_\C$  (valued in a common Cartan subalgebra of the flavor symmetry) enter the same way as $\sigma,\varphi$. We will come back to them later in Section \ref{sec:comp}.

\subsubsection{Rewriting 3d $\CN=4$ as 1d quantum mechanics}
\label{sec:QMaction}

Another useful step in preparation for describing vortex lines is to rewrite a bulk 3d ${\CN=4}$ gauge theory as 1d $\SQM_A$ supersymmetric quantum mechanics \cite{VV}.

By ``rewriting a 3d theory as a 1d theory,'' we mean to reinterpret all the fields of the 3d theory on $\R_t\times \C_z$ as fields on $\R_t$ valued in functions (or sections of various bundles) on $\C_z$. 
Given a 3d gauge group $G$ and representation $R\oplus R^*$, the 3d $\CN=4$ multiplets decompose under the 1d $\SQM_A$ subalgebra%
\footnote{Note that the 1d $\CN=4$ multiplets used here are sometimes denoted ``$\CN=(2,2)$'' multiplets in the literature. This is because they are the multiplets one obtains by reducing 2d $\CN=(2,2)$ chiral and vectormultiplets to 1d.} %
as follows:
\begin{itemize}
\item The 3d hypermultiplets split into pairs of 1d chiral multiplets, with bottom components $X$ and $Y$.
 More precisely, the bottom components are maps $X(z,\bar z)$, $Y(z,\bar z)$ from the $\C_z$ plane into the original target space $R\oplus R^*$ of the 3d theory. 
\item The 1d gauge group consists of all $G$-valued gauge transformations $g(z,\bar z)$ in the $\C_z$ plane. We will denote this infinite-dimensional group as $\CG$.
\item The 3d vectormultiplet splits into 1) a 1d vectormultiplet for the gauge group $\CG$, containing the connection $A_t$ and the triplet of scalars $\sigma,\varphi,\varphi^\dagger$; and 2) a 1d chiral multiplet with bottom component $A_{\bar z}$.
\end{itemize}

The supersymmetric Lagrangian for this 1d $\CN=4$ quantum mechanics includes an important superpotential term
\be W = \int_{\C_z} d^2z\, \text{Tr}\,X D_{\bar z} Y\,, \label{A-W} \ee
which captures the kinetic terms for $X$ and $Y$ in the $\C_z$ plane.
Note that the superpotential involves the chiral multiplet $A_{\bar z}$ (in $D_{\bar z} = \pd_{\bar z}-iA_{\bar z}$) as well as $X$ and $Y$. 

In this 1d $\CN=4$ quantum mechanics, the half-BPS equations \eqref{A-eqs} may now be interpreted as familiar equations for SUSY vacua (\emph{i.e.} full-BPS equations). In particular, the F-term equations coming from $W$ reproduce most of \eqref{A-vortex}\,:
\be \frac{\delta W}{\delta X} = D_{\bar z}Y = 0\,,\qquad \frac{\delta W}{\delta Y} \sim D_{\bar z}X = 0\,,\qquad \frac{\delta W}{\delta A_{\bar z}} \sim \mu_\C = 0\,.\ee
The remaining vortex equation $F_{z\bar z}-\mu_\R=0$ appears as a D-term in quantum mechanics.

\subsubsection{Holomorphic gauge}
\label{sec:hol}

A common technique for analyzing the vortex equations \eqref{A-vortex} involves trading the real D-term equation $F_{z\bar z}=\mu_\R$ for a complexification of the gauge group. In mathematics, this is often called a Kobayashi-Hitchin correspondence (with a prototypical realization in the Donaldson-Uhlenbeck-Yau Theorem \cite{Donaldson-stab, UhlenbeckYau}). 
This ultimately allows a complex-analytic or (even better) an algebraic description of the moduli space of solutions.
We briefly recall the basic ideas, aiming to provide intuition rather than mathematical rigor.

Recall that the first three vortex equations $D_{\bar z}X=D_{\bar z}Y = \mu_\C=0$ are critical-point equations for the superpotential $W$ in \eqref{A-W}, whereas $F_{z\bar z}=\mu_\R$ is a real D-term constraint for the infinite-dimensional gauge group $\CG$ of all $G$-valued gauge transformations on $\C_z$. The space of solutions to the vortex equations on $\C_z$ is thus a real symplectic quotient 
\begin{align} \CM &=  \{\text{$A,X,Y$ s.t. $\delta W=0$}\}/\!/\CG  \label{M-real} \\
  &=  \{\text{$A,X,Y$ s.t. $\delta W=0$ and $F_{z\bar z}=\mu_\R$}\}/\CG\,.  \notag \end{align}
By comparison with the finite-dimensional setting \cite{KempfNess, Kirwan-quotient}, we \emph{expect} to be able to ignore the D-term constraint while at the same time complexifying the gauge group $\CG \leadsto \CG_\C$, and possibly imposing some stability conditions. Roughly, we should have
\be \CM \approx \{\text{$A,X,Y$ s.t. $\delta W=0$}\}/\CG_\C\,, \label{M-complex} \ee
where $\CG_\C$ is the group of all $G_\C$-valued gauge transformations on $\C_z$. 

In \eqref{M-complex}, we can further use complexified gauge transformations to gauge-fix $A_{\bar z}=0$, so that the covariant derivative $D_{\bar z}=\pd_{\bar z}$ becomes an ordinary derivative.
We are left with a residual gauge group consisting of holomorphic gauge transformations $\CG_\C^{\rm hol} = \{g(z) \in \CG_\C\;\text{s.t.}\; \pd_{\bar z}g = 0\}$,
and a complex-analytic moduli space
\be \CM \approx \left\{\begin{array}{c}\text{holomorphic $G_\C$ bundles on $\C_z$} \\[.1cm]
\text{w/ hol'c sections $X(z),Y(z)$ of an associated $R\oplus R^*$ bundle} \\[.1cm]
\text{s.t. $\mu_\C(X,Y)=0$} \end{array}\right\} \Big/ \CG_\C^{\rm hol}\,. \label{M-hol} \ee

Making the equivalence of \eqref{M-real} and \eqref{M-hol} precise can be a subtle and difficult endeavor.
 One must specify boundary conditions as $z\to \infty$, as well as stability conditions for the $G_\C$-bundles and holomorphic sections appearing in \eqref{M-hol}. Some of the mathematical history of this endeavor, starting with \cite{Taubes-vortex, JaffeTaubes} for abelian $G$, was reviewed in the introduction. In Section \ref{sec:comp} we will consider moduli spaces with a vacuum boundary condition at $z\to \infty$, whose holomorphic/algebraic formulation was established relatively recently by \cite{VW-vortices}.

For the moment, we are not interested in moduli spaces \emph{per se}, but rather in the structure of singularities in the BPS equations. There is a large body of mathematical work on singularities and their algebraic data for the case of trivial or adjoint $R$ (the latter leading to Hitchin's equations), \emph{e.g.} \cite{MehtaSeshadri, Simpson, Sabbah, BiquardBoalch}, used in characterizing surface operators in 4d $\CN=4$ SYM \cite{GW-surface, Witten-wild}. Some recent work on singularities for abelian $G$ and general $R$ appears in \cite{BaptistaBiswas}. However, there does not seem to exist a classification of singularities of our half-BPS equations for general $G$ and $R$.

In the remainder of this section, we attempt to build up a partial classification, focusing primarily on holomorphic/algebraic data, as it will feed directly into algebraic definitions of moduli spaces. The classification is motivated both by \cite{GW-surface, Witten-wild} and recent work on 3d $\CN=4$ Coulomb branches \cite{BDGH, BFN-lines, Web2016}.

\subsection{Free matter}
\label{sec:VL-matter}

Even a theory with free hypermultiplet matter can have interesting, nontrivial half-BPS vortex lines.
Indeed, they illustrate most of the main features of vortex lines in gauge theories, while avoiding subtleties such as the equivalence of real and holomorphic moduli spaces \eqref{M-real}-\eqref{M-hol} above.
We discuss free hypermultiplets in this section, and then add gauge interactions in Section \ref{sec:VL-gauge}.
 
Consider the 3d $\CN=4$ theory of a single free hypermultiplet. (In this case, $G=1$ and the hypermultiplet scalars $(X,Y)$ are just valued in $R\oplus R^*=\C\oplus \C$.) The $\SQM_A$ BPS equations \eqref{A-eqs} simply require $X,Y$ to be constant in time, and holomorphic in the $\C_z$ plane,
\be \pd_{\bar z}X = \pd_{\bar z}Y = 0\,. \label{hyper-SQMA} \ee
A large family of solutions with a singularity at the origin come from allowing $X$ and $Y$ to have poles of some order, say
\be X(z) = \frac{a}{z^k} + \frac{b}{z^{k-1}} + \ldots\,,\qquad Y(z) = \frac{a'}{z^{k'}} + \frac{b'}{z^{k'-1}} + \ldots\,. \label{XYpoles} \ee
Given such a solution, we can attempt to define a ``disorder'' line operator $\V_{k,k'}$ using a standard prescription: we excise the line $\{z=0\}$ from spacetime, and restrict the path integral on $\C^*_z\times \R_t$ to field configurations that approach \eqref{XYpoles} near $z=0$.

There is actually some choice in how to interpret \eqref{XYpoles}. The vortex-line operators $\V_{k,k'}$ that we define in this paper will \emph{allow} poles of order $\leq k$, $\leq k'$ in $X,Y$ at $z=0$, but do not \emph{require} poles. In other words, we do not fix the coefficients of singular terms, such as $a,b,a',b',...$, above.%
\footnote{This contrasts with the surface operators defined by Gukov-Witten \cite{GW-surface}, which did give the adjoint matter fields a first-order pole with fixed residue.} %
A qualitative feature of this choice is that the $U(1)_m$ flavor symmetry that rotates $X$ and $Y$ with opposite charge is preserved. As we shall verify later, vortex lines defined in this manner turn out to be naturally dual to B-type Wilson lines.

\subsubsection{Lagrangians in the loop space}
\label{sec:VL-Lags}

There is an important additional constraint on the values of $k$ and $k'$ appearing in \eqref{XYpoles} that we must discuss. In order for \eqref{XYpoles} to be a half-BPS field configuration, it is not quite sufficient to just satisfy the bosonic BPS equations \eqref{hyper-SQMA}; we must also consider the fermionic fields. Equivalently,
we must make sure that a singularity of the form \eqref{XYpoles} makes sense for entire 1d $\SQM_A$ multiplets.

From the superpotential \eqref{A-W}, it is clear that the 1d chiral multiplet with bottom component $X$ has an F-term $\pd_z\ol Y$. Similarly, the multiplet with bottom component $Y$ has an F-term $-\pd_z\ol X$. This structure is ultimately governed by the holomorphic symplectic form $\Omega= dX\wedge dY$ on the 3d target space.

Suppose then that we work on the ``punctured'' spacetime $\C^*_z\times \R_t$, and expand $X$ and $Y$ into modes as
\be X = \sum_{n\in \Z} x_n z^n = \sum_{n\in \Z} x_n(r,t) r^ne^{in\theta}\,,\qquad Y = \sum_{n\in \Z} y_n z^n = \sum_{n\in \Z} y_n(r,t) r^ne^{in\theta}\,.\ee
The respective F-terms in the $X$ and $Y$ multiplets are
\be \pd_z \ol Y = \sum_{n\in \Z} \pd_r\bar y_{-n-1}r^{-n-1}e^{in\theta}\,,\qquad -\pd_z \ol X = - \sum_{n\in \Z} \pd_r \bar x_{-n-1}r^{-n-1}e^{in\theta} \,.\ee
Therefore, the pairs of modes $( r^nx_n, \frac{1}{r^{n+1}}\pd_r \bar y_{-n-1})$ all lie in the same multiplet, as do the pairs $(r^n y_n,\frac{1}{r^{n+1}}\pd_r \bar x_{-n-1})$. If we think about a putative singularity at $z=0$ as a boundary condition on the modes, we encounter a familiar structure: a ``Dirichlet'' boundary condition that sets any mode $x_n\big|_{r=0} = 0$ must be accompanied by a ``Neumann'' boundary condition that leaves its conjugate $y_{-n-1}\big|_{r=0}$ unconstrained. 
 
For example, we would describe the trivial (\ie\ empty) vortex line $\id$ in this language as the boundary condition
\be \id\,:\quad \tfrac{1}{r^n}x_{-n}\big|_{r=0} = r^{n-1}\pd_r\bar y_{n-1}\big|_{r=0} = 0\,,\quad
\tfrac{1}{r^n}y_{-n}\big|_{r=0} = r^{n-1}\pd_r\bar x_{n-1}\big|_{r=0} = 0
\quad\forall\,n>0\,, \label{trivVL} \ee
which simply says that all negative modes $x_{-n},y_{-n}$ vanish at the origin, while all positive modes are unconstrained. In other words, $X,Y$ are regular on $\C_z$.

Alternatively, we could ``flip'' a mode from $Y$ to $X$, allowing $X$ to have a first-order pole, while constraining $Y$ to have a first-order zero. Then the boundary condition sets 
\be y_0\big|_{r=0} = \tfrac{1}{r}\pd_r\bar x_{-1}\big|_{r=0} = 0\,, \ee
which is effectively Dirichlet for $y_0$ and Neumann for $x_{-1}$.

There is a natural geometric characterization of the sorts of singularities that are preserved by the $\SQM_A$ subalgebra. Let
\be \Omega_L = \frac{1}{2 \pi i}\oint dz\, dX\wedge dY = \sum_{n\in \Z} d x_n \wedge d y_{-n-1} \ee
be the holomorphic symplectic form on the \emph{loop space} $L(R\oplus R^*)$ of the original 3d target, parameterized by the modes of $X$ and $Y$. Then the above analysis of multiplets amounts to saying that half-BPS singularities must be supported on \emph{holomorphic Lagrangian submanifolds} in the loop space, with respect to $\Omega_L$.%
\footnote{Such holomorphic Lagrangian submanifolds appear naturally as half-BPS boundary conditions for 2d $\CN=(4,4)$ sigma-models. They were studied extensively in \cite{KapustinWitten} and many subsequent papers, and are often referred to as (B,A,A) branes. The connection between (B,A,A) branes and line operators in 3d $\CN=4$ theories comes from reduction of the 3d theories along a circle linking the line --- which turns the line into a boundary condition for an effective 2d $\CN=(4,4)$ theory. We elaborate on this construction in \cite{lineops}.}
 
For example, from this geometric perspective, the trivial line operator is the Lagrangian
\be \id\,:\quad \{x_n=y_n=0\}_{n<0}\,. \ee
The vortex line that we first described in \eqref{XYpoles}, with free coefficients $a,b,a',b',...$, corresponds to a holomorphic Lagrangian if and only if $k+k'=0$. In this case, we get the vortex line
\be \V_k\,:\quad \left\{ \begin{array}{l@{\qquad}l} x_n=0 & n < -k \\
 y_n=0 & n < k \end{array}\right\}\,. \label{def-Vk} \ee
If $k+k'\neq 0$, the singularity \eqref{XYpoles} is not half-BPS.

\subsubsection{Flipping modes with 1d chirals}
\label{sec:flip}

An alternative definition of the vortex line $\V_k$ comes from coupling the 3d theory of a free hypermultiplet to additional purely 1d degrees of freedom --- namely, to free 1d chiral multiplets.

Consider, for example, a single 1d $\CN=4$ chiral multiplet $q$, localized on the line $\ell$ at $\{z=0\}$.%
\footnote{The sort of 1d $\CN=4$ multiplets that can be coupled to the bulk theory must be of ``1d $\CN=(2,2)$'' type. This is because the bulk multiplets themselves reduce to ``1d $\CN=(2,2)$'' type multiplets under the subalgebra $\text{SQM}_A$, as discussed in Section \ref{sec:QMaction}.} %
We will denote the scalar component of this supermultiplet by $q$. If we couple the 1d chiral to the bulk hypermultiplet fields with a superpotential $W_{1d} = q\, X\big|_{z=0}$, then the total superpotential (including \eqref{A-W}) becomes
\be W = \int d^2z \big[ X\pd_{\bar z}Y + q X \delta^{(2)}(z,\bar z)\big]\,. \label{Wflip} \ee
The F-term equation $\pd_{\bar z}Y=0$ gets modified to
\be \pd_{\bar z} Y + q\delta^{(2)} = 0\quad\Rightarrow\quad Y = -\frac{q}{z}+\text{regular, holomorphic}\,, \ee 
allowing $Y$ to have a pole with (undetermined) coefficient $-q$.
Dually, there is a new F-term equation for $q$, namely
\be \frac{\delta W}{\delta q}=0\quad\Rightarrow\quad X \delta^{(2)}(z,\bar z) =0 \quad\Rightarrow\quad X = z \cdot(\text{regular, holomorphic})\,,\ee
which requires $X$ to have a first-order zero. Altogether, coupling to the 1d chiral $q$ provides an equivalent definition of the vortex line $\V_{-1}$.

This sort of procedure, using 1d matter to ``flip'' a mode from $X$ to $Y$, is analogous to ``flips'' of supersymmetric boundary conditions from Neumann to Dirichlet and vice versa. Such flips were introduced in \cite{DGG}, in the context of  3d $\CN=2$ boundary conditions for 4d $\CN=2$ theories, as a generalization of Witten's $SL(2,\Z)$ action on boundary conditions \cite{Witten-SL2}.

It is easy to generalize the coupling $W_{1d}=qX$ to produce other vortices $\V_k$. If $k<0$, we can introduce $|k|$ 1d chiral multiplets $q_1,...,q_{|k|}$ with scalar components $q_1,...,q_{|k|}$, and a superpotential coupling
\be W_{1d} = \big[ q_1X + q_2\pd_z X + \ldots + q_{|k|}\pd_z^{|k|-1} X \big]\big|_{z=0}\,, \label{Wk} \ee
so that the F-terms of the total superpotential effectively set
\be Y = -(|k|-1)!\frac{q_{|k|}}{z^k} - \cdots - \frac{q_2}{z^2}-\frac{q_1}{z}+\text{regular}\,,\qquad X = z^k\cdot(\text{regular})\,.\ee
Note that the 1d chirals $q_i$ must have nontrivial charges under the $U(1)_E$ group of spacetime rotations in the $\C_z$ plane, in order for the coupling \eqref{Wk} to preserve this symmetry. From the point of view of the 1d SQM along the line, $U(1)_E$ (mixed with the bulk $U(1)_H$ R-symmetry) is an ordinary flavor symmetry.

Dually, to produce $\V_k$ with $k>0$, we could introduce 1d chiral multiplets $q_1,...,q_k$ and a coupling
\be W_{1d} = \big[ q_1Y + q_2\pd_z Y + \ldots + q_{k}\pd_z^{k-1} Y \big]\big|_{z=0}\,, \label{WkY} \ee
which effectively sets
\be X = (k-1)!\frac{q_{k}}{z^k} + \cdots + \frac{q_2}{z^2}+\frac{q}{z}+\text{regular}\,,\qquad Y = z^k\cdot(\text{regular})\,.\ee

\subsubsection{Multiple hypermultiplets}

For trivial gauge group and $N$ hypermultiplets, \ie\ $R\oplus R^*\simeq \C^N\oplus \C^N$, the family of vortex lines described above generalizes in a straightforward way. Let $(X^i,Y_i)_{i=1}^N$ be the complex hypermultiplet scalars. Then the holomorphic symplectic form on loop space is
\be \label{OmegaL-gen} \Omega_L = \frac{1}{2\pi i} \oint dz \sum_{i=1}^N dX^i\wedge dY_i = \sum_{i=1}^N\sum_{n\in \Z} dx^i_n\wedge dy_{i,-n-1}\,, \ee
and a general half-BPS vortex should correspond to a holomorphic Lagrangian in the space of modes $x^i_n,y_{i,n}$. The simplest holomorphic Lagrangians are just products of \eqref{def-Vk}; they define vortex lines
\be \V_{k_1,...,k_N}\,:\quad \left\{ \begin{array}{l@{\qquad}l} x^i_n=0 & n < -k_i \\
 y_{i,n}=0 & n < k_i \end{array}\right\}\,, \label{def-VkN} \ee
for which each $X^i$ is allowed a pole of order $k_i$ (and $Y_i$ is required to have a zero of order $k_i$) or vice versa. However, many more intricate configurations are possible as well.

As before, any vortex line \eqref{def-VkN} can equivalently be engineered by coupling the bulk 3d $\CN=4$ theory to free 1d chiral matter, with appropriate $U(1)_E$ charges.

\subsection{Algebraic reformulation}
\label{sec:VL-alg}

We now introduce some standard algebraic notation that will be useful in the remainder of the paper. In an infinitesimal neighborhood of the origin in $\C_z$, the holomorphic functions may be described as formal Taylor series. The ring of formal Taylor series is denoted
\be \CO = \C[\![z]\!]\,. \ee
Similarly, the holomorphic functions in an infinitesimal punctured neighborhood of the origin --- with a possible meromorphic singularity at the origin --- are formal Laurent series, denoted
\be \CK = \C(\!(z)\!)\,. \ee
The ring $\CK$ is an algebraic version of the loop space $L\C$.

Above, we encountered the loop space $L(R\oplus R^*) \simeq T^*(LR)$. Its algebraic version is
\be R(\CK)\oplus R^*(\CK) \simeq T^*R(\CK)\,, \label{RK} \ee
where $R(\CK) = R\otimes \CK$ denotes formal Laurent series whose coefficients are elements of $R$, or (equivalently) vectors in $R$ whose entries are formal Laurent series. For example, if $R=\C^2$, an element of $R(\CK)\oplus R^*(\CK)$ looks like a vector and a covector of formal series,
\begin{align} &X(z)\oplus Y(z) = \bp X^1(z) \\ X^2(z) \ep\oplus \bp Y_1(z) \\ Y_2(z)\ep^T \in \bp \CK \\ \CK \ep \oplus \bp \CK\\ \CK\ep^T \,.
\end{align}
Geometrically, we may think of $R(\CK)\oplus R^*(\CK)$ as the space of holomorphic sections of a holomorphic $R\oplus R^*$ bundle on an infinitesimal punctured disc.

The holomorphic symplectic form on the algebraic loop space $R(\CK)\oplus R^*(\CK)$ is still given by the residue formula \eqref{OmegaL-gen}.
A general half-BPS vortex-line operator in a theory of free hypermultiplets is labeled by a choice of holomorphic Lagrangian submanifold
\be \CL_0 \subset T^*R(\CK)\,. \label{L0-hypers} \ee
We can think of this Lagrangian as specifying how sections of a holomorphic $R\oplus R^*$ on an infinitesimal punctured disc are allowed to extend over the origin. In a theory with $N$ hypermultiplets, $R\simeq \C^N$, the simple holomorphic Lagrangians \eqref{def-VkN} described above, labeled by an $N$-tuple of integers $\k=(k_1,...,k_N)$, may be expressed algebraically as
\be \V_\k\,:\quad  \CL_0 = \bp z^{k_1}\CO \\ z^{k_2}\CO \\ \vdots \\ z^{k_N}\CO \ep \oplus \bp z^{-k_1}\CO \\ z^{-k_2}\CO \\ \vdots \\ z^{-k_N}\CO\ep^T\,. \label{L0k} \ee

\subsubsection{An algebro-geometric category}

As prefaced in the Introduction, we expect vortex lines preserved by the A-twist to be objects of a braided tensor category. In the case of a 3d theory of free hypermultiplets, the category turns out to have a description in algebraic geometry as
\be \CC_A = \text{D-mod}(R(\CK))\,, \label{cat-free} \ee
namely, the derived category of D-modules on the algebraic loop space $R(\CK)$. The physics and mathematics of this category (and its gauge-theory analogues) will be explored in \cite{lineops}. For now, we just observe that holomorphic Lagrangians $\CL_0 \subset T^*R(\CK)$, such as \eqref{L0k}, naturally correspond to objects in $\CC_A$. The Lagrangian is the micro-local support of a particular D-module.

\subsection{Adding gauge interactions}
\label{sec:VL-gauge}

We would like to extend the characterizations of vortex lines in theories of free hypermultiplets (Section \ref{sec:VL-matter}) to gauge theories. As before, we expect to have several different but highly overlapping  descriptions of vortex lines, as
\begin{itemize} \setlength{\itemsep}{-.1cm}
\item[1)\,] singular solutions to the physical BPS equations \eqref{A-eqs} 
\item[1')] singularities in holomorphic or algebraic data, such as \eqref{M-hol}
\item[2)\,] coupled 3d-1d systems
\item[3)\,] objects of a geometrically defined category.
\end{itemize}

In the case of free hypermultiplets, there was no distinction between (1) and (1'), since the BPS equations were automatically holomorphic. This is no longer true of gauge theories. We saw in Section \ref{sec:hol} that, in gauge theory, rewriting the BPS equations in terms of holomorphic data amounts to replacing an infinite-dimensional symplectic quotient by an infinite-dimensional holomorphic quotient. The precise relation can be quite subtle.

Nevertheless, there are some natural physical expectations for how the correspondence should work. The most practical approach (which we will follow, motivated by \cite{GW-surface}) is to use a quantum-mechanics description (2) of a given line operator as a link between real-analytic (1) and holomorphic (1') regimes.
In this section we will build up our intuition with several important classes of examples, and then combine them to describe a general class of A-type line operators in gauge theories in Section \ref{sec:VL-GR}.

\subsubsection{Trivial line}
\label{sec:VL-triv}

In gauge theory with any $G$ and $R$, a canonical example of an A-type line operator is given by the trivial line  $\id$.

As a 3d-1d coupled system, we would say that $\id$ is defined by doing nothing: coupling the bulk 3d theory to the trivial 1d quantum mechanics with Hilbert space $\C$.

The $\SQM_A$ BPS equations are just the standard ones \eqref{A-eqs} in the bulk. In the presence of the trivial line, they must have ordinary, nonsingular solutions. In particular, near $z=0$ the hypermultiplet fields are nonsingular and the gauge group is unbroken.

It is useful to give a holomorphic characterization of the trivial line, at least for purposes of establishing some notation. Since the hypermultiplet scalars are nonsingular, they belong to the subspace
\be (X,Y) \in \CL_0 =  R(\CO)\oplus R^*(\CO) \,\subset\, R(\CK)\oplus R^*(\CK)\,, \ee
in the algebraic notation of Section \ref{sec:VL-alg}. Moreover, in an infinitesimal neighborhood of the origin, the group of complexified, holomorphic gauge transformations (preserved in holomorphic gauge) is
\be \CG_0 = G(\CO)\,, \ee
where
\be G(\CO) := \{\text{the algebraic group $G_\C$ defined over formal Taylor series $\CO$}\}\,  \label{def-GO} \ee
is an algebraic version of the positive loop group. In the case of $G=U(n)$, the group $G(\CO)$ simply consists of invertible $n\times n$ matrices whose entries are formal Taylor series in $z$.

Note that the algebraic group $G(\CO)$ acts naturally on $R(\CK)\oplus R^*(\CK) \simeq T^*R(\CK)$ (multiplying a Taylor-series entry of some $g(z)\in G(\CO)$ with a formal Laurent series in $T^*R(\CK)$ gives another formal Laurent series). Moreover, $G(\CO)$ preserves the Lagrangian subspace $\CL_0 \subset T^*R(\CK)$. Altogether, the trivial line is associated to the holomorphic data
\be \id\,:\quad \CL_0=T^*R(\CO)\,,\;\CG_0=G(\CO)\,,\qquad \text{with $\CG_0$ preserving $\CL_0$}\,. \ee

\subsubsection{Abelian vortex lines and screening}
\label{sec:VL-abel}

Consider $G=U(1)$ gauge theory with a single hypermultiplet $(X,Y)\in T^*\C$, where $X,Y$ have charges $+1,-1$.

Working with holomorphic data, we can try to define a vortex line the same way as in Section \ref{sec:VL-matter}: we allow $X$ to have a pole of order $k$ near $z=0$, and dually require $Y$ to have a zero of order $k$, \emph{i.e.}
\be X \in z^{-k}\CO\,, \qquad Y \in z^k\CO\,.  \label{Xeg-U1} \ee
(Note that the holomorphic-Lagrangian constraint of Section \ref{sec:VL-matter} must still be satisfied.) More succinctly, $(X,Y) \in \CL_0 = z^{-k}\CO \oplus  z^k\CO\,.$ This sort of singularity in the hypermultiplets does not require any breaking of gauge symmetry; we can still have full, nonsingular, holomorphic gauge transformations near the origin,
\be \CG_0 = \CG(\CO)  =  \{a+z\C[\![z]\!]\,,\;a\neq 0\}\,.  \label{Geg-U1} \ee

The vortex lines defined by \eqref{Xeg-U1}--\eqref{Geg-U1} can actually be screened, by dynamical vortex particles. (This was discussed from a physical, analytic perspective in \cite{HKT}.) From a holomorphic perspective, we can act with a gauge transformation $g(z)=z^k$, which is well defined in a formal \emph{punctured} neighborhood of $z=0$, to make $X,Y$ nonsingular:
\be  g(z) = z^k\,: \quad  z^{-k}\CO \oplus  z^k\CO\,\mapsto\, \CO\oplus \CO\,. \label{g-sing-U1} \ee
Physically, \eqref{g-sing-U1} corresponds to a ``large'' gauge transformation in the complement of the line operator, \emph{i.e.} on $\C_z^*$. Line operators related by such gauge transformations are physically equivalent; here we find that the vortex line \eqref{Xeg-U1}--\eqref{Geg-U1} is equivalent to the trivial line $\id$.

In order to define nontrivial vortex lines in $G=U(1)$ gauge theory, we must add more hypermultiplets. Thus, let us now consider $N$ fundamental hypers $(X^i,Y_i)_{i=1}^N\in T^*\C^N$, where the charges of $X^i,Y_i$ are all $+1,-1$ as before.

Choosing a vector of integers $\k= (k_1,...,k_n)\in \Z^N$, we define a putative vortex line in terms of the holomorphic data
\be \label{XY-sing-gen0}  \V_\k\,:\quad (X^i,Y_i) \in \CL_0 = \bigoplus_{i=1}^N z^{-k_i}\CO\oplus z^{k_i}\CO\,, \qquad \CG_0 = G(\CO)\,. \ee
In other words, we allow each $X^i$ to have a pole of order $k_i$ and require that $Y_i$ have a zero of order $k_i$ (or vice versa when $k_i<0$); and we again leave the gauge group unmodified.

Now these vortex lines are only partly screened. A singular gauge transformation $g=z^m$ (for $m\in \Z$) can be used to shift all integers $k_i$ simultaneously, but not individually. Thus there are equivalences of vortex lines
\be \V_\k \sim \V_{\k'}\quad\text{if}\quad \k-\k'= m(1,...,1) \;\;\text{for}\;\;m \in \Z\,. \ee
Physical vortex charge becomes an element of the quotient lattice $\k \in \Z^N/\Z$. \medskip

Let us also explain how to engineer these vortex lines by coupling to quantum mechanics, providing a more physical definition from which one can recover the holomorphic data above. For simplicity, we focus on $N=1$ hypermultiplets and ignore screening.

To obtain the vortex line \eqref{Xeg-U1} with $k=1$, we follow the same procedure as for free matter. Namely, we  introduce a 1d chiral multiplet $q$ of gauge charge $+1$, and a superpotential coupling $qY\big|_{z=0}$. The total superpotential, in 1d $\CN=4$ terms, becomes
\be W  = \int d^2z \big[ - YD_{\bar z}X + q Y \delta^{(2)}(z,\bar z)\big]\,, \ee
generalizing \eqref{Wflip}.%
\footnote{We have used an integration by parts to replace $XD_{\bar z}Y\leadsto -YD_{\bar z}X$, which is more convenient for introducing singularities in $X$ (as opposed to $Y$).}
The F-term for $Y$ sets $D_{\bar z}X=q\,\delta^{(2)}$. After complexifying the gauge group and passing to a holomorphic gauge with $A_{\bar z}=0$, this implies $X = \frac{q}{z}+$(regular, holomorphic), in other words $X\in z^{-1}\CO$ near $z=0$. Dually, the F-term for $q$ sets $Y\big|_{z=0}=0$, and the F-term for $X$ sets $D_{\bar z}Y=0$; after passing to holomorphic gauge, these together imply $Y\in z\,\CO$ near $z=0$.

The generalization to higher $k$ gets more interesting. Suppose $k=2$. To get $X\in z^{-2}\CO$, we introduce a pair of 1d chirals $q_1,q_2$, and a higher-derivative coupling
\be W  = \int d^2z \big[ - YD_{\bar z}X + \big(q_1 Y + q_2\pd_z Y) \delta^{(2)}(z,\bar z)\big]\,. \label{q12gauge} \ee
Note that the covariant $D_z$ derivative cannot enter $W$, because $A_z$ is not a chiral field. One may therefore be worried about gauge invariance. It turns out that \eqref{q12gauge} \emph{can} be made invariant under the group of real (physical) gauge transformations $g(z,\bar z)\in U(1)$ near the origin of $\C_z$, if we give $q_1$ and $q_2$ a transformation rule
\be q_1 \to g\big|_0 q_1 + \pd_z g\big|_0 q_2\,,\qquad q_2 \to g\big|_0 q_2\,, \label{q12trans} \ee
for $g(z,\bar z)\in U(1)$. (Here $\big|_0$ is shorthand for evaluation at $z=\bar z=0$.)

To recover the holomorphic data from \eqref{q12gauge} we note that in holomorphic gauge the F-terms $\delta W/\delta Y=0$ and $\delta W/\delta q_i=0$ set
\be X = \frac{q_2}{z^2}+\frac{q_1}{z}+\text{regular}\;\in\; z^{-2}\CO\,,\qquad  Y\big|_0=\pd_zY\big|_0=0 \quad\Rightarrow\quad Y\in z^2\CO\,,\ee
as desired. Moreover, the gauge transformation \eqref{q12trans} of $q_1,q_2$ is just right to ensure that, in holomorphic gauge, the polar terms in $X$ transform as expected:
\be X(z)\to g(z)X(z) \quad\text{with}\; g(z) = g\big|_0+ z \pd_z g\big|_0+... \ee

The pattern is now clear. For any $k>0$, we may introduce 1d chirals $q_1,...,q_k$ with
\be W  = \int d^2z \big[ - YD_{\bar z}X + \big(q_1 Y + q_2\pd_z Y + ...+q_k\pd_z^{k-1} Y) \delta^{(2)}(z,\bar z)\big]\,. \ee
This is gauge-invariant if the $(q_1,...,q_k)$ are given an appropriate linear gauge transformation that involves the first $k-1$ derivatives of $g$ at $z=\bar z=0$. In holomorphic gauge, the F-terms will restrict $Y\in z^k\CO$, and allow $X\in z^{-k}\CO$ as desired.

Similarly, for a $U(1)$ gauge theory with $N\geq 1$ hypermultiplets, we can engineer the vortex-line operators $\V_\k$ from \eqref{XY-sing-gen0} by coupling to a collection of $|k_1|+|k_2|+...|k_N|$ 1d chiral multiplets, and using them to ``flip'' the required modes from $X$ to $Y$ or vice versa. The gauge group near the origin will remain unmodified (in other words, $\CG_0=G(\CO)$) as long as the 1d chirals are given an appropriate gauge transformations, involving derivatives of $g$.

\subsubsection{Pure gauge theory}
\label{sec:VL-pureG}

Next, we recall (and generalize) ways to define an A-type line operator in terms of gauge-symmetry breaking. To avoid additional constraints related to hypermultiplets, we focus on pure gauge theory (meaning general $G$ and $R = 0$). The main interesting examples require $G$ to be nonabelian.

A class of line operators associated to gauge-symmetry breaking that is now quite standard was introduced in \cite{GW-surface} and generalized (as surface operators) in \cite{AGGTV,DGH}. An operator in this class is characterized by choosing a Levi subgroup $\mathbb L \subset G$, which becomes the unbroken physical gauge group at $z=0$. In additional, there are some continuous parameters involved. For A-type line operators in a 3d $\CN=4$ theory, a relevant continuous parameter is the holonomy $\alpha$ of the gauge connection around an infinitesimal loop linking the line operator. This holonomy must be $\mathbb L$-invariant, and can be conjugated to take values in the real torus $T$ of $G$ (modulo the Weyl group of $\mathbb L$). Unlike the case of surface operators in 4d $\CN=2$ theories, the parameter $\alpha$ does \emph{not} get complexified.

In holomorphic terms, the data $(\mathbb L,\alpha)$ gets replaced by a single parabolic subgroup ${P\subset G_\C}$. The subgroup $P$ is a minimal parabolic subgroup containing $\mathbb L$, and is the subgroup of $G_\C$ preserved by the line operator after passing to holomorphic gauge.
As explained carefully in \cite{GW-surface}, the continuous parameter $\alpha$ determines \emph{which} $P$ to take.  
In general, there are finitely many discrete choices of $P$'s, corresponding to finitely many chambers in which $\alpha$ can lie. For example, if $G=U(2)$ and $\mathbb L = T= U(1)^2$, the generic holonomy is
\be \alpha = \bp  \alpha_1 & 0 \\ 0 & \alpha_2 \ep \in \mathfrak t/\Lambda_{\rm cochar}\simeq T\,. \ee
There are two possible parabolic subgroups containing $T$, namely the lower and upper Borels
\be  P = B = \bp *&* \\ 0&* \ep\,,\quad\text{or}\quad P = B_- = \bp * & 0 \\ * & * \ep \qquad (B,B_-\subset GL(2,\C))\,.\label{U2B} \ee
If $\alpha_i$ are small and $\alpha_1>\alpha_2$ then the holomorphic data contains $P=B$; whereas if $\alpha_2>\alpha_1$ the holomorphic data contains $P = B_-$.

Notably, most of the information in $\alpha$ gets lost in the translation to holomorphic data. In later sections, we will calculate spaces of local operators bound to A-type lines by taking $Q_A$-cohomology of certain moduli spaces of solutions to BPS equations. These calculations depend only on the holomorphic data. Stated more generally, the A-twist of 3d $\CN=4$ gauge theory is locally insensitive to real continuous parameters.

We may reformulate and generalize the holomorphic data in algebraic terms. The group of holomorphic gauge transformations in an infinitesimal neighborhood of the origin is $G(\CO)$, as in \eqref{def-GO}. Breaking $G_\C$ to a parabolic $P$ right at the origin $z=0$ means that, in an infinitesimal neighborhood, we break $G(\CO)$ to
\be \CI_{P} = \{g(z)\in G(\CO)\;\text{s.t.}\; g(0)\in P\} \label{def-Iwahori} \ee
This is called a ``parahoric'' subgroup of $G(\CO)$. When $P=B$ is a Borel, then $\CI_B$ is called an ``Iwahori'' subgroup. For example, if $G=U(2)$ and $P=B$ as in \eqref{U2B}, we have
\be \CI_B = \left\{ g(z)\in GL(2,\CO)\;\text{s.t.}\; g(z)= \bp a(z) & b(z) \\ z\,c(z) & d(z) \ep \right\}\,. \label{def-IB} \ee

So far, \eqref{def-Iwahori} describes a ``zeroth-order'' breaking of gauge symmetry on the support of a line operator. We would also like to consider higher-order symmetry breaking, in a neighborhood of $z=0$. In algebraic terms, this is easily characterized by choosing a general subgroup
\be \CG_0 \subseteq G(\CO) \label{G0GO} \ee
to remain unbroken.%
\footnote{\label{foot:fin}In this paper, we will only consider symmetry breaking up to some finite order around $z=0$, which means that $\CG_0$ has finite codimension inside $G(\CO)$. In principle one could consider subgroups of infinite codimension as well.
Choices of $\CG_0$ with infinite codimension are relevant for line-like operators constructed by wrapping boundary conditions on a circle, and will be discussed further in \cite{lineops}.

One may generalize in yet another direction, and choose the group $\CG_0$ of holomorphic gauge transformations near the origin to be a subgroup of the full algebraic loop group $G(\CK)$, rather than a subgroup of $G(\CO)$. This is possible because, once the origin is excised from the plane $\C_z$, all ``singular gauge transformations'' in $G(\CK)$ become available. We will not need such choices in this paper, but they will be part of the general categorical setup of \cite{lineops}.
}
For example, if $G=U(2)$ we could take any
\be \CG_0 = \CI_B^k :=  \left\{ g(z)\in GL(2,\CO)\;\text{s.t.}\; g(z)= \bp a(z) & b(z) \\ z^k\,c(z) & d(z) \ep \right\}\,, \quad k\geq 0 \,. \label{IBk} \ee
As in the case of zeroth-order symmetry breaking, the algebra/holomorphic data \eqref{G0GO} should be supplemented by additional real parameters, when describing a breaking of the real gauge group $G$ and a singularity of the real physical fields. For example, there may be higher-order poles in the real gauge connection. Such parameters were discussed in \cite{Witten-wild} in the context of wild ramification. We will not need them for computations in the A-twist.

Finally, we recall from \cite{GW-surface} that line operators characterized by a breaking of gauge symmetry have a natural construction by coupling to quantum mechanics. For zeroth order breaking, we may construct the line operator labeled by $(\mathbb L,\alpha)$ --- or holomorphically by $P$ --- by introducing a 1d $\SQM_A$ SQM sigma-model with K\"ahler target
\be \CX = G/\mathbb L \simeq G_\C / P\,. \label{coset-fin} \ee
The space $\CX$ is a homogeneous $G$-space, with a left $G$-action that manifests as a flavor symmetry in the 1d quantum mechanics.
This 1d theory is coupled to the 3d $\CN=4$ bulk by gauging the $G$ flavor symmetry with the bulk gauge symmetry. Since the stabilizer of any point of $\CX$ is (conjugate to a copy of) $\mathbb L$, the effect is to break $G$ to $\mathbb L$.

The continuous parameters $\alpha$ enter as real K\"ahler parameters in the 1d sigma-model to $\CX$. This makes it quite clear that the A-twist (whose supercharge acts as de Rham differential in 1d) will be locally insensitive to them. Many examples of line operators of this type are discussed in \cite{AsselGomis}, by realizing the coset space $\CX$ as a 1d gauged linear sigma model (GLSM). In the 1d GLSM's, the parameters $\alpha$ entered as real FI parameters.

More generally, we expect to be able to realize a line operator with holomorphic data $\CG_0$ by coupling to a 1d sigma-model with target
\be \CX = G(\CO)/\CG_0\,. \label{coset-inf} \ee
Coupling to the 3d bulk is again done by gauging the 1d flavor symmetry. In this case, however, the flavor symmetry group is $\CG(\CO)$, acting by left multiplication on $\CX$; or, in real/physical terms, the symmetry group is the group $\CG$ of gauge transformations on the disc that appeared in Section \ref{sec:QMaction}. Though it may look exotic, gauging this infinite-dimensional flavor symmetry is a perfectly reasonable operation! As discussed in Section \ref{sec:QMaction}, when we rewrite the 3d $\CN=4$ bulk theory as 1d $\SQM_A$ quantum mechanics, the decomposition of the bulk $G$ gauge multiplet contains a 1d vectormultiplet for the infinite-dimensional gauge group $\CG$. This 1d $\CG$ vectormultiplet can be used canonically to gauge the $\CG$ flavor symmetry of quantum-mechanics with target $\CX$.

As mentioned briefly in Footnote \ref{foot:fin}, we will only consider symmetry breaking up to some finite order around $z=0$, which implies that $\CX = G(\CO)/\CG_0$ is a finite-dimensional space. Thus, \emph{most} of the infinite-dimensional flavor symmetry group $G(\CO)$ (or $\CG$ in the real case) acts trivially on $\CX$. In turn, the coupling between 1d and 3d theories induced by gauging will only involve a finite number of derivatives. For example, if we chose $\CG_0 = \CI_P$ as in \eqref{def-Iwahori}, we would find
\be \CX = G(\CO)/\CI_P = G_\C/P\,,  \ee
and recover the well-known setup \eqref{coset-fin}, and a coupling with no derivatives at all.

\subsection{General A-type line operators}
\label{sec:VL-GR}

For a 3d $\CN=4$ theory with general gauge group $G$ and hypermultiplet representation $T^*R$, we may combine the various ingredients described above to define vortex-line operators.

In terms of holomorphic/algebraic data, we characterize a vortex line by choosing
\begin{itemize}
\item[1)] a holomorphic Lagrangian subspace $\CL_0 \subset T^*R(\CK)$, encoding the meromorphic singularity in the hypermultiplet scalars
\item[2)] a subgroup $\CG_0\subseteq G(\CO)$ of the group of holomorphic gauge transformations in an infinitesimal neighborhood of $z=0$, encoding the breaking of gauge symmetry.
\end{itemize}
These two choices must be compatible, in the sense that $\CG_0$ must preserve $\CL_0$. Moreover, as we saw in Section \ref{sec:VL-abel}, there are redundancies in this data, as some vortex-line operators can be related by ``screening.'' In algebraic terms, two pairs of data $(\CL_0,\CG_0)$ and $(\CL_0',\CG_0')$ are physically equivalent if there exists an element $g(z) \in G(\CK)$ such that
\be \hspace{-1in}\text{screening equivalence}\,:\quad (g\cdot \CL_0\,,\;g\,\CG_0\,g^{-1}) = (\CL_0',\CG_0')\,. \ee
Here
\be G(\CK) := \{\text{the algebraic group $G_\C$ defined over formal Laurent series $\CK$}\}\,  \label{def-GK} \ee
is the group of algebraic%
\footnote{Naively, one may want to consider here the group of \emph{holomorphic} gauge transformations in an infinitesimal punctured neighborhood of $z=0$. However, there is now a big difference between holomorphic and algebraic: the former contain gauge transformations with essential singularities, whereas the latter only contain meromorphic gauge transformations. We refer the reader to a careful discussion in \cite{Witten-wild} on how to interpret the distinction physically, and why a restriction to algebraic gauge transformations is sensible.} %
gauge transformations in an infinitesimal punctured neighborhood of $z=0$. Informally, elements of $G(\CK)$ are often called singular gauge transformations.

When defining an A-type line operator in the full, physical 3d $\CN=4$ theory, this data should be accompanied by additional real parameters, associated to a $\CG_0$-invariant singularity in the holomorphic connection $A_z$. We will not need them for analyses in the A-twist. In the quantum-mechanics definition of vortex-line operators (further below), the real parameters are K\"ahler parameters of $G(\CO)/\CG_0$.

\subsubsection{Example: U(2) with matter}

Let us give a simple example of line operators in the general class above, in the case of nonabelian gauge theory with matter. We take $G=U(2)$ and $R=\C^2$ the fundamental representation.

Since $G_\C=GL(2,\C)$, the group of holomorphic gauge transformations in an infinitesimal neighborhood of $z=0$ is
\be G(\CO) = \left\{g(z)=\bp a(z) & b(z) \\ c(z) & d(z) \ep\text{s.t.}\; a,b,c,d\in \CO\,,\; \det g\big|_{z=0}\neq 0 \right\}\,. \ee
Suppose that we require the hypermultiplets $X = \bsp X^1\\ X^2 \esp$ and $Y =  \bsp Y_1 \\ Y_2 \esp^T$ to take the form
\be (X,Y) \,\in\, \CL_0 = \bp z^{-k_1}\CO \\ z^{-k_2}\CO \ep \oplus \bp z^{k_1}\CO \\ z^{k_2}\CO  \ep^T\,. \label{sing-U2}  \ee
for some $k_1,k_2\geq 0$.

If $k_1=k_2$, the holomorphic Lagrangian subspace $\CL_0$ is preserved by the full $G(\CO)$ gauge symmetry, so we may simply choose $\CG_0=G(\CO)$ to define a vortex-line operator.

 If $k_1\neq k_2$, the gauge group must be broken.
A simple case is $(k_1,k_2)=(1,0)$. Then we are looking at $X^1\in z^{-1}\CO$, $X^2\in \CO$.
The largest subgroup of $G(\CO)$ that preserves this singularity is the standard Iwahori $\CI_B$ from \eqref{def-IB}, containing elements of the form
\be g(z) = \bp a(z) & b(z) \\ z\,c(z) & d(z) \ep\,,\quad a,b,c,d\in \CO\,. \label{I-U2} \ee
Then we can choose $\CG_0 = \CI_B$ together with \eqref{sing-U2} to define a vortex line.

Many other interesting options are possible. For example, if $k_1>k_2$, a maximal subgroup of $G(\CO)$ that preserves the meromorphic singularity $X^1\in z^{-k_1}\CO,\, X^2\in z^{-k_2}\CO$ is the ``higher'' Iwahori subgroup $\CI_B^{k_1-k_2}$ from \eqref{IBk}, containing elements of the form 
\be g(z)= \bp a(z) & b(z) \\ z^{k_1-k_2}c(z) & d(z) \ep\,. \ee
For fixed $k_1\geq k_2$, we can define a vortex-line operator by supplementing \eqref{sing-U2} with $\CG_0=\CI_B^k$ for any $k\geq k_1-k_2$.

\subsubsection{Coupling to quantum mechanics}
\label{sec:VL-genQM}

The vortex-line operators characterized by holomorphic data $\CL_0$ and $\CG_0$ can be systematically engineered by coupling the 3d $\CN=4$ theory to a 1d $\CN=4$ sigma-model (with multiplets of ``1d $\CN=(2,2)$'' type). The procedure for doing so combines the quantum-mechanics construction of singularities in free-matter theories (Section \ref{sec:flip}) and in pure gauge theories (Section \ref{sec:VL-pureG}).

Many examples of this construction are known in the literature, usually involving 1d GLSM's and brane constructions (\emph{e.g.} many appear in \cite{AsselGomis}). Here we give a general geometric description.

We consider general $G$ and $R$, but assume for simplicity that $\CL_0 \subset R(\CK)\oplus R^*(\CK)$ is a subspace of the form
\be \CL_0\simeq (z^{-k_1}\CO,z^{-k_2}\CO,...,z^{-k_N}\CO)^T\oplus (z^{k_1}\CO,z^{k_2}\CO,...,z^{k_N}\CO)\,, \label{L0QM} \ee
for some integers $\k=(k_1,...,k_N)$.

Let us ignore the gauge group for the moment. We learned in Section \ref{sec:flip} that the singularity \eqref{L0QM} can be engineered by introducing  $|\k| := |k_1|+|k_2|+...+|k_N|$ 1d chiral multiplets $q_i$, and a superpotential
\be W = \int d^2z \,XD_{\bar z}Y + W_0(q;X,\pd_zX,...;Y,\pd_z Y,...)\big|_{z=\bar z=0}\,, \label{WW0} \ee
where $W_0$ contains quadratic couplings between the $q$'s and appropriate $\pd_z$ derivatives of the bulk hypermultiplets $X$ and $Y$. These quadratic couplings effectively ``flip'' non-negative modes of $X$ into negative modes of $Y$ and vice versa, to recover $\CL_0$.

Formally, we may think of $W_0\big|_{z=\bar z=0}$ as a function
\be W_0\big|_{z=\bar z=0}\,: V\times (R(\CO)\oplus R^*(\CO)) \to \C\,, \ee
where $R(\CO)\oplus R^*(\CO)$ is parameterized by $\pd_z$ derivatives of $X$ and $Y$, and
\be V = \bigoplus_{i=1}^N \begin{cases} z^{-k_i}\C & k_i > 0 \\
  z^{k_i}\C & k_i < 0 \end{cases} \quad \simeq\; \C^{|k_1|+...+|k_N|} \ee
is the finite-dimensional vector space parameterized by the $q$'s.

Now, the fact that $\CL_0$ is only invariant under $\CG_0$ rather than all of $G(\CO)$ means that the superpotential $W_0\big|_{z=\bar z=0}$ is invariant only under $\CG_0$.
Thus, a coupled 3d-1d system with total superpotential \eqref{WW0} only makes sense if we break gauge symmetry explicitly near $z=0$. We would rather like to break gauge symmetry through a coupling to a 1d sigma-model.

In pure gauge theory, we broke gauge symmetry by coupling to the coset space $\CX = G(\CO)/\CG_0$. In the presence of matter, we enhance this construction as follows. The vector space $V$ is a finite-dimensional representation of group $\CG_0$.%
\footnote{Explicitly, the 1d chirals $q_i$ discussed above correspond to the negative modes appearing in $\CL_0$. They transform linearly under an element $g(z)\in \CG_0$, in a way that depends on $g$ and its $\pd_z$ derivatives at $z=0$. See, for example, \eqref{q12trans}.} %
It can therefore be used to define a holomorphic, homogeneous, associated vector bundle $\CE$ over $\CX$,
\be \CE = (\CG(\CO)\times V)/\CG_0\,, \ee
whose points are pairs $(g,q)\in \CG(\CO)\times V$ modulo the equivalence relation
\be (gh,q) \sim (g,hq) \quad \forall\,h\in \CG_0\,. \ee
The map $\CE\to \CX$ just forgets $q$; so all fibers of $\CE$ are isomorphic to $V$. The $G(\CO)$ action on $\CX$ lifts to the total space of the bundle, with an element $g'\in G(\CO)$ sending $(g,q)\mapsto (g'g,q)$.

In order to engineer our desired vortex line by coupling to quantum mechanics in a gauge-invariant way, we introduce a 1d $\CN=4$ sigma-model whose target is the total space of $\CE$. We couple to the 3d bulk theory (also rewritten as a 1d $\CN=4$ theory) by
\begin{itemize}
\item Gauging the flavor symmetry of the sigma-model with the bulk gauge symmetry (exactly as in \eqref{coset-inf}).
\item Introducing a $G(\CO)$-invariant superpotential $\int d^2z\, XD_{\bar z}Y+\wt W_0$, where $\wt W_0: \CE\times (R(\CO)\oplus R^*(\CO))\to \C$ is defined by
\be \wt W_0\big( (g,q);X;Y)\big) = W_0(q;g^{-1}\cdot X;g^{-1}\cdot Y)\big|_{z=\bar z=0}\,. \label{tW0} \ee
\end{itemize}
Here on the RHS we suppressed potential $\pd_z$ derivatives of $X$ and $Y$ in order to simplify the notation. We also schematically write $g^{-1}\cdot X$, $g^{-1}\cdot Y$ to denote the action of $g(z)^{-1}\in G(\CO)$ on $X$ and $Y$.

To check that $\wt W_0$ is well defined on the quotient space $\CE$, note that $\wt W_0((gh^{-1},hq);X;Y) = W_0(hq;hg^{-1}X;hg^{-1}Y)\big|_{z=\bar z=0}=W_0(q;g^{-1}\cdot X;g^{-1}\cdot Y)\big|_{z=\bar z=0} = \wt W_0((g,q);X;Y)$ due to $\CG_0$-invariance of $W_0$. Moreover, $\wt W_0$ is invariant under the left action of $G(\CO)$, since $\wt W_0((g'g,q);g'\cdot X;g'\cdot Y) = W_0(q;g^{-1}g'{}^{-1}g'\cdot X;...)\big|_{z=\bar z=0} = W_0(q;g^{-1}\cdot X,g^{-1}\cdot Y)\big|_{z=\bar z=0} = \wt W_0((g,q);X;Y)$\,. 

Finally, we emphasize that the gauge-fixed form of \eqref{tW0} looks just like the simpler \eqref{WW0}. In holomorphic terms, we use the bulk $G(\CO)$ action to bring any point $(g,q)\in \CE$ to $(1,q)$. The stabilizer of $(1,q)$ is $\CG_0$, and the superpotential over this point is manifestly $\wt W_0((1,q);X;Y)=W_0(q;X;Y)\big|_{z=\bar z=0}$.
From \eqref{WW0}, we recover the original Lagrangian $\CL_0$.

\subsubsection{Category}

In \cite{lineops}, we will propose that the category of line operators in the A-twist of a 3d $\CN=4$ gauge theory is
\be \CC_A = \text{D-mod}_{G(\CK)}(R(\CK))\,.  \label{cat-gauge}\ee
This is the derived category of D-modules on the loop space $R(\CK)$, equivariant for the loop group $G(\CK)$; it generalizes \eqref{cat-free} to gauge theories. It turns out that half-BPS A-type line operators characterized by the algebraic data $(\CL_0,\CG_0)$ naturally define objects in \eqref{cat-gauge}. There are much more general objects in \eqref{cat-gauge} as well, which will be explored in \cite{lineops}. 

A version of the category \eqref{cat-gauge} recently appeared in work of Costello-Creutzig-Gaiotto on chiral boundary conditions for 3d $\CN=4$ theories \cite{CCG}. There, it was the category of modules for a boundary VOA. These modules are naturally associated to bulk line operators that end on the boundary, much as in the classic relation between 3d Chern-Simons and WZW \cite{Witten-Jones, EMSS}.

\subsubsection{Mass parameters and quantization}
\label{sec:masses1}

As we shall see in the examples below, it is possible to deform the above vortex lines by turning on complex masses and/or an Omega background. After rewriting our 3d $\CN = 4$ gauge theories as 1d $\SQM_A$ quantum mechanics, both of these deformations are interpreted as turning on twisted masses for flavor symmetries, \emph{cf.} \cite[Sec 2.5]{VV}. In particular, rotations of the $\C_z$ plane, which are involved in the Omega background, simply become symmetries of the target space of the quantum mechanics.

Such twisted masses do not affect the vortex-line operators \emph{per se}. Rather, they deform the spaces of local operators at junctions of vortex lines, and the algebraic structure of local operators coming from collision. Local operators will be the subject of the next section. We shall see, just like in \cite{VV}, that complex masses and the Omega background deform cohomology to equivariant cohomology in various constructions.


It is also worth noting that, in the presence of an Omega background, turning on quantized mass parameters $m_\C = \lambda \varepsilon$ (where $\lambda$ is an integral cocharacter of the 3d Higgs-branch flavor symmetry $F$) is equivalent to introducing a flavor vortex for a subgroup $U(1)_\lambda \subseteq F.$ This is mirror to the phenomenon mentioned at the end of Section \ref{sec:Wilson-Omega}, relating abelian Wilson lines to quantized FI parameters. See \cite{BDGH} for further discussion.

\section{Junctions of vortex lines}
\label{sec:comp}

Given a pair $\CL,\CL'$ of half-BPS A-type line operators in a 3d $\CN=4$ gauge theory, we would like to be able to compute the $Q_A$-cohomology of the space of local operators at their junction. 
In categorical terms, we seek $\text{Hom}_A(\CL,\CL')$. We would also like to find the OPE induced from collision of junctions, \emph{i.e.} composition of Homs.

As prefaced in the Introduction, junctions of A-type line operators are not nearly as easy to access as junctions of B-type line operators. Even the simplest case, where $\CL = \CL'=\id$ are both the trivial line, the algebra
\be \text{End}_A(\id) = \text{Hom}_A(\id,\id)  \supseteq \C[\CM_C] \ee
contains the Coulomb-branch chiral ring. The ring  $\C[\CM_C]$  includes monopole operators, whose OPE's famously receive perturbative and (in nonabelian gauge theories) nonperturbative quantum corrections, making them difficult to compute with a semi-classical approach.

Fortunately, the last few years have seen remarkable progress in developing exact, TQFT-based methods to compute the Coulomb-branch chiral ring, \emph{e.g.} \cite{CHZ-Hilbert, BDG, VV, Nak, BFNII, BFN-lines, HananySperling-fans, Web2016, DG-star, BPR-defq, Pufu-Coulomb, Pufu-bubbling, CCG}.
Many of these methods can be adapted to exact computations of local operators at junctions of more general vortex lines as well. This was already done in limited contexts in \cite{BDGH, BFN-lines, Web2016}.

In this paper, we adapt the approach of \cite{VV} to compute spaces $\text{Hom}_A(\CL,\CL')$ and their OPE. 
Physically, this requires choosing a half-BPS boundary condition $\CB$ and embedding line operators and their junctions in a solid cylinder, with $\CB$ wrapped on the outside.
We will also attempt to put the approach of \cite{VV} in a more algebraic and categorical framework, explaining in particular how a choice of boundary condition $\CB$ furnishes a \emph{representation of the category} of line operators.

We will quickly restrict our focus to boundary conditions $\CB$ labeled by massive vacua of a bulk 3d gauge theory, as was done in \cite{VV}. Such boundary conditions --- when available --- allow for \emph{relatively} simple computations of spaces of local operators. Even so, mathematically, we will need to employ equivariant intersection cohomology or Borel-Moore homology.
 Many other interesting boundary conditions can be studied. A particular class that directly generalizes the construction of Braverman-Finkelberg-Nakajima is discussed in Appendix \ref{app:BFN}.
 
The algebraic definitions of moduli spaces in this section --- in particular, their equivalence with analytic definitions, via a Kobayashi-Hitchin correspondence --- are almost all conjectural. 

\subsection{Cylinder setup}
\label{sec:cyl}

The basic idea of \cite{VV} was to ``probe'' bulk local operators with a boundary condition. Let us explain how this idea extends to line operators. 

Just like line operators, BPS boundary conditions for 3d $\CN=4$ theories are classified by the 2d SUSY subalgebras that they preserve. We are interested in half-BPS boundary conditions $\CB$ that preserve 2d $\CN=(2,2)$ SUSY and $U(1)_C\times U(1)_H$ R-symmetry. (These were studied in \cite{KRS} for 3d $\CN=4$ sigma-models, and in \cite{BDGH, ChungOkazaki} for gauge theories.) Such boundary conditions are compatible with both the A and B twists of the bulk.

\begin{figure}[htb]
\centering
\includegraphics[width=5in]{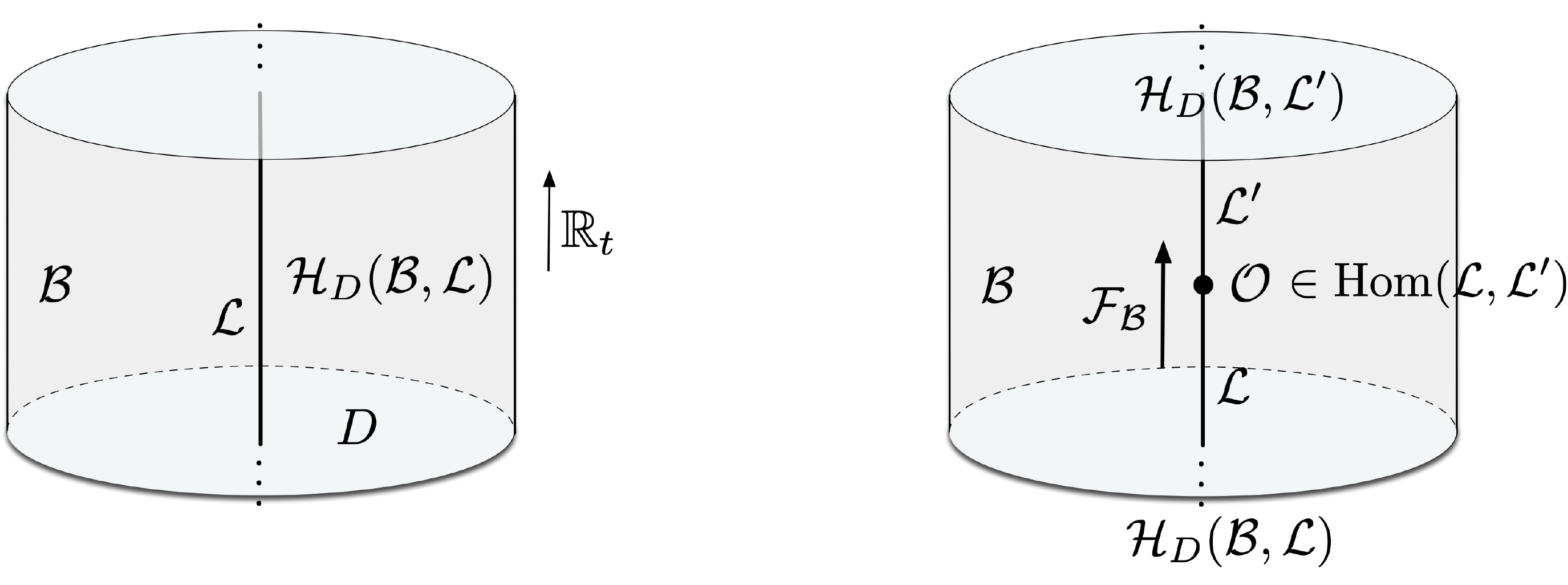}
\caption{Left: the SUSY Hilbert space $\CH_D(\CB,\CL)$ on a disc punctured by $\CL$, with boundary condition $\CB$. Right: using this setup to define a representation of the category of line operators on the vector spaces $\CH_D(\CB,\CL)$, for various $\CL$.}
\label{fig:cyl-rep}
\end{figure}

Given such a boundary condition $\CB$ and an A-type half-BPS line operator $\CL$, we can place the bulk 3d theory in a solid cylinder $D\times \R_t$, with $\CB$ wrapped around the boundary of the disc $D$, and $\CL$ supported at the origin of the disc (Figure \ref{fig:cyl-rep}). The physical Hilbert space $\text{Hilb}_D(\CB,\CL)$ on the disc has \vspace{-.1cm}
\begin{itemize} \setlength{\itemsep}{-.1cm}
\item an action of $Q_A$, since the entire setup preserves this supercharge; and
\item a $\Z$-valued cohomological grading given by charge under the $U(1)_C$ R-symmetry.
\end{itemize}
Taking $Q_A$-cohomology, we define the ``supersymmetric Hilbert space''
\be \CH_D(\CB,\CL) = H^\bullet_{Q_A}(\text{Hilb}_D(\CB,\CL))\,. \ee 

By using a state-operator correspondence, the Hilbert space $\CH_D(\CB,\CL)$ may also be interpreted as the $Q_A$-cohomology of the space of local operators, at a point where the line $\CL$ intersects the boundary $\CB$. This perspective shows that the space $\CH_D(\CB,\CL)$ will be empty (zero-dimensional) unless the $\CL$ can end on the boundary in a way that preserves $Q_A$.

For fixed $\CB$ and varying $\CL$, the spaces $\CH_D(\CB,\CL)$ give us a way to access information about the junctions of line operators. Specifically, the $Q_A$-cohomology of the space of local operators at a junction $\text{Hom}_A(\CL,\CL')$ acts on $\CH_D(\CB,\CL)$, by mapping it to $\CH_D(\CB,\CL')$, as on the right of Figure \ref{fig:cyl-rep}. In the special case that $\CL=\CL'$, we expect to find an action of the algebra $\text{End}_A(\CL) = \text{Hom}_A(\CL,\CL)$ on $\CH_D(\CB,\CL)$ alone.

More abstractly, given any $\CB$, the cylinder setup sends every object $\CL$ in the category of A-type line operators to the vector space $\CH_D(\CB,\CL)$, and sends any morphism $\CO\in \text{Hom}_A(\CL,\CL')$ to an ordinary linear map of corresponding vector spaces,
\be \label{repB} \hspace{-.5in} \text{probe with $\CB$}\,:\quad \begin{array}{ccc}
\CL & \overset{\CF_\CB}\longmapsto & \CH_D(\CB,\CL) \\[.1cm]
\text{Hom}_A(\CL,\CL') & \overset{\CF_\CB}{\longrightarrow} & \text{Hom}_\C\big(\CH_D(\CB,\CL),\CH_D(\CB,\CL')\big) \end{array}
\ee
The maps $\CF_\CB$ are functorial, meaning that they preserve composition of morphisms. Altogether, \eqref{repB} is a functor from the category $\CC_A$ of line operators to the category $\text{Vect}$ of vector spaces.%
\footnote{More precisely, one should think of $\CC_A$ as a dg-category, and $\CF_\CB$ as a functor the category of dg vector spaces. In this paper, we pass to cohomology, and the distinction will not be important.} %
This is usually called a representation of the category.

An important question is how close the maps in \eqref{repB} are to being isomorphisms. We can make a few general remarks.

A necessary condition for the functor $\CF_\CB$ to be faithful --- meaning that all $\CH_D(\CB,\CL)$ are nonzero and all the maps $\text{Hom}_A(\CL,\CL')\to \text{Hom}_\C\big(\CH_D(\CB,\CL),\CH_D(\CB,\CL')\big)$ are injective --- is that all lines $\CL$ can end on the boundary condition $\CB$, in a way that preserves $Q_A$. Otherwise, some $\CH_D(\CB,\CL)$ will clearly be zero. The vacuum boundary conditions that we use further below seem to have this property, at least for the sort of half-BPS line operators defined in Section \ref{sec:VL}. If a single boundary condition $\CB$ is \emph{not} sufficient to faithfully probe all the line operators of interest, then could try to analyze the maps \eqref{repB} for multiple boundary conditions at once.

The functor \eqref{repB} will almost never be full --- meaning that all maps $\CF_\CB:\text{Hom}_A(\CL,\CL')\to \text{Hom}_\C\big(\CH_D(\CB,\CL),\CH_D(\CB,\CL')\big)$ are surjective. Indeed, we would not want this! Linear transformations of vector spaces, $\text{Hom}_\C\big(\CH_D(\CB,\CL),\CH_D(\CB,\CL')\big)$  form a large, boring matrix algebra. The local operators $\text{Hom}_A(\CL,\CL')$ at junctions of lines should be embedded inside in an interesting way. 

Physically, $\text{Hom}_\C\big(\CH_D(\CB,\CL),\CH_D(\CB,\CL')\big)$ simply consists of all operators acting on the (supersymmetric) cylinder Hilbert space. This includes not only local operators at the junction of $\CL$ and $\CL'$, but \emph{e.g.} surface operators extended along $D$ (at a fixed time) and line operators that wrap the boundary $\pd D\times \{t_0\}$ at fixed time, and look like interfaces along~$\CB$. 
Below, we will attempt to characterize the image of $\CF_\CB:\text{Hom}_A(\CL,\CL')\to \text{Hom}_\C\big(\CH_D(\CB,\CL),\CH_D(\CB,\CL')\big)$ in a precise way that excludes these other non-local operators.

\subsection{Quantum mechanics and cohomologies}
\label{sec:QM-Hom}

The representation \eqref{repB} of Hom spaces, associated to a boundary condition $\CB$, is useful because the RHS tends to be vastly more computable than the LHS.

In \cite{VV}, a three-step procedure was proposed for computing $\CH_A(\CB,\id)$. It generalizes easily to any $\CH_A(\CB,\CL)$. The basic idea is to
\begin{itemize}
\item[1)] Rewrite the 3d $\CN=4$ theory on $D\times \R_t$ as a 1d $\SQM_A$ quantum mechanics on $\R_t$, as in Section \ref{sec:BPSA}. This 1d theory has an infinite-dimensional target, roughly consisting of maps from $D$ to the original 3d target, subject to appropriate boundary conditions near $0\in D$ (coming from $\CL$) and $\pd D$ (coming from $\CB$). Additional degrees of freedom supported on $\CL$ or $\CB$ may further enhance the target of this quantum mechanics.

\item[2)] Solve the BPS equations \eqref{A-eqs} along $D$ to localize the theory from (1) to an effective 1d $\SQM_A$ sigma-model with a vastly smaller target $\CM_D(\CB,\CL)$.

\item[3)] Compute the $Q_A$-cohomology of the Hilbert space of the effective 1d quantum mechanics (\emph{a.k.a.} the space of SUSY ground states) by taking  cohomology,
\be \CH_A(\CB,\CL) \simeq H^\bullet(\CM_D(\CB,\CL))\,. \label{H-coh} \ee

\end{itemize}

Note that the supercharge $Q_A$ acts as an ``A-type'', or ``de-Rham-type'' supercharge in the 1d $\SQM_A$ quantum mechanics that describes this system on a cylinder. Thus, just as in Witten's classic work  \cite{Witten-Morse}, the $Q_A$-cohomology of the full Hilbert space of states should be given by a form of de Rham cohomology. (We will comment further on the precise cohomology being used in Section \ref{sec:cohomologies} below.)  Step (2) is based on the premise that taking cohomology gives equivalent results before or after localizing to the solutions of BPS equations.

A nice simplification arises in this framework.
Cohomology is intrinsically topological, and cannot have local dependence on the K\"ahler structure of $\CM_D$ in Step (2). Thus we expect to be able to compute the SUSY Hilbert space \eqref{H-coh} by using an algebraic description of $\CM_D$.

\subsection{Local operators and convolution}
\label{sec:conv}

The action of local operators $\CO\in \text{Hom}_A(\CL,\CL')$ on disc Hilbert spaces acquires a natural description in terms of a \emph{convolution product} in cohomology. This again generalizes constructions of \cite{VV}. Earlier, such convolution products were used to define the OPE in the Coulomb-branch chiral ring defined by Braverman-Finkelberg-Nakajima \cite{Nak, BFNII} (see also \cite[App. A]{DG-star} for related discussion).

\subsubsection{Convolution in quantum mechanics}
\label{sec:convQM}

The convolution product is simply an implementation of a state-operator correspondence in A-type quantum mechanics. Let's briefly review this idea.
 To keep thing simple, consider A-type (de Rham type) $\CN=2$ quantum mechanics with a smooth, compact target $\CX$ and nilpotent supercharge $Q$.%
 \footnote{We are ultimately interested in analyzing a 1d $\CN=4$ theory. However, the relevant structure of convolution products shows up already for 1d $\CN=2$, which is why we consider a more general $\CN=2$ setup here. The fact that we actually have 1d $\CN=4$ SUSY leads to additional features, such as the ability to compute Hilbert spaces via fixed-point localization.} %
 The $Q$-cohomology of the Hilbert space is $\CH=H^\bullet(\CX)$. The state-operator correspondence in topological quantum mechanics says that the ($Q$-cohomology of the) space of local operators $\text{Ops}$ at a point is isomorphic to the Hilbert space on the sphere $S^0$ linking the point. Since $S^0$ is just two points (with opposite orientations), our theory on $S^0\times \R$ is just two non-interacting copies of the theory on $\R$,
\be \raisebox{-.3in}{\includegraphics[width=2in]{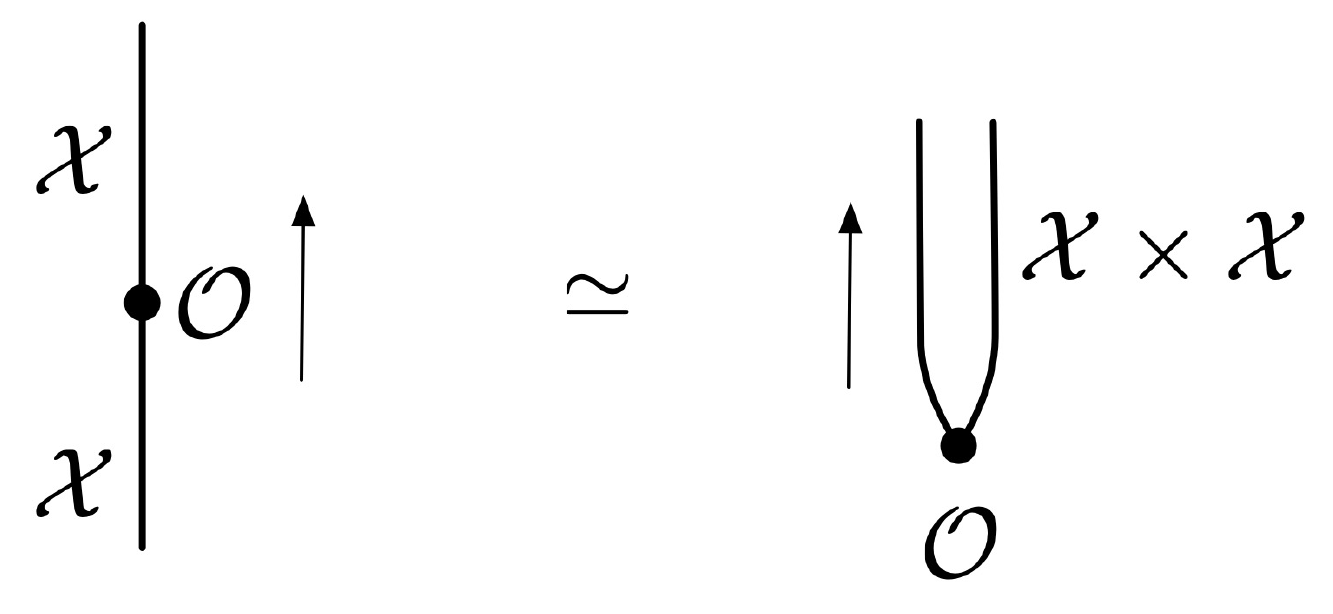}}
\ee
 In other words, it's quantum mechanics with target $\CX\times \CX$. We deduce that local operators~are
\be H^\bullet_Q(\text{Ops}) = H^\bullet(\CX\times \CX)  \,. \label{stateop-1d} \ee

From one perspective, this result is hardly surprising. Using the K\"unneth formula and Hodge duality, we have an isomorphism
\be \label{EndH} H^\bullet_Q(\text{Ops})  \simeq  H^\bullet(\CX)\otimes H^\bullet(\CX) \overset{\text{Hodge}}\simeq H^\bullet(\CX)\otimes H^\bullet(\CX)^* \, \simeq \text{End}_\C(H^\bullet(\CX))\,. \ee
This is just the full set of linear transformations acting on the complex vector space $H^\bullet(\CX)$. 

However, there is also an intrinsic geometric description of the OPE on $H^\bullet_Q(\text{Ops})$ and its action on $\CH$ that avoids (or rather, repackages) Hodge duality, and which we will generalize momentarily. To see the action of $H^\bullet_Q(\text{Ops})$ on $\CH$, we use the two maps from $\CX\times \CX$ to $\CX$, coming from projection onto the first and second factors
\be \label{conv-action} \begin{array}{ccccc}
 &&\CX_2\times \CX_1  \\
  & \hspace{-.5cm}\raisebox{.1cm}{$\pi_2$} \swarrow && \searrow \raisebox{.1cm}{$\pi_1$} \\
 \CX_2 &&&&  \hspace{-.5cm}   \CX_1 \end{array} \ee
This is called a convolution diagram.
Given a local operator $\CO \in H^\bullet(\CX_2\times \CX_1)$, we can define an action on $\CH =  H^\bullet(\CX_1)=  H^\bullet(\CX_2)$ by
\be \CO:  \begin{array}{ccc} H^\bullet(\CX_1) & \longrightarrow & H^\bullet(\CX_2)  \\
  v &\mapsto& (\pi_2)_*(\CO\wedge \pi_1^*(v))\,.  \end{array}
\ee
The Hodge duality above has been repackaged in the push-forward map $(\pi_2)_*$, which involves an integration along the fibers of $\pi_2$, \emph{i.e.} an integration along $\CX_1$.%
\footnote{The whole story may be even more familiar in standard, non-supersymmetric quantum mechanics. Consider a particle moving on $\CX=\R$, with Hilbert space $\CH = L^2(\R)$. Linear operators $\CO:\CH\to\CH$ can be represented by their integral kernel $K_\CO(x,y)$ (in a manner made precise by the Schwartz kernel theorem) so that $\CO|x\rangle = \int dx\, K_\CO(x,y)|y\rangle$. This is the convolution product of \eqref{conv-action}, in the infinite-dimensional setting. Similarly, the product of two operators is represented as convolution of their kernels $K_{\CO'\cdot \CO}(x,y) = \int dz\, K_{\CO'}(x,z)K_{\CO}(z,y)$, analogous to \eqref{conv-prod}.
}

Similarly, the product (the algebra structure) on  $H^\bullet_Q(\text{Ops})$ comes geometrically from considering three copies of $\CX$, and projections onto pairs of factors
\be \label{conv-prod} \begin{array}{ccccc}
 &&\CX_3\times \CX_2\times \CX_1  \\
  & \hspace{-.8cm}\raisebox{.1cm}{$\pi_{31}$} \swarrow &  \hspace{.5cm}  \downarrow \raisebox{.1cm}{$\pi_{32}$} & \searrow \raisebox{.1cm}{$\pi_{21}$} \\
 \CX_3\times \CX_1 &&   \CX_3\times \CX_2&&  \hspace{-.8cm}  \CX_2\times \CX_1 \end{array} \ee
Given any $\CO \in H^\bullet(\CX_2\times \CX_1)$ and $\CO'\in H^\bullet(\CX_3\times \CX_2)$, there is now a ``convolution product''  defined by
\be \CO' \cdot \CO = (\pi_{31})_*\big( \pi_{32}^*(\CO')\wedge \pi_{21}^*(\CO)\big) \ee
Ignoring the indices and identifying $ H^\bullet(\CX_2\times \CX_1) = H^\bullet(\CX_3\times \CX_2) = H^\bullet(\CX_3\times \CX_1) = H^\bullet_Q(\text{Ops})$, we see that this is a product $H^\bullet_Q(\text{Ops})\times H^\bullet_Q(\text{Ops})\to H^\bullet_Q(\text{Ops})$. Working through the various push-forwards and pull-backs involved, one can show that it is the \emph{same} as the more naive product resulting from the identification $H^\bullet_Q(\text{Ops}) \simeq \text{End}_{\C}(H^\bullet(\CX))$ in \eqref{EndH}.

Note that the same sort of analysis could have been used to describe local operators at half-BPS \emph{junctions} of $\CN=2$ quantum mechanics theories, with different targets. At a junction of SQM with target $\CX_1$ and SQM with target $\CX_2$, the local operators are $H^\bullet_Q(\text{Ops}(\CX_1,\CX_2)) = H^\bullet(\CX_2\times \CX_1)$. They act on states in the Hilbert space $H^\bullet(\CX_1)$ to produce states in $H^\bullet(\CX_2)$. Geometrically, the action comes from the same convolution diagram \eqref{conv-action}, now with $\CX_1$ and $\CX_2$ interpreted as (potentially) different spaces.

The OPE coming from collision of two different junctions --- say between SQM with targets $\CX_1,\CX_2$ and SQM with targets $\CX_2,\CX_3$ --- is also encoded geometrically in the convolution diagram \eqref{conv-prod}, with $\CX_1,\CX_2,\CX_3$ interpreted as potentially different spaces.

\subsubsection{Generalization to line operators}
\label{sec:catrep}

Now consider a junction of two A-type line operators $\CL_1,\CL_2$ in a 3d $\CN=4$ theory. Placing the junction inside a solid cylinder with boundary condition $\CB$ on the outside, we expect to obtain localized descriptions of the theory above and below the junction, as quantum mechanics with target $\CM_D(\CB,\CL_2)$ and $\CM_D(\CB,\CL_1)$, respectively. By analogy with \eqref{stateop-1d}, a naive guess would be that the space of local operators at the junction, as represented in the presence of the boundary condition, is
\be H^\bullet\big(\CM_D(\CB,\CL_2)\times   \CM_D(\CB,\CL_1)\big) \,\simeq\, \text{Hom}_\C\big(H^\bullet(\CM_D(\CB,\CL_1)),H^\bullet(\CM_D(\CB,\CL_2))\big)\,. \label{hom-huge} \ee

This is not quite right. The problem is that the space in \eqref{hom-huge}, which also appeared on the RHS of \eqref{repB}, is too large. It certainly contains local operators at the junction of $\CL_1$ and $\CL_2$; however, as discussed below \eqref{repB}, it also contains various extended operators. It captures all ways to interpolate between solutions of the BPS equations on a disc below the junction and a disc above the junction.

In order to capture only local operators at the junction itself, we should restrict our attention to solutions of the BPS equations on two discs that \emph{only} differ in a neighborhood of the origin $z=0$.
One way to formalize this constraint is to introduce a moduli space of solutions to BPS equations on a punctured disc $D^*$,
\be \CM_{D^*}(\CB) = \left\{ \begin{array}{c}\text{sol's to BPS equations on a punctured disc $D^*$} \\
\text{subject to the boundary condition $\CB$ at $\pd D$} \end{array}\right\}\,,
\ee
with a boundary condition $\CB$ near the ``outer'' $S^1$ boundary of the punctured disc, but with free boundary conditions near the origin.  Both $\CM_D(\CB,\CL_1)$ and $\CM_D(\CB,\CL_2)$ map to $\CM_{D^*}(\CB)$, just by forgetting the data of a solution near the origin:
\be \begin{array}{ccccc} \CM_D(\CB,\CL_2) &&&&  \hspace{-.5cm} \CM_D(\CB,\CL_1) \\
 & \hspace{-.5cm} p_2\searrow && \swarrow p_1 \\
 && \CM_{D^*}(\CB) \end{array} \ee
Then we consider a space
\be  \label{Mfiber} \boxed{ \begin{array}{rcl} \CM_{\rm rav}(\CB;\CL_2,\CL_1)  &=& \CM_D(\CB,\CL_2) \times_{\CM_{D^*}(\CB)} \CM_D(\CB,\CL_1)   \\
&:=&  \big\{ (x,y) \in \CM_D(\CB,\CL_2)\times   \CM_D(\CB,\CL_1) \;\text{s.t.}\; p_2(x)=p_1(y) \big\}\,. \end{array}}
\ee

Mathematically, the construction in \eqref{Mfiber} is called a fiber product; the product is ``fibered over'' $\CM_{D^*}(\CB)$. Note that the fiber product is a subspace of the ordinary product, $\CM_{\rm rav}(\CB;\CL_2,\CL_1) \subseteq \CM_D(\CB,\CL_2) \times \CM_D(\CB,\CL_1)$.%
\footnote{We use the notation ``rav'' because, schematically, $\CM_{\rm rav}(\CL_2,\CL_1;\CB)$ is the space of solutions to BPS equations on $D\cup_{D^*}D$, \emph{i.e.} on two copies of the disc $D$ (punctured by $\CL_2$ and $\CL_1$, respectively), identified over the punctured disc $D^*$. The union of discs $D\cup_{D^*}D$ looks like a ``raviolo.''} %
Physically, the space $\CM_{\rm rav}(\CB;\CL_2,\CL_1)$ is the right place to look for operators that are localized at the junction of $\CL_1$ and $\CL_2$. Thus, we expect the image of the map \eqref{repB} to be the cohomology of $\CM_{\rm rav}$, 
\begin{align} & \boxed{ \CF_\CB\,:\; \text{Hom}_A(\CL_1,\CL_2) \; \to\hspace{-.3cm}\to \;  H^*\big( \CM_{\rm rav}(\CB;\CL_2,\CL_1)\big) } \label{Brav} \\ & \hspace{1.5in}  \subseteq  \text{Hom}_\C\big(\CH_D(\CB,\CL_1),\CH_D(\CB,\CL_2)\big)\,. \notag \end{align}
Moreover, for sufficiently generic $\CB$, we expect \eqref{Brav} to be an isomorphism.

Cohomology classes $\CO \in H^*\big( \CM_{\rm rav} \big)$ have natural convolution products and/or actions, just like classes in the bigger space \eqref{hom-huge} did. 
The most direct way to see this is to note that there are convolution diagrams involving $\CM_{\rm rav}$ alone:
\be \label{action-Mrav} \begin{array}{ccccc}
 && \hspace{-.5cm}\CM_D(\CB,\CL_2) \times_{\CM_{D^*}(\CB)} \CM_D(\CB,\CL_1) = \CM_{\rm rav}(\CB;\CL_2,\CL_1) \\
  & \hspace{-.5cm}\raisebox{.1cm}{$\pi_2$} \swarrow && \hspace{-6.4cm} \searrow \raisebox{.1cm}{$\pi_1$} \\
 \CM_D(\CB,\CL_2) &&& &  \hspace{-3.2cm} \hspace{-.5cm}  \CM_D(\CB,\CL_1) \end{array} \ee
computing the action $H^*\big(\CM_{\rm rav}(\CB;\CL_2,\CL_1)\big):\, H^*\big(\CM_D(\CB,\CL_1)\big) \to H^*\big(\CM_D(\CB,\CL_2)\big) $\,; and
\be \label{prod-Mrav} \begin{array}{ccccc}
 && \hspace{-2cm} \CM_D(\CB,\CL_3)\times_{\CM_{D^*}(\CB)} \CM_D(\CB,\CL_2) \times_{\CM_{D^*}(\CB)} \CM_D(\CB,\CL_1) \\
  & \hspace{-.8cm}\raisebox{.1cm}{$\pi_{31}$} \swarrow &  \hspace{-2cm}  \hspace{.5cm}  \downarrow \raisebox{.1cm}{$\pi_{32}$} &   \hspace{-2cm} \searrow \raisebox{.1cm}{$\pi_{21}$} \\
\CM_{\rm rav}(\CB;\CL_3,\CL_1) &&   \hspace{-2cm} \CM_{\rm rav}(\CB;\CL_3,\CL_2)&&   \hspace{-1cm}  \hspace{-.8cm}  \CM_{\rm rav}(\CB;\CL_2,\CL_1)\end{array} \ee
computing a product $H^*\big(\CM_{\rm rav}(\CB;\CL_3,\CL_2)\big)\otimes H^*\big(\CM_{\rm rav}(\CB;\CL_2,\CL_1)\big)\to H^*\big(\CM_{\rm rav}(\CB;\CL_3,\CL_1)\big)$ from collision of junctions. This product is the representation of the intrinsic product $\text{Hom}_A(\CL_2,\CL_3)\otimes \text{Hom}_A(\CL_1,\CL_2)\to \text{Hom}_A(\CL_1,\CL_3)$ of local operators at junctions, on the left side of \eqref{Brav}.

We finish with two remarks. Mathematically, restricting convolution algebras to fiber products rather than ordinary direct products, as we did here, is a common operation. A thorough discussion of such products and their use in geometric representation theory is contained in \cite{ChrissGinzburg}.
In Section \ref{sec:convQM} we already saw that convolution coming from ordinary products was a little boring: it just reproduced the full algebra of linear transformations (a matrix algebra) acting on a vector space.
In contrast, convolution with fiber products can define interesting and highly nontrivial subalgebras.

Physically, the space $\CM_{\rm rav}$ may also be interpreted as solutions to BPS equations on a ``Gaussian pillbox'' surrounding the location of a putative local operator, as in Figure \ref{fig:cyl-intro} of the Introduction. Topologically, the pillbox is a sphere $S^2$.
 From this perspective, identifying local operators with the cohomology of $\CM_{\rm rav}$ is a reflection of the state-operator correspondence in 3d, which would relate local operators with the Hilbert space on $S^2$. We have described a modification of the 3d state-operator correspondence in the presence of a boundary condition.

\subsection{Algebraic reformulation}
\label{sec:Mrav}

We would like to do concrete computations of algebras of local operators at junctions of lines. To this end, we propose an algebraic reformulation of the moduli spaces $\CM_D$ and $\CM_{\rm rav}$ that appeared above. Mathematically, this formulation is (as yet) conjectural; it generalizes the Kobayashi-Hitchin correspondence of \cite{Taubes-vortex, JaffeTaubes, VW-vortices} in the case of the trivial line.

Suppose that a bulk 3d $\CN=4$ theory has gauge group $G$ and hypermultiplets in a representation $T^*R$. We saw in Section \ref{sec:VL} that a line operator $\CL$ (as seen by the A-twist) is characterized by algebraic data consisting of 1) a subgroup $\CG_0\subseteq G(\CO)$ of algebraic gauge transformations in an infinitesimal neighborhood of $z=0$; and 2) a $\CG_0$-invariant Lagrangian subspace $\CL_0$ of the algebraic loop space $T^*R(\CK)$. 

A large class of boundary conditions $\CB$, described in detail in \cite{BDGH}, also admit an algebraic characterization. The boundary conditions we will use in the remainder of this paper are a special subset of those considered in \cite{BDGH}: they are labeled by \emph{massive vacua} $\nu$ of the bulk 3d theory. A vacuum boundary condition $\CB_\nu$ is not quite a boundary condition in the ordinary sense (\emph{i.e.} at finite distance), but rather requires the theory to be in a fixed supersymmetric vaccum $\nu$ asymptotically near spatial infinity.%
\footnote{One \emph{can} associate a finite-distance half-BPS boundary condition to a given vacuum $\nu$ as well, which has the same effect as requiring fields to approach $\nu$ at infinity. See \cite{BDGH} for its definition. Such vacuum-mimicking boundary conditions were studied by \cite{HoriIqbalVafa} in the analogous context of 2d $\CN=(2,2)$ theories. They are sometimes called Lefschetz branes. More recently, their physics (in 2d) was revisited in \cite{GMW, GMW-rev}, and they are related mathematically to Fukaya-Seidel categories \cite{Seidel-book}. We will \emph{not} need finite-distance boundary conditions in this paper, aside from Appendix \ref{app:BFN}.}

Returning to the solid cylinder setup of \ref{sec:cyl}, we consider the bulk 3d theory on $D\times \R_t$ with line operator(s) at the origin of $D$ and a vacuum boundary condition $\CB_\nu$, associated to a vacuum $\nu$, at the outer boundary of $D$. We will interpret $\CB_\nu$ as an asymptotic condition, requiring the fields of the theory to approach the vacuum $\nu$ as $|z|\to \infty$.

We should explain what a massive vacuum $\nu$ is. To characterize $\nu$ physically, we would turn on real FI parameters $t_\R$ to make the Higgs branch (or some region thereof) smooth, then turn on generic complex mass parameters $m_\C$ to induce a potential on the Higgs branch with isolated, nondegenerate, zero-energy minima. The vacuum $\nu$ is chosen to be one of these minima. All of this translates nicely to algebra. The algebraic data of a massive vacuum consists of a point $\boxed{\nu\in T^*R}$ such that
\begin{itemize}
\item the complex moment map evalued at $\nu$ vanishes, $\nu \in \mu_\C^{-1}(0)$
\item $\nu$ lies in the stable part of $\mu_\C^{-1}(0)\subseteq T^*R$, for a \emph{choice} of stability condition; \\
(the stability condition, used in defining the smooth Higgs branch $\CM_H \simeq \mu_\C^{-1}(0)^{\rm stab}/G_\C$, is correlated with the physical choice of FI parameters)
\item the stabilizer of $\nu$ under the $G_\C$ action is trivial, \emph{i.e.} the orbit $G_\C\cdot \nu \simeq G_\C$ is maximal \\
(physically, this means gauge symmetry is broken, so the Coulomb branch is massive)
\item $\nu$ is an isolated fixed point of a torus of the flavor symmetry group $F_\C$ \\
(physically, this means the Higgs branch will become massive around $\nu$ for generic $m_\C$)
\end{itemize}

Note that not all theories have isolated massive vacua. A theory needs sufficient matter content and abelian factors in the gauge group for massive vacua to exist (after a mass and FI deformation). A huge class of interesting theories --- including most abelian theories, and ``good'' and ``ugly'' A-type quivers \cite{GW-Sduality} --- do have this property. Having massive vacua makes the analysis of spaces $\CM_D$ and $\CM_{\rm rav}$ particularly simple, so we will use them in this paper. Nevertheless, it is important to mention that even without massive vacua there are \emph{many} other choices of boundary conditions available, see \emph{e.g.} Appendix \ref{app:BFN}.

In order to algebraically describe fields that ``approach $\nu$ as $|z| \rightarrow \infty$" we define
\be \CO_\infty := \C[\![z^{-1}]\!]\,,\qquad \CK_\infty := \C(\!(z^{-1})\!)  \ee
to be the rings of formal Taylor and Laurent series in $z^{-1}$, respectively. These are the algebraic functions in a formal neighborhood of $z=\infty$, and a formal punctured neighborhood of $z=\infty$. The hypermultiplet fields $X(z)$, $Y(z)$ must be nonsingular near infinity, meaning they take values in $T^*R(\CO_\infty)$ there, and we can define an evaluation map
\be \text{val}_\infty\,:\; T^*R(\CO_\infty)\to T^*R\,,\quad X(z),Y(z)\, \mapsto\, X(\infty),Y(\infty) \ee
that sets $z^{-1}\mapsto 0$. The hypermultiplet fields that are gauge-equivalent to a configuration approaching $\nu$ at infinity are precisely those in the orbit
\be \label{infinity-orbit} X,Y \,\in\, G(\CK_\infty)\cdot \text{val}_\infty^{-1}(\nu)\, \subseteq\, T^*R(\CK_\infty)\,,\ee
 where $G(\CK_\infty)$ is the algebraic group $G_\C$ defined over $\CK_\infty = \C(\!(z^{-1})\!)$, and represents the gauge transformations near infinity. As indicated, the orbit \eqref{infinity-orbit} is a subscheme of $T^*R(\CK_\infty)$.

Near the origin, we must implement the constraint defining the line operator, that the hypermultiplets lie in $\CL_0$. We do this by taking the intersection $\CL_0 \, \cap\, \big[  G(\CK_\infty)\cdot \text{val}_\infty^{-1}(\nu)\big]$. Since $\CL_0$ is a subscheme of $T^*R(\CK)$, the intersection lies in $T^*R(\CK)\cap T^*R(\CK_\infty) = T^*R[z,z^{-1}]$, and can thus be described entirely in terms of 
  \emph{Laurent-polynomial-valued}  $X(z),Y(z)$.
  
 Finally, we must quotient by gauge transformations near the origin. The group of gauge transformations that preserves the line operator is (by definition) $\CG_0 \subseteq G(\CO)$. The intersection $\CL_0 \, \cap\, \big[  G(\CK_\infty)\cdot \text{val}_\infty^{-1}(\nu)\big]$ is preserved by the subgroup of $\CG_0$ containing \emph{polynomial-valued} gauge transformations. To describe this, we define
\be G[z]:=G_\C(\C[z]) \subset G(\CO)\,,\qquad \CG_0[z] := \CG_0 \cap G[z]\,, \ee 
\emph{i.e.} $G[z]$ is polynomial-valued holomorphic gauge transformations on all of $\C$ (not just a formal neighborhood of the origin), and $\CG_0[z]$ is the subgroup of $G[z]$ preserving the line operator.

Now let us put everything together. Given a line operator $\CL$ (with data $\CG_0,\CL_0$) and a vacuum boundary condition $\CB_\nu$, the algebraic version of the moduli space of solutions to BPS equations on the disc is
\be \begin{array}{ccl}
\CM_D(\CB_\nu,\CL) &=& {\CG_0}[z] \big\backslash  \wt \CM_D(\CB_\nu,\CL)\,,\qquad 
 \wt \CM_D(\CB_\nu,\CL) :=  \CL_0 \,\cap\, \big(G(\CK_{\infty})\cdot \text{val}_\infty^{-1}(\nu)\big)\,. \end{array} \label{MD-alg} \ee
In general, a space such as $\CM_D = \CG_0[z]\backslash \wt\CM_D$ should be interpreted as a derived stack. However, the fact that $\nu$ is a massive vacuum ensures that $\CG[z]$ acts freely in \eqref{MD-alg}, and that $\CM_D$ has the much simpler structure of a (potentially singular) variety. In fact, as discussed in Section \ref{sec:vxnumber}, it is a disjoint union of finite-dimensional varieties.

It is also instructive to recast the space \eqref{MD-alg} geometrically. It is a moduli space of bundles and sections on $\cp^1$, thought of as the compactification of the infinite disc $D$,
\be \label{MD-geom} \CM_D(\CB_\nu,\CL) = \left\{ \begin{array}{l}
 \text{$E,(X,Y)$ s.t. $E$ is a principal algebraic $G_\C$ bundle on $\cp^1$} \\
 \text{\quad with structure reduced to $\CG_0$ near $z=0$ and trivialized at $z=\infty$,} \\
 \text{and $(X(z),Y(z))$ is a section of an associated $T^*R$ bundle} \\
 \text{\quad satisfying $\mu(X,Y)=0$} \\
 \text{\quad with $(X,Y)\in \CL_0$ near $z=0$ and $(X,Y)\in G_\C \cdot \nu$ at $z=\infty$}
\end{array} \right\}\Big/\text{iso}
\ee
This moduli space must be considered modulo isomorphisms, \emph{i.e.} gauge transformations.

We again emphasize that the moduli spaces \eqref{MD-alg}, \eqref{MD-geom} generalize spaces studied in \cite{VV, Braverman-W, BFFR} (for the case $\CL=\id$). They are generalized vortex moduli spaces, encountered in many places in math and physics, as reviewed in the Introduction and Section \ref{sec:VL}. The proposed (yet unproven) equivalence of \eqref{MD-alg}, \eqref{MD-geom} with physical solutions to the vortex equations is a natural extension of \cite{VW-vortices} to incorporate a potential singularity at $z=0$.

The ``raviolo space'' $\CM_{\rm rav}$, used for computing local operators at a junction of lines, should admit a similar algebro-geometric description. Given a pair of line operators $\CL',\CL$, we propose that
\begin{align} \notag 
\CM_{\rm rav}(\CB_\nu;\CL',\CL) &= \left\{\begin{array}{l}
 \text{pairs $E',(X',Y')$ and $E,(X,Y)$ of bundles/sections on $\cp^1$,} \\
 \text{\;\; each satisfying constraints as in \eqref{MD-geom} for (resp.) $\CL'$ and $\CL$ } \\
 \text{together with an isomorphism $g: \big(E,(X,Y)\big)\overset{\sim}{\to} \big(E',(X',Y')\big)$} \\
 \text{\quad away from $z=0$, \emph{i.e.} $g\in G[z, z^{-1}]$}
 \end{array} \right\} \Big/\text{iso} \\
 & = \raisebox{-.2cm}{${\CG_0}'[z]$} \big\backslash \wt \CM_{\rm rav}(\CB_\nu;\CL',\CL)  \big/ \raisebox{-.2cm}{${\CG_0}[z]$}\,, \label{Mrav-alg}
\end{align}
\be \notag \begin{array}{l}  \wt \CM_{\rm rav}(\CB_\nu;\CL',\CL)  :=  \big(\CL_0' \,\cap\, (G(\CK_\infty)\cdot \text{val}_\infty^{-1}(\nu))\big) \times  G[z, z^{-1}] \times \big(\CL_0 \,\cap\, (G(\CK_\infty)\cdot \text{val}_\infty^{-1}(\nu))\big)\big|_{(*)}  \\ 
 \hspace{1.9in} X',Y' \hspace{1.1in} g \hspace{1.1in} X,Y
 \end{array} \ee
with a constraint $(*)$ requiring $(X',Y')=(gX,Yg^{-1})$. Here again, having a massive vacuum guarantees that $X,Y$ and $X',Y'$ are Laurent polynomials in $z$. The element
\be g(z)\in G[z,z^{-1}] := G_\C(\C[z,z^{-1}]) \ee
is a gauge transformation valued in Laurent polynomials that relates $X,Y$ on the ``bottom'' disc with $X',Y'$ on the ``top" disc, away from $z=0$. The remaining gauge transformations $(g_0',g_0)\in {\CG_0}'[z]\times {\CG_0}[z]$ on the top and bottom discs act on the algebraic data as
\be X',Y',g,X,Y\;\mapsto\; g_0'X,\,Y'g_0'{}^{-1},\, g_0' g g_0^{-1},\, g_0 X,\, Y g_0^{-1}\,. \ee

\subsubsection{Coupling to quantum mechanics}
\label{sec:Mrav-QM}

In Section \ref{sec:VL} we also reviewed how A-type line operators could be engineered by coupling to $\SQM_A$ quantum mechanics. Such a definition can also be incorporated fairly easily into the algebraic moduli spaces above, either replacing singularity data given by $\CG_0,\CL_0$, or further enhancing it.

We'll just describe the case where a line operator is \emph{entirely} defined by coupling to 1d degrees of freedom, with no other singularity present in the bulk fields. Suppose that we define $\CL$ by introducing a 1d sigma-model with K\"ahler target $\CE$ as in Section \ref{sec:VL-GR}, thought of as an algebraic variety with complexified flavor symmetry $G(\CO)$. (All but a finite part of $G(\CO)$ is assumed to act trivially.) In an algebraic formulation, the sigma-model is coupled to the bulk by gauging $G(\CO)$.

We may also introduce an algebraic $G(\CO)$-invariant superpotential $\wt W_0:\CE\times T^*R(\CO)\to \C$, as in \eqref{tW0}.
Let  $W = \int d^2z X D_{\bar z} Y + \wt W_0$, and note that the critical locus $\delta W = 0$ is algebraic. Explicitly, if $\alpha$ are local coordinates on $\CE$, and $x_n,y_n$ are the modes of $X$ and $Y$, then the critical locus is equivalent to
\be \delta W=0\,:\quad \mu_\C(X,Y)=0\,;\qquad   \frac{\pd \wt W_0}{\pd \alpha}=0\,,\quad \frac{\pd \wt W_0}{\pd y_n} = x_{-n-1}\,,\quad \frac{\pd \wt W_0}{\pd x_n} = -y_{-n-1}\quad (n\geq 0)\,. \label{tWcrit} \ee
The equations involving $\frac{\pd \wt W_0}{\pd y_n}$ and $\frac{\pd \wt W_0}{\pd x_n}$ are not really constraints on the space $\CE\times T^*R(\CO)$, since the negative modes $x_{-n-1}$ and $y_{-n-1}$ are not part of $T^*R(\CO)$ to begin with. Instead, one can view \eqref{tWcrit} as equations on $\CE\times T^*R(\CK)$. For example, in the extreme case of vanishing superpotential $\wt W_0=0$, last two equations in \eqref{tWcrit} set all negative modes to zero, so that the critical locus $\delta W=0$ is precisely $\CE\times T^*R(\CO)$ inside $\CE\times T^*R(\CK)$.

In the presence of a line operator $\CL$ with quantum-mechanics data $\CE,W_0$, and a vacuum boundary condition $\CB_\nu$, we expect that the moduli space $\CM_D$ can be described as
\begin{align}\notag  \CM_D(\CB_\nu,\CL) &= \left\{\begin{array}{l}
\text{$\alpha,E,(X,Y)$ s.t. $E$ is a $G_\C$ bundle on $\cp^1$ trivialized at $\infty$;} \\
\text{\quad $(X,Y)$ is a section of an associated bundle on $\cp^1\backslash\{0\}$} \\
\text{\quad with $(X,Y)\in G_\C\cdot \nu$ at $z= \infty$;} \\
\text{\quad and $\alpha \in \CE$; all subject to $\delta W =0$ }
\end{array} \right\} \Big/ \text{iso} \\
&= G[z]\big\backslash  \wt \CM_D(\CB_\nu,\CL) \,,  \label{MD-QM}
\end{align}
\be \hspace{-.6in} \wt \CM_D(\CB_\nu,\CL) :=  \CE\times \big[T^*R(\CK)\; \cap\; (G(\CK_{\infty})\cdot \text{val}_\infty^{-1}(\nu))\big]\big|_{\delta W=0}\,. \notag \ee
Similarly, given a pair of line operators with data $\CE',W_0'$ and $\CE,W_0$, the raviolo space is
\begin{align}  \notag \CM_{\rm rav}(\CB_\nu;\CL',\CL) &=  \left\{\begin{array}{l}
 \text{$\alpha', E',(X',Y');g;\alpha,E,(X,Y)$} \\ \text{\quad  s.t. each tuple $\alpha', E',(X',Y')$ and $\alpha, E,(X,Y)$ is as above } \\
 \text{\quad (in particular, $\alpha'\in \CE'$, $\alpha\in \CE$ and $\delta W' = \delta W=0$)} \\
 \text{\quad and $g: \big(E,(X,Y)\big)\overset{\sim}\to \big(E',(X',Y')\big)$ on $\cp^1\backslash\{0\}$}
\end{array} \right\} \Big/ \text{iso} \\
& = \raisebox{-.2cm}{$G[z]$} \big\backslash \wt \CM_{\rm rav}(\CB_\nu;\CL',\CL) \big/ \raisebox{-.2cm}{$G[z]$}\,,  \label{Mrav-QM}\end{align}
\be \notag \hspace{-.3in}  \begin{array}{l} \wt \CM_{\rm rav}(\CB_\nu;\CL',\CL)  =
\CE'\times \big[T^*R(\CK)\, \cap\, (G(\CK_{\infty})\cdot \text{val}_\infty^{-1}(\nu))\big] \times  G[z, z^{-1}]   \times \CE \times \big[T^*R(\CK)\, \cap\, (G(\CK_{\infty})\cdot \text{val}_\infty^{-1}(\nu))\big]\Big|_{(*)} \\
\hspace{1.2in} \alpha' \hspace{.7in} X',Y' \hspace{1.55in} g  \hspace{.35in} \alpha \hspace{.7in} X,Y  \end{array} \ee
with constraints $(*)$ given by $\delta W'=\delta W=0$ and $(X',Y') = (gX,Yg^{-1})$.

\subsection{Vortex number}
\label{sec:vxnumber}

A key feature of massive-vacuum boundary conditions is that the spaces $\CM_D$ and $\CM_{\rm rav}$ break up into a disjoint union of finite-dimensional components. This is the main reason we use them here. It makes the cohomology of these moduli spaces much easier to analyze by elementary methods. It also endows the cohomology with an additional grading.

In the case of $\CM_D$, components are labeled by \emph{vortex number} $\n\in \pi_1(G)$,
\be \label{D-conn} \CM_D(\CB_\nu,\CL) = \bigsqcup_{\n \in \pi_1(G)} \CM_D^\n(\CB_\nu,\CL)\,, \ee
and correspondingly the cohomology $\CH_D(\CB_\nu,\CL) = \bigoplus_{\n\in \pi_1(G)} H^*\big( \CM_D^\n(\CB_\nu,\CL)\big)$ is graded by $\pi_1(G)$.
Physically, vortex number is usually interpreted as a first Chern class, and expressed as an integral of the curvature of the $G$-bundle on the disc
\be \n = \frac{1}{2\pi} \int_D \text{Tr}\,F\,, \ee
which is well defined because the vacuum $\nu$ at infinity trivializes the structure of the bundle (and in particular ensures that $\text{Tr}F\to 0$ sufficiently fast).

Topologically, vortex number arises because the group $G(\CK_\infty)$ that appears in \eqref{MD-alg} is a version of the loop group $LG_\C$, which has connected components labeled by $\pi_1(G)$.  Viewed as an algebraic ind-scheme, $G(\CK_\infty)$ is stratified rather than disconnected, with strata labeled by elements $\n \in \pi_1(G)$. However, after passing to the quotient by $\CG_0[z]$ in \eqref{MD-alg}, one again finds connected components labeled by $\n \in \pi_1(G)$.

The most direct way to understand vortex number algebraically is as a \emph{degree}. The basic example (and the only one relevant for us) is 
$G=U(N)$. In this case $G_\C = GL(N,\C)$, and $G(\CK_\infty)$ is the group of invertible $N\times N$ matrices whose entries are formal Laurent series in $z^{-1}$. Given any $g(z)\in G(\CK_\infty)$, the determinant
\be \text{det}\, g(z) = a_\n z^\n + a_{\n-1}z^{\n -1} + ... \quad \in \C(\!(z^{-1})\!)  \ee
is a nonzero formal series, and has a well-defined degree $\n \in \Z \simeq \pi_1(U(N))$ given by the highest  power of $z$ that appears with nonzero coefficient.

When a line operator $\CL$ breaks gauge symmetry near the origin from $G(\CO)$ to $\CG_0$, the notion of vortex number and the corresponding decomposition \eqref{D-conn} may be refined. We will see this happening in nonabelian examples. Nevertheless, there is \emph{always} a decomposition by at least the vortex numbers $\n\in \pi_1(G)$, which is what we are discussing here.

In a similar way, the raviolo spaces used to construct local operators at junctions break up into connected components labeled by pairs of vortex numbers
\be \label{rav-conn} \CM_{\rm rav}(\CB_\nu;\CL',\CL) = \bigsqcup_{\n',\n\in \pi_1(G)} \CM_{\rm rav}^{\n',\n}(\CB_\nu;\CL',\CL)\,.\ee
In the algebraic formulation of \eqref{Mrav-alg}, $\n'$ and $\n$ are the degrees of the two $G(\CK_\infty)$ elements used to relate the vacuum at infinity to sections $X',Y'$ and $X,Y$ (respectively) near the origin.

The decomposition \eqref{rav-conn} implies that the cohomology of $\CM_{\rm rav}$ will be graded by pairs of vortex numbers $\n',\n$. The difference $\n'-\n$ corresponds to the physical monopole charge of a local operator. It is the charge under the $U(1)_t$ topological flavor symmetries dual to the center of the group $G$.

\subsection{Summary and interpretation}
\label{sec:Asum}

The final approach used to compute local operators at junctions of lines looks as follows.

Given a 3d $\CN=4$ gauge theory with data $G,R$, we choose a massive supersymmetric vacuum $\nu$ (assuming the theory has massive vacua). We use $\nu$ to define an asymptotic boundary condition $\CB_\nu$, which allows us to represent the entire category of line operators. We expect the representation --- in particular, the maps $\CF_{\CB_\nu}$ from \eqref{Brav} on spaces of local operators --- to be injective for a large set of line operators.

For every line operator $\CL$ defined by algebraic data $\CG_0,\CL_0$, we construct the algebraic moduli space $\CM_D(\CB_\nu,\CL)$ as in \eqref{MD-alg}. If $\CL$ is defined by coupling to quantum mechanics, we can use the definition \eqref{MD-QM} instead. In our examples, $\CM_D(\CB_\nu,\CL)$ will break up into infinitely many finite-dimensional components, labeled by vortex numbers $\n$.
We take cohomology to (conjecturally) construct the $Q_A$-cohomology of the physical Hilbert space on the disc,
\be \CH_D(\CB_\nu,\CL) = H^*\big(\CM_D(\CB_\nu,\CL)\big)
 = \bigoplus_{\n} H^*\big( \CM_D^\n(\CB_\nu,\CL)\big)\,. \label{MD-sum} \ee
The Hilbert space is graded by vortex number.

For every pair of line operators $\CL,\CL'$, we construct the raviolo space $\CM_{\rm rav}(\CB_\nu;\CL',\CL)$, using algebraic data as in \eqref{Mrav-alg} or quantum-mechanics data as in \eqref{Mrav-QM} (or some combination thereof). Again, the raviolo spaces break up into finite-dimensional
components labeled by pairs of vortex numbers. We expect the $Q_A$-cohomology $\text{Hom}_A(\CL,\CL')$ of the space of local operators at a junction of lines to be represented by the cohomology
\be
H^*\big(\CM_{\rm rav}(\CB_\nu;\CL',\CL)\big) =
 \bigoplus_{\n',\n} H^*\big(\CM_{\rm rav}^{\n',\n}(\CB_\nu;\CL',\CL)\big)\,. \label{Mrav-sum}  \ee
The product of local operators at junctions (\emph{a.k.a.} composition of Hom's) and their action on the Hilbert spaces $\CH_D(\CB_\nu,\CL)$ are both given by convolution, as in Section \ref{sec:conv}.

\subsubsection{Identifying monopole operators}

The sort of local operators we expect to find at junctions of A-type lines are a generalization of operators in the bulk Coulomb-branch chiral ring. In particular, we should see operators formed out of bulk vectormultiplet scalars, as well as monopole operators. The vectormultiplet scalars $\varphi$ will appear in a straightforward way as equivariant parameters (see Section \ref{sec:cohomologies}). We recall how monopole operators are identified, following \cite[Sec 10]{KapustinWitten} and \cite{BFNII, VV}.

A  physical monopole operator is labeled by a ``monopole charge'' $A$.  Mathematically, this is an element of the cocharacter lattice $A \in \text{cochar}(G) \simeq  \text{Hom}(\C^*,T_\C)$. The charge $A$ thus determines a group homomorphism from $\C^*$ to the maximal torus $T_\C\subseteq G_\C$.  (In the physical definition of a monopole operator, $A$ is literally used to embed a fundamental Dirac singularity for $U(1)$ into gauge theory with group $G$.) Let  $z^A \in G[z, z^{-1}]$ denote the image of $z\in \C^*[z, z^{-1}]$ under this homomorphism. For example, if $G=U(N)$, cocharacters are $N$-tuples of integers $A = (A_1,...,A_N)\in \Z^N$, and
\be z^A = \text{diag}(z^{A_1},z^{A_2},...,z^{A_N}) \in G[z, z^{-1}]\,. \ee

At a junction of lines $\CL$ and $\CL'$, we may use any element
\be w z^A \in G[z, z^{-1}]\,,\qquad w\in \text{Weyl}(G) \label{wzA} \ee
to try to define a monopole operator. Note, these are elements of the extended affine Weyl group $\text{Weyl}(G)\ltimes\text{cochar}(G) \simeq W_{\rm aff}(G)\simeq \text{Weyl}(G(\CK))$.
 When $\CL=\CL'=\id$, only the orbit of $w z^A$ under $\text{Weyl}(G)\times \text{Weyl}(G)$ acting on the left and right matters; this action can be used to remove $w$ and to conjugate $A$ to a dominant cocharacter $(A_1\geq A_2\geq...\geq A_N)$, whence one usually says that the charges of bulk monopole operators are dominant cocharacters. However, if $\CL$ and $\CL'$ break the bulk gauge symmetry to  $\CG_0$ and $\CG_0'$, respectively, we may only act on \eqref{wzA} with $\text{Weyl}(\CG_0)\times \text{Weyl}(\CG_0')$. Then monopole charges take values in
\be \text{for $\text{Hom}_A(\CL,\CL')$\,:}\quad \text{Weyl}(\CG_0') \backslash \text{Weyl}(G(\CK)) / \text{Weyl}(\CG_0)\,. \label{LLcharges} \ee

Now consider a space $\CM_{\rm rav}(\CB_\nu;\CL',\CL)$. In the algebraic formulation \eqref{Mrav-alg}, points of $\CM_{\rm rav}$ are labeled in part by singular gauge transformations $g(z)\in G[z, z^{-1}]$. We expect that a putative monopole operator $M_{w,A}$ of ``charge'' $(w,A)\in \text{Weyl}(G)\ltimes\text{cochar}(G) \simeq  \text{Weyl}(G(\CK))$ is represented by the fundamental class of a subvariety of $\CM_{\rm rav}(\CB_\nu;\CL',\CL)$ consisting of all points that are gauge-equivalent to a configuration with $g(z) = wz^A$. 
In other words, given the map that forgets the hypermultiplets
\footnote{To make contact with BFN-like constructions, we could further map $\CG_0[z]'\backslash   G[z, z^{-1}]/\CG_0[z]\hookrightarrow \CG_0[z]'\backslash   G(\CK)/\CG_0[z] \to\!\!\!\!\to \CG_0'\backslash   G(\CK)/\CG_0$. Then we can pull back classes from $\CG_0'\backslash   G(\CK)/\CG_0$, which is an equivariant affine Grassmannian or affine (partial) flag variety, depending on the choice of $\CG_0$}
\be \label{GrGz}  \CM_{\rm rav}(\CB_\nu;\CL',\CL)   \overset{\pi_g}\longrightarrow  \CG_0[z]'\backslash   G[z, z^{-1}]/\CG_0[z]\,, \ee
a monopole operator $M_{w,A}$ should correspond to the pullback
\be \label{mon-GrG} M_{w,A} \;\sim\; \pi_g^*[wz^A] \in H^*\big( \CM_{\rm rav}(\CB_\nu;\CL',\CL)\big)\,, \ee
where $[wz^A]$ is the fundamental class of the closure of the double-orbit $\CG_0[z]' \cdot wz^A \cdot \CG_0[z]$ in $\CG_0[z]'\backslash   G[z, z^{-1}]/\CG_0[z]$. 
Similarly, we expect ``dressed'' monopole operators corresponding to Chern classes of line bundles over the $wz^A$ orbit. 

Note that the formula \eqref{mon-GrG} does \emph{not} imply that a junction of line operators $\CL,\CL'$ will have monopole operators of all possible charges! It may well be that, for given $(w,A)$, the pull-back $\pi_g^*[wz^A]$ is zero. For example, this would happen if the condition imposed on hypermultiplet fields by $\CL_0'$ and $\CL_0$ made it impossible to have points with $g(z)$ in the orbit of $w,z^A$, in the space $\CM_{\rm rav}(\CB_\nu;\CL',\CL)$.

When we decompose $ \CM_{\rm rav} =\bigsqcup_{\n',\n}  \CM_{\rm rav}^{\n',\n}$ into components labeled by vortex numbers,  $\pi_g^*[wz^A]$ can only be supported on components with $\n'-\n$ equal to the topological type of $A$ (as a cycle in $\pi_1(G)$). We write this relation as $\n'-\n \sim A$, noting that it only depends on the $\text{Weyl}(G)\times \text{Weyl}(G)$ orbit of $w z^A$. Then we expect a dressed or undressed monopole operator of charge $(w,A)$ to be represented as a diagonal sum, schematically
\be M_{w,A} = \sum_{\n'-\n\sim  A} M_{w,A}^{\n',\n}\,,\qquad M_{w,A}^{\n',\n} \in H^*\big(\CM_{\rm rav}^{\n',\n}(\CB_\nu;\CL',\CL)\big) \label{mon-diag} \ee

\subsubsection{Idempotents}
\label{sec:idem}

We claimed in \eqref{Brav} that the map $\text{Hom}_A(\CL,\CL') \to H^*\big(\CM_{\rm rav}(\CB_\nu;\CL',\CL)\big)$ should always be surjective. This would let us relate any class in $H^*\big(\CM_{\rm rav}(\CB_\nu;\CL',\CL)\big)$ to a physical operator. The surjectivity statement requires a slightly technical modification when dealing with vacuum boundary conditions. Namely, due to the decomposition of $\CM_D$ into disjoint components $\CM_D^\n$, there are extra operators acting on (and among) the cohomologies $H^*\big( \CM_D^\n(\CB_\nu,\CL)\big)$, which have nothing to do with local operators at junctions of lines.
These extra operators are projections to summands of \eqref{MD-sum} with fixed $\n$.  If $\CL$ breaks gauge symmetry, so that vortex number is refined, then even more projections will appear.

Mathematically, these projections are ``orthogonal idempotents'' $e_\n$. They are represented as classes
\be e_\n = \pi^*[1] \cap H^*\big(\CM_{\rm rav}^{\n,\n}(\CB_\nu;\CL,\CL)\big) \ee
for each fixed $\n$, and they satisfy $e_\n e_{\n'} = \delta_{\n,\n'}e_\n$. 

There are two ways to correct the surjectivity statement \eqref{Brav} to account for these spurious operations. One option (\emph{cf.} \cite[Sec 4.4.1]{VV}) is to enhance $\text{Hom}_A(\CL,\CL')$ on the LHS, by throwing in all possible idempotents, acting by multiplication on both the left and right. With this enhancement of the LHS, surjectivity should be regained.

Alternatively, we may focus on operators in $H^*\big(\CM_{\rm rav}(\CB_\nu;\CL',\CL)\big)$ that act in a way that is independent of decomposition by vortex number. (In particular, we want operators whose convolution products are independent of vortex number.)
 We would expect such operators to come from actual elements of $\text{Hom}_A(\CL,\CL')$. They will necessarily be represented as infinite diagonal sums over graded components $H^*\big(\CM_{\rm rav}^{\n',\n}(\CB_\nu;\CL,\CL)\big)$, just like in \eqref{mon-diag}.

\subsection{Equivariant intersection cohomology}
\label{sec:cohomologies}

We have been vague so far about precisely \emph{which} cohomology theory we should be using to compute $H^*(\CM_D)$ and $H^*(\CM_{\rm rav})$. The moduli spaces in question split into finite-dimensional components, which helps. However, the components are typically noncompact and singular.
Physically, we should ask ourselves how to interpret supersymmetric quantum mechanics with a noncompact and singular target space. We consider these potential problems one at a time.

To handle noncompactness, we introduce equivarance. This standard technique is familiar from classic work on localization of supersymmetric path integrals \cite{AtiyahJeffrey, CMR, BT-2d, MNS, Nek-Omega}. 
Given A-type SQM with a Riemannian target $\CX$, and an abelian isometry group $T$ that acts on $\CX$ (a flavor symmetry), one may turn on twisted masses for $T$. The twisted masses $m$ take values in the complexified Lie algebra $\mathfrak t_\C$. Physically, they introduce a scalar potential $|m V|^2$, where $V\in \mathfrak t^*\otimes T\CX$ is the vector field generating the $T$ action. For generic $m$, the potential localizes physical wavefunctions to a neighborhood of the fixed locus of $T$. In particular, if $\CX$ happens to be noncompact but the $T$ action has a compact fixed locus, low-energy wavefunctions will decay exponentially near infinity. The SUSY ground states are well defined, and become identified with classes in $T$-equivariant cohomology.

In the case at hand, we introduce
\begin{itemize}
\item[1)] Twisted masses $m_\C$ for a torus of the Higgs-branch flavor symmetry $T_F$ that preserves the massive vacuum $\nu$; these twisted masses are the complex masses of the bulk 3d $\CN=4$ theory.
\item[2)] A twisted mass $\varepsilon$ for the diagonal $U(1)_\varepsilon$ subgroup of $U(1)_E\times U(1)_H$ that includes rotations in the $z$-plane and leaves the supercharge $Q_A$ invariant. As discussed in Section \ref{sec:Wilson-Omega}, this amounts to turning on the A-type Omega background.
\end{itemize}
Then we work with the equivariant cohomologies
\be  \label{H-eq} H^*_{T_F\times U(1)_\varepsilon}\big(\CM_D(\CB_\nu,\CL)\big)\,,\qquad H^*_{T_F\times U(1)_\varepsilon}\big(\CM_{\rm rav}(\CB_\nu;\CL',\CL)\big)\,. \ee
In the examples we study, the $T_F\times U(1)_\varepsilon$ action on $\CM_D$ and $\CM_{\rm rav}$ spaces actually has isolated fixed points. Then, by localization \cite{AtiyahBott, GKM} we will be able to describe the full equivariant cohomology in terms of appropriate linear combinations of fixed-point classes, \emph{i.e.} delta-function forms at the fixed points. We also note that our moduli spaces $\CM_D,\CM_{\rm rav}$ are K\"ahler, and the $T_F\times U(1)_\varepsilon$ metric isometry extends holomorphically to a complex isometry $T_{F,\C}\times \C^*\varepsilon$, with the same fixed-point set. In an algebraic context, we will consider $H^*_{T_{F,\C}\times \C^*_\varepsilon}\big(\CM_D(\CB_\nu,\CL)\big)$, $H^*_{T_{F,\C}\times \C^*_\varepsilon}\big(\CM_{\rm rav}(\CB_\nu;\CL',\CL)\big)$ instead of \eqref{H-eq}; however, the two are completely equivalent.

Mathematically, the spaces \eqref{H-eq} are modules for the polynomial algebra $\C[m_\C,\varepsilon]$ of equivariant parameters. Since the $T_F\times U(1)_\varepsilon$ (or $T_{F,\C}\times \C^*_\varepsilon$) action has fixed points, they are free modules: no constraints are imposed on $m_\C,\varepsilon$. Moreover, physically, $m_\C$ and $\varepsilon$ are fixed complex numbers, so $\C[m_\C,\varepsilon]\simeq \C$. Thus, there is no interesting structure in the $\C[m_\C,\varepsilon]$ action, and we will usually leave it implicit.

We may go a step further. Since the disc moduli spaces take the form $\CM_D = \CG_0[z]\backslash \wt \CM_D$, there is an equivalence with $\CG_0[z]$-equivariant cohomology of $\wt \CM_D$,
\be \label{H-eq-phi} H^*_{T_{F,\C}\times \C^*_\varepsilon}\big(\CM_D(\CB_\nu,\CL)\big) \simeq H^*_{\CG_0[z]\times T_{F,\C}\rtimes \C^*_\varepsilon}\big(\wt \CM_D(\CB_\nu,\CL)\big)\,. \ee
The RHS is naturally a module for $\C[\varphi]^{\text{Weyl}(\CL)}$, the polynomial algebra in equivariant parameters for the constant gauge transformations in $\CG_0[z]$, invariant under the part of the Weyl group preserved by $\CL$. Physically, the $\varphi$'s are the bulk vectormultiplet scalars. The $\CG_0[z]$ action on $\wt \CM_D(\CB_\nu,\CL)$ is free (it does \emph{not} have fixed points), so the corresponding action of $\C[\varphi]^{\text{Weyl}(\CL)}$ on equivariant cohomology is interesting. It is literally the action of the Coulomb-branch $\varphi$ operators in the disc Hilbert space.
 Roughly, we will find that the $\varphi$'s act on each component $H^*_{\CG_0[z]\times T_F \rtimes U(1)_\varepsilon}\big(\wt \CM_D^\n(\CB_\nu,\CL)\big)$ by measuring the vortex number $\n$, generalizing an analogous structure from \cite{VV}.

Similarly, raviolo spaces are of the form $\CM_{\rm rav} = \CG_0[z]'\backslash \wt\CM_{\rm rav}/\CG_0[z]$, so equivariant cohomology can be lifted
\be H^*_{T_{F,\C}\times \C^*_\varepsilon}\big(\CM_{\rm rav}(\CB_\nu;\CL',\CL)\big) \simeq H^*_{\CG_0[z]'\times \CG_0[z]\times T_{F,\C}\rtimes \C^*_\varepsilon}\big(\wt\CM_{\rm rav}(\CB_\nu;\CL',\CL)\big)\,. \ee
The RHS is a module for $\C[\varphi',\varphi]^{\text{Weyl}(\CL')\times \text{Weyl}(\CL)}$, where $\varphi'$ and $\varphi$ are the vectormultiplet scalars acting above and below the junction. Neither $\CG_0[z]'$ nor $\CG_0[z]$ has fixed points, so both $\varphi'$ and $\varphi$ are set to constants. The values of $\varphi',\varphi$ are related to $\n',\n$ on the component $H^*_{\CG_0[z]'\times \CG_0[z]\times T_{F,\C}\rtimes \C^*_\varepsilon}\big(\wt\CM_{\rm rav}^{\n',\n}\big)$, and $\varphi'-\varphi$ measures the monopole charge of local operators at the junction.

Equivariant cohomology takes care of noncompactness. However, we must still deal with the fact that the moduli spaces $\CM_D$ and $\CM_{\rm rav}$ may be singular. In \cite{VV}, this issue was deftly avoided, because (in the examples studied there) the bulk Coulomb-branch chiral ring $\text{End}_A(\id)$ was generated by monopole operators of minuscule charge, which could be captured by $\CM^{\n',\n}_{\rm rav}$ spaces for very small $\n',\n$, which turned out to be smooth. In the presence of nontrivial line operators, the spaces $\CM^{\n',\n}_{\rm rav}$ are almost \emph{never} smooth, even for minuscule monopole charge. So there is a genuine and practical difficulty to overcome. 

The singularities of $\CM_{\rm rav}$ are an artefact of our simplifications from Section \ref{sec:QM-Hom}. In particular, they arise from restricting to solutions of BPS equations in Step 2, which propagates to the definition of $\CM_{\rm rav}$ as a fiber product \eqref{Mfiber}. A physically rigorous analysis would return to the very-infinitely-dimensional space of \emph{all} field configurations in the presence of a monopole singularity, and then impose BPS equations by turning on a suitable potential.
We will shortcut such an analysis with a well-motivated guess: when $\CM_{\rm rav}$ or $\CM_D$ spaces turn out to be singular, we will take their equivariant \emph{intersection cohomology}.

The guess is motivated in part by the relation between intersection cohomology and $L^2$ cohomology \cite{Zucker, Goresky}. More practically, mathematical constructions using intersection cohomology have been known to match physical expectations in many setups similar to ours. This includes the identification of the Satake category \cite{MirkovicVilonen} (generated by intersection cohomology sheaves) with 't Hooft lines of 4d super-Yang-Mills \cite[Sec. 10]{KapustinWitten}. More directly: for particular classes of 3d $\CN=4$ theories whose bulk Coulomb-branch chiral rings are expected to be finite W-algebras, the spaces $\CM_{\rav}(\CB_\nu;\id,\id)$ appeared in \cite{Braverman-W, BFFR}; it was shown there that their equivariant cohomology reproduces the desired W-algebras. (This mathematical work was an important inspiration for \cite{VV}.)

Motivated in particular by \cite{Braverman-W, BFFR}, we will use equivariant intersection cohomology whenever we encounter singular spaces. This is how all cohomologies \eqref{MD-sum}, \eqref{Mrav-sum} in this paper are to be interpreted. Important examples of spaces with unavoidable singularities will appear in Section \ref{sec:conifold}.

There is another option for handling singularities and noncompactness, which might be deemed equally reasonable from a physical perspective: instead of equivariant intersection cohomology, one might use equivariant Borel-Moore homology. This is a topological homology theory that is Poincar\'e-dual to cohomology with compact support; a thorough review, relevant for convolution constructions, is contained in \cite{ChrissGinzburg}. Notably, Borel-Moore homology was used in the Braverman-Finkelberg-Nakajima constructions of Coulomb-branch chiral rings \cite{Nak, BFNII}.

In our actual examples, we will only encounter relatively mild singularities, modeled locally on transverse intersections such as $\{xy=0\}\subset \C^2$. Equivariant intersection cohomology and equivariant Borel-Moore homology give exactly the same answers in these cases. (Both are computed using a normalization of the singularity, \emph{e.g.} pulling $\{xy=0\}$ apart to $\{x=0\}\sqcup \{y=0\}$.)  Thus, so far, both seem equally good for matching physical expectations. Strictly speaking, the pull-back maps \eqref{mon-GrG} and infinite sums \eqref{mon-diag} that appear in the representations of monopole operators only make sense in Borel-Moore homology, so the latter may well be a better mathematical model to use in the future.

\section{Line operators in abelian theories}
\label{sec:abel}

We begin to describe examples of A-type and B-type line operators, in increasingly interesting theories. In this section, we focus on abelian 3d $\CN=4$ gauge theories, with
\be \hspace{.5in} G=U(1)^r\,,\qquad R = \C^N\quad\text{($N$ hypers)}\,. \ee

We will start by looking at the A and B twists of a free hypermultiplet ($r=0$, $N=1$), and its 3d mirror, which is SQED with a hypermultiplet ($r=1$, $N=1$). This gives us the simplest examples of nontrivial A-type and B-type line operators, and we check that our prescriptions for local operators at their junctions are consistent with 3d mirror symmetry.
We will also encounter ``flavor'' Wilson lines and ``screenable'' vortex lines, exchanged by mirror symmetry, which are both trivial in flat space.

Then in Section \ref{sec:tsu2}, we will consider SQED with two hypermultiplets ($r=1$, $N=2$). This theory, also known as $T[SU(2)]$ \cite{GW-Sduality}, is self-3d-mirror. 
We will again describe how line operators and their junctions map across mirror symmetry. 
We will then use this example to make contact with work of Assel and Gomis \cite{AsselGomis}, involving quivers and brane constructions. We will need to extend the proposal of  \cite{AsselGomis} slightly, by adding higher-derivative couplings to superpotentials.

In Section \ref{sec:gen-abel} we will outline the mirror map of half-BPS line operators and their junctions in general abelian theories, extending and complementing the work of \cite{HKT, TongWong}. We follow the philosophy of Kapustin and Strassler \cite{KapustinStrassler}, that any abelian mirror symmetry breaks up into iterations of the basic $r=0,N=1$ $\leftrightarrow$ $r=1,N=1$ duality.

Finally, in Section \ref{sec:flavor}, we include some general remarks about ``flavor vortex'' lines, \emph{i.e.} A-type line operators created by a singular gauge transformation for a flavor symmetry. 
 In abelian theories, all vortex lines appear as flavor vortices, making them particularly relevant. However, they play an important role in nonabelian theories as well, and were studied extensively in the mathematical works \cite{BFNII, BFN-ring, BFN-lines}, and discussed in the lectures \cite{BZ-lectures}.

\subsection{Basic mirror symmetry}
\label{sec:basic}

Let $\CT_{\rm hyper}$ be the theory of a free hypermultiplet ($G=1$, $R=\C$), and let $\SQED_1$ be a $G=U(1)$ gauge theory with a hypermultiplet $R=\C$ of charge +1. We will consider the A and B twists of these theories in turn, and confirm the expected mirror map of line operators.

\subsubsection{Free hyper}
\label{sec:hyperAB}

Let's first consider the B-twist. Since there is no gauge group, the only standard half-BPS Wilson line is the trivial line operator $\id \simeq \W_\C$. The endomorphisms of the trivial line are polynomials in the complex hypermultiplet scalars $X,Y$ --- \emph{i.e.} element of the bulk Higgs-branch chiral ring --- which get quantized to a Heisenberg algebra in the presence of Omega background:
\be \text{End}_B(\id) = \C[X,Y] \;\text{(commutative)}\,,\qquad \text{End}_B^\varepsilon(\id) = \C[X,Y]/([X,Y]=\varepsilon)\,. \label{Heis} \ee

There are also B-type line operators corresponding to direct sums of the trivial line. Given any finite-dimensional vector space $V$ (which can be thought of as a trivial representation of $G=1$), there is an associated half-BPS Wilson line $\W_V$. However, since $G$ acts trivially, it is equivalent to 
\be \W_V \simeq V\otimes \id \simeq \id^{\oplus \text{dim}\,V}\,. \label{WV-triv} \ee
Its endomorphism algebra is just the algebra of matrices with polynomial entries,
\be \text{End}_B^{\varepsilon}(\W_V) = \text{End}(V)\otimes \big(\C[X,Y]/([X,Y]=\varepsilon)\big)\,. \ee
At junctions we find matrices $\text{Hom}_B^\varepsilon(\W_V,\W_{V'}) = \text{Hom}(V,V') \otimes \big(\C[X,Y]/([X,Y]=\varepsilon)\big)$, etc.

There do exist flavor Wilson lines, but as long as we work in flat spacetime $\C_z\times \R_t$ (or just work locally in the category of lines), they are all of the form \eqref{WV-triv}. Concretely, the full flavor symmetry is $F=USp(1)$, and we could take $V$ to be any representation of $F$, or any representation of a subgroup of $F$. We can define a flavor Wilson line $\W_V = \text{Hol}_{\R_t}(\rho_V(\CA_f))$ as in \eqref{def-W}, where $\CA_f$ is a background superconnection for the flavor symmetry. However, since $G$ acts trivially, we find $\W_V\simeq V\otimes \id$ in the category of B-type lines, with all morphisms consisting of matrices as above.%
\footnote{Partition functions involving flavor Wilson lines can be very slightly nontrivial. On a closed 3-manifold, one can introduce a nonzero background connection with nontrivial holonomies. Then the flavor Wilson lines measure these holonomies, and their insertion modifies partition functions by overall prefactors. Such prefactors appeared frequently in index computations of \cite{AsselGomis}.}

So far, it sounds like the category of B-type line operators $\CC_B$ is fairly boring. This is approximately correct. Since the theory $\CT_{\rm hyper}$ is a 3d $\CN=4$ sigma-model to the flat hyperk\"ahler target $T^*\C$, the complete category $\CC_B$ was identified by \cite{RW, KRS} as $\CC_B = D^b\text{Coh}(T^*\C)$. This category is generated by the structure sheaf $\id\simeq \CO_{T^*\C}$. It does contain more interesting objects, such as the structure sheaf of a subspace $\CO_\C$ or a skyscraper at the origin $\CO_0$. These sheaves can be constructed as complexes of $\id$ (that's what it means for $\id$ to generate); but they are not simple half-BPS Wilson lines themselves. Physically, $\CO_\C$ seems to corresponds to a wrapped $\CN=(2,2)$ boundary condition (of the type studied in \cite{BDGH}); and $\CO_0$ seems to correspond to a wrapped $\CN=(0,4)$ boundary condition (of the type studied in \cite{CostelloGaiotto} and very recently in \cite{Okazaki04}.). They will be further discussed in \cite{lineops}.

The A-twist of a free hypermultiplet is much richer in objects. A basic family of half-BPS vortex-line operators was described in detail in Section \ref{sec:VL-matter}. These are the operators
\be \V_k\,: \quad \CL_0 = z^{-k} \CO \oplus z^{k}\CO \,\subseteq \, T^*\C(\CK)\,,\qquad k\in \Z\,, \label{hyper-V} \ee
with a pole of order $k$ in $X(z)$ and a zero of order $k$ in $Y(z)$, or vice versa, depending on the sign of $k$. Also recall that $\V_0 = \id$ is the trivial line.

We can use the methods of Section \ref{sec:comp} to find local operators at junctions. The computation looks a little trivial, but we will see using mirror symmetry that the result is actually correct!

To apply Section \ref{sec:comp} we should choose a massive vacuum $\nu$. There is only one. The free hypermultiplet admits a complex mass deformation by $m_\C \in \mathfrak f_{\C}$ in the Cartan of the flavor symmetry $F = USp(1)$. The corresponding torus $U(1)_F\subset F$ rotates $X,Y$ with charges +1,-1; and the massive vacuum is at the fixed point $(X,Y)=(0,0)$ of this rotation.

Having fixed the vacuum $\nu$ at $z\to \infty$, the moduli spaces
\be \begin{array}{rl} \CM_D(\CB_\nu,\V_k) &= \{(X,Y) = (0,0)\}\,,\\[.2cm]
 \CM_{\rm rav}(\CB_\nu;\V_{k'},\V_{k}) &=  \{(X',Y')=(X,Y) = (0,0), \, g=1\} \end{array} \ee
are all just isolated points. Taking cohomology, we find that the Hilbert spaces
\be \CH(\CB_\nu,\V_k) = \C \ee
are all one-dimensional, as are the spaces of local operators at junctions
\be H^*\big( \CM_{\rm rav}(\CB_\nu;\V_{k'},\V_{k})\big) = \C\,. \label{hyper-A-ops}\ee
Recall form Section \ref{sec:comp} that, in general, \eqref{hyper-A-ops} is only guaranteed to be a representation of the actual space $\text{Hom}_A(\V_k,\V_{k'})$ of local operators at a junction of $\V_k$ and $\V_{k'}$. In this example (and in fact in all abelian theories), 3d mirror symmetry will confirm that the map $\text{Hom}_A(\V_k,\V_{k'})\to\hspace{-.3cm}\to H^*\big( \CM_{\rm rav}(\CB_\nu;\V_{k'},\V_{k})\big)$ is an isomorphism. For now we simply claim that
\be \text{Hom}_A(\V_k,\V_{k'}) = \C\quad (\text{any $k,k'\in \Z$}) \label{hyper-A-ops2} \ee

A special case of \eqref{hyper-A-ops2} should be unsurprising. Namely, for $k=k'=0$ we expect
\be  \text{End}_A(\V_0,\V_0) = \text{End}_A(\id,\id) \simeq \C[\CM_C] \ee
to be the bulk Coulomb-branch chiral ring. However, since the gauge group $G=1$ is trivial, the only operator in the chiral ring is the identity, which generates the `$\C$' appearing in \eqref{hyper-A-ops2}.

Finally, we note that \eqref{hyper-A-ops2} encodes an important qualitative statement about half-BPS vortex lines in the free hypermultiplet theory. Namely, since $\text{Hom}_A(\V_k,\id)$ and $\text{Hom}_A(\id,\V_k)$ are nonzero,  all vortex lines can end! More generally, any two vortex lines can be joined together, in a unique way. 

Just as in the B-twist, we expect the complete category of line operators in the A-twist to contain more exotic objects, which preserve $Q_A$ but are not easily described as half-BPS vortex lines. Some of these exotic objects are already known, in a different guise. The full category $\CC_A$ for a free hypermultiplet, mentioned in \eqref{cat-free}, is equivalent to modules for a beta-gamma VOA \cite{CostelloGaiotto, CCG}. This is a very rich category (\emph{cf.} the recent \cite{Creutzig-braid}), which contains the simple objects $\V_k$. In VOA terms, $\V_0$ is the vacuum module and the $\V_k$ are spectral-flow modules thereof.

\subsubsection{B-twist of SQED$_1$}

We expect the categories of line operators in the B and A twists of a free hypermultiplet to match the categories in the A and B twists of $\SQED_1$,
\be \CC_B(\text{hyper}) \simeq \CC_A(\SQED_1)\,,\qquad \CC_A(\text{hyper}) \simeq \CC_B(\SQED_1)\,. \ee
We will check this match for the half-BPS objects and Hom spaces computed above.

$\SQED_1$ is a $G=U(1)$ gauge theory with hypermultiplet scalars $(X,Y)\in T^*\C$ of charges $(+1,-1)$. In the B-twist, there are basic half-BPS Wilson-line operators $\W_k$, $k\in \Z$, each associated to the 1-dimensional representation of $U(1)$ of charge $k$. We expect these Wilson lines to be the mirrors of the vortex lines $\V_k$ \eqref{hyper-V} for the free hypermultiplet.

Given two Wilson lines $\W_k$, $\W_{k'}$, the local operators at their junction are polynomials in the hypermultiplet scalars $X,Y$ whose total charge is equal to $k'-k$. But for any $k,k'$, there is exactly one such polynomial:
\begin{align} \text{Hom}_B(\W_{k},\W_{k'}) &= \C[X,Y]\big|_{\text{charge}=k'-k}   \notag \\
 &= \begin{cases} \C\langle X^{k'-k} \rangle & k'-k \geq 0 \\
  \C\langle Y^{k-k'} \rangle & k'-k < 0 \end{cases} \notag \\
 & \simeq \C\,, \label{Hom-W-SQED} \end{align}
perfectly consistent with \eqref{hyper-A-ops2}. In particular, every Wilson line can end, because there are hypermultiplet operators of the right charge to make the endpoint gauge-invariant.

As a special case, we find that $\text{Hom}_B(\id,\id) = \C\langle 1\rangle \simeq \C[\CM_H]$ reproduces the Higgs-branch chiral ring, which is trivial in $\SQED_1$. Indeed, the Higgs branch is the hyperk\"ahler quotient $T^*\C/\!/\!/U(1)$, which is a point.

\subsubsection{A-twist of SQED$_1$}
\label{sec:SQED1-VL}

We already described the basic A-type line operators $\SQED_1$ in Section \ref{sec:VL-abel}. We found a one-parameter family of vortex-line operators $\V_k$ with unbroken gauge symmetry $\CG_0=G(\CO)$ and $\CL_0  = z^{-k}\CO\oplus z^{k}\CO$ exactly as in \eqref{hyper-V}. Moreover, we claimed that $\V_k\simeq \id$ for all $k$, because each of these line operators can be screened by dynamical vortices.

So far, this nicely matches Wilson lines in the B-twist of a free hyper. The vortex line $\V_k$ in $\SQED_1$ is mirror to a flavor Wilson line $\W_k$ of charge $k$ for the torus $U(1)\subset F$ of the flavor symmetry. But $\V_k\simeq \id$ due to screening, while $\W_k\simeq \id$ because it has no gauge charge.

Just as in the B-twist of a free hypermultiplet, the A-twist of $\SQED_1$ does have more exotic line operators that are not easily described as vortex lines. For example, we might expect a line operator that breaks the group $G(\CO)\to 1$ completely. This would come from a wrapped Dirichlet boundary condition. See \cite{lineops} for further details.

We will now proceed to derive the algebras $\text{End}_A(\V_k)$ of local operators bound to the basic vortex lines in $\SQED_1$. This computation mildly generalizes results of \cite[Sec 3.4, 4.1]{VV}.
We go through it partly to remind the reader of some technical aspects (which will later be suppressed).
We will find that $\text{End}_A(\V_k)$ recovers the Coulomb-branch chiral ring for any $k$, and will see explicitly the isomorphism $\V_k\simeq \V_0=\id$ as objects in the category $\CC_A$.

We choose a massive vacuum $\nu$ given in algebraic terms by $(X,Y)=(1,0)$. Note that this satisfies the various requirements from Section \ref{sec:Mrav}; in particular, the complex moment $\mu=XY$ vanishes, and the stabilizer of $\nu$ under the complexified $\C^*$ gauge group is trivial. Physically, we find $\nu$ by first deforming the theory with a real FI parameter $t_\R<0$, so that the real moment-map constraint $\mu_\R =|X|^2-|Y|^2 +t_\R =0$ forces $X\neq 0$, thus breaking $U(1)$ gauge symmetry and ensuring that the Coulomb branch will be massive. After quotienting by $U(1)$, the physical vacuum can be described as $(X,Y)=(|t_\R|^{\frac12},0)$. It lies in the complexified $\C^*$ orbit of the algebraic vacuum $(X,Y)=(1,0)$.

To find the algebraic moduli space \eqref{MD-alg} of solutions to BPS equations on the disc, we require $(X(z),Y(z))$ to be in the intersection of $G(\CK_\infty)\cdot\text{val}_\infty{}^{-1}(\nu)$ and $\CL_0$. This forces $Y\equiv 0$, and restricts $X(z)$ to be $z^{-k}$ times a polynomial in $z$. We then quotient by the action of $\CG_0 \cap G(\C[z]) = G(\C[z])$. However, this group of invertible polynomials in $z$ is very simple: the only invertible polynomials are constants, so $G(\C[z])=\C^*$. (These are literally the constant gauge transformations on the disc.) Thus,
\be \CM_D(\CB_\nu;\V_k) = \C^* \big\backslash \{z^{-k}\C[z]\}\,. \ee
We may further decompose $\CM_D(\CB_\nu;\V_k) = \bigsqcup_\n \CM_D^\n(\CB_\nu;\V_k)$, where the connected components of fixed vortex number $\n\in \Z$ are polynomials of top degree $\n$,
\be \label{SQED1-MDn}  \CM_D^\n(\CB_\nu,\V_k) = \C^* \big\backslash \{X(z)=x_\n z^\n + x_{\n-1}z^{\n-1}+\ldots + x_{-k} z^{-k},\; x_\n\neq 0\} \simeq \C^{\n+k}\,. \ee
The $\C^*$ action can be used to fix $x_\n=1$, whence each component is simply an affine space $\C^{\n+k}$, with coordinates $x_{\n-1},...,x_{-k}$.
Note that $\CM_D^\n$ will be empty unless $\n\geq -k$.

The cohomology of each component \eqref{SQED1-MDn} is just $\C$. Therefore, the $Q_A$-cohomology of the Hilbert space on the disc becomes
\be \CH_D(\CB_\nu,\V_k) = \bigoplus_{\n\geq -k} H^*\big(\CM_D^\n(\CB_\nu,\V_k)\big) \simeq \bigoplus_{\n\geq -k} \C\,. \ee
This is not a terribly enlightening description, and we can do better by working equivariantly.

Let us write $\CM_D^\n(\CB_\nu,\V_k) = \C^*\backslash \wt \CM_D^{\n,k}$, where $\wt \CM_D^{\n,k} = \{X(z)=x_\n z^\n + x_{\n-1}z^{\n-1}+\ldots + x_{-k} z^{-k},\; x_\n\neq 0\}$ as in \eqref{SQED1-MDn}. We will also turn on an Omega background, \ie\ work equivariantly with respect to the $\C^*_\varepsilon$ action of loop rotations. We consider the equivariant cohomology $H^*_{\C^*_\varepsilon}\big(\CM_D^\n(\CB_\nu,\V_k)\big) = H^*_{\C^*\times \C^*_\varepsilon}(\wt \CM_D^{\n,k}) $. Mathematically, this space must be a module for
\be H^*_{\C^*\times \C^*_\varepsilon}(pt) = \C[\varphi,\varepsilon]\,. \ee
Physically, $\varepsilon$ is some fixed complex number (the parameter of the Omega background), so we will not distinguish between polynomials in $\varepsilon$ and constants, \emph{i.e.} $\C[\varepsilon]\simeq \C$.
On the other hand, $\varphi$ is a bulk local operator corresponding to the complex vectormultiplet field, and the structure of equivariant cohomology as a $\C[\varphi]$ module is important.

$H^*_{\C^*\times \C^*_\varepsilon}(\wt\CM_D^{\n,k})$ can easily be computed by localization \cite{AtiyahBott, GKM}. It is generated by a single fixed-point class $|\n\rangle$, associated to the $\C^*_\varepsilon$ fixed point $\{X(z) = z^\n\} \in \CM_D^\n(\CB_\nu,\V_k)$ (or to its $\C^*$ orbit $\{X(z) = x_\n z^\n\}\subset \wt \CM_D^{\n,k}$).
Explicitly, $\C^*_\varepsilon$ acts on $z$ with weight $-1$ and on both $X,Y$ with weight $\frac12$ (due to mixing with the $U(1)_H$ R-symmetry). 
Under an infinitesimal $\C^*\times \C^*_\varepsilon$ rotation with parameters $\varphi,\varepsilon$, the coefficient $x_\n$ transforms as
\be \delta\, x_\n = \big[\varphi+(\n+\tfrac12)\varepsilon\big]x_\n\,, \ee
and is thus fixed under a subgroup of $\C^*\times \C^*_\varepsilon$ with $\varphi =-(\n+\frac12)\varepsilon$. This tells us that
\be \CH_D^\varepsilon(\CB_\nu,\V_k) =  \bigoplus_{\n\geq -k} H_{\C^*\times\C^*_\varepsilon}^*\big(\wt \CM_D^{\n,k}\big) \simeq \bigoplus_{\n\geq -k} \C\,|\n\rangle \ee
with an action of Coulomb-branch scalars given by
\be \varphi |\n\rangle = -(\n+\tfrac12)\varepsilon|\n\rangle\,. \label{phiSQED1} \ee
Thus, the local operator $\varphi$ acts on the disc Hilbert space by measuring vortex number.

Throughout the examples in this paper, we will normalize equivariant cohomology classes $|p\rangle$ corresponding to fixed points $p\in \CM$ as
\be \label{norm-p} |p\rangle = \frac{1}{\omega_{p}}\delta_{p}\,,\qquad \delta_{p} :=  i_*(\mb 1_{p})\,, \ee
where $\mb 1_{p}$ is the fundamental class of an isolated point $p$, $i:p\hookrightarrow \CM$ is the inclusion of the fixed point, and $\omega_p=e(N_p)$ is the Euler class of the normal bundle to $p$ in $\CM$.
The Euler class is a product of equivariant weights of $N_p$.
In the present case, for the fixed point $p=z^\n$, the coordinates on $N_p$ are the remaining coefficients $x_{\n-1},x_{\n-2},...,x_{-k}$ of $X(z)$, whence
\begin{align} \omega_{z^\n} &= (\varphi+(\n-\tfrac12)\varepsilon)(\varphi+(\n-\tfrac32)\varepsilon)\cdots (\varphi+(\tfrac12-k)\varepsilon) \\
 &= (-1)^{\n+k}\varepsilon^{\n+k}(\n+k)! \notag\end{align}

Now, the local operators bound to $\V_k$ come from cohomology classes in the raviolo space $\CM_{\rm rav}(\CB_\nu;\V_k,\V_k)$. Specializing \eqref{Mrav-alg}, we find that
\be \CM_{\rm rav}(\CB_\nu;\V_k,\V_k) = \bigsqcup_{\n',\n\geq-k} \CM_{\text{rav},k}^{\n',\n}\,, \ee
with each component given by $\CM_{\text{rav},k}^{\n',\n} = \C^*{}'\big\backslash \wt \CM_{\text{rav},k}^{\n',\n} \big/\C^*$  with
\be \wt \CM_{\text{rav},k}^{\n',\n}  = \begin{cases} \{ X(z)'=g(z)X(z),\; g(z)=\alpha z^{\n'-\n},\;
   X(z) = x_{\n}z^{\n}+...+x_{-k}z^{-k} \} & \n'\geq \n \\[.2cm]
   \{ X(z)'=x_{\n'}z^{\n'}+...+x_{-k}'z^{-k},\; g(z)=\alpha z^{\n'-\n},\;
   X(z) = g(z)^{-1}X(z)' \} & \n' < \n \end{cases}
\ee
where $\alpha,x_\n,x_{\n'}\neq 0$. Again, there is a unique fixed-point class in equivariant cohomology
\be H^*_{\C^*{}'\times \C^*\times \C^*_\varepsilon}\big(\wt \CM_{\text{rav},k}^{\n',\n} \big) = \C|\n',\n\rangle \ee
associated to the $\C^*{}'\times \C^*$ orbit of $\{X'=z^{\n'},g=z^{\n'-\n},X=z^{\n}\}$. With the normalization \eqref{norm-p}, $|\n',\n\rangle = [\CM_{\text{rav},k}^{\n',\n}]$ is equivalent to the fundamental class of the entire raviolo space.
 This class represents an abelian monopole operator $v_{\n'-\n}$ of charge $\n'-\n$, acting on a state of vortex number $\n$ to produce a state of vortex number $\n'$.

As discussed in Section \ref{sec:idem}, the actual local operators bound to a line operator must be represented in a way that is independent of the vortex moduli spaces they act on. In this case, the monopole operator $v_n$ of charge $n$ is the diagonal sum
\be v_n\, = \sum_{\n'-\n=n} |\n',\n\rangle\,. \ee
A short calculation (\emph{cf.} \cite[Sec 4.1]{VV}) shows that the convolution action of $v_n$ on Hilbert spaces is given uniformly for all $n\geq \Z$ by
\be  \label{SQED-EndVk}  \begin{cases}
  v_n|\n\rangle = \prod_{i=1}^n (\varphi+(i-\tfrac12-k)\varepsilon)|\n+n\rangle & n \geq 0 \\
  v_{n}|\n\rangle = |\n+n\rangle &  n < 0\,,\quad  \n+n\geq -k \\
  v_{n}|\n\rangle = 0 & n < 0\,,\quad \n + n < -k \end{cases}
\ee
The prefactor $\prod_{i=1}^n (\varphi+(i-\tfrac12-k)\varepsilon)$ appearing in the action of positive monopole operators is a product of the equivariant weights of coefficients of $X'$ that are \emph{missing} when we set $X(z)' = z^n X(z)$; in other words, weights of the coefficients $x_{-k+n-1},...,x_{-k}$.

From the action \eqref{SQED-EndVk} we extract an algebra that is completely independent of vortex number, which we expect to be $\text{End}^\varepsilon_A(\V_k)$. The algebra is generated by just two monopole operators $v_1,v_{-1}$, subject to the simple relation $[v_1,v_{-1}]=\varepsilon$,
\be \label{SQED-EndVk-2}  \text{End}^\varepsilon_A(\V_k) \simeq \C[v_1,v_{-1}]/([v_1,v_{-1}]-\varepsilon)\,. \ee
The higher monopole operators and the scalar $\varphi$ are obtained from $v_{\pm 1}$ as
\be \label{SQED-vphi} v_n = \begin{cases} (v_1)^n & n\geq 0 \\
 (v_{-1})^{|n|}  & n < 0 \end{cases}\,,\qquad \varphi = v_1v_{-1}+(k-\tfrac12)\varepsilon = v_{-1}v_1 + (k+\tfrac12)\varepsilon\,. \ee
Notably, the scalar satisfies $[\varphi,v_{\pm}]=\mp \varepsilon v_{\pm 1}$, counting monopole number.

Written in the form \eqref{SQED-EndVk-2}, the algebra $\text{End}^\varepsilon_A(\V_k)$ is clearly isomorphic to the Heisenberg algebra $\C[X,Y]\big/([Y,X]-\varepsilon)$ that arose in the B-twist of a free hypermultiplet. In particular, for all $k$, $\text{End}^\varepsilon_A(\V_k) \simeq \text{End}^\varepsilon_A(\V_0=\id) \simeq \C_\varepsilon[\CM_C]$ is the Omega-deformed Coulomb-branch chiral ring. (The undeformed chiral ring of SQED was described in \cite{BKW-II} and follows from the Coulomb-branch geometry of \cite{SW-3d, IS}, while its deformation quantization appeared in \cite{BDG}.) The only (indirect) dependence on $k$ appears in the relation between $\varphi$ and $v_1v_{-1}$.  In particular, to relate $\text{End}_A^\varepsilon(\V_k)$ to $\text{End}_A^\varepsilon(\V_{k'})$, we simply send $\varphi\mapsto \varphi-(k'-k)\varepsilon$. This is precisely the effect of screening a vortex line of charge $k$ by a dynamical vortex of charge $k'-k$.

To describe the isomorphism $\V_k\simeq \V_0$ (say) more directly in the category of line operators, we should look for local operators at the junction of $\V_k$ and $\V_0$ (and vice versa) that implement it. Consider the raviolo space
\be \CM_{\rm rav}^{\n-k,\n}(\CB_\nu;\V_k,\V_0) \simeq \big\{X'=z^{\n-k}+x_{\n-1}z^{\n-k-1}+...+x_{0}z^{-k},\, g=z^{-k},\, X=z^{\n}+x_{\n-1}z^{\n-1}+...+x_{0}\big\} \ee
There are \emph{no} missing coefficients in either $X$ or $X'$; thus by forgetting $X',g$ the space maps bijectively to  $\CM_D^\n(\CB_\nu,\V_0)$, and by forgetting $X,g$ the space maps bijectively to $\CM_D^{\n-k}(\CB_\nu,\V_k)$.  An analogous statement holds for $\CM_{\rm rav}^{\n,\n-k}(\CB_\nu;\V_0,\V_k)$. The diagonal sums of fundamental classes of these spaces define $\n$-independent local operators
\be \begin{array}{c} \CO_k=\sum_{\n} \big[\CM_{\rm rav}^{\n-k,\n}(\CB_\nu;\V_k,\V_0)\big] \in \text{Hom}_A(\V_0,\V_k) \\[.2cm]
\wt \CO_k=\sum_{\n} \big[\CM_{\rm rav}^{\n,\n-k}(\CB_\nu;\V_0,\V_k)\big] \in \text{Hom}_A(\V_k,\V_0) \end{array} \ee
that are inverses in the sense that $\wt \CO_k\CO_k=\text{id}_{\V_0}$, $\CO_k\wt\CO_k=\text{id}_{\V_k}$. They explicitly implement the isomorphisms of objects
\be  \CO_k:\V_0\overset{\sim}\to\V_k\,,\qquad \wt\CO_k:\V_k \overset{\sim}\to \V_0\,. \ee
By more carefully looking at the right and left $\C^*$ gauge actions on $\CM_{\rm rav}^{\n-k,\n}(\CB_\nu;\V_k,\V_0)$ (which we already used to fix $x_\n=1$ above), we also find that $\varphi\CO_k = \CO_k(\varphi-k\varepsilon)$ and $\varphi\wt\CO_k = \wt\CO_k(\varphi+k\varepsilon)$, thus recovering the relation between $\varphi$ on $\V_k$ and $\varphi$ on $\V_0$.

\subsection{SQED$_2$}
\label{sec:tsu2}

We extend the analysis of the basic mirror pair above to $\SQED_2$, \emph{a.k.a.} $T[SU(2)]$. This theory has gauge group $G=U(1)$ and two hypermultiplets, \emph{i.e.} $R=T^*\C^2$ parameterized by $X=(X_1,X_2)^T$ of charge $+1$ and $Y=(Y^1,Y^2)$ of charge $-1$.

$\SQED_2$ is self-mirror, so we expect the categories of A-type and B-type lines to be (nontrivially) equivalent to each other. We begin by explaining how this happens. In the A-twist, there are basic vortex lines $\V_k$ labeled by a pair of integers $k=(k_1,k_2)$, with algebraic data
\be \V_k\,:\quad \CG_0 = G(\CO)\,,\qquad \CL_0 = \big\{ X \in (z^{-k_1}\CO,z^{-k_2}\CO)^T,\, Y\in (z^{k_1}\CO,z^{k_2}\CO)\big\}\,. \label{SQED2-Vk}\ee
However, screening implies that $\V_k$ and $\V_{k'}$ are equivalent if $k-k'\in (1,1)\Z$. Thus the nontrivial, inequivalent vortex lines are labeled by the quotient lattice
\be \label{SQED2-Vcharge} \frac{\Z^2}{(1,1)\Z} \simeq \Z\,. \ee

To give a dual description of Wilson lines, it is helpful to think of $G=U(1)$ as embedded in the maximal torus $\wt T=U(1)_1\times U(1)_2$ of the full isometry group $USp(2)$ acting on two hypermultiplets. Each $U(1)_a$ acts on $X_i$ with charge $\delta_{ai}$, and $G\subset \wt T$ is embedded as the diagonal, generated by cocharacter $\gamma=(1,1)$. $\SQED_2$ has a basic set of Wilson lines
\be \W_n\,:\quad  \text{1d rep $\C_n$ of $\wt T$ with charge $n=(n_1,n_2)$} \ee
labeled by the 1-dimensional representations of $\wt T$ with charges $n=(n_1,n_2)$.However, the Wilson lines with zero gauge charge --- meaning $\gamma\cdot n=0$ --- are flavor Wilson lines, and should be equivalent to the $\id$, just as they were in the theory of a free hypermultiplet.

More generally, we may make use of the monoidal/tensor structure in the category of Wilson lines, coming from collisions of line operators supported on parallel lines. Here we expect the tensor product of Wilson lines to simply correspond to tensoring representations, so that $\W_m\otimes \W_n = \W_{m+n}$.  Then, whenever $\gamma\cdot n=0$, $\W_{m+n} = \W_m\otimes \W_n \simeq \W_m\otimes \id  = \W_m$. The sublattice of $\wt T$ charges satisfying $\gamma\cdot n=0$ is generated by $(1,-1)$, so, altogether, the inequivalent Wilson lines are labeled by elements of the quotient lattice
\be \label{SQED2-Wcharge} \frac{\Z^2}{(1,-1)\Z} \simeq \Z\,\,. \ee
The lattices \eqref{SQED2-Vcharge} and \eqref{SQED2-Wcharge} are swapped under 3d mirror symmetry.

\subsubsection{Junctions of Wilson lines}

The general structure of local operators bound to Wilson lines and their junctions was laid out in Section \ref{sec:WL}.  The general prescription has a particularly simple manifestation in abelian theories, which we'd like to review here in the case of $\SQED_2$.

We turn on a B-type Omega background. Following Section \ref{sec:Wilson-Omega}, the algebra of bulk local operators $\text{End}_B^\varepsilon(\id)$ (the quantized Higgs-branch chiral ring) is a quantum symplectic reduction of a tensor product of two Heisenberg algebras, one for each hypermultiplet:
\be \text{End}_B^\varepsilon(\id) = \C[X_1,Y_1,X_2,Y_2]^{G}/(X_1Y_1+X_2Y_2-\varepsilon+t_\C)\,,  \label{SQED2-id-alg1} \ee
where we have left implicit the commutation relations $[X_i,Y_j]=\varepsilon \delta_{ij}$. Also note that $\mu_\C = X_1Y_1+X_2Y_2-\varepsilon = Y_1X_1+Y_2X_2+\varepsilon$ is the normal-ordered moment map, and we have introduced a complex FI parameter $t_\C$.

There is a simpler description of this algebra (see \cite[Sec 2]{BDGH}). Let $h=X_1Y_1-X_2Y_2$, $e=X_2Y_1$, and $f=X_1Y_2$. Then $e,f,h,$ and $\mu_\C$ generate the $G$-invariant subalgebra  $\C[ X, Y]^{G}\subset \C[X,Y]$. 
In abelian theories, the moment map $\mu_\C$ is always $G$-invariant, and is a central element of $\C[ X, Y]^{G}$.
After imposing $\mu_\C+t_\C=0$, we are left with
\be \text{End}_B^\varepsilon(\id)  = \C[e,f,h]/(*) \ee
with relations
\be 
\qquad
(*)\,:\quad \begin{array}{c}  [h,e]=2\varepsilon e\,,\qquad [h,f]=-2\varepsilon f\,,\qquad [e,f]=\varepsilon\,h \\[.2cm]
	ef+fe+\tfrac12 h^2=\tfrac12(t_\C^2-\varepsilon^2) \end{array}
\ee
This is just a central quotient of the enveloping algebra $U(\mathfrak{sl}_2)$, with the quadratic Casimir $C_2= ef+fe+\tfrac12 h^2$ set equal to $\tfrac12(t_\C^2-\varepsilon^2)$. The appearance of $\mathfrak{sl}_2$ is due to the full bulk flavor symmetry group $PSU(2)$, acting on the pair of hypermultiplets. The operators $e,f,h$ above are components of the moment map for this flavor symmetry.

On a Wilson line of charge $n=(n_1,n_2)$, the general prescription of Section \ref{sec:Wilson-Omega} implies that the algebra of local operators is modified to
\be \text{End}_B^\varepsilon(\W_n) = \C[X_1,Y_1,X_2,Y_2]^{G}/(X_1Y_1+X_2Y_2-\varepsilon+t_\C + \varepsilon (\gamma\cdot n))\,,  \label{SQED2-W-alg1}\ee
where $\gamma\cdot n=n_1-n_2$ is the gauge charge. Alternatively,
\be \text{End}_B^\varepsilon(\W_n) \simeq \C[e,f,h]/(*)\,,\qquad
 (*)\,:\quad \begin{array}{c}  [h,e]=2\varepsilon e\,,\qquad [h,f]=-2\varepsilon f\,,\qquad [e,f]=\varepsilon\,h \\[.2cm]
 ef+fe+\tfrac12 h^2=\tfrac12(t_\C+(\gamma\cdot n)\varepsilon)^2-\tfrac12\varepsilon^2 \end{array} \label{SQED2-W-alg2}\ee
 Note that modifying \eqref{SQED2-id-alg1} to \eqref{SQED2-W-alg1} is equivalent to shifting the complex FI parameter $t_\C \mapsto t_\C + \varepsilon(\gamma\cdot n)$.
This shift does \emph{not} imply an equivalence of Wilson lines for different $\gamma\cdot n$, since in a given 3d theory the bulk FI parameter $t_\C$ is fixed once and for all.%
\footnote{In contrast, the screening transformations $\varphi\mapsto \varphi-k\varepsilon$ from Section \ref{sec:SQED1-VL} \emph{were} isomorphisms, because $\varphi$ was a dynamical operator.} %
 However, it does give us an easy way to find any $\text{End}_B^\varepsilon(\W_n)$ once we know $\text{End}_B^\varepsilon(\id)$. 

In the special case that the bulk FI parameter is set to zero, each algebra $\text{End}_B^\varepsilon(\W_n)$ is isomorphic to a central quotient of $U(\mathfrak{sl}_2)$ with the Casimir fixed to $C_2=\tfrac12((\gamma\cdot n)^2-1)\varepsilon^2$. These quantized values of $C_2$ are precisely where interesting representations appear, such as finite-dimensional irreducibles. This played an important role in the physics of symplectic duality \cite{BDGH}.

Finally, we consider the junction of two different Wilson lines $\W_n$ and $\W_{n'}$. Now the local operators at the junction are not gauge-invariant, but must have gauge charge $\Delta n:= \gamma\cdot n'-\gamma \cdot n$. Abstractly, the prescription of Section \ref{sec:Wilson-Omega} gives
\begin{align} \label{SQED2-B-Hom} \text{Hom}^\varepsilon_B(\W_n,\W_{n'}) &= \C[X_1,Y_1,X_2,Y_2]^{\text{charge}=\Delta n} / (\mu_\C+t_\C+\varepsilon(\gamma\cdot n)) \\
 &\simeq  (\mu_\C+t_\C+\varepsilon(\gamma\cdot n'))\backslash \C[X_1,Y_1,X_2,Y_2]^{\text{charge}=\Delta n} \notag\,. \end{align}
When $\Delta n\neq 0$, the moment map $\mu_\C$ is no longer central, so we must be careful about quotienting by left vs. right ideals.

The vector space $\text{Hom}^\varepsilon_B(\W_n,\W_{n'})$ is naturally a bimodule for the two algebras $\text{End}_B^\varepsilon(\W_{n'})$ and $\text{End}_B^\varepsilon(\W_{n})$, acting on the left and right. While \eqref{SQED2-B-Hom} is infinite-dimensional, it is finitely generated as a bimodule. In other words, all operators at the junction can be obtained by starting from a finite set, and acting with arbitrary elements of $\text{End}_B^\varepsilon(\W_{n'})\times\text{End}_B^\varepsilon(\W_{n})$.
 It is illuminating to describe this structure from the perspective of $U(\mathfrak{sl}_2)$.
 
Let $\rho_{\Delta n}$ denote the $(|\Delta n|+1)$-dimensional vector space
\be \label{SQED2-rho} \rho_{\Delta n} := \begin{cases} \C\big\langle X_1^{\Delta n},\, X_1^{\Delta n-1}X_2,\,X_1^{\Delta n -2}X_2^2,..., X_2^{\Delta n}\big\rangle & \Delta n  \geq  0 \\
 \C\big\langle Y_1^{|\Delta n|},\, Y_1^{|\Delta n|-1}Y_2,\,Y_1^{|\Delta n| -2}Y_2^2,..., Y_2^{|\Delta n|}\big\rangle & \Delta n < 0\,, \end{cases}
\ee
which in each case contains local operators of charge $\Delta n$. Then it is easy to see that
\be \text{Hom}^\varepsilon_B(\W_n,\W_{n'}) \simeq \text{Hom}_B^\varepsilon(\W_{n'}) \cdot \rho_{\Delta_n} \cdot \text{Hom}_B^\varepsilon(\W_{n})\,. \ee
Moreover, each $\rho_{\Delta n}$ may be identified with an irreducible representation of $\mathfrak{sl}_2$, under the ``adjoint'' action coming from difference of left and right actions. For example, with $e=X_2Y_1$ and $f=X_1Y_2$ (the same identification holds in both the left and right algebras), we have
\be \begin{array}{c}  [e,Y_1^{\Delta n}] = 0\,,\quad [e,Y_1^{\Delta n-1}Y_2]=\varepsilon Y_1^{\Delta n}\,,\quad\ldots,\quad [e,Y_2^{\Delta n}] = \varepsilon(\Delta n)Y_1Y_2^{\Delta_n-1} \\[.2cm]
 [f,Y_1^{\Delta n}] = \varepsilon(\Delta n) Y_1^{\Delta n-1}Y_2\,,\quad [f, Y_1^{\Delta n-1}Y_2] = \varepsilon(\Delta n-1)Y_1^{\Delta n-2}Y_2^2\,,\quad \ldots,\quad [f,Y_2^{\Delta n}] = 0 \end{array}
\label{SQED2-Hom-rep} \ee
for $\Delta n < 0$, and similarly for the dual representation when $\Delta n > 0$. Thus, the local operators at a junction of Wilson lines are generated by the finite-dimensional $\mathfrak{sl}_2$ representation whose highest or lowest weight is the difference of Wilson-line charges.

We can actually simplify the presentation of Hom spaces a bit further.
All the vectors in $\rho_{\Delta n}$ can be obtained as in \eqref{SQED2-Hom-rep} by acting on a highest-weight vector $X_1^{\Delta n}$ or $Y_2^{|\Delta n|}$ with a difference of left and right $\mathfrak{sl}_2$'s. Thus, as a bimodule, the entire Hom space can be generated from a single local operator of charge $\Delta n$,
\be  \text{Hom}^\varepsilon_B(\W_n,\W_{n'}) \simeq \begin{cases} \text{Hom}_B^\varepsilon(\W_{n'}) \cdot X^{\Delta n} \cdot \text{Hom}_B^\varepsilon(\W_{n}) & \Delta n \geq 0 \\
\text{Hom}_B^\varepsilon(\W_{n'}) \cdot Y^{|\Delta n|} \cdot \text{Hom}_B^\varepsilon(\W_{n}) & \Delta n < 0\,. \end{cases} \ee

\subsubsection{Junctions of vortex lines}

We can derive the dual structures for vortex lines, in the A-twist, by using the methods of Section \ref{sec:comp}. We choose a massive vacuum whose algebraic description is
\be \nu\,:\quad (X_1,X_2;Y_1,Y_2) = (1,0;0,0)\,, \ee
and a corresponding asymptotic boundary condition $\CB_\nu$.

The bulk Coulomb-branch algebra $\text{End}^\varepsilon_A(\id) \simeq \C_\varepsilon[\CM_C]$ was constructed in \cite{VV}. It can be generalized to any vortex line $\V_k$ from \eqref{SQED2-Vk}, labeled by charges $k=(k_1,k_2)$, as follows. First, each disc moduli space $\CM_D(\CB_\nu,\V_k)$ has connected components
\be \CM_D^\n(\CB_\nu,\V_k) = \C^*\big\backslash \left\{ \bp X_1(z) \\ X_2(z) \ep = \bp x_{1,\n} z^\n + x_{1,\n-1}z^{\n-1} + ...+ x_{1,-k_1}z^{-k_1} \\
\phantom{x_{1,\n} z^\n +} x_{2,\n-1}z^{\n-1} + ...+ x_{2,-k_2}z^{-k_2} \ep\,,\; x_{1,\n}\neq 0 \right\}\,,
\ee
where the action of complexified, constant $\C^*$ gauge transformations can be used to fix ${x_{1,\n}=1}$. Thus, the moduli space is just an affine space $\CM_D^\n(\CB_\nu,\id)\simeq \C^{2\n+k_1+k_2}$, parameterized by the remaining coefficients.
It is empty unless $\n\geq -k_1$.
 Under loop rotation $\C^*_\varepsilon$,  there is a single fixed point at the origin, giving rise to a state $|\n\rangle$ in equivariant cohomology. Thus,
\be \CH(\CB_\nu,\V_k) \simeq \bigoplus_{\n} \C|\n\rangle\,, \ee
and the equivariant parameters $\varphi,\varepsilon$ for $\C^*\times \C^*_\varepsilon$ obey
\be (\varphi+(\n+\tfrac12)\varepsilon)|\n\rangle = 0\,. \label{SQED2-pe}\ee

A new feature that did not arise in the $\SQED_1$ example of Section \ref{sec:SQED1-VL} is that there is additional flavor symmetry. We may work equivariantly for the entire torus $\wt T = U(1)_1\times U(1)_2$, where each $U(1)_a$ acts on $X_i$ with charge $\delta_{ai}$. The gauge group is embedded as the diagonal, and we may identify a torus $U(1)_m$ of the flavor group with $U(1)_1$. (This is a choice.) In equivariant cohomology with respect to $\C^*_1\times \C^*_2\times \C^*_\varepsilon \simeq \C^*\times\C^*_{m}\times \C^*_\varepsilon$ there are now three equivariant parameters $\varphi,\, m_\C,\,\varepsilon$. The flavor parameter $m_\C$ is complex mass of the bulk theory. The relation \eqref{SQED2-pe} becomes $\big(\varphi+(n+\tfrac12)\varepsilon+m_\C\big)|\n\rangle=0$.

Next, for every pair $\n',\n$ there is a raviolo space
\be \CM_{\rm rav}^{\n',\n}(\CB_\nu,\V_k,\V_k) = \C^*\big\backslash \{X(z)',\,g(z) = \alpha z^{\n'-\n},\, X(z)\,,\; \alpha\neq 0 \}\big/\C^*\,,\ee
with
\be \notag X(z)' = g(z) X(z)\,,\quad X(z) =  \bp x_{1,\n} z^\n + x_{1,\n-1}z^{\n-1} + ...+ x_{1,-k_1}z^{-k_1} \\
\phantom{x_{1,\n} z^\n +} x_{2,\n-1}z^{\n-1} + ...+ x_{2,-k_2}z^{-k_2} \ep\,,\quad x_{1,\n}\neq 0 \ee
if $\n'-\n \geq 0$, and
\be\notag X(z)' =  \bp x_{1,\n'} z^{\n'} + x_{1,\n'-1}z^{\n'-1} + ...+ x_{1,-k_1}z^{-k_1} \\
\phantom{x_{1,\n'} z^{\n'} +} x_{2,\n'-1}z^{\n'-1} + ...+ x_{2,-k_2}z^{-k_2} \ep\,,\quad X(z) = g(z)^{-1} X(z)'\,,\quad x_{1,\n'}\neq 0 \ee
if $\n'-\n < 0$. The $\C^*\times \C^*$ gauge action can be used to fix $\alpha=1$ and either $x_{1,\n}=1$ or $x_{1,\n'}=1$ (depending on whether $\n'-\n \geq 0$ or $< 0$), showing that each $\CM_{\rm rav}^{\n',\n}(\CB_\nu,\V_k,\V_k)$ is also an affine space. Its equivariant cohomology contains a single fixed-point class, which is also the fundamental class $|\n',\n\rangle = \big[\CM_{\rm rav}^{\n',\n}(\CB_\nu,\V_k,\V_k)\big]$. Taking a diagonal sum, we identify monopole operators bound to $\V_k$ as
\be v_n = \sum_{\n'-\n=n} \big[\CM_{\rm rav}^{\n',\n}(\CB_\nu,\V_k,\V_k)\big]\,. \ee

A careful calculation of convolution products shows that $v_n = (v_1)^n$ if $n\geq 0$ and $v_n=(v_{-1})^{|n|}$ if $n<0$, and that the basic monopoles $v_{\pm 1}$ act on vortex states as
\be \text{on $\V_k$}\,:\quad \begin{array}{l}  v_1|\n\rangle = \big(\varphi-(k_1-\tfrac12)\varepsilon\big)\big(\varphi-(k_2-\tfrac12)\varepsilon+m_\C\big)|n+1\rangle \\[.1cm]
 v_{-1}|\n\rangle = |\n-1\rangle\,. \end{array} \ee
The prefactors that appear in the action of $v_1$ are the equivariant weights of the coefficients $x_{1,-k_1}$ and $x_{2,-k_2}$ that are \emph{missing} in $X(z)'$, when we set $X(z)'=z\cdot X(z)$. More abstractly, we find that $\text{End}_A^\varepsilon(\V_k)$ is generated by $v_{\pm1}$ and $\varphi$, with relations
\be \begin{array}{l} v_1v_{-1} = \big(\varphi-k_1\varepsilon+\tfrac12\varepsilon\big)\big(\varphi-k_2\varepsilon+\tfrac12\varepsilon+m_\C\big) \\[.1cm]
v_{-1}v_{1} =\big(\varphi-k_1\varepsilon-\tfrac12\varepsilon\big)\big(\varphi-k_2\varepsilon-\tfrac12\varepsilon+m_\C\big) \end{array}\,,\qquad [\varphi,v_{\pm 1}]=\mp \varepsilon\,v_{\pm 1}\,.\ee

In order to compare with the algebras on Wilson lines, it is helpful to write these algebras more universally in terms of $\mathfrak{sl}_2$. (In this case, $\mathfrak{sl}_2$ appears due to the \emph{topological} $SU(2)$ flavor symmetry enjoyed by $\SQED_2$ in the infrared.) On each vortex line $\V_k$, we define
\be e=-v_{-1}\,,\quad f = v_1\,,\quad h = 2(\varphi+m_\C-(k_1+k_2)\varepsilon)\,. \ee
(The latter equation redefines the dynamical operator $\varphi$ in terms of $h$.) Then
\be \text{End}_A^\varepsilon(\V_k) \simeq \C[e,f,h]/(*) \ee
with familiar relations
\be (*)\qquad  \begin{array}{c} [h,e]=2\varepsilon e\,,\qquad [h,f]=-2\varepsilon f\,,\qquad [e,f]=\varepsilon h\,; \\[.2cm]  ef+fe+\tfrac12 h^2 = \tfrac12(m_\C-(\tilde\gamma\cdot k)\varepsilon)^2-\tfrac12\varepsilon^2\,, \end{array}\ee
where $\tilde\gamma=(1,-1)$. The algebra manifestly depends only on $\tilde \gamma\cdot k = k_1-k_2$, as expected due to screening. Moreover, comparing with \eqref{SQED2-W-alg2}, we clearly have
\be \text{End}^\varepsilon_B(\V_k) \simeq \text{End}^\varepsilon_A(\W_n) \ee
due to 3d mirror symmetry, with the identifications $\tilde\gamma \cdot k \leftrightarrow \gamma\cdot n$ and $m_\C\leftrightarrow t_\C$.

As for $\SQED_1$ in Section \ref{sec:SQED1-VL}, screening isomorphisms correspond to raviolo spaces that map bijectively to vortex moduli spaces on both sides. In this case, given any $k',k$ such that $k'-k= a(1,1)$ is an integer multiple of $(1,1)$, we observe that
\be \CM_{\rm rav}^{\n+a,\n}(\CB_\nu;\V_{k'},\V_k) = \big\{ X(z)',\, g(z) = z^a,\, X(z)\big\}\,, \ee
\be X(z)' = g(z)X(z)\,,\qquad X(z) =  \bp x_{1,\n} z^\n + x_{1,\n-1}z^{\n-1} + ...+ x_{1,-k_1}z^{-k_1} \\
\phantom{x_{1,\n} z^\n +} x_{2,\n-1}z^{\n-1} + ...+ x_{2,-k_2}z^{-k_2} \ep \notag \ee
maps bijectively to both $\CM_D^{\n+a}(\CB_\nu,\V_{k'})$ and $\CM_D^{\n}(\CB_\nu,\V_{k})$. (There are no missing coefficients in either $X'$ or $X$.) The operators $\CO = \sum_\n \big[ \CM_{\rm rav}^{\n+a,\n}(\CB_\nu;\V_{k'},\V_k)\big]$ and $\wt \CO = \sum_{\n} \big[ \CM_{\rm rav}^{\n-a,\n}(\CB_\nu;\V_{k},\V_{k'})\big]$
obey $\CO\wt \CO = \wt \CO\CO=id$ and implement isomorphisms $\CO:\V_k\overset\sim\to \V_{k'}$, $\wt\CO:\V_{k'}\overset\sim\to \V_k$.

The operators at junctions of inequivalent vortex lines are (of course) not isomorphisms. Consider, for example, the junction of $ \V_{(0,0)}=\id$ and $\V_{(a,0)}$ with $a > 0$. There are $a+1$ basic ways to map the vortex state $|\n\rangle$ on $\id$ to a vortex state $|\n'\rangle$ on $\V_{(a,0)}$. Namely, for every $0\leq i \leq a$,
 we can consider correspondences
\be \CM_{\rm rav}^{\n+i,\n}(\CB_\nu;\V_{(a,0)},\V_{(0,0)}) = \big\{ X(z)',\, g(z) = z^i,\, X(z) \big\}\,,\ee
\be X' = \bp z^{\n+i}+x_{1,\n-1}z^{\n+i-1}+...+x_{1,a-i}z^a \\
    \phantom{z^\n+} x_{2,\n-1}z^{\n+i-1}+...+z_{2,0}z^i \ep\,,\quad
X = \bp z^\n +x_{1,\n-1}z^{\n-1}+...+x_{1,a-i}z^{a-i} \\
      \phantom{z^\n+} x_{2,\n-1}z^{\n-1}+...+z_{2,0} \ep \notag \ee
Letting $w_i = \sum_\n \big[ \CM_{\rm rav}^{\n+i,\n}(\CB_\nu;\V_{(a,0)},\V_{(0,0)})\big]$ denote the diagonal sum of fundamental classes, one finds by working through the convolution algebra that
\be w_i |\n\rangle = (\varphi+\tfrac12\varepsilon)(\varphi+\tfrac32\varepsilon)\cdots (\varphi+(i-\tfrac12)\varepsilon)|\n+i\rangle\,, \ee
with a prefactor from the missing coefficients in $X_2'$. 
We claim that the $a+1$ operators
\be w_0,w_1,...,w_a \,\in\, \text{Hom}_A^\varepsilon(\V_{(0,0)},\V_{(a,0)}) \ee
generate the entire space $\text{Hom}_A^\varepsilon(\V_{(0,0)},\V_{(a,0)})$ as a bimodule for $\text{End}_A^\varepsilon(\V_{(a,0)})\otimes \text{End}_A^\varepsilon(\V_{(0,0)})$, and that they
span the 3d mirror of the $\mathfrak{sl}_2$ representation $\rho_{\Delta n}$ that appeared in \eqref{SQED2-rho}, with $\Delta n=a$. 
The claim follows from the universal construction in Section \ref{sec:gen-abel} below; but we also invite the interested reader to verify it by a direct computation.

\subsection{SQED$_2$ via 1d quiver}
\label{sec:quiver}

As mentioned previously, \cite{AsselGomis} proposed a mirror map of half-BPS line operators in 3d $\CN=4$ quiver gauge theories based on brane constructions  \cite{HananyWitten}. Some abelian theories have a quiver realization, in which case the analysis of this section should be compatible with \cite{AsselGomis}. In particular, $\SQED_2$ is such a theory.

Here we will briefly comment on how the vortex lines of $\SQED_2$ defined above match the construction of \cite{AsselGomis}, after a mild but interesting modification.

Recall that $\SQED_2$ has the quiver description
\be 
	\raisebox{-.3in}{ \includegraphics[width=3in]{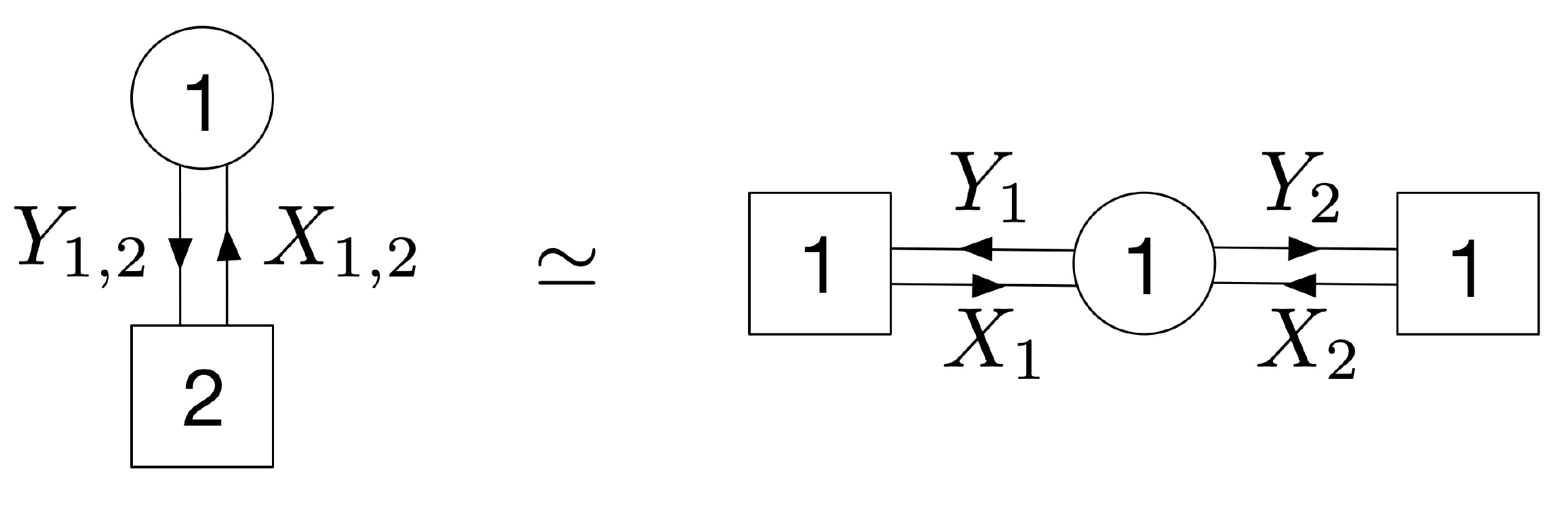}} 
\ee
where the circular node denotes the $U(1)$ gauge group and the square node denotes the two flavors of hypermultiplets. We consider a vortex line $\V_{(k,0)}$, which is 3d-mirror to a Wilson line of gauge charge $k$. It was proposed in \cite{AsselGomis} that, for $k\geq 0$, this vortex line can be engineered by coupling the 3d quiver to a 1d $\SQM_A$ quiver quantum mechanics in the following way:
\be \label{fig:diagram}
	\raisebox{-.5in}{\includegraphics[width=1.6in]{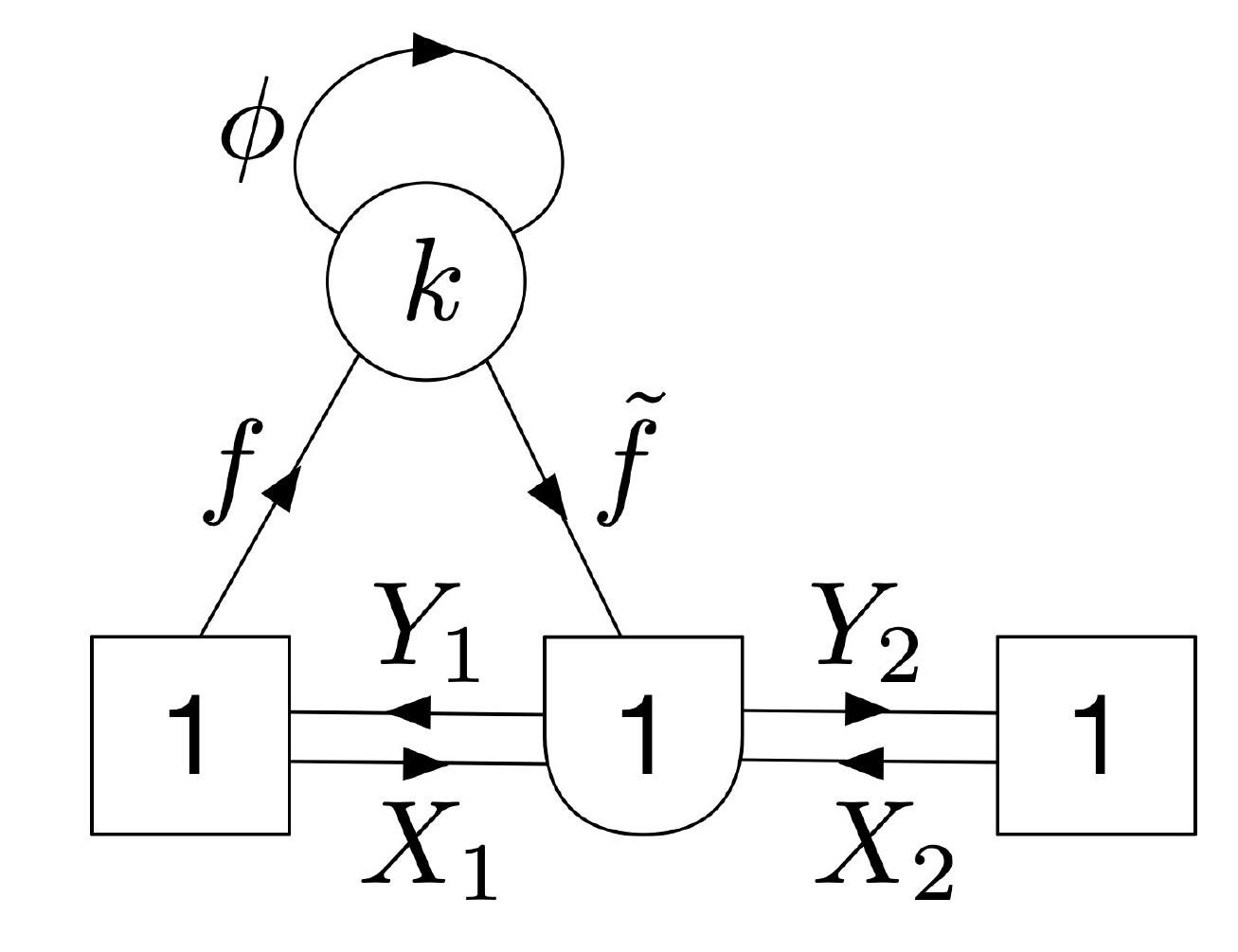}}
\ee
The quantum mechanics is a 1d $U(k)$ gauge theory with chiral multiplets
\be f = \bp f^1 \\ \vdots \\ f^k\ep\,,\qquad \tilde f = (\tilde f_1,...,\tilde f_k)\,,\qquad \phi =\bp \phi^1{}_1 & \cdots &  \phi^1{}_k \\ &\ddots \\ \phi^k{}_1 & \cdots & \phi^k{}_k \ep \ee
in the fundamental, antifundamental, and adjoint representations, respectively. The 1d FI parameter $t_{1d}$ is given a positive value, so that
\be \label{1dFI} ff^\dagger-\tilde f^\dagger \tilde f + [\phi,\phi^\dagger] = t_{1d} > 0\,. \ee
The 1d and 3d fields are then coupled through a gauge-invariant superpotential
\be W_{AG} =   \tilde f f Y_1 \big|_{z=0}\,, \label{WAG1} \ee

On the other hand, following Sections \ref{sec:VL-abel}, we would expect to find a realization of $\V_{(k,0)}$ from coupling to $k$ free 1d chiral multiplets $q_1,...,q_k$, with a superpotential
\be \label{WAGus} W_0 = q_1Y_1+ q_2\pd_z Y_1+ \ldots + q_k \pd_z^{k-1}Y_1\,\big|_{z=0}\,, \ee
so that the F-term equations set $X_1(z) = (k-1)!\frac{q_k}{z^k}+...+\frac{q_2}{z^2}+\frac{q_1}{z}+...\in z^{-k}\CO$ and $Y_1(z) \in z^k\CO$.

How can the descriptions in \eqref{WAG1} and \eqref{WAGus} match? We claim that they are equivalent after modifying the superpotential \eqref{WAG1} to
\be \label{WAG2} W_{AG}' = \tilde f f Y_1 + (\tilde f\phi f) \pd_z Y_1 + \ldots + (\tilde f\phi^{k-1} f) \pd_z^{k-1} Y_1\,\big|_{z=0}\,.\ee
One motivation for this comes from observing that the quantum-mechanics in \eqref{fig:diagram} has an additional flavor symmetry $U(1)_\varepsilon$ that rotates $\phi$ with charge $-1$. The brane construction of \cite{AsselGomis} identifies $U(1)_\varepsilon$ with rotations in the $z$ plane of the 3d theory, consistent with the higher-order terms in \eqref{WAG2}.

A stronger argument comes from analyzing the K\"ahler quotient
\be \label{M1d} \CM_{\rm 1d} = \big\{ f,\tilde f,\phi \,\big|\, \eqref{1dFI} \big\}\big/U(k) \simeq \big\{f,\tilde f,\phi\big\}^{\rm stab}\big/GL(k,\C)\,. \ee
When the 1d FI parameter is nonzero, this turns out to be a smooth space, so the 1d quantum mechanics is equivalent to a sigma-model with target $\CM_{\rm 1d}$. To see that $\CM_{\rm 1d}$ is smooth, we first translate the K\"ahler quotient to an algebraic quotient on the RHS of \eqref{M1d}. The real moment-map constraint \eqref{1dFI} imposes a stability condition that freely acting on $f$ with polynomials in $\phi$ generates the entire vector space $\C^k$. This is equivalent to saying that the vectors
\be f\,,\quad \phi f\,,\quad \phi^2 f\,,\;\ldots\;, \phi^{k-1}f \ee
are a basis of $\C^k$. We may then gauge-fix all of $GL(k,\C)$ by setting
\be f = \bp 1 \\ 0 \\0  \\\vdots \\ 0 \ep\,,\quad \phi f = \bp 0 \\ 1 \\0  \\\vdots \\ 0 \ep\,,\quad \phi^2 f =  \bp 0 \\ 0 \\ 1  \\\vdots \\ 0 \ep\,, \;\ldots \quad \phi^{k-1} f = \bp 0\\ 0 \\ 0 \\\vdots \\ 1 \ep \ee
to be the standard basis. Only $2k$ degrees of freedom remain, in the entries of $\tilde f$ and the final column of $\phi$,
\be \tilde f = (*,*,\ldots,*)\,,\qquad \phi = \bp 0 & 0 & \cdots & 0 & * \\
  1 & 0 &\cdots & 0 & * \\
   && \ddots \\
  0 & 0 &\cdots & 1 & * \ep \ee
These entries are completely unconstrained, and give global coordinates on $\CM_{\rm 1d}$, identifying
\be \CM_{\rm 1d} \simeq \C^{2k}\,. \ee
These $2k$ degrees of freedom are captured in the gauge-invariant operators 
\be q_i :=  \tilde f\phi^{i-1} f\qquad  p_i =\Tr(\phi^i)\,,\qquad i=1,...,k\,.\ee
Thus, the ring of functions $\CM_{\rm 1d}$ (the chiral ring of this quantum mechanics) is the free polynomial algebra $\C[\CM_{\rm 1d}] = \C[q_1,...,q_k,p_1,...,p_k]$.

Now we are very close! We have found that the quiver quantum mechanics of \eqref{fig:diagram} is equivalent to $2k$ free 1d chiral multiplets $q_i,p_i$. The modified superpotential $W_{AG}'$ in \eqref{WAG2} couples the $q_i$ to derivatives of $Y_1$, reproducing our expected superpotential \eqref{WAGus}. The remaining 1d chirals $p_i$ don't do anything at all, and they don't need to: the space of SUSY ground states in $\SQM_A$ quantum mechanics with $k$ free chirals is one-dimensional (after a small deformation to regularize the physical theory, it may be identified with de Rham cohomology of $\C^k$), so adding the $p_i$ without coupling them to 3d fields does not affect the line operator.

Finally, we recall that \cite{AsselGomis} introduced equivalencies among different vortex lines, called ``hopping dualities,'' that came from Hanany-Witten-like operations on brane configurations. In abelian theories, hopping dualities turn out to be the same as screening. For example, in $\SQED_2$, a hopping duality relates \eqref{fig:diagram} to
\be \label{fig:diagram2}
	\raisebox{-.5in}{\includegraphics[width=1.6in]{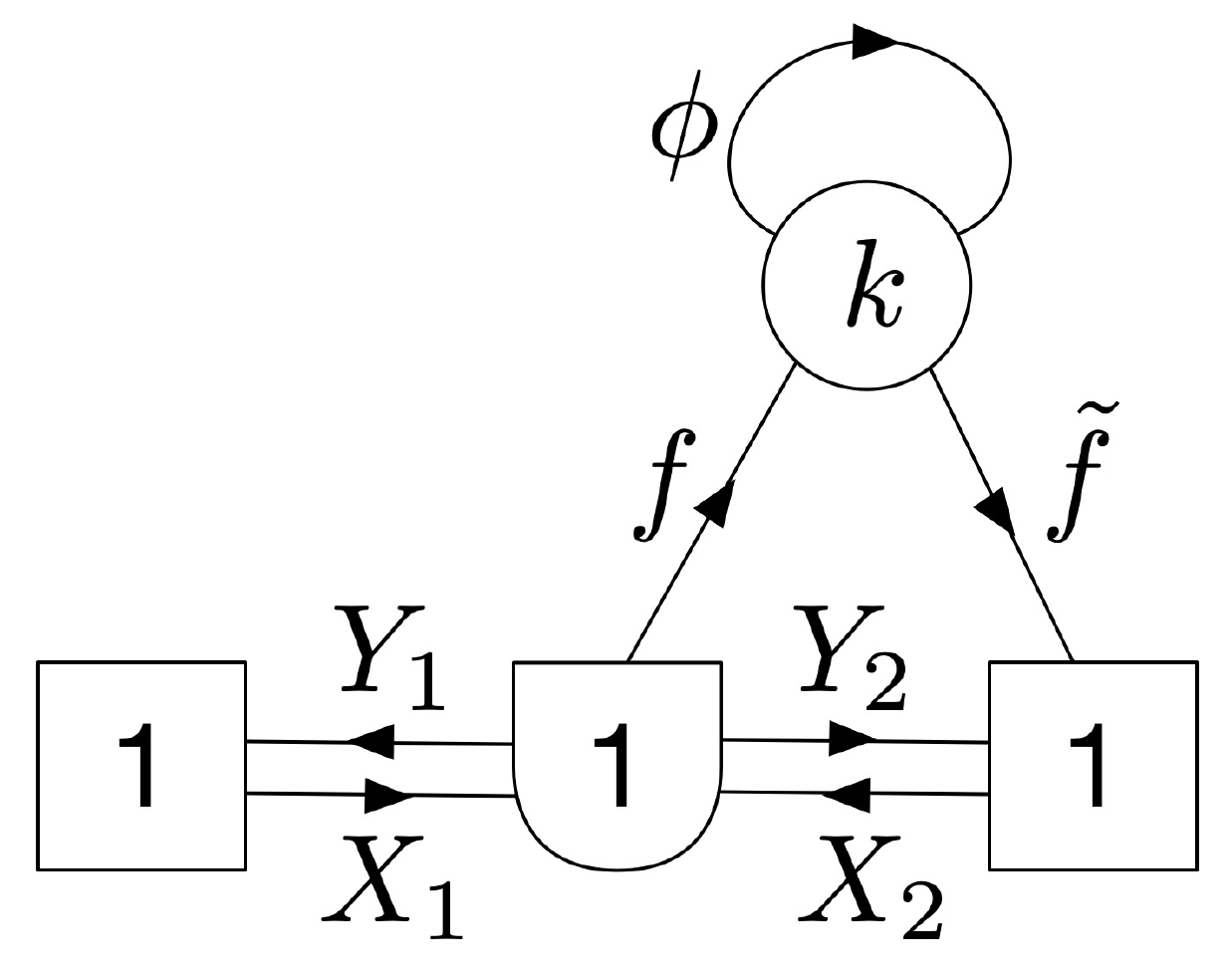}}
\ee
After a modification analogous to \eqref{WAG2}, we find that the superpotential couplings corresponding to this quiver are
\be \wt W_{AG}' = (\tilde f f) X_2 + (\tilde f\phi f) \pd_z X_2 + \ldots + (\tilde f\phi^{k-1} f) \pd_z^{k-1} X_2\,\big|_{z=0}\,.\ee
After recognizing that $q^i=\tilde f \phi^i f$ are all equivalent to free 1d chirals, the F-terms now set
\be X_2(z) \in z^k\CO\,,\quad Y_2(z) \in z^{-k}\CO\,, \ee
which defines the vortex line $\V_{(0,-k)}$, related to $\V_{(k,0)}$ by screening.

\subsection{General abelian theories}
\label{sec:gen-abel}

Fully characterizing the half-BPS Wilson lines and vortex lines (and their junctions) in a general abelian gauge theory is no harder than the previous examples. We now summarize the general structure, using the mirror symmetry between $\SQED_1$ and a free hypermultiplet as a key ingredient.

Here we focus only on half-BPS Wilson lines and vortex lines, as we did in the examples above, rather than the more general categories of B-type and A-type line operators. We suspect that half-BPS Wilson and vortex lines generate the full categories.

\subsubsection{Wilson lines via gauging}
\label{sec:abel-WL}

Suppose we have a 3d theory $\CT$ with gauge group $G=U(1)^r$ and $N\geq r$ hypermultiplets in a faithful representation of $G$.%
\footnote{Faithfulness, which implies $N\geq r$, ensures that the 3d mirror is also an abelian gauge theory and precludes gauginos from appearing in spaces of local operators in the B-twist, \emph{cf.} \ref{sec:Wilson-bulk}.} %
It is useful to think of constructing this theory by starting with $N$ \emph{free} hypermultiplets $(X_i,Y_i)$, whose flavor symmetry $USp(N)$ has a maximal torus $\widehat T= \widehat U(1)_1\times \widehat U(1)_2 \times\cdots\times \widehat U(1)_N$, where each $\widehat U(1)_i$ acts on $(X_j,Y_j)$ with charges $(\delta_{ij},-\delta_{ij})$. Then we form $\CT$ by taking the free theory $(\CT_{\rm hyper})^{\otimes N}$ and gauging a $G=\prod_{a=1}^r U(1)_a$ subgroup of~$\widehat T$. The subgroup is defined by a collection of $r$ linearly independent cocharacters
\be \label{T-cochar} q_a\, \in\, \text{cochar}(\widehat T) = \Z^N \ee
that define how each $U(1)_a$ is embedded in $\widehat T$. In more physical terms, the $i$-th hypermultiplet has charges $(q_{a,i},-q_{a,i})$ under $U(1)_a$.

We also recall that, for every $U(1)_a$ factor in $G$, the theory $\CT$ acquires 1) a $U(1)_a^{\rm top}$ topological flavor symmetry, which rotates the $U(1)_a$ dual photon; and 2) a triplet of FI parameters $t^a_\R,t^a_\C$, which are the scalars in a background $U(1)_a^{\rm top}$ vectormultiplet. 

The free matter theory $(\CT_{\rm hyper})^{\otimes N}$ has no nontrivial Wilson lines (just as in our $N=1$ example from Section \ref{sec:hyperAB}). However, the endomorphism algebra of the trivial line is gigantic:
\be \text{End}_B^{\varepsilon}(\id) =   \C[X,Y]  := \bigotimes_{i=1}^N \C[X_i,Y_i] \qquad\text{with}\quad [X_i,Y_j]=\varepsilon \delta_{ij}\,. \ee
Another way to say this is that $(\CT_{\rm hyper})^{\otimes N}$ has many flavor Wilson lines $\W_n$ labeled by points $n\in \Z^N$ in the character lattice of $\widehat T$; but for every $n,n'$,
\be \label{W-hyperN} \text{Hom}_B^\varepsilon(\W_n,\W_{n'}) = \C[X,Y]\,. \ee  
In particular, the operator $1\in \C[X,Y]$ provides isomorphisms $\W_n\overset{\sim}\to \W_{n'}$ among every pair of Wilson lines; so for all $n$, $\W_n\simeq \W_0 = \id$.

Heuristically, the effect of gauging $G$ is to produce \emph{more} nontrivial equivalence classes of line operators, with \emph{fewer} Homs. These go hand in hand: restricting the Homs removes some of the isomorphisms among different Wilson lines. 
In the extreme case that the entire torus is gauged ($G=\widehat T$), we expect every Wilson line $\W_n$ to become distinct, but each junction $\text{Hom}_B(\W_n,\W_{n'})$ to support just a single local operator.

To describe more explicitly what happens, we note that the algebra $\C[X,Y]$ is $\Z^N$ graded, by $\widehat T$ charge. We can decompose it into graded components by defining the combinations
\be z_i := X_iY_i -\tfrac12\varepsilon = Y_iX_i + \tfrac12\varepsilon\,,\qquad  w_n \,:=\, \prod_{i=1}^N \begin{cases} X_i^{n_i} & n_i \geq  0 \\
  Y_i^{|n_i|} & n_i < 0 \end{cases}\,, \ee
for $i\in [1,N]$ and $n\in \Z^N$. The $z$'s commute with each other, and the $z$'s and $w$'s satisfy
\be \label{rel-zw} (*)\quad  [z_i,w_n]=-\varepsilon n_i w_n\,,\qquad w_n w_{n'} = \Big( \prod_{\substack{i ~s.t.~ n_i n_i' < 0 \\ |n_i'| \geq |n_i|}} [z_i]^{-n_i}\Big) w_{n+n'} \Big( \prod_{\substack{i ~s.t.~ n_i n_i' < 0 \\ |n_i| > |n_i'|}} [z_i]^{n_i'}\Big)\,, \ee
where the $\varepsilon$-shifted products are
\begin{align}
	&[z]^k :=
	\begin{cases}
		\prod^{k-1}_{l=0} \left( z - \left( l + \frac 1 2 \right) \varepsilon \right)		&k>0\,, \\
		\prod^{|k|-1}_{l=0} \left( z + \left( l + \frac 1 2 \right) \varepsilon \right)	&k<0\,, \\
		1		&k=0\,. \\
	\end{cases}
\end{align}
As an algebra, we have%
\footnote{Equivalence follows because the RHS is manifestly a subalgebra of the LHS, and the generators of the LHS can be expressed as $X_i = w_{(0...010...0)}$, $Y_i=w_{(0...0(-1)0...0)}$ with a $\pm1$ in the $i$-th slot.} %
$\C[X,Y]  \simeq \C[w_n,z_i]_{n\in \Z^N,i\in[1,N]}\big/(*)$. More so, the commutator $\frac{1}{(-\varepsilon)}[z,-]$ measures $\widehat T$ charge, so the graded components of $\C[X,Y]$ are the eigenspaces of $\frac{1}{(-\varepsilon)}[z,-]$. Each eigenspace contains an operator $w_n$ and arbitrary polynomials in the $z$'s. Thus, as a graded vector space
\be \C[X,Y] \simeq \bigoplus_{n\in \Z^N} 
 \C[z]\langle w_n\rangle\,. \ee

Now, gauging a subgroup $G\subset \widehat T$ specified by cocharacters $q_1,...,q_r$  reduces the space of local operators at a junction of Wilson lines from \eqref{W-hyperN} to operators whose gauge charge agrees with $n'-n$, modulo relations imposed by the moment maps. In other words,
\be \label{abel-Whom}  \text{Hom}_B^\varepsilon(\W_n,\W_{n'}) \;\;=  \hspace{-.2in} \bigoplus_{\scriptsize\begin{array}{c}m\in \Z^N \\ q_a\cdot (n'-n-m)=0\;\;\forall\, a\end{array} } \hspace{-.5in}  \C[z]\langle w_m\rangle \,\Big/\big(q_a\cdot(z+n\varepsilon)+t^a_\C\big)_{a=1}^r\,,\ee
where $q_a\cdot z = \sum_i q_{a,i}z_i = \mu_\C^a$ is the moment map for the $U(1)_a$ factor of $G$.
The space \eqref{abel-Whom} contains the operator $1$, giving an isomorphism $\W_n\simeq \W_{n'}$, if and only if $q_a\cdot(n'-n)=0$ $\forall a$. Thus, if we view the $q_{a}$ as defining  a map $q:\Z^r\to \Z^N$, with a dual map $q^T:\Z^N\to \Z^r$, we find that inequivalent Wilson lines are labeled by elements of the quotient lattice
\be \label{abel-Wlattice}  n\in \frac{\Z^N}{\ker q^T}\,. \ee

Finally, we observe that the endomorphism algebras
\be \text{End}_B^\varepsilon(\W_n) = \Big( \bigoplus_{m\in \Z^N\;\text{s.t.}\; q_a\cdot m=0} \hspace{-.2in}  \C[z]\langle w_m\rangle \Big) \,\Big/\big(q_a\cdot(z+n\varepsilon)+t^a_\C\big)_{a=1}^r \ee
can all be obtained from the algebra  $\text{End}_B^\varepsilon(\id)$ of bulk local operators by simply shifting $t_\C^a \mapsto t_\C^a + (q_a\cdot n)\varepsilon$.

\subsubsection{Vortex lines via ungauging}
\label{sec:abel-VL}

We can dualize the analysis of Wilson lines above to obtain a universal description of half-BPS A-type vortex lines in abelian theories as well! 

When analyzing Wilson lines, it was useful to think of a theory $\CT$ as obtained from a free matter theory $(\CT_{\rm hyper})^{\otimes N}$ by gauging a $G$ subgroup of the flavor symmetry. For the propose of analyzing vortex lines, we espouse a dual perspective.

Let $\SQED_1$ be the $U(1)$ gauge theory with a single hypermultiplet, as in Section \ref{sec:basic}. Then $(\SQED_1)^{\otimes N}$ is a theory with $N$ hypermultiplets $(X_i,Y_i)$ and gauge group $\prod_{i=1}^N U(1)_i$, such that $(X_i,Y_i)$ has charge $(\delta_{ij},-\delta_{ij})$. It is the 3d mirror of $N$ free hypers:
\be \label{SQEDN}  (\SQED_1)^{\otimes N} \; \overset{\text{3d MS}}{\longleftrightarrow} \; (\CT_{\rm hyper})^{\otimes N}\,. \ee
The theory $(\SQED_1)^{\otimes N}$ has \emph{topological} flavor symmetry with maximal torus $\widehat T = \prod_{i=1}^N \widehat U(1)_i$, mirror to the ordinary flavor symmetry of $(\CT_{\rm hyper})^{\otimes N}$. Each $\widehat U(1)_i$ rotates a dual photon for the $i$-th factor in the gauge group of  $(\SQED_1)^{\otimes N}$. 

Now, suppose we gauge a subgroup $G=U(1)^r$ of $\widehat T$ on both sides, with cocharacters $\{q_a\}_{a=1}^r$. On the right we recover $\CT$. On the left, gauging a topological flavor symmetry is equivalent (after flowing to the infrared) to \emph{ungauging} the associated gauge symmetry \cite{KapustinStrassler, Witten-SL2}. Specifically, gauging $G\subset \widehat T$ effectively ungauges a $U(1)^r$ part of the gauge symmetry, such that the remaining gauge group $\wt G\simeq U(1)^{N-r}$ is specified by cocharacters $\{\wt q_\alpha\}_{\alpha=1}^{N-r}$ that satisfy
\be \text{im}(\wt q:\Z^{N-r}\to \Z^N) = \text{ker}(q^T:\Z^N\to \Z^r)\,. \ee 
(In other words, the cocharacters $\wt q_\alpha$ form a basis for the sublattice $\ker q^T$.) We thus obtain a $U(1)^{N-r}$ gauge theory $\wt\CT$, which is the 3d mirror of $\CT$:
\be \label{SQEDN2} \begin{array}{cccc} &  (\SQED_1)^{\otimes N} & \overset{\text{3d MS}}{\longleftrightarrow} & (\CT_{\rm hyper})^{\otimes N} \\
\text{gauge $G$:\quad } & \downarrow && \downarrow \\
& \wt\CT & \overset{\text{3d MS}}{\longleftrightarrow} &  \CT\,. \end{array}
 \ee

For example: suppose we wanted to produce the quiver gauge theory $\wt\CT = \raisebox{-.03in}{\includegraphics[width=1in]{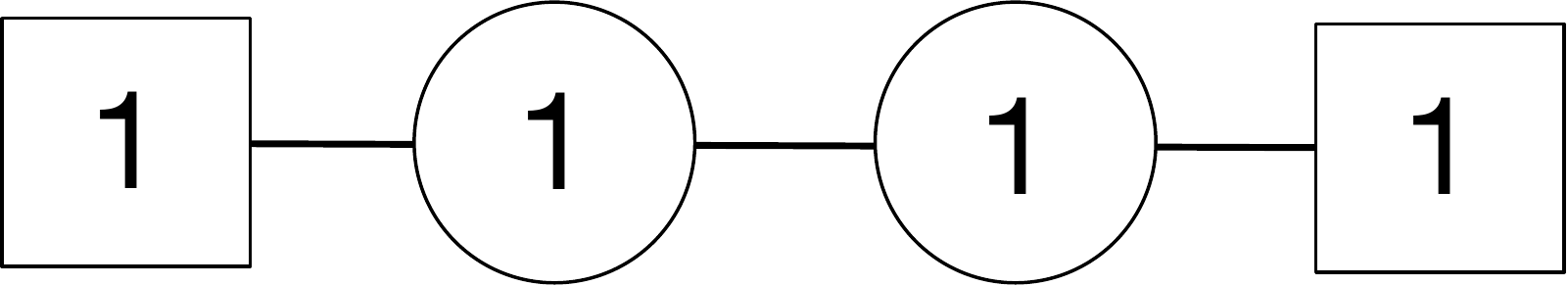}}$. We start with $(\SQED_1)^{\otimes 3}$, such that the charge vectors for each factor of $U(1)^3$ are $(1,0,0),$ $(0,1,0),$ $(0,0,1)$. Then we gauge the diagonal $\widehat U(1)$ subgroup of the topological flavor symmetry. This leaves behind a $\wt G=U(1)^2$ theory for which the $X$'s have charges $(1,-1,0),\,(0,1,-1)$ (since these vectors span orthogonal complement of $(1,1,1)$). This is the quiver gauge theory $\wt \CT$. Dually, starting with $(\CT_{\rm hyper})^{\otimes 3}$ and gauging the diagonal flavor symmetry produces $\CT = \SQED_3$, which is the 3d mirror of $\wt\CT$.

Given \emph{any} theory $\wt \CT$ with $N$ hypermultiplets and gauge group $\wt G=U(1)^{N-r}$ acting faithfully, we can always view it as the mirror of a theory $\CT$ with $N$ hypermultiplets and gauge group $G=U(1)^r$ (also acting faithfully), such that the cocharacters $\wt q,q$ that specify $\wt G$ and $G$ satisfy $ \text{im}\, \wt q = \ker q^T$ and $ \text{im}\, q = \ker \wt q^T$. In particular, any such $\wt \CT$, can be constructed from $(\SQED_1)^{\otimes N}$ by gauging a $G$ subgroup of the topological flavor symmetry.

We may use this perspective to construct the half-BPS vortex lines of $\wt \CT$. First, note that the basic half-BPS vortex lines of $(\SQED_1)^{\otimes N}$ are obtained just as in Section \ref{sec:SQED1-VL}. For every $k\in \Z^N$, there is a vortex line $\V_k$ with algebraic data
\be \label{abel-Vk} \V_k\,: \quad \CG_0=G(\CO)\,,\qquad \CL_0 = \big\{ (X_i,Y_i)\in z^{-k_i}\CO\oplus z^{k_i}\CO\big\}\,. \ee
These line operators are all equivalent due to screening, and every space $\text{Hom}_A^\varepsilon(\V_k,\V_{k'})$ is isomorphic to $\text{End}_A^\varepsilon(\id)$.

To be explicit, the bulk algebra in $(\SQED_1)^{\otimes N}$ is obtained as $N$ copies of \eqref{SQED-EndVk-2}, namely
\be \text{End}_A^\varepsilon(\id) = \bigotimes_{i=1}^N \C[v_{i,1},v_{i,-1}]/([v_{i,1},v_{i,-1}]-\varepsilon)\,. \ee
A more natural set of generators is given by the monopole operators $v_n$ of charge $n\in \Z^N$ and the Coulomb-branch scalars $\varphi_i$, which are identified as
\be \varphi_i = v_{i,1}v_{i,-1}-\tfrac12\varepsilon\,,\qquad v_n = \prod_{i=1}^n \begin{cases} (v_{i,1})^{n_i} & n_i\geq 0 \\
 (v_{i,-1})^{|n_i|} & n_i< 0\,. \end{cases} \ee
Then we may rewrite
\be  \text{End}_A^{\varepsilon}(\id) = \C[\varphi_i,v_n]_{i\in[1,N],n\in \Z^N}\big/(*)\,, \ee
\be  (*):\quad  [\varphi_i,v_n]=-\varepsilon n_i v_n\,,\qquad v_n v_{n'} = \Big( \prod_{\substack{i ~s.t.~ n_i n_i' < 0 \\ |n_i'| \geq |n_i|}} [\varphi_i]^{-n_i}\Big) v_{n+n'} \Big( \prod_{\substack{i ~s.t.~ n_i n_i' < 0 \\ |n_i| > |n_i'|}} [\varphi_i]^{n_i'}\Big)\,, \notag \ee
identical to \eqref{rel-zw}.

In $(\SQED_1)^{\otimes N}$, the endomorphism algebra of any vortex line is isomorphic to
\be \label{abel-endV} \text{End}_A^\varepsilon(\V_k) \simeq  \text{End}_A^\varepsilon(\id) = \C[\varphi_i,v_n]_{i\in[1,N],n\in \Z^N}\big/(*)\,. \ee
However, one must be a little careful when comparing this statement to equivariant cohomology computations. As we saw in Section \ref{sec:SQED1-VL}, the natural equivariant parameters $\varphi_i^{eq}$ for the gauge action in the presence of a vortex line $\V_k$ are related to \eqref{abel-endV} by a redefinition $\varphi_i^{eq} = \varphi_i+k_i\varepsilon$. Similarly, every junction space $\text{Hom}_A^\varepsilon(\V_k,\V_{k'})$ is isomorphic to \eqref{abel-endV} as a $\text{End}_A^\varepsilon(\V_{k'})\otimes \text{End}_A^\varepsilon(\V_{k})$ bimodule. The operator $1\in \text{Hom}_A^\varepsilon(\V_k,\V_{k'})$ that obviously implements the isomorphism $\V_k\overset\sim\to \V_{k'}$ in this description is identified in equivariant cohomology with the fundamental class of a raviolo space containing the singular gauge transformation $g(z) = z^{k'-k}$ (as opposed to $g(z)=1$). These shifts/identifications are a direct consequence of screening --- in particular, the fact that the line operator $\V_{k}$ only looks like $\V_{k'}$ after being decorated by a dynamical vortex of charge $k'-k$.

When we gauge a $G$ subgroup of the topological flavor symmetry to obtain our desired theory $\wt \CT$, there will be more inequivalent line operators, and there will be fewer local operators (hence fewer ways to screen).
Altogether, the inequivalent vortex lines of $\wt \CT$ are labeled by elements of the quotient
\be \label{abel-Vlattice} k \in \frac{\Z^N}{\text{im}\,\wt q}\,, \ee
where $\text{im}\, \wt q$ corresponds to screening by dynamical vortices. Of course, since $\text{im}\,\wt q=\text{ker}\,q^T$, this matches the mirror lattice \eqref{abel-Wlattice}.
Every Hom space also gets reduced exactly as in \eqref{abel-Whom}\,:
\be \label{abel-Vhom} \text{Hom}_A^\varepsilon(\V_k,\V_{k'}) \;\; = \hspace{-.2in} \bigoplus_{\scriptsize\begin{array}{c}n\in \Z^N \\ q_a\cdot (k'-k-n)=0\;\;\forall\, a\end{array} } \hspace{-.5in}  \C[\varphi]\langle v_n\rangle  \,\Big/\big(q_a\cdot(\varphi+k\varepsilon)+m^a_\C\big)_{a=1}^r\,.\ee
In particular,
\be \label{abel-Vend} \text{End}_A^\varepsilon(\V_k) \; = \hspace{-.1in} \bigoplus_{n\in \Z\,\text{s.t.}\, q_a\cdot n=0} \hspace{-.2in}  \C[\varphi]\langle v_n\rangle  \,\Big/\big(q_a\cdot(\varphi+k\varepsilon)+m^a_\C\big)_{a=1}^r\,.\ee

We interpret \eqref{abel-Vend} as follows. The monopole operators in our $\wt G$ gauge theory can only have charges $n$ corresponding to cocharacters of $\wt G$, \emph{i.e.} linear combinations $n= \sum_\alpha c_\alpha \wt q_\alpha \in \text{im}\,\wt q$. But  $\text{im}\,\wt q = \ker q^T$, whence the constraint $q_a\cdot n=0$. Moreover, the scalars $\varphi$ in vectormultiplets of $(\SQED_1)^{\otimes N}$ that have been ``ungauged'' are no longer dynamical. They should be set equal to complex mass parameters $m_\C^a$, associated to a new, ordinary $U(1)^r$ flavor symmetry of $\wt \CT$. This is precisely implemented by setting $q_a\cdot(\varphi+k\varepsilon) = -m_\C^a$, where the shift by $k\varepsilon$ accounts for the screening affect above!

Note that when $q_a\cdot k\neq 0$, the line operator $\V_k$ itself may be interpreted as a \emph{flavor vortex}. It originates from a dynamical vortex in $(\SQED_1)^{\otimes N}$, which gets frozen out when part of the gauge symmetry is rendered non-dynamical. Thus, physically, $\V_k$ can be defined by introducing a singular profile for the background gauge connection associated to the $U(1)^r$ flavor symmetry of $\wt T$. This is consistent with hypermultiplets having a profile $(X_i,Y_i)\sim(z^{k_i},z^{-k_i})$, which is related to the vacuum by a singular flavor transformation $g(z) = z^k$. Unlike singular gauge transformations (which lead to dynamical vortices), singular flavor transformations are nontrivial and define new operators.
Correspondingly, the algebra \eqref{abel-Vend} is simply related to $\text{End}_A^\varepsilon(\V_k)$ by shifting $m_\C^a \mapsto m_\C^a+(q_a\cdot k)\varepsilon)$.

Formula \eqref{abel-Vhom} has a similar interpretation. 
The monopole operators $v_n$ that are allowed to appear in $\text{Hom}_A^\varepsilon(\V_k,\V_{k'})$ satisfy $q_a\cdot n=q_a\cdot(k'-k)$ rather than $q_a\cdot n=0$. These operators are thus labeled by cocharacters that involve both the gauge group $\wt G$ and the new $U(1)^r$ flavor symmetry. They are \emph{flavor monopole} operators, defined in the presence of a point-like singularity of the background connection for the flavor symmetry.%
\footnote{Such operators played an important role in the generalized superconformal index of \cite{KW-index}.}
When two vortex lines $\V_k,\V_{k'}$ have different values of $q_a\cdot k$, such a point-like singularity is necessarily created at their junction.

\subsection{Flavor vortices, abelian and nonabelian}
\label{sec:flavor}

We learned an interesting lesson in Section \ref{sec:abel-VL}: in any abelian gauge theory where the gauge group $\wt G$ acts faithfully, every half-BPS vortex line operator defined as in \eqref{abel-Vk} is either a flavor vortex, or isomorphic to a flavor vortex after screening by dynamical vortices. Under abelian mirror symmetry,
\be \hspace{-.6in} \begin{array}{lccc} 
\text{trivial:\qquad} & \text{screenable (gauge) vortices} & \leftrightarrow & \text{flavor Wilson lines} \\
\text{nontrivial:\qquad } & \text{flavor vortex lines} & \leftrightarrow & \text{gauge Wilson lines}\,.
\end{array} \ee

Flavor vortices are somewhat more versatile than what we just found, and we take a moment to discuss them. Given any 3d $\CN=4$ theory $\CT$ (abelian, nonabelian, sigma-model...) with flavor symmetry $F$ acting on the Higgs branch, a half-BPS A-type flavor vortex $\V_\lambda$ can be defined in algebraic terms by
\begin{itemize}
\item[1)] choosing a cocharacter $\lambda \in \text{cochar}(T_F)$ of a torus of $F$; and
\item[2)] acting with the singular flavor transformation $g(z) =z^\lambda$ to specify a singular profile for the matter fields.
\end{itemize}
Note that if $\CT$ has gauge group $G$, this process will leave the $G(\CO)$ unbroken.

Alternatively, we may first gauge the subgroup $U(1)_\lambda\subset T_F\subseteq F$ generated by the cocharacter $\lambda$. If $\CT$ has gauge group $G$, we obtain a new theory $\CT'$ with gauge group $G\times U(1)_\lambda$. The theory $\CT'$ has a dynamical vortex $\CV_\lambda$ of charge $\lambda$, which is created and destroyed by ordinary monopole operators. Subsequently, we un-gauge the $U(1)_\lambda$ symmetry, either by sending its gauge coupling to zero, or by gauging its dual topological symmetry as in \eqref{SQEDN2}. After ungauging, the vortex $\CV_\lambda\leadsto \V_\lambda$ gets frozen out, define a vortex line operator in the original theory $\CT$. The endomorphism algebra of $\V_\lambda$ and its Hom's with other line operators all descend from dynamical operations in $\CT'$; for example, elements of $\text{Hom}_A(\V_\lambda,\id)$ and $\text{Hom}_A(\id,V_\lambda)$ come from ordinary monopole operators of $\CT'$. From a more algebraic perspective, Hom spaces of $\CT$ (including $\V_\lambda$) will be Hamiltonian reductions of Hom spaces of $\CT'$, just like in \eqref{abel-Vhom}--\eqref{abel-Vend}.

Abelian theories $\CT$ (with a faithful $G$ action) are special only insofar as they have such a large flavor group $F$ that \emph{all} nontrivial half-BPS vortex lines can be interpreted as flavor vortices. Then all Homs are obtained as in Section \ref{sec:abel-VL}.

Braverman-Finkelberg-Nakajima introduced flavor vortices from a geometric, mathematical perspective in \cite{BFN-lines}, and used them to construct partial resolutions of Coulomb branches. Physically, one would resolve the Coulomb branch by introducing real mass parameters (associated with abelian subgroups of the flavor symmetry). In the mathematical definition of the Coulomb branch, real masses are unavailable; but one can look at flavor vortices instead. The space $\text{Hom}_A(\id,\V_\lambda)$ for a flavor vortex is interpreted as the space of sections of a line bundle over the Coulomb branch. (This statement is dual to the correspondence between Wilson lines and bundles on the Higgs branch from Section \ref{sec:Wilson-sheaves}). Looking at these spaces of sections for different choices of $\lambda$ allows one to probe parts of the Coulomb branch (such as exceptional divisors) that are invisible to ordinary global functions. Mathematically, allows one to build a resolution of the Coulomb branch by using a standard ``Proj'' construction.

Flavor vortices also play a spacial role in the correspondence of \cite{CostelloGaiotto, CCG} between the category of A-type lines and modules for a boundary VOA. The flavor symmetry $F$ of a 3d theory $\CT$ also becomes a flavor symmetry of its boundary VOA, and flavor vortices $\V_\lambda$ are spectral flow modules (of the vacuum module) in the VOA.


\section{Nonabelian theories: the abelianizing line}\label{sec:IwahoriLine}


In the final two sections of the paper, we study two particular examples of half-BPS vortex line operators in $U(2)$ SQCD with four flavors, \emph{i.e.} $G=U(2)$ and $R = (\C^2)^{\oplus 4}$.
Our initial motivation for looking at these examples was to put the proposed computational methods of Section \ref{sec:comp} through a more rigorous test. However, in both cases, the computations turned out to reveal some beautiful structure, and taught us some interesting lessons.

The current section considers an ``abelianizing'' or ``Iwahori'' A-type line operator $\V_\CI$, which can be defined for any nonabelian gauge theory as in Section \ref{sec:VL-pureG}. Namely, $\V_\CI$ breaks the gauge group $G$ to its maximal torus $T$ along a line, and may be accompanied by a monodromy defect for the connection;
 but it does \emph{not} introduce any singularity in the hypermultiplet fields. In terms of algebraic data, $\V_\CI$ breaks $G(\CO)$ to the Iwahori subgroup $\CI=\CI_B$ from \eqref{def-IB}, while retaining the standard Lagrangian $\CL_0$,
\be \V_\CI\,:\quad \CG_0 = \CI\,,\qquad \CL_0 = R(\CO)\oplus R^*(\CO)\,. \label{VI-triv} \ee
Alternatively, $\V_\CI$ may be defined by introducing a 1d $\SQM_A$ sigma-model whose target is the flag manifold $\CX=G/T \simeq G(\CO)/\CI$, and coupling it to the vectormultiplets of the bulk 3d theory by gauging its flavor symmetry.

In any gauge theory, we actually expect the Iwahori line operator $\V_\CI$ to be equivalent in $Q_A$-cohomology to a direct sum of trivial lines 
\be \V_\CI \,\simeq\, \id^{\oplus N}\,,\qquad N = |\text{Weyl}(G)| = \text{rank}\,H^\bullet(G/T)\,. \ee
We will explain the mathematical basis for this expectation in Section \ref{sec:Deligne}. The upshot is that the algebra of local operators bound to $\V_\CI$ should look like $N\times N$ matrices whose entries are bulk Coulomb-branch operators; and the space of local operators at any junction with $\V_\CI$ should look like an $N$-component vector whose entries are local operators at a corresponding junction with $\id$. More succinctly, letting $V= H^\bullet(G/T)= \C^N$,
\be \label{VIV} \begin{array}{c} \text{End}_A^\varepsilon(\V_\CI) \simeq \C_\varepsilon[\CM_C]\otimes V\otimes V^*\,,\\[.2cm]
\text{Hom}_A^\varepsilon(\CL,\V_\CI) \simeq \text{Hom}_A^\varepsilon(\CL,\id)\otimes V\,,\qquad
\text{Hom}_A^\varepsilon(\V_\CI,\CL) \simeq \text{Hom}_A^\varepsilon(\id,\CL)\otimes V^*\,. \end{array}\ee

In the remainder of this section, we will implement the computational methods of Section~\ref{sec:comp} to actually find $\text{End}_A^\varepsilon(\V_\CI)$, $\text{Hom}_A^\varepsilon(\V_\CI,\id)$, and $\text{Hom}_A^\varepsilon(\id,\V_\CI)$, for the case of $G=U(2)$ and $R=(\C^2)^{\oplus 4}$. The \emph{way} in which these spaces get arranged into matrix algebras or vectors \eqref{VIV} turns out to be highly nontrivial, and will provide a good test of our methods.

The computation of $\text{End}_A^\varepsilon(\V_\CI)$ reveals an additional piece of structure. We find that $\text{End}_A^\varepsilon(\V_\CI)$ most directly takes the form of an abelianized version of the bulk Coulomb-branch algebra (closely related to the abelianization construction of \cite{BDG}), tensored with a copy of the \emph{nil-Hecke algebra} $\mb H_2$ for $GL(2)$ \cite{KostantKumar}. Abstractly, the nil-Hecke algebra may be defined as the $G$-equivariant cohomology of a product of flag varieties. Here $G/B\simeq \cp^1$, and
\be \mb H_2  = H_{GL(2)}^*(\cp^1\times \cp^1)\,, 
 \ee
with a product from convolution. Explicit relations for $\mb H_2$ will be given in Section \ref{sec:nilHecke}. In Section \ref{sec:mxalg} we relate this presentation of $\text{End}_A^\varepsilon(\V_\CI)$ with the expected matrix algebra.

We note that the structure of $\text{End}^\varepsilon_A(\V_\CI)$, both as a $2\times 2$ matrix algebra over $\C_\varepsilon[\CM_C]$, 
and as a semidirect product of an abelianized $\C_\varepsilon[\CM_C]$ with the nil-Hecke algebra, was discussed by Webster in \cite{Web2016}. The analysis of \cite{Web2016} was performed in a BFN-like setup (as in Appendix~\ref{app:BFN}) rather than by choosing a massive vacuum boundary condition $\CB_\nu$ as we do here. Of course, the structure of the line operator $\V_\CI$ should be independent of how it is probed by boundary conditions. 
Happily, the final results of our computation here agree with \cite{Web2016}.

We also recall that a version of the nil-Hecke algebra (in fact, a categorification of thereof) appeared in physics in the work of Gukov and Witten \cite{GW-surface}. They considered a surface operator in 4d $\CN=4$ SYM that broke gauge symmetry $G\to T$. These operators were not trivial as in \eqref{VI-triv}, because in 4d $\CN=4$ SYM the breaking of gauge symmetry is accompanied by a singularity in the adjoint-valued matter fields. Nevertheless, the breaking of gauge symmetry was sufficient to introduce a copy of the nil-Hecke algebra, sitting inside a larger affine Hecke algebra that described line operators bound to the surface.

\subsection{Triviality of line operators with nonsingular matter} \label{sec:Deligne}

Consider 3d $\CN=4$ gauge theory with arbitrary group $G$ and hypermultiplets in $T^*R$.
Suppose that we define an A-type line operator $\V_\CX$ by coupling to 1d $\SQM_A$ quantum mechanics with K\"ahler target $\CX$, by gauging a flavor symmetry of $\CX$. In algebraic terms, we think of $\CX$ as a (possibly singular) variety or stack with an action of $G(\CO)$ that is identified with holomorphic gauge transformations in a neighborhood of the line.  We propose that
\be \V_\CX \simeq \id\otimes H^*(\CX)\,, \ee
as an object in the category of A-type line operators. Taking $G=U(2)$ and $\CX=\cp^1$ recovers the main example \eqref{VI-triv}.

Evidence for the proposal comes from considering the spaces of local operators $\text{End}_A^\varepsilon(\V_\CX)$,  $\text{Hom}_A^\varepsilon(\V_\CX,\CL)$, and  $\text{Hom}_A^\varepsilon(\CL,\V_\CX)$, for any other line operator $\CL$. We explained in Section~\ref{sec:comp} that these spaces can be computed --- or more precisely, represented --- by probing with a boundary condition, \emph{e.g.} a boundary condition $\CB_\nu$ labeled by a massive vacuum. 
For example, $\text{End}_A^\varepsilon(\V_\CX)$ is represented in the  cohomology $H^*_{\C^*_\varepsilon}\big(\CM_{\rm rav}(\CB_\nu;\V_\CX,\V_\CX)\big)$, acting by convolution on the disc Hilbert space $H^*_{\C^*_\varepsilon}\big(\CM_D(\CB_\nu,\V_\CX)\big)$.

The key point is that for a line operator $\V_\CX$ of the type above, the spaces $\CM_D$ and $\CM_{\rm rav}$, defined algebraically in \eqref{MD-QM}--\eqref{Mrav-QM} (where `$\CX$' was called `$\CE$'), acquire the structure of fibrations
\be \label{fib} \begin{array}{ccc} \CX &\to& \CM_D(\CB_\nu,\V_\CX) \\ && \downarrow \\&&  \CM_D(\CB_\nu,\id)\,, \end{array} \qquad 
\begin{array}{ccc} \CX\times \CX &\to& \CM_{\rm rav}(\CB_\nu;\V_\CX,\V_\CX) \\ && \downarrow \\&&  \CM_{\rm rav}(\CB_\nu;\id,\id)\,, \end{array}  \qquad
\begin{array}{ccc} \CX &\to& \CM_{\rm rav}(\CB_\nu;\CL,\V_\CX) \\ && \downarrow \\&&  \CM_{\rm rav}(\CB_\nu;\CL,\id)\,, \end{array} 
\ee
etc. 
The vertical maps are obtained by forgetting the data of (one or two copies of) $\CX$.  These maps are fibrations (meaning the pre-image of \emph{every} point is exactly $\CX$ or $\CX\times \CX$)  precisely because there is no superpotential to impose any constraints on $\CX$, such as relations between $\CX$ and the bulk hypermultiplets.
The fact that symmetries of $\CX$ are gauged in coupling to the bulk will (in general) simply lead to the fibrations \eqref{fib} being nontrivial, meaning \emph{e.g.} that $\CM_D(\CB_\nu,\V_\CX)$ is not necessarily a direct product $  \CM_D(\CB_\nu,\id)\times \CX$.

Now, a classic result of Deligne \cite{Deligne} says that the cohomology of a smooth K\"ahler fibration will always factorize into cohomology of the base times cohomology of the fiber, regardless of whether the fibration is trivial or not. More precisely, the topological Leray spectral sequence degenerates. The result was generalized to intersection cohomology (or Borel-Moore homology) of singular varieties or stacks by the Beilinson-Bernstein-Deligne-Gabber decomposition theorem \cite{BBDG}. The implication is that all cohomologies factorize%
\footnote{The tensor products on the RHS of \eqref{fact} should be taken over polynomials in equivariant parameters $\C[\varepsilon,...]$ meaning the parameters act the same way in each factor. Since we treat $\varepsilon$ as a number anyway, we have not written this explicitly.}
\be \label{fact} \begin{array}{rcl} H_{\C^*_\varepsilon}^*(\CM_D(\CB_\nu,\V_\CX)) &\simeq& H_{\C^*_\varepsilon}^*(\CM_D(\CB_\nu,\id)) \otimes V \,,  \\[.1cm]
 H_{\C^*_\varepsilon}^*(\CM_{\rm rav}(\CB_\nu;\V_\CX,\V_\CX)) &\simeq& H_{\C^*_\varepsilon}^*(\CM_{\rm rav}(\CB_\nu;\id,\id)) \otimes V \otimes V^*\,, \\[.1cm]
 H_{\C^*_\varepsilon}^*(\CM_{\rm rav}(\CB_\nu;\CL,\V_\CX)) &\simeq& H_{\C^*_\varepsilon}^*(\CM_{\rm rav}(\CB_\nu;\CL,\id))  \otimes V^*\,,
\end{array}
\ee
etc., 
with $V\simeq V^*\simeq H_{\C^*_\varepsilon}^*[\CX]$. In turn, the convolution products from Section \ref{sec:comp}  factorize, into convolutions involving the moduli spaces for $\id$, and simple quantum-mechanics convolution algebras for $V$ as in Section \ref{sec:convQM}.

Assuming that there exist boundary conditions such that the representations of local operators at various junctions are faithful, we are led to conclude that there are isomorphisms
\be \label{fact-End} \begin{array}{ccc} \text{End}^\varepsilon_A(\V_\CX) &\simeq& \text{End}^\varepsilon_A(\id)\otimes \text{End}_\C(V)\,, \\[.1cm]
 \text{Hom}_A^\varepsilon(\CL,\V_\CX) &\simeq& \text{Hom}_A^\varepsilon(\CL,\id) \otimes V \quad \forall\,\CL\,, \\[.1cm]
 \text{Hom}_A^\varepsilon(\V_\CX,\CL) &\simeq& \text{Hom}_A^\varepsilon(\id,\CL) \otimes V^*\quad \forall \,\CL\,.
 \end{array}
 \ee
Moreover, in any piece of the category of line operators that can be represented faithfully by a particular boundary condition, the isomorphisms will be \emph{functorial}. Under the assumption that functoriality can be extended to the entire category, the isomorphisms \eqref{fact-End} imply that the line operators $\V_\CX$ and $\id\otimes V$ are equivalent.

\subsection{The Iwahori line and boundary condition}

Now we move on to our main example.
Consider SQCD with gauge group $G=U(2)$ and $N_f=4$ fundamental hypermultiplets, \emph{i.e} $R=(\C^2)^{\oplus 4}$. Explicitly, we denote the hypermultiplet scalars $(X,Y)\in T^*R$ as
\be X^a{}_i\,,\quad Y^i{}_a\,,\ee
where $a=1,2$ indexes the fundamental/antifundamental representation of $G=U(2)$ and $i=1,...,4$ indexes the antifundemental/fundamental representation of the flavor symmetry $F = PSU(4)$. Complexified gauge transformations $g\in G_\C = GL(2,\C)$ act as $X_i\to g X_i$, $Y^i\to Y^i g^{-1}$.

We introduce the Iwahori line operator $\V_\CI$, with algebraic data
\be \CG_0 = \CI\,,\qquad \CL_0 = R(\CO)\oplus R^*(\CO)\,, \ee
where the Iwahori subgroup is
\begin{align}
	\CI = \left \{ 
	g(z) = \begin{pmatrix}
		a(z) & b(z) \\
		z\,c(z) & d(z)
	\end{pmatrix} \in G(\CO) \right \} = \big\{g\in G(\CO)\;\big|\; g(0)\in B\big\}\,.
\end{align}
Equivalently, $\V_\CI$ is defined by coupling to $\SQM_A$ quantum mechanic with target $\CX = \cp^1$ by gauging the $U(2)$ flavor symmetry (note only $PSU(2)$ acts nontrivially); or, algebraically, by gauging $G_\C=GL(2,\C)$.

We will compute $\text{End}_A^\varepsilon(\V_\CI)$, $\text{Hom}_A^\varepsilon(\id,\V_\CI)$, and $\text{Hom}_A^\varepsilon(\V_\CI,\id)$ in the formalism of Section~\ref{sec:comp}. We choose a massive vacuum $\nu$, and assume that the representation of these spaces in the presence of the vacuum boundary condition $\CB_\nu$ is faithful. The fact that we eventually recover all the expected structure of a matrix algebra on $\V_\CI$ confirms that the assumption is reasonable.

For the massive vacuum $\nu$ we take
\be \label{vacU2}
\nu: \hspace{0.5cm} X^a{}_i = \delta^a{}_i = \bp 1&0&0&0 \\ 0&1&0&0\ep\,, \hspace{1cm} Y^i{}_a \equiv 0\,.
\ee
Note that this satisfies the requirements of Section \ref{sec:Mrav}: the complex moment map vanishes, $(\mu_\C)^i{}_j(\nu)  = \sum_a Y^i{}_aX^a{}_j \big|_\nu=0$; the complexified gauge group $G_\C$ is completely broken by the choice of $X$; and a torus $T_F=(\C^*)^4/\C^*$ of the flavor group $F_\C$ is preserved, after compensating with gauge transformations.

Physically, $\nu$ becomes a massive vacuum when generic complex masses and a real FI parameter are turned on. The FI parameter should be negative, $t_\R<0$, in order for a point on the $G_\C$ orbit of $\nu$ to satisfy the real moment-map constraint $X^\dagger X-Y Y^\dagger+t_\R=0$.

\subsection{Disc Hilbert spaces}\label{sec:IwahoriVortexHilbertSpace}

Given the asymptotic boundary condition $\CB_\nu$, we can construct vortex moduli spaces $\CM_D(\CB_\nu,\id)$ and $\CM_D(\CB_\nu,\V_\CI)$, and take cohomology to get the corresponding Hilbert spaces $\CH(\CB_\nu,\id)$ and $\CH(\CB_\nu,\V_\CI)$. Recall that local operators on $\V_\CI$ and at junctions between $\V_\CI$ and $\id$ will then be represented as maps among the Hilbert spaces.

\subsubsection{Trivial line}

From Section \ref{sec:Mrav}, we find that the space of solutions to BPS equations in the presence of the trivial line is
\be \label{MDid} \CM_D(\CB_\nu,\id) = G[z]\big\backslash \big(\CL_0\,\cap\, [G(\CK_\infty)\cdot \text{val}_\infty^{-1}(\nu)] \big)\,.\ee
This is a vortex moduli space that was also analyzed in \cite{VV}, following a careful physical construction in \cite{Eto1,Eto2}. We recall some relevant details here, and then generalize to $\V_\CI$.

In $\CL_0\,\cap\, [G(\CK_\infty)\cdot \text{val}_\infty^{-1}(\nu)]$, the $Y$ hypermultiplets are identically zero, while the $X$ hypermultiplets contain polynomial entries
\be X(z) = \bp X^1{}_1(z) & X^1{}_2(z) & X^1{}_3(z) & X^1{}_4(z)  \\ X^2{}_1(z) & X^2{}_2(z) & X^2{}_3(z) & X^2{}_4(z) \ep\,. \ee
If we denote by $X(z)_I$ the $2\times 2$ submatrix of $X$ with columns $I=(i_1,i_2)$, then being in the inverse image of the vacuum $\nu$ imposes the constraint
\be \label{MDidX}  \text{det} X(z)_{(1,2)}\neq 0\,,\qquad \text{deg}\,\text{det} X(z)_I < \text{deg}\,\text{det} X(z)_{(1,2)} \quad \forall\, I\neq (1,2)\,.\ee
The gauge group $G[z] := GL(2,\C[z])$, consisting of invertible $2\times 2$ entries with polynomial entries, acts as $X(z)\to g(z) X(z)$. After quotienting by $G[z]$, the moduli space decomposes into components
\be \CM_D(\CB_\nu,\id) = \bigsqcup_{\n\in  \Z_{\geq 0}}  \CM_D^\n(\CB_\nu,\id)\,,\ee
where the $\n$-th component contains the matrices $X(z)$ with $\text{deg}\,\text{det} X(z)_{(1,2)} =\n$\,.

As discussed in \cite{Eto1, Eto2}, each component $\CM^{\n}_D(\CB_\nu,\id)$ is smooth, and covered by open affine charts (or ``patches'') that are labeled by nonnegative cocharacters $k=(k_1,k_2) \in \Z^2$ with $k_1 + k_2=\n$. Each chart, denoted $\CM_D^{(\n;k)}(\CB_\nu,\id)$, is of the form
\be \label{bulkGaugeFixed}
	X(z) =  \begin{pmatrix}
		z^{k_1} + \sum_{d=0}^{k_1 - 1} x^1{}_{1 , d} z^d &  \sum_{d=0}^{k_2 - 1} x^1{}_{2 , d} z^d & \sum_{d=0}^{k_1 - 1} x^1{}_{3 , d}z^d & \sum_{d=0}^{k_1 - 1} x^1{}_{4 , d}z^d\\
		\sum_{d=0}^{k_1-1} x^2{}_{1 , d} z^d  & z^{k_2} + \sum_{d=0}^{k_2 - 1} x^2{}_{2 , d} z^d & \sum_{d=0}^{k_2 - 1} x^2{}_{3 , d}z^d & \sum_{d=0}^{k_2 - 1} x^2{}_{4 , d} z^d
	\end{pmatrix},
\ee
and is freely parameterized by the coefficients $x^i{}_{a,d}$ of the various polynomials.

This description makes it easy to describe the equivariant cohomology of $\CM_D^\n(\CB_\nu,\id)$. The combined action of the flavor torus $T_F$ and loop rotation $\C^*_\varepsilon$ has a unique fixed point at the origin of each chart $\CM_D^{(\n;k)}$, \emph{i.e.} at $x^i{}_{a,d}\equiv 0$. 
As usual, a compensating torus-valued gauge transformation is required to keep the point fixed. Denoting the equivariant parameters%
\footnote{As equivariant parameters for $F=PSU(4)$, we use four complex masses $m_i$ $(i=1,...,4)$.
Technically, they are defined up to a simultaneous translation $m_i\mapsto m_i+c$. We fix this ambiguity by imposing the constraint $m_1+m_2+m_3+m_4=0$, as it turns out to simplify several formulas.} %
for $T$ (gauge), $T_F$, and $\C^*_\varepsilon$ as $\varphi_a$, $m_i$, and $\varepsilon$, we find that $x^a{}_{i,d}$ transforms as
\be \label{torusAction} \delta x^a{}_{i,d} \sim \big(\varphi_a+m_i+(d+\tfrac12)\varepsilon\big)x^a{}_{i,d}\,. \ee
Thus, the origin of the chart $\CM_D^{(\n;k)}$ is fixed when $\delta x^a_{a,k_a}=0$, \emph{i.e.}
for $\varphi_a=-m_a-(k_a+\tfrac12)\varepsilon$.
Moreover, the Euler class $\omega_{\n,k}$ of the normal bundle to the fixed point is a product of equivariant weights of the remaining $x^a{}_{i,d}$ in the chart,
\be \label{weight-id}  \omega_{\n,k} = \prod\limits_{a = 1}^2\bigg[\prod\limits_{i=1}^2 \prod\limits_{d=0}^{k_i-1}(m_i - m_a + (d-k_a)\varepsilon)\prod\limits_{i=3}^4 \prod\limits_{d=0}^{k_a-1}(m_i - m_a + (d-k_a)\varepsilon)\bigg]\,. \ee

Using fixed-point localization to compute the cohomology, we find that the disc Hilbert space has graded components $\CH(\CB_\nu,\id) = \bigoplus_{\n\geq 0} \CH^\n(\CB_\nu,\id)$, with the $\n$-th component generated by $\n+1$ fixed-point classes
\be  \CH^\n(\CB_\nu,\id) = H^*_{T_F\times \C^*_\varepsilon}\big( \CM^\n_D(\CB_\nu,\id)\big) \simeq \bigoplus_{k_1,k_2\geq 0\atop k_1+k_2=\n}\C|\n,k\rangle\,.\ee
We normalize the fixed-point states the same way as in abelian theories \eqref{norm-p}, namely
\be |\n,k\rangle = \frac{1}{\omega_{\n,k}} \delta_{\n,k}\,,\ee
where $\delta_{\n,k}$ denotes the (Poincar\'e dual of the) fundamental class of the origin on the $\n,k$ chart.
(We assume that $m,\varepsilon$ take generic values, and invert them at will.)
The equivariant parameters $\varphi$, representing operators formed from complex vectormultiplet scalars, act as
\be \varphi_a |\n,k\rangle = -(m_a+(k_a+\tfrac12)\varepsilon)|\n,k\rangle\,, \ee
In particular, $\text{Tr}\,\varphi=\varphi_1+\varphi_2$ measures vortex number $\n$.

\subsubsection{Iwahori line}

Generalizing the moduli space and Hilbert space to a disc punctured by the Iwahori line $\V_\CI$ is fairly straightforward. From \eqref{MD-alg} and \eqref{MD-QM} we now have
\begin{align} \CM_D(\CB_\nu,\V_\CI) &= G[z]\big\backslash \big(\cp^1 \times \CL_0\,\cap\, [G(\CK_\infty)\cdot \text{val}_\infty^{-1}(\nu)] \big) \\
&= \CI[z]\big\backslash \big(\CL_0\,\cap\, [G(\CK_\infty)\cdot \text{val}_\infty^{-1}(\nu)] \big)\,.
\end{align}
In the first description, the moduli space consists of matrix of polynomials $X(z)$ satisfying constraints \eqref{MDidX}, exactly as for the trivial line (because these constraints come from $\CB_\nu$); together with a choice of point $p = \bsp p_1\\p_2\esp \in \cp^1$. Polynomial-valued gauge transformations $g(z)\in G[z]$ act as
\be p\mapsto g(0)p\,,\qquad X(z)\mapsto g(z)X(z)\,. \ee
In the second description, we have gauge-fixed $p = \bsp 1 \\ 0 \esp$, thereby breaking the gauge group to polynomial-valued elements of the Iwahori subgroup, \emph{i.e.} $g(z)\in G[z]$ satisfying $g(0)\in B$.

Just as in the case of the trivial line, the moduli space decomposes into connected components $\CM_D^\n(\CB_\nu,\V_\CI)$ labeled by vortex number $\n\geq 0$.
More so, it is clear that $\CM_D(\CB_\nu,\V_\CI)$ is a $\cp^1$ fibration over $\CM_D(\CB_\nu,\id)$ (the map $\CM_D(\CB_\nu,\V_\CI)\to \CM_D(\CB_\nu,\id)$ just forgets $p\in \cp^1$). Thus we expect each $\CM_D^\n(\CB_\nu,\V_\CI)$ to be labeled by affine charts that are \emph{products} of the charts \eqref{bulkGaugeFixed}, and two standard charts on $\cp^1$ (omitting either the north or south poles). Abstractly, the charts on $\cp^1=G_\C/B$ are labeled by elements of the Weyl group $\sigma \in \text{Weyl}(G) = \Z_2 = \{ 1, w\}$.

We can describe the new affine charts quite explicitly. For given $\n$, the charts are labeled by nonnegative cocharacters $k=(k_1,k_2)$ satisfying $k_1+k_2=\n$ \emph{and} by $\sigma\in \{1,w\}$.
After fixing $p = \bsp 1\\0 \esp$, we can subsequently gauge-fix the Iwahori $\CI[z]$ by setting
\be \label{Iwahori-charts}
\begin{array}{rl}
	\CM_D^{(\n,k,1)}: & X(z) = \begin{pmatrix}
		z^{k_1} + \sum_{d=0}^{k_1 - 1} x^1{}_{1 , d} z^d &  \sum_{d=0}^{k_2 - 1} x^1{}_{2 , d} z^d & \sum_{d=0}^{k_1 - 1} x^1{}_{3 , d}z^d & \sum_{d=0}^{k_1 - 1} x^1{}_{4 , d}z^d\\
		\sum_{d=0}^{k_1} x^2{}_{1 , d} z^d  & z^{k_2} + \sum_{d=0}^{k_2 - 1} x^2{}_{2 , d} z^d & \sum_{d=0}^{k_2 - 1} x^2{}_{3 , d}z^d & \sum_{d=0}^{k_2 - 1} x^2{}_{4 , d} z^d
	\end{pmatrix}\\
	\text{or} &\\
	\CM_D^{(\n,k,w)}: & X(z) = \begin{pmatrix}
		\sum_{d=0}^{k_1 - 1} x^1{}_{1 , d} z^d &  z^{k_2} + \sum_{d=0}^{k_2 - 1} x^1{}_{2 , d} z^d & \sum_{d=0}^{k_2 - 1} x^1{}_{3 , d}z^d & \sum_{d=0}^{k_2 - 1} x^1{}_{4 , d} z^d\\
		z^{k_1} + \sum_{d=0}^{k_1-1} x^2{}_{1 , d} z^d  & \sum_{d=0}^{k_2} x^2{}_{2 , d} z^d & \sum_{d=0}^{k_1 - 1} x^2{}_{3 , d}z^d & \sum_{d=0}^{k_1 - 1} x^2{}_{4 , d} z^d
	\end{pmatrix}\\
\end{array}
\ee 
One can determine transition functions on the overlaps by solving for a $g\in \CI[z]$ that sends the gauge-fixed form of one chart to that of another.

Again, there is a unique fixed point for $T_F\times \C^*_\varepsilon$ (up to a compensating torus-valued gauge transformation) at the origin of each chart. In terms of equivariant parameters, the compensating action is fixed by
\begin{align}\label{gaugeFixingIwahori}
	\varphi_a + m_{\sigma(a)} + (k_{\sigma(a)} + \tfrac12) \varepsilon = 0\,,,
	\qquad \sigma \in \{ 1 , w \} .
\end{align}
The Euler class of the normal bundle to each fixed point is given by
\begin{align}
	 \omega_{\n,k,\sigma} = (-1)^\sigma (m_1 - m_2 + ( k_1 - k_2 ) \varepsilon )\, \omega_{\n,k}\,,
\end{align}
where $(-1)^\sigma$ is the signature of the permutation $\sigma$ and  $\omega_{\n,k}$ is the corresponding Euler class for the trivial line \eqref{weight-id}.
Normalizing fixed-point classes as
\be |\n,k,\sigma\rangle = \frac{1}{\omega_{\n,k,\sigma}}\delta_{\n,k,\sigma}\,, \ee
we find a Hilbert space
\be \label{lineHilbert} \CH(\CB_\nu,\V_\CI) \simeq \bigoplus_{\n\geq 0} \bigoplus_{\scriptsize\begin{array}{c}k_1,k_2\geq 0\,, \\ k_1+k_2=\n\,; \\ \sigma\in \{1,w\}\end{array}} \C|\n,k,\sigma\rangle\; \simeq\; \CH(\CB_\nu;\id) \otimes_{\C[\varphi]} H^\bullet_{\C^*}(\cp^1)\,.  \ee
(Here $\C[\varphi]=\C[\varphi_1,\varphi_2]$ is the ring of polynomials in gauge equivariant parameters, and tensoring over it on the RHS means that its actions on $\CH(\CB_\nu;\id)$ and $H^*_{\C^*}(\cp^1)$ are compatible.)

\subsection{Local operators on the Iwahori line}

We are ready to begin describing local operators in our SQCD example.

In \cite{VV, Braverman-W, BFFR}, it was shown that the space of bulk local operators $\text{End}_A^\varepsilon(\id) = \C_\varepsilon[\CM_C]$ is faithfully represented in the cohomology of the raviolo space $\CM_{\rm rav}(\CB_\nu;\id,\id)$. We will review this result momentarily. The algebra structure on local operators, and their compatible action on the disc Hilbert space
\be H_{T_F\times \C^*_\varepsilon}^*\big(\CM_{\rm rav}(\CB_\nu;\id,\id)\big) :\; \CH(\CB_\nu,\id) \to \CH(\CB_\nu,\id) \label{act-id} \ee
both come from convolution, as discussed in Section \ref{sec:conv}. Since the representation \eqref{act-id} is faithful, the algebra structure of local operators may be fully reconstructed from it.

In this section, we will focus on the local operators $\text{End}_A^\varepsilon(\V_\CI)$ bound to the Iwahori line, which are represented in the cohomology of the raviolo space $\CM_{\rm rav}(\CB_\nu;\V_\CI,\V_\CI)$.  Now there is a convolution action
\be H_{T_F\times \C^*_\varepsilon}^*\big(\CM_{\rm rav}(\CB_\nu;\V_\CI,\V_\CI)\big):\; \CH(\CB_\nu,\V_\CI)\to  \CH(\CB_\nu,\V_\CI)\,,\ee
from which the algebra structure of local operators may be reconstructed.

\subsubsection{Summary of bulk local operators}
\label{sec:SQCD-bulk}

The bulk algebra $\text{End}_A^\varepsilon(\id) \simeq \C_\varepsilon[\CM_C]$, quantized in the Omega background, is now well known from many perspectives. The Coulomb branch $\CM_\C$ itself is a moduli space of two $PSU(2)$ monopoles in the presence of a Dirac singularity \cite{HananyWitten}, \emph{a.k.a.} a slice in the $PGL(2)$ affine Grassmannian \cite{BDG}; its quantization produces a finite W-algebra, or, equivalently, a central quotient of a Yangian \cite{KWWY, BDG}. This algebra was computed using the particular methods of Section \ref{sec:comp} in \cite{Braverman-W, VV}. It was also computed using `BFN' methods in \cite{BFN-quiver}. Moreover, the Coulomb-branch algebra $\C_\varepsilon[\CM_C]$ is equivalent to the quantized Higgs-branch algebra of a mirror quiver that we will discuss in greater detail in Section \ref{sec:FundamentalWilsonLine} and Appendix \ref{app:Wilson}.

Using any of these approaches, one finds that $\text{End}_A^\varepsilon(\id)$ is generated (as an algebra) by six operators $V^0,V^+,V^-,W^0,W^+,W^-$. Our 3d $\CN=4$ theory has a topological $U(1)_{\rm top}$ flavor symmetry acting on the Coulomb branch, whose charge is monopole number. This symmetry is enhanced in the infrared to $PSU(2)_{\rm top}$ \cite{GW-Sduality}, whose complexification acts on the Coulomb-branch chiral ring. The operators $V^0,V^+,V^-$ are the components of the moment map that generates this action, and the $W$ operators transform as an additional adjoint representation:
\be \label{SQCD-comm} \begin{array}{c}
[\Phi,V^\pm] = \pm 2\varepsilon V^\pm\,,\qquad  [V^+,V^-]=\varepsilon \Phi\,, \\[.2cm]
[\Phi,W^\pm] = \pm 2\varepsilon W^\pm\,, \qquad [\Phi,W^0]=0\,,\qquad
 [V^+,W^+]=[V^-,W^-]=0\,, \\[.2cm]
[V^\pm,W^0]=  \mp 2\varepsilon W^\pm\,,\qquad [V^+,W_-]=-[V^-,W^+]=\varepsilon W^0\,.
\end{array}
\ee

More physically, the operator $V^0 = -2(\varphi_1+\varphi_2) =-2\text{Tr}\varphi$ is the trace of the vectormultiplet scalar, and $V^\pm$ are fundamental nonabelian monopole operators of charge $\pm 1$, defined by the dominant cocharacters $(1,0)$ and $(0,-1)$, respectively.
The operators $W^\pm$ are dressed monopoles of charge $\pm1$; while $W^0$ is a mixture of $\text{Tr}(\varphi^2)$ and a monopole operator defined by cocharacter $(1,-1)$, of total monopole charge zero. We also note that, in units where $\varepsilon$ has charge $+1$ under the $U(1)_C$ R-symmetry, $V^0,V^\pm$ have R-charge $+1$ and $W^0,W^\pm$ have R-charge $+2$.

The algebra $\text{End}_A^\varepsilon(\id)$ is generated by these six operators, subject to \eqref{SQCD-comm} and two additional relations. The two extra relations conveniently described by defining the $PSU(2)_{\rm top}$-invariant combinations
\be
\begin{array}{rl}
	V^2 &:= 2 V^+ V^- + 2 V^- V^+ + (V^0)^2\,,\\
	VW &:= 2 V^+ W^- + 2 V^- W^+ + V^0 W^0\,,\\
	W^2 &:= 2 W^+ W^- + 2 W^- W^+ + (W^0)^2\,.\\
\end{array}
\ee
With all complex masses set to zero, the two relations are simply $VW=0$ and $W^2 = \frac14(V^2)^2+3\varepsilon V^2+3\varepsilon^2$. At generic mass parameters, the relations are deformed to
\be \notag
\begin{array}{rl}
	VW & = \tfrac{1}{2}(m_R - m_L)\big[V^2 + (m_R+m_L+2 \varepsilon)(m_R+m_L-2 m_C-2 \varepsilon)\big]\,,\\[.2cm]
	W^2 & = \tfrac{1}{4} (V^2)^2 - \tfrac{1}{2}(2 m_C^2 - 2 m_C (m_L + m_R) + (m_L + m_R)^2 + 4 \varepsilon m_C -2 \varepsilon^2) V^2\\
	&\quad  + (2 m_C^2 (m_L^2 + m_R^2) - m_C (m_L + m_R) (3 m_L^2 - 2 m_L m_R + 3 m_R^2) \\ &\quad + \tfrac{1}{4}(m_L + m_R)^2 (5 m_L^2 - 6 m_L m_R + 5 m_R^2)\\
	&\quad + 2 \varepsilon m_C (m_L + m_R) (2 m_C - m_L - m_R) - \varepsilon^2 ((m_L - m_R)^2 - 8 m_C (m_L + m_R) + 4 (m_L + m_R)^2))\,,
\end{array}
\ee
with $m_R := m_4 - m_3 - \varepsilon$, $m_L := m_2 - m_1 - \varepsilon$, and $m_C = m_4 - m_1 - 2\varepsilon$.

Finally, let us describe the algebra $\text{End}_A^\varepsilon(\id)$ in terms of its representation in the equivariant cohomology of $\CM_{\rm rav}(\CB_\nu;\id,\id)$ --- since this is the approach that will generalize.
The raviolo space itself simplifies to
\be \CM_{\rm rav}(\CB_\nu;\id,\id) = G[z]' \big\backslash \{ (X',g,X) \} \big/ G[z]\,, \label{Mrav-idid} \ee
where $X(z)$ and $X(z)'$ are each of the form \eqref{MDidX}, $g(z)\in G[z,z^{-1}]$, and $X'=gX$.
The space breaks up into components $\CM_{\rm rav}^{\n,\n'}(\CB_\nu;\id,\id)$ labeled by pairs of vortex numbers $\n,\n'\geq 0$, with each component containing matrices $X',X$ such that $\text{deg}\,\text{det} X_{(1,2)}'=\n'$ and $\text{deg}\,\text{det} X_{(1,2)}=\n$. Notably, within each component, we can fully gauge-fix both $G[z]'$ and $G[z]$, by requiring both $X'$ and $X$ to be of the form \eqref{bulkGaugeFixed}. 

Now, for any cocharacter $A = (A_1,A_2)$ of $G$, let $\CO_A$ denote the double orbit
\be \CO_A := {G[z]'\,z^A\,G[z]} \,\subset G[z,z^{-1}] \ee
under the left and right $G[z]$ actions. The orbit only depends on the Weyl-conjugacy class of~$A$. Similarly, let $\CS_A$ denote the subvariety of $\CM_{\rm rav}$ with $g$ restricted to lie in $\CO_A$,
\be \CS_A := G[z]' \big\backslash \{ (X',g\in \CO_A,X) \} \big/ G[z] \,\subset \CM_{\rm rav}(\CB_\nu;\id,\id)\,. \ee
Note that $\CS_A$ only intersects components $\CM_{\rm rav}^{\n',\n}$ with $\n'-\n=A_1+A_2$.
Also recall that there are maps
\be \CM_D(\CB_\nu,\id) \overset{\pi'}{\longleftarrow} \CM_{\rm rav}(\CB_\nu;\id,\id) \overset\pi\longrightarrow \CM_D(\CB_\nu,\id)\,, \ee
with $\pi'$, $\pi$ forgetting $(g,X)$ and $(X',g)$, respectively. The map $\pi|_{\CS_A}$ need not surject onto $\CM_D(\CB_\nu, \id)$; a given singular gauge transformation won't map a generic non-singular field configuration to a non-singular field configuration. Nonetheless, the fiber of $\pi|_{\CS_A}$ over a point in the image of $\pi|_{\CS_A}$ is exactly a copy of $G[z]' \backslash \CO_A \subset G[z]' \backslash G[z,z^{-1}].$

Operators of monopole charge $\mathfrak m$ come from cohomology classes supported on the closure of $\CS_A$ with $A_1+A_2=\mathfrak m$. We only need to consider $\mathfrak m=0,1,-1$, as operators of higher monopole number are generated by successive convolutions. For trivial cocharacter $A=(0,0)$, we find that both $\pi$ and $\pi'$ are isomorphisms when restricted to $\CS_{(0,0)}$,
\be   \CM_D(\CB_\nu,\id) \underset{\sim}{\overset{\pi'}{\longleftarrow}} \CS_{(0,0)} \underset{\sim}{\overset\pi\longrightarrow} \CM_D(\CB_\nu,\id)\,, \ee
and that the fundamental class $[\CS_{(0,0)}]$ is the identity operator $\mb 1\in \text{End}_A^\varepsilon(\id)$.
The operator $V^0=-2\text{Tr}\,\varphi$, which is a ``dressed'' version of the identity, is the equivariant class $-2(\varphi_1+\varphi_2)[\CS_{(0,0)}]$, where $\varphi_a$ are equivariant parameters for $G[z]$. This class evaluates to $2\varepsilon \n+2(m_1 + m_2+\varepsilon)$ within each component $\CM_{\rm rav}^{\n,\n}$.

For the fundamental cocharacter $A=(1,0)$, we find that $ \pi :\CS_{(1,0)} \to\hspace{-2.7ex}\to  \CM_D(\CB_\nu,\id)$ is a fibration, with fiber $G[z]'\backslash \CO_{(1,0)} \simeq \cp^1$. This $\cp^1$ is a familiar orbit in the affine Grassmannian of $GL(2)$. The monopole operator $V^+$ is the fundamental class $[\CS_{(1,0)}]$, while the dressed monopole $W^+$ is a linear combination of the equivariant volume form on the $\cp^1$ fiber, tensored with the fundamental class of the base, a $(V^0+ \rm{masses})V^+$. For antifundamental cocharacter $A=(-1,0)$, we find that $\pi' :\CS_{(-1,0)} \to\hspace{-2.7ex}\to  \CM_D(\CB_\nu,\id)$ is a $\cp^1$ fibration, with $V^-$ coming from its fundamental class $V^-$ and $W^-$ coming from a linear combination of the volume form and $(V^0+ \rm{masses})V^-$. Finally, the operator $W^0$ gets contributions from the classes $((V^0)^2+aV^0+b)[\CS_{(0,0)}]$, $\text{Tr}(\varphi^2)[\CS_{(0,0)}]$ and $[\CS_{(1,-1)}]$, where $a,b$ are functions of the masses $m_i$ and $\varepsilon$.

\subsubsection{Monopole number 0}\label{sec:IwahoriMono0}

Now consider the raviolo space in the presence of the Iwahori line,
\begin{align} \label{Mrav-VIVI} \CM_{\rm rav}(\CB_\nu;\V_\CI,\V_\CI) &= \CI[z]'\big\backslash \{(X',g,X)\} \big/ \CI[z]\,, \end{align}
with $X(z)',g(z),X(z)$ exactly the same as in \eqref{Mrav-idid}. This space splits into components $\CM_{\rm rav}^{\n',\n}(\CB_\nu;\V_\CI,\V_\CI)$, labeled by the degrees $\n',\n\geq 0$ of the determinant of $X_{(1,2)}'$ and $X_{(1,2)}$.
We would like to describe some equivariant cohomology classes in these components.

For fixed monopole number $\mathfrak m=\n'-\n$, we should consider $g(z)$ with $\det g(z)=\text{const}\cdot z^{\mathfrak m}$. 
We focus on small monopole number $|\mathfrak m|\leq 1$, because the bulk algebra was generated by operators with $|\mathfrak m|\leq 1$, and we expect that the algebra on the Iwahori line will behave similarly. We begin here with $\n'=\n$, \emph{i.e.}  operators of monopole number zero.
There are also natural maps
\be  \CM_D(\CB_\nu,\V_\CI) \overset{\pi'}{\longleftarrow}  \CM_{\rm rav}(\CB_\nu;\V_\CI,\V_\CI) \overset{\pi}{\longrightarrow} \CM_D(\CB_\nu,\V_\CI) \ee
that forget either $(g,X)$ or $(X',g)$.

The most basic local operators of monopole number zero come from orbits of the identity. Let
\be \CO_{id} = \CI[z]'\bsp 1&0\\0&1 \esp \CI[z] \simeq \CI[z]\;\subset G[z, z^{-1}] \ee
be the double (Iwahori) orbit of the identity, and let
\be \CS_{id} = \CI[z]'  \backslash \big\{\big(X',g\in \CO_{id},X\big)\big\}/\CI[z]
 \;\subset\; \bigsqcup_{\n\geq 0} \CM_{\rm rav}^{\n,\n} \;\subset\; \CM_{\rm rav}\, \ee
be the subvariety of $\CM_{\rm rav}(\CB_\nu;\V_\CI,\V_\CI)$ with $g$ restricted to lie in $\CI_{id}$. Then both maps $\pi:\CS_{id} \overset{\sim}\to \CM_D$, $\pi':\CS_{id} \overset{\sim}\to \CM_D$ are isomorphisms; indeed, we may identify $\CS_{id}$ with the diagonal
\be \CS_{id} \simeq \Delta \subset \CM_D(\CB_\nu,\V_\CI)\times \CM_D(\CB_\nu,\V_\CI)\,. \ee
Thus the fundamental class $[\CS_{id}]$  in equivariant cohomology represents the identity operator
\be [\CS_{id}]= \mb 1:\,\CH(\CB_\nu,\V_\CI) \overset{\sim}\longrightarrow \CH(\CB_\nu,\V_\CI)\,.\ee

We also find operators corresponding to the bulk vectormultiplet scalars, which may be considered dressed versions of the identity (or Chern classes of bundles on $\CS_{id}$). In the bulk, we had to take $G$-invariant combinations  $\text{Tr}(\varphi) = \varphi_1+\varphi_2$ and $\text{Tr}(\varphi^2)$. On the Iwahori line, we have access to $\varphi_1$ and $\varphi_2$ independently. They are the equivariant cohomology classes
\be \varphi_a := \varphi_a'[\CS_{id}] =  \varphi_a[\CS_{id}] \,\in\, H^*_{\CI[z]'\times \CI[z] \times T_F\rtimes \C^*_\varepsilon}(\wt\CM_{\rm rav})\,, \ee
where $\CM_{\rm rav} = \CI[z]'\backslash \wt \CM_{\rm rav}/ \CI[z]$. We already know from \eqref{gaugeFixingIwahori} that their action on a fixed-point basis of $\CH(\CB_\nu,\V_\CI)$ is given by
 \be
\varphi_a \ket{\n,k,\sigma} = -(m_{\sigma(a)}+(k_{\sigma(a)}+\tfrac{1}{2})\varepsilon)\ket{\n,k,\sigma}.
\ee

More interestingly, we may consider the orbit of $\bsp 1&0\\0&1\esp$ under $G[z]'\times G[z]$,
\be \CO_{(0,0)} := G[z]' \bsp 1&0\\0&1\esp G[z]\,\subset G[z, z^{-1}]\,,\ee
and the corresponding subvariety
\begin{align} \CS_{(0,0)} &= \CI[z]'\backslash  \big\{\big(X',g\in \CO_{(0,0)},X\big)\big\}/\CI[z]
 \;\subset\; \bigsqcup_{\n\geq 0} \CM_{\rm rav}^{\n,\n} \;\subset\; \CM_{\rm rav} \\
  &\simeq \CI[z]' \backslash   \big\{\big(g\in \CO_{(0,0)},X\big)\big\}/\CI[z]\,. \notag \end{align}
For the second line, note that no information is carried in $X'$, since the constraint $X'=gX$ determines it uniquely; moreover, as long as $g\in \CO_{(0,0)}$, requiring $X'$ to be regular at the origin imposes no additional conditions on $X$. The map $\pi$ now gives us a smooth fibration
\be \pi : \CS_{(0,0)} \to \CM_D(\CB_\nu,\V_\CI)\,, \label{mon0-fib} \ee
whose fibers are isomorphic to the orbits $\CI[z]'\backslash\CO_{(0,0)} \subset \CI[z]'\backslash G[z^\pm]$. The space  $\CI[z]'\backslash G[z^\pm]$ is a version of the affine flag variety, and it is well known that the orbit $\CI[z]'\backslash\CO_{(0,0)}$ inside it is a copy of $\cp^1$, covered by two affine charts
\be\begin{pmatrix}1&0\\c&1\end{pmatrix} \qquad  \begin{pmatrix}0&1\\1&d \end{pmatrix},\ee
with transition function $d = 1/c$. We expect to find local operators in $\text{End}_A^\varepsilon(\V_\CI)$ labeled by cohomology classes of this $\cp^1$ fiber.

A more universal way to analyze the setup is the following. Under the map
\be \pi_g: \CM_{\rm rav}(\CB_\nu;\V_\CI,\V_\CI) \to \CI[z]' \backslash G[z,z^{-1}]/\CI[z] \ee
that forgets $X'$ and $X$, the subvariety $\CS_{(0,0)}$ maps to the substack $\CI[z]' \backslash \CO_{(0,0)}/\CI[z]  \subset \CI[z]' \backslash G[z,z^{-1}]/\CI[z]$. We can construct classes in $H^*_{\CI[z]'\times \CI[z]\times T_F\rtimes \C^*_\varepsilon}(\wt\CM_{\rm rav})$ that are supported on $\CS_{(0,0)}$ by pulling them back from classes in $H^*_{\CI[z]'\times \CI[z]}(G[z,z^{-1}])$ that are supported on $\CO_{(0,0)}$. If we write $H^*_{\CI[z]'\times \CI[z]}(G[z,z^{-1}]) \simeq H^*_{\CI[z]}(\CI[z]'\backslash G[z,z^{-1}])$, then we are precisely looking at $\CI[z]$-equivariant cohomology classes on $\CI[z]'\backslash \CO_{(0,0)} \simeq \cp^1$, which is equivalent to $\C^*$-equivariant cohomology classes on $\cp^1$.

Recall some general elementary facts about the $\C^*$-equivariant cohomology of $\cp^1$. Let `$n$' and `$s$'
denote the points at the poles, fixed by the $\C^*$ action, and let $\alpha$ denote the equivariant parameter for $\C^*$. Then, as a module over $H^*_{\C^*}(pt) = \C[\alpha]$, the equivariant cohomology $H^*_{\C^*}(\cp^1)$ is freely generated by the fundamental class $[\cp^1]$ and the equivariant volume form $[\omega]$. The Atiyah-Bott localization formula relates these to fixed-point classes via
\begin{align}\label{GOmodLeftIClasses}
	[\cp^1] =  \frac{\delta_n - \delta_s }{\alpha}\,,
	&&[\omega] = \delta_n + \delta_s\,.
\end{align}

In the case at hand, the equivariant parameter for the $\C^*$ action on $\cp^1\simeq \CI[z]'\backslash \CO_{(0,0)}$ is identified as
\be \alpha = \varphi_2-\varphi_1\,. \ee
Define the pullbacks of the various $\cp^1$ classes to be
\be \delta_1 := \pi_g^* \delta_n\,,\qquad \delta_w := \pi_g^*\delta_s\,,\qquad \pd := \pi_g^* [\cp^1]\,,\qquad s:= \mb 1-\pi_g^*[\omega]\,.\ee
They are cohomology classes supported on $\CS_{(0,0)}$, which are constant along the base of the fibration \eqref{mon0-fib}. Moreover, they satisfy the same relations as in \eqref{GOmodLeftIClasses}, \emph{e.g.} $\pd = (\delta_1-\delta_w)/\alpha$.

Finally, we would like to determine the convolution product among these classes. We will do this by computing the action of $\alpha=\varphi_2-\varphi_1$, $\delta_1$, and $\delta_w$ on the fixed-point basis of $\CH_D(\CB_\nu,\V_\CI)$. From  \eqref{gaugeFixingIwahori}  we easily find that
\be \label{phiAction} \alpha \ket { \n , k , \sigma} = (-1)^\sigma (m_1 - m_2 + ( k_1 - k_2 ) \varepsilon ) \ket { \n , k , \sigma}\,. \ee
The action of $\delta_\sigma$ (for $\sigma=1,w$) is given by convolution
\begin{align}
	\delta_{\sigma'} \ket {\n , k , \sigma }
	= \pi'_* ( \delta_{\sigma'} \wedge \pi^* |\n,k,\sigma\rangle)
	 = \frac{1}{\omega_{\n , k,\sigma}} \pi'_* ( \delta_{\sigma'} \wedge \pi^* \delta_{\n , k ,\sigma} )\,.
\end{align} 
The pullback $\pi^* \delta_{\n , k , \sigma}$ gives the fundamental class of the entire $\cp^1$ fiber
that lies over the fixed point $(k, \sigma)$.
Wedging with $\delta_{\sigma'}$ we obtain
the fundamental class of the fixed point $(k, \sigma)$ in the base times the $\sigma'$ fixed point in the fiber. Let's denote this as $\delta_{\n, k ; \sigma ; \sigma'}$. Pushing forward by $\pi'$ simply sends this to the corresponding fixed-point class in the space of $X'$, supported on 
$X' = g X$, so that $\pi'_* \delta_{\n, k ;\sigma ; \sigma'} = \delta_{\n , k ; \sigma' \sigma}$. Accounting for normalizations of states by Euler classes, we finally obtain
\be \label{fixedPtAction1}
	\delta_{\sigma'} \ket { \n , k , \sigma} = \frac{\omega_{\n , k,\sigma' \sigma}}{\omega_{\n , k,\sigma}} \ket { \n , k , \sigma' \sigma} = (-1)^{\sigma'} \ket { \n , k , \sigma' \sigma} .
\ee

From \eqref{fixedPtAction1}, we easily see that $\delta_1$ acts the same way as the identity operator $\mb 1$ found earlier. Moreover, $s=-\delta_w$ acts as a Weyl reflection
\be s \ket{\n , k , \sigma } = \ket{ \n, k , w \sigma }\,,  \label{sAction} \ee
and the fundamental class of the fiber $\delta = (\delta_1-\delta_w)/\alpha$ acts as
\be \partial \ket { \n , k , \sigma } = \frac{(-1)^\sigma}{m_1 - m_2 + ( k_1 - k_2 ) \varepsilon} \left( \ket{ \n , k , 1} + \ket {\n , k , w }\right)\,. \label{pdAction} \ee
It is also useful to note that $s\,\alpha=-\alpha\,s$ (where now the two sides denote the convolution product, \emph{not} the wedge/cup product in cohomology); intuitively, this is because the gauge transformation $g=\bsp 0&1\\1&0\esp$ associated to $s=\delta_w$ swaps the equivariant parameters $\varphi_1,\varphi_2$, whose difference appears in $\alpha$.

\subsubsection{The nil-Hecke algebra}
\label{sec:nilHecke}

The operators $\mb 1$, $\alpha=\varphi_2-\varphi_1$, $s$, and $\pd$ found above generate a copy of the nil-Hecke algebra $\mb H_2$ for $SL(2)$ \cite{KostantKumar}. Indeed, it is easy to verify the standard algebra relations
\be \begin{array}{c} s^2=\mb 1\,, \qquad \pd^2 = 0\,,\qquad s\pd = -\pd s = \pd\,, \\[.2cm]
 s\alpha = -\alpha s\,,\qquad  \{\pd,\alpha\} = 2\cdot\mb 1 \,,\qquad [\pd,\alpha] = 2s\,,\end{array} \ee
where \emph{all} products come from convolution, and are induced from the action on the disc Hilbert space in \eqref{phiAction}, \eqref{sAction}, \eqref{pdAction}. Abstractly, the nil-Hecke algebra is obtained from the polynomial algebra $\C[\alpha]$ by first adjoining the Weyl reflection $s$ and then the ``Demazure operator''
\be \pd = \frac{1}{\alpha}(1-s)\,. \ee 

Geometrically, the nil-Hecke algebra is defined as the equivariant cohomology
\be \mb H_2 = H_{SL(2,\C)}^*(\cp^1\times \cp^1)\,, \ee
with its natural convolution product. By writing $\cp^1 = SL(2,\C)/B$, one also obtains the equivalent description $\mb H_2 \simeq H^*(B\backslash SL(2,\C)/B)$. As discussed in the introduction to this section, we expected the nil-Hecke algebra to appear on the Iwahori line due to the appearance of $\cp^1\times \cp^1$ in the raviolo space
\be \CM_{\rm rav}(\CB_\nu;\V_\CI,\V_\CI) \simeq G[z]'\big\backslash \big(\cp^1\times \{X',g,X\}\times \cp^1\big)\big/ G[z]\,.\ee

\subsubsection{Monopole Number 1}\label{sec:IwahoriMono1}
Next, let's identify the basic operators of monopole number 1. We will find that they have the structure of a product of the nil-Hecke algebra $\mb H_2$ and an abelianized monopole algebra, along the lines of \cite{BDG, Web2016}.

We must consider cohomology classes supported on subvarieties with $\det g(z) = \text{const}\cdot z$. The most basic such classes come from the full $G[z]'\times G[z]$ orbit of $g(z)=\bsp z&0\\0&1\esp  = z^{(1,0)}$, 
\be \CO_{(1,0)} := G[z]'\cdot \bsp z&0\\0&1 \esp \cdot G[z] \,\subset\, G[z,z^{-1}]\,, \ee
and the corresponding subvariety 
\be \CS_{(1,0)} = \pi_g^{-1}\big( \CI[z]'\backslash \CO_{(1,0)}/\CI[z]\big) = 
 \CI[z]'\backslash  \big\{\big(X',g\in \CO_{(1,0)},X\big)\big\}/\CI[z] \;\subset\; \bigsqcup_{\n\geq 0} \CM_{\rm rav}^{\n+1,\n}\,. 
 \ee
We are interested in cohomology classes supported on $\CS_{(1,0)}$, constructed by pulling back $\CI[z]$-equivariant cohomology classes of $\CI[z]' \backslash G[z,z^{-1}]$ that  are supported on $\CI[z]'\backslash \CO_{(1,0)}$.

The cycle $\CI[z]'\backslash \CO_{(1,0)}$ in the affine flag variety is well known to be a $\cp^1$ bundle over $\cp^1$; more precisely, it is the projectivization of the rank-two bundle $\CO(1)\oplus \CO(-1)\to \cp^1$.
Let's briefly recall why. The affine flag variety $\CI[z]' \backslash G[z,z^{-1}]$ is itself a $\cp^1$ fibration over (a polynomial version of) the affine Grassmannian, $\CI[z]'\backslash G[z,z^{-1}]\overset{f}\to G[z]'\backslash G[z,z^{-1}]$. An explicit way to see the fibration is to first write $\CI[z]'\backslash G[z,z^{-1}] \simeq G[z]\backslash \big(G[z,z^{-1}]\times \cp^1\big)$, introducing a copy of $\cp^1$ with homogeneous coordinates $p=(p_1,p_2)$. Then the map
\be \CI[z]'\backslash G[z,z^{-1}]  \simeq  G[z]'\backslash \big(G[z,z^{-1}]\times \cp^1\big) \overset{f}\to G[z]'\backslash G[z,z^{-1}]\ee
simply forgets the $\cp^1$. In this description, $\CI[z]'\backslash\CO_{(1,0)}$ takes the form
\be \CI[z]'\backslash\CO_{(1,0)} \simeq G[z]'\backslash \big\{ h'(z)\cdot \bsp z&0\\0&1\esp\cdot h(z),p \big\}\,,\qquad h'(z)\in G[z]', h(z)\in G[z]\,, \ee
which by forgetting $p$ is sent to the corresponding $G[z]'\times G[z]$ orbit in the affine Grassmannian,
\be f(\CI[z]'\backslash\CO_{(1,0)}) = G[z]'\backslash \big\{ h'(z)\cdot \bsp z&0\\0&1\esp\cdot h(z) \big\} \subseteq G[z]'\backslash G[z,z^{-1}]\,,\qquad h'(z)\in G[z]', h(z)\in G[z]\,.\ee
The space $f(\CI[z]'\backslash\CO_{(1,0)})$ is a $\cp^1$ covered by two affine charts of the form
\be \bp z&0 \\ c&1\ep\,,\qquad \bp 0&z\\1&d \ep\,,\ee
with transition function $d=1/c$. The fiber of $f:\CI[z]'\backslash\CO_{(1,0)}\to f(\CI[z]'\backslash\CO_{(1,0)})$ is parameterized by $(p_1,p_2)$, which transforms as $(p_1,p_2)\mapsto (-c p_1,p_2/c)$ under the $G[z]'$ transformation required to relate the two charts on the base. This identifies $\CI[z]'\backslash\CO_{(1,0)}$ itself as the total space of the bundle $\mathbb P(\CO(1)\oplus \CO(-1))$.

Now, the subvariety $\CS_{(1,0)}$ turns out to surject onto the space of allowed $X$'s, just like in the case of monopole number zero. (This is because requiring $X'=gX$ to be regular at the origin $z=0$ does not impose any constraints on $X$ when $g\in \CO_{(1,0)}$.) Therefore, we have a fibration
\be \CI[z]'\backslash \CO_{(1,0)} \to \CS_{(1,0)} \overset{\pi}{\to} \CM_D(\CB_\nu,\V_\CI)\,. \label{fib-mon1} \ee

Under the action of (the maximal torus of) $\CI[z]$, the space $\CI[z]'\backslash \CO_{(1,0)}$ has four fixed points
\be \sigma z^{e_a}\,=\; \bp z&0\\0&1\ep\,,\qquad \bp 1&0\\0&z\ep\,,\qquad \bp 0&z\\1&0\ep\,,\qquad \bp 0&1\\z&0\ep\,, \ee
which we can label by an element $\sigma\in \{1,w\}$ of the Weyl group and a cocharacter $e_1=(1,0)$ or $e_2=(0,1)$.  The pull-backs of these fixed-point classes under $\pi_g^*$ are cohomology classes in $\CM_{\rm rav}$ supported at points on the fibers of the fibration \eqref{fib-mon1} and constant along the base. We denote the pulled-back fixed-point classes $\delta_{e_a,\sigma}$. Any equivariant cohomology class can be expressed in terms of these fixed point classes after inverting the equivariant parameters, so suffices to understand the convolution action of the fixed-point classes $\delta_{e_a,\sigma}$ on fixed-point states in $\CH(\CB_\nu,\V_\CI)$. To do so, we will state and use a general formula.

Suppose we want to understand the action of cohomology classes arising from the pullback of a cycle in $\CG_0[z]'\backslash G[z, z^{-1}]$, \eg\, from an orbit $\CG_0[z]'\backslash \CO_A$.
The pullback will sit in some subvariety $\CS \subset \CM_{\rav}(\CB;\CL',\CL)$ that, in general, does \emph{not} fiber uniformly over the space of all $X$'s. This is not a problem, and is deftly handled by intersection cohomology.
Furthermore, this cycle will have some fixed points $g_i$ under the maximal torus of $\CG_0[z]$, denote the pulled-back fixed point class by $\delta_{g_i}$, and cohomology classes coming from the pulled-back cycle can be expressed in terms of the $\delta_{g_i}$ using localization. For a state $\ket{\psi}$ associated to the fixed point $X_\psi \in \CM_D(\CB,\CL)$ we have, \cf\, (4.45), (4.48) in \cite{VV},%
\footnote{The classes $v^{\pm}_b$ appearing in (4.45) and (4.48) of \cite{VV} differ from $\delta_{g_i}$ here, in that they include extra denominators from the Atiyah-Bott localization formula.
 The $v's$ of \cite{VV} are not honest classes in $H^\bullet_{T_F\times \C^*_\varepsilon}(\CM_{\rav})$; they only make sense after equivariant parameters have been inverted. Here and below, we only work with actual cohomology classes.}
\begin{align}\label{fixedPointActions}
	\delta_{g_i} \ket{\psi} = \begin{cases} \frac{\omega_{g_i \cdot \psi}}{\omega_\psi} e(N_\CS|g_i;\psi) \ket{g_i \cdot \psi} & g_i \cdot \psi \in \CM_D(\CB, \CL')\\ 0 & \rm else \end{cases},
\end{align}
where $e(N_\CS|g_i;\psi)$ is the equivariant weight of (the normal bundle to) $\CS$ at the fixed point $(g_i \cdot \psi; g_i; \psi)\in \CM_\rav(\CB_\nu; \CL', \CL)$, $\omega_{\psi}$ (resp. $\omega_{g_i \cdot \psi}$) is the equivariant weight of (the normal bundle to) $\psi$ in $\CM_{D}(\CB_\nu, \CL)$ (resp. $g_i \cdot \psi$ in $\CM_{D}(\CB_\nu, \CL')$).

Using \eqref{fixedPointActions}, we can compute the convolution action of the $\delta_{e_a,\sigma}$ on states $|\n,k,\sigma\rangle$ in the disc Hilbert space. Namely, we know that $\CS_{(1,0)}$ fits into correspondences over all $X$'s and the gauge transformation $\sigma z^{e_a}$ corresponding to each fixed point on $\CI[z]' \backslash \CO_{(1,0)}$ will send $X\mapsto X'= \sigma z^{e_a}X$. We find
\begin{align}
	\delta_{e_a; \sigma'} |\n , k , \sigma \rangle &= \frac{\omega_{\n , k,\sigma'\sigma}}{\omega_{\n , k,\sigma}} |\n , k + e_{\sigma(a)} , \sigma' \sigma \rangle \\
	&= (-1)^{\sigma'} P \left( - m_{\sigma (a)} - (k_{\sigma(a)}+1) \varepsilon\right) |\n , k + e_{\sigma(a)} , \sigma' \sigma \rangle\,, \notag
\end{align} where $P(x) = (x+m_1)(x+m_2)(x+m_3)(x+m_4)$.

The \emph{actual} equivariant cohomology of $\CI[z]' \backslash \CO_{(1,0)}$ is related to fixed-point classes by localization. It differs from the fixed-point classes themselves by introducing a few (but not arbitrary!) denominators, analogous to \eqref{GOmodLeftIClasses}. Using the global structure of $\CI[z]' \backslash \CO_{(1,0)}$, as the bundle $\mathbb P(\CO(1)\oplus \CO(-1))$, and pulling back global cohomology classes to $\CM_{\rm rav}$, we may finally identify four basic monopole operators of charge 1:
\begin{align}\label{IwahoriMonopoles+}
	V^+_1 &= \frac{\delta_{e_1; 1} - \delta_{e_1 ; w}}{\alpha ( \alpha + \varepsilon )} - \frac{\delta_{e_2;1} - \delta_{e_2;w}}{\alpha ( \alpha - \varepsilon ) },\nonumber \\
	V^+_2 &=  \frac{\delta_{e_1; 1} - \delta_{e_1 ; w}}{ \alpha + \varepsilon } + \frac{\delta_{e_2;1} - \delta_{e_2;w}}{ \alpha - \varepsilon } , \nonumber \\
	V^+_3 &=\frac{\delta_{e_1; 1}+ \delta_{e_1;w} - \delta_{e_2;1} - \delta_{e_2;w}}{\alpha} , \nonumber \\
	V^+_4 &=\delta_{e_1; 1}+ \delta_{e_1;w} + \delta_{e_2;1} + \delta_{e_2;w},
\end{align}
where $\alpha =  \varphi_2 - \varphi_1.$ The operator $V^+_1$ is (the pullback of) the fundamental class of $\P(\CO(1) \oplus \CO(-1))$, $V^+_4$ is the equivariant volume form, and $V^+_2,V^+_3$ are fundamental classes of the base/fiber $\cp^1$'s.

An enlightening way to repackage this data is the following. Let us define
\be u_a^+ := \delta_{e_a,1}\,.\ee
These are ``abelianized'' monopole operators discussed in \cite{Web2016, DG-star}, slight renormalizations of the  abelianized monopole operators of \cite{BDG, VV}. They act on states as
\begin{align} \label{posAbelianizedMonopolesAction}
u^+_a \ket{ \n , k ; \sigma } = P \left( - m_{\sigma (a)} - (k_{\sigma(a)}+1) \varepsilon\right) \ket{ \n +1, k + e_{\sigma(a)} ; \sigma }.
\end{align} 
Then the four monopole operators $V_i^+$ above can all be expressed as convolution products of $u_a^+$ with elements of the nil-Hecke algebra:
\begin{align}
	&V^+_1 = \partial u^+_1 \partial,
	&&V^+_2 =\partial ( u^+_1 + u^+_2 ) , \nonumber \\
	&V^+_3 = \alpha V^+_1,
	&&V^+_4 = \alpha V^+_2 .
\end{align}
Conversely, we have
\begin{align}
	u^+_1 &= \frac 1 4 \left( V^+_1 \alpha ( \alpha + \varepsilon ) + V^+_2 ( \alpha + \varepsilon ) + V^+_3 \alpha + V^+_4 \right) , \nonumber \\
	u^+_2 &= \frac 1 4 \left(- V^+_1 \alpha ( \alpha + \varepsilon ) + V^+_2 ( \alpha + \varepsilon ) - V^+_3 \alpha + V^+_4 \right).
\end{align}
Some other simple relations among operators of monopole numbers $\mathfrak m=0,1$ are 
\be \label{posAbelianizedMonopolesRels}
s u^+_a = u^+_{w(a)} s \qquad u^+_a \varphi_b = (\varphi_b - \delta_{ab} \varepsilon) u^+_a \qquad u^+_a u^+_b = u^+_b u^+_a.\\
\ee

\subsubsection{Monopole Number $-1$}\label{sec:IwahoriMono-1}
Now let's consider the negative monopole sector. We expect the algebra to be symmetric under $\n \rightarrow - \n$ (and $\varepsilon \to -\varepsilon$), however due to our choice of boundary conditions, the computation at monopole number $-1$ looks somewhat different from that of the previous section. A generic configuration $X$ on the bottom disk does not fit into any correspondences, seeming to indicate that negative monopole operators do not exist. However, this is only over generic configurations and for special $X$ a larger class of $g$'s is allowed. 

Consider the $G'[z]\times G[z]$ orbit $\CO_{(0,-1)}$. This orbit is obtained by simply multiplying the elements of the above double orbit by $z^{-1}$ and so must be another copy of $\P ( \CO (1) \oplus \CO (-1))$. There should be monopole operators labeled by the $\CI[z]$-equivariant cohomology classes of this $\P ( \CO (1) \oplus \CO (-1) )$. Just as with the positive monopole operators, we get four operators
\begin{align}\label{IwahoriMonopoles-}
	V^-_1 &= \frac{\delta_{-e_1; 1} - \delta_{-e_1 ; w}}{\alpha ( \alpha - \varepsilon )} - \frac{\delta_{-e_2;1} - \delta_{-e_2;w}}{\alpha ( \alpha + \varepsilon ) },\nonumber \\
	V^-_2 &=  \frac{\delta_{-e_1; 1} - \delta_{-e_1 ; w}}{ \alpha - \varepsilon } + \frac{\delta_{-e_2;1} - \delta_{-e_2;w}}{ \alpha + \varepsilon } , \nonumber \\
	V^-_3 &=\frac{\delta_{-e_1; 1}+ \delta_{-e_1;w} - \delta_{-e_2;1} - \delta_{-e_2;w}}{\alpha} , \nonumber \\
	V^-_4 &=\delta_{-e_1; 1}+ \delta_{-e_1;w} + \delta_{-e_2;1} + \delta_{-e_2;w}.
\end{align}  where $\alpha = \varphi_2 - \varphi_1$. The shifted denominated from \eqref{IwahoriMonopoles+} are due to the different equivariant weights of the fixed points.

It is important to note that the gauge transformations in $\CO_{(0,-1)}$ do not fit into correspondences with all $X$, \ie\, $\CS_{(0,-1)}$ does not surject onto to the space of $X$'s. This is not a problem. To evaluate the action of the fixed points, $\delta_{-e_a, \sigma}$ we use that the equivariant weights of the normal bundles to the fixed points are given by
\begin{align}
	e(N_{\CS_{(0,-1)}}|\sigma' z^{-e_a}; k,\sigma) = P ( - m_{\sigma(a)} - k_{\sigma(a)} \varepsilon ).
\end{align} Thus, using fixed point action from (\ref{fixedPointActions}), we have
\begin{align}
	\delta_{- e_a , \sigma'} \ket{ \n , k , \sigma } &=\begin{cases}
		(-1)^{\sigma'}\ket{ \n-1 , k - e_{\sigma' (a)}, \sigma' \sigma} & k_{\sigma'(a)} < 1\\
		0 & \text{else}\\
		\end{cases}.
\end{align} The right side vanishes if $k_{\sigma'(a)} < 1$ since there are no vortex configurations with $k_a < 0$. Just as with the positive monopole operators, the $V^-_i$ can all be expressed in terms of abelianized monopoles $u^-_a = \delta_{- e_a , 1}$. They satisfy relations similar to \ref{posAbelianizedMonopolesRels}: \be \label{negAbelianizedMonopolesRels}
	s u^-_a = u^-_{w(a)} s \qquad u^-_a \varphi_b = (\varphi_b + \delta_{ab} \varepsilon) u^-_a \qquad u^-_a u^-_b = u^-_b u^-_a.\\
\ee And the following relation with the positive abelianized operators, \cf\, equation (3.43) of \cite{BDG}, \be \label{posnegAbelianizedMonopolesRels}
u^+_a u^-_a = P(\varphi_a - \tfrac{1}{2}\varepsilon) \qquad u^-_a u^+_a = P(\varphi_a + \tfrac{1}{2}\varepsilon).\\
\ee

\subsection{Junctions between $\V_{\CI}$ and $\mathds{1}$}

Now consider the junctions $\Hom^\varepsilon_A(\V_{\CI}, \mathds{1})$ and $\Hom^\varepsilon_A(\mathds{1}, \V_{\CI})$, which are naturally bi-modules for the algebras $\End^\varepsilon_A(\V_{\CI})$ and $\End^\varepsilon_A(\mathds{1})$. Just as with the computations of operators in the previous section, we consider the following (equivalent) spaces of correspondences
\be \CM_{\rm rav}(\CB_\nu; \mathds{1},\V_{\CI}) = G[z]' \bigg \backslash \left \{ (X',g,X) \in \wt{\CM_D}(\CB_\nu,\mathds{1}) \times G[z, z^{-1}] \times \wt{\CM_D}(\CB_\nu,\V_{\CI}) ~\bigg |~ X\in X'=gX  \right \} \bigg /  \CI[z] \ee
and
\be \CM_{\rm rav}(\CB_\nu;\V_{\CI}, \mathds{1}) = \CI'[z] \bigg \backslash \left \{ (X',g,X) \in \wt{\CM_D}(\CB_\nu,\V_{\CI}) \times G[z, z^{-1}] \times \wt{\CM_D}(\CB_\nu,\mathds{1}) ~\bigg |~ X\in X'=gX  \right \} \bigg /  G[z]\ee
As usual, these spaces have projection maps to $\CM_D(\CB_\nu,\mathds{1})$ and $\CM_D(\CB_\nu,\V_{\CI})$, and decompose into a disjoint union with components labeled by the vortex numbers $\n', \n$ of the top and bottom disks. Bimodule elements will correspond to (diagonal sums of) cohomology classes of these components and they are represented, via convolutions using the above spaces and maps, as linear operators between $\CH(\CB_\nu, \V_{\CI})$ and $\CH(\CB_\nu,\mathds{1})$. We will only consider bimodule elements corresponding to spaces with $\n' = \n$. We expect, but do not prove, that they generate the bimodules.

Let us start with the space $\CM_{\rm rav}(\CB_\nu; \mathds{1}, \V_{\CI})$. We consider $\CO_{(0,0)}\subset G[z,z^{-1}]$, the $G[z]'\times G[z]$ orbit of $g = \bsp 1&0\\0&1\\\esp$, and the corresponding subvariety 
\be \CS^{\mathds{1}, \V_{\CI}}_{(0,0)} \subset \bigsqcup \limits_{\n\geq0} \CM_{\rm rav}^{\n, \n}(\CB_\nu; \mathds{1}, \V_{\CI})\,.\ee
The space $G[z]' \backslash \CO_{(0,0)} \subset G[z]'\backslash G[z,z^{-1}]$ is a single point and $\pi: \CS^{\mathds{1}, \V_{\CI}}_{(0,0)} \to \CM_{D}(\CB_\nu,\V_\CI)$ is a surjection onto the space of $X$'s. Using the same procedure as in the previous subsections, we then have a single operator, call it $B_-$, that acts as
\be
	B_- |\n , k, \sigma \rangle = \frac{(-1)^\sigma}{m_1 - m_2 + ( k_1 - k_2 ) \varepsilon } | \n , k \rangle .
\ee
We expect $B_-$ to generate the bimodule. Another (expected) generator is given by
\be
	B_+ := B_- (\varphi_2-\varphi_1) \Rightarrow B_+ |\n , k , \sigma \rangle = |\n , k \rangle,
\ee
which is related to $B_-$ via $B_+ \partial = 2B_-$. It is worth noting that $B_- \partial = \tfrac 1 2 B_+ \partial^2 = 0$.

Finally, we move on to the space $\CM_{\rm rav}(\CB_\nu;\V_{\CI}, \mathds{1})$. We again consider $\CO_{(0,0)}\subset G[z,z^{-1}]$ and the corresponding subvariety 
\be \CS^{\V_{\CI}, \mathds{1}}_{(0,0)} \subset \bigsqcup \limits_{\n\geq0} \CM_{\rm rav}^{\n, \n}(\CB_\nu; \V_{\CI}, \mathds{1})\,.\ee
Just as in Section \ref{sec:IwahoriMono0}, $\CI[z]' \backslash \CO_{(0,0)}$ is a copy of $\cp^1.$ Furthermore, $\pi:\CS^{\V_{\CI}, \mathds{1}}_{(0,0)}\to\CM_{D}(\CB_\nu, \id)$ surjects onto the space of $X$'s. We find two operators, call them $b_+$ and $b_-$, with actions
\be
	b_+ |\n, k \rangle = \frac 1 2 \left(| \n , k ; 1\rangle + |\n , k ; w\rangle \right)
	\qquad 
	b_- |\n , k \rangle = \frac 1 2 (m_1 - m_2 + ( k_1 - k_2 ) \varepsilon )\left( |\n , k ; 1\rangle - |\n , k ; w\rangle \right).
\ee
Again, we expect that the entire bimodule is generated by either $b_+$ or $b_-$. In particular one may pass from one to the other via
\begin{align}
	(\varphi_2 - \varphi_1) b_+ = b_-,
	\qquad \partial b_- = 2 b_+ .
\end{align}

The transformations $B_+$ and $b_+$ are Weyl symmetric, in the sense that
\begin{align}
	B_+ s = B_+,
	\qquad
	s b_+ = b_+,
\end{align} while the operators $B_-$ and $b_-$ are Weyl antisymmetric \begin{align}
B_- s = -B_-,
\qquad
s b_- = -b_-.
\end{align}
They furthermore satisfy
\begin{align}\label{transitionIdentities}
	B_+b_+ = 1 \qquad B_- b_+ = 0 \qquad B_+ b_- = 0 \qquad B_- b_- = 1.
\end{align}

Then, given an operator $\CO^{\pm} \in \End^\varepsilon_A(\V_{\CI})$ of definite parity $s \CO^\pm s = \pm \CO^\pm$, we find
\begin{align}
	B_+ \CO^\pm b_+ = B_+ s \CO^\pm s b_+ = \pm  B_+ \CO^\pm b_+.
\end{align}
Since any $\CO$ can be written as $\CO = \CO^+ + \CO^-$, the associated $\End_A(\mathds{1})$ operator only knows about the symmetric part
\begin{align}
	B_+ \CO b_+ = B_+ \CO^+ b_+.
\end{align}
Sandwiching $B_+ ... b_+$ is also an algebra homomorphism $(\End^\varepsilon_A(\V_{\CI}))^W \rightarrow \End^\varepsilon_A(\mathds{1})$. Given two Weyl invariant operators $\CO_1$ and $\CO_2$, we have
\begin{align}
	(B_+ \CO_1 b_+)(B_+ \CO_2 b_+) = \frac 1 2 B_+ \CO_1 ( 1 + s ) \CO_2 b_+ = B_+ \CO_1 \CO_2 b_+.
\end{align}It is worth noting that we could sandwich with any combination of $B_+, B_-$ and $b_+,b_-$. Inside a sandwich $B_+... b_+$ or $B_-...b_-$ (resp. $B_+...b_-$ or $B_-...b_+$), only Weyl symmetric (resp. antisymmetric) operators survive.

\subsection{$\text{End}^{\varepsilon}_A(\V_{\CI})$ as a matrix algebra over $\text{End}^{\varepsilon}_A(\mathds{1})$}
\label{sec:mxalg}

There were many hints in the previous sections that the algebra $\text{End}^{\varepsilon}_A(\V_{\CI})$ factors into a product $\text{End}^{\varepsilon}_A(\mathds{1}) \otimes \End(H^*_{\C^*}(\cp^1))$. In particular, we found that the positive and negative monopole operators could all be expressed in terms of products of abelianized monopoles, which only acted on $\n, k$, and the nil-Hecke algebra, which only acted on $\sigma$. Furthermore, we found two ``inclusion" maps $b_{\pm}$ and two ``projection" maps $B_{\pm}$ that relate $\text{End}^{\varepsilon}_A(\mathds{1})$ and $\text{End}^{\varepsilon}_A(\V_{\CI})$.

Let us introduce a new basis of $\CH(\CB_\nu, \V_{\CI})$ under which the action of $s$ is diagonal: $\ket{\n, k; \pm} := b_\pm \ket{\n, k}$. By abuse of notation, we will write $b_\pm$ as $2 \times 1$ matrices, \ie\ column vectors \be
b_+ = \begin{pmatrix}
	1\\0\\
\end{pmatrix} \qquad b_- = \begin{pmatrix}
0\\1\\
\end{pmatrix}.
\ee In light of \eqref{transitionIdentities}, the operators $B_\pm$ have a natural representation as row vectors\be
B_+ = \begin{pmatrix}
	1&0\\
\end{pmatrix} \qquad B_- = \begin{pmatrix}
	0&1\\
\end{pmatrix}.
\ee The matrix elements of the various $\End_A^\varepsilon(\V_{\CI})$ operators found above can be determined by sandwiching between $B_\pm$ and $b_\pm$. For example, the monopole number 0 operators can be represented by matrices over $\End_A^\varepsilon(\mathds{1})$ as\footnote{The operator $(\varphi_2 - \varphi_1)^2$ can be expressed in terms of $V^2, W^0, V^0$ and complex masses but simply represents the Coulomb branch operator $\Tr \varphi^2$.}\be
s = \begin{pmatrix}
	1 & 0\\
	0 & -1\\
\end{pmatrix} \qquad \varphi_1 + \varphi_2 = \begin{pmatrix}
-\tfrac 1 2 V^0 & 0\\
0 & -\tfrac 1 2 V^0\\
\end{pmatrix} \qquad \partial = \begin{pmatrix}
0 & 2\\
0 & 0\\
\end{pmatrix} \qquad \varphi_2 - \varphi_1 = \begin{pmatrix}
0 & (\varphi_2 - \varphi_1)^2\\
1 & 0\\
\end{pmatrix}.
\ee The operators of nonzero monopole charge can be expressed similarly. 

Conversely, we can find generators of this matrix algebra by sandwiching elements of $\End_A^\varepsilon(\mathds{1})$ between $b^\pm$ and $B^\pm$. A nice set of them come in two flavors. The first flavor contains products of the $b_\pm$ and $B_\pm$ and is related to the nil-Hecke algebra:
\be \label{nilHeckeOps}
\begin{array}{lc@{\quad}l}
	\begin{pmatrix}
		1 & 0\\ 0 & 0
	\end{pmatrix}  = b_+ B_+ = \tfrac 1 2 (\one + s) &\hspace{0.5 cm}& \begin{pmatrix}
	0 & 1\\ 0 & 0
\end{pmatrix}  = b_+ B_- = \tfrac 1 2 \partial\\
	 \begin{pmatrix}
		0 & 0\\ 1 & 0
	\end{pmatrix} = b_- B_+ = (\varphi_2 - \varphi_1) - \tfrac 1 2 (\varphi_2 - \varphi_1)^2 \partial &\hspace{0.5 cm}&\begin{pmatrix}
	0 & 0\\ 0 & 1
\end{pmatrix}  = b_- B_-= \tfrac1 2 (\one - s)\\
\end{array}.
\ee The second corresponds to the spherical subalgebra \be
\begin{array}{l}
	\begin{pmatrix}
		V^0 & 0\\ 0 & V^0
	\end{pmatrix} = b_+ V^0 B_+ + b_- V^0 B_- = -2(\varphi_1+\varphi_2),\\
	\begin{pmatrix}
		V^{\pm} & 0\\ 0 & V^{\pm}
	\end{pmatrix} = b_+ V^\pm B_+ + b_- V^\pm B_-= \pm\tfrac1 2 \big\{\partial,u^\pm_1-u^\pm_2\big\},\\
	\begin{pmatrix}
		W^0 & 0\\ 0 & W^0
	\end{pmatrix} = b_+ W^0 B_+ + b_- W^0 B_- = \begin{array}{l}\{\partial,u^+_1\partial u^-_2+u^+_2\partial u^-_1\}+\big(2(\varphi_1^2+ \varphi_2^2)+2(m_2+m_3+\varepsilon)(\varphi_1 + \varphi_2)\\ 3 (m_1 + m_4) (m_2 + m_3)+(m_2+m_3)^2+m_1 m_4 + \varepsilon^2\big)\end{array},\\
	\begin{pmatrix}
		W^{\pm} & 0\\ 0 & W^{\pm}
	\end{pmatrix} = b_+ W^\pm B_+ + b_- W^\pm B_- = \begin{array}{l}\pm \tfrac 1 2 \big\{\partial,(2\varphi_2-m_2-m_3) u^{\pm}_1-(2\varphi_1-m_2-m_3) u^{\pm}_2\big\}\end{array},
\end{array}
\ee where $\{\cdot, \cdot \}$ is the anticommutator.


\section{The conifold vortex line and its mirror}
\label{sec:conifold}


In the previous section we found that the simplest nonabelian A-type line operators --- those that break the gauge group but introduce no singularity in the hypermultiplets --- are a little \emph{too} simple. In the category  of A-type line operators, they are equivalent to direct sums of the trivial line. In this section we study a particular example involving singularities in the hypermultiplets as well.

The 3d $\CN=4$ theory we consider on the A-side is the same one from Section \ref{sec:IwahoriLine}: SQCD with $G=U(2)$ and four flavors of hypermultiplets, $R = (\C^2)^{\oplus 4}$. This theory has a standard quiver presentation shown on the LHS:
\be \raisebox{-.5in}{\includegraphics[width=5.2in]{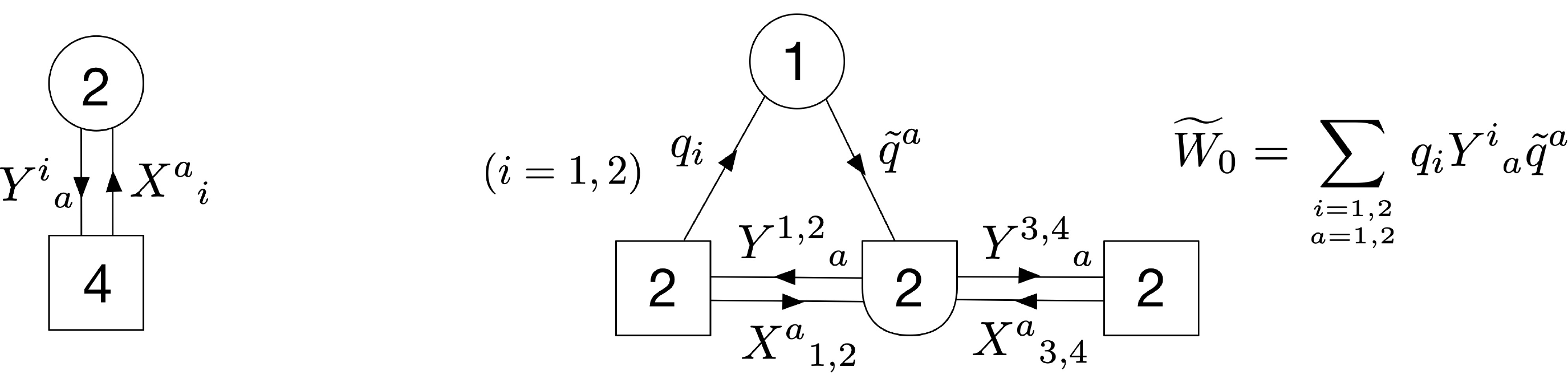}} \label{fig:quiver24} \ee
The line operator $\V_{\rm con}$ that we are interested in can be defined by coupling the bulk theory to 1d $\SQM_A$ quiver quantum mechanics as shown on the RHS of \eqref{fig:quiver24}.

The 1d/3d superpotential $\wt W_0$ induces the desired singularity in the bulk hypermultiplets at $z=0$. One way to understand this is to observe that, with nonzero 1d FI parameter, the Higgs branch of the $\SQM_A$ quantum mechanics is a resolved conifold $\CE = \big[\CO(-1)\oplus \CO(-1)\to\cp^1\big]$. This is a homogeneous bundle over the flag manifold $G_\C/B=\cp^1$.
Moreover, the matrix $p^a{}_i := \tilde q^aq_i$ that appears in $\wt W_0$ generates the global functions on $\CE$. We end up with the general setup described in Section \ref{sec:VL-genQM} coupling to the conifold $\CE$ breaks gauge symmetry $G\to T$ (due to the base $\cp^1$) \emph{and} introduces a singularity
\be X^a{}_i(z) = \frac1z p^a{}_i + \text{regular}\qquad (a,i=1,2)\,, \ee
together with a dual constraint on the zero-mode of $Y$. We will give further details in Section~\ref{sec:FundamentalVortexLine}. We denote the resulting A-type line operator $\V_{\rm con}$ and sometimes call it the ``conifold line'' because it comes from ``coupling to a conifold.''

This is a useful example for testing our computational methods because the conifold line has a simple 3d mirror. The 3d mirror of the bulk SQCD theory is a quiver gauge theory with $G=U(1)\times U(2)\times U(1)$ \cite{dBHOO, dBHOOY, HananyWitten}, as in \eqref{dualQuiver} below.
More so, it was shown by \cite{AsselGomis} that the vortex line $\V_{\rm con}$ maps under mirror symmetry to a Wilson line $\W_{\mb 2}$ in the fundamental representation of the central $U(2)$ node.

It is straightforward (albeit  tedious) to compute the algebra of local operators $\text{End}_B^\varepsilon(\W_{\mb 2})$ bound to the $\W_{\mb 2}$ Wilson line, since the entire algebra can be realized in terms of free hypermultiplet scalars, as in Section \ref{sec:Wilson-Omega}.
We will summarize some results of this computation in Section \ref{sec:FundamentalWilsonLine}, leaving further details to Appendix \ref{app:Wilson}.

Then, in Section \ref{sec:FundamentalVortexLine} and Appendix \ref{app:conifoldops}, we will compute elements of the algebra of local operators $\text{End}_A^\varepsilon(\V_{\rm con})$ on the conifold line, using the methods of Section \ref{sec:comp}. We expect of course that
\be \text{End}_A^\varepsilon(\V_{\rm con}) \simeq \text{End}_B^\varepsilon(\W_{\mb 2})\,.\ee
We will verify this by identifying the A-side mirrors of various generators of $\text{End}_B^\varepsilon(\W_{\mb 2})$, and matching relations among their products --- now computed via convolution.

\subsection{Fundamental Wilson line}\label{sec:FundamentalWilsonLine}

We consider the 3d mirror to $U(2)$ SQCD with four flavors, given by the quiver 
\be \label{dualQuiver}
\raisebox{-.5in}{\includegraphics[width=2in]{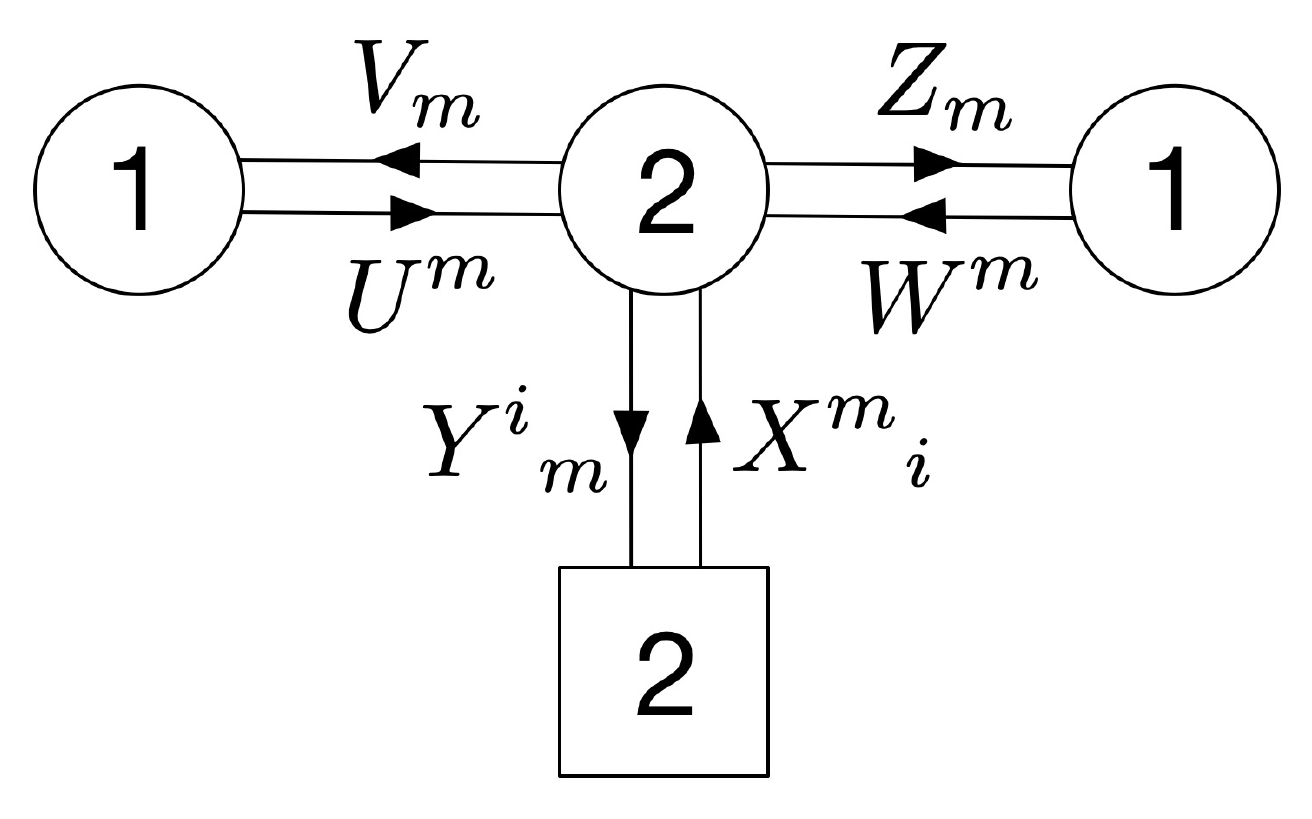}}
\ee
This is a gauge theory with gauge group $G=U(1)_L \times U(2)_C \times U(1)_R$, where the subscripts refer to the ``left,'' ``center,'' and ``right'' nodes. There is also a $PSU(2)_F$ flavor symmetry, corresponding to the square node, which is mirror to the $PSU(2)_{\rm top}$ topological symmetry of SQCD. The hypermultiplets come in bifundamental representations associated to each edge. We use indices $m=1,2$ and $i=1,2$ for the (anti)fundamental representations of $U(2)_C$ and $PSU(2)_F$, respectively. We then label the hypermultiplet scalars as $(U^m,V_m)$ in the $(1, \mathbf 2,0) \oplus (-1,\bar{ \mathbf 2 },0)$ representation of $U(1)_L\times U(2)_C\times U(1)_R$;  $(W^m , Z_m )$ in the $(0,\mathbf 2,1) \oplus (0,\bar{ \mathbf 2 },-1)$; and  $(X^m{}_i , Y^i{}_m)_{i=1,2}$ in the $(0,\mathbf 2,0) \oplus (0,\bar{ \mathbf 2 },0)$.%
\footnote{The charges for the two $U(1)$ factors differ by a sign from standard quiver conventions. This does not affect any final results.}

\subsubsection{Bulk algebra}

We will first describe local operators in the bulk, and then generalize to the Wilson line. \emph{All} these operators are built from polynomials in the hypermultiplet scalars, which obey
\begin{align}
	[ U^m , V_n ] = \varepsilon \delta^m{}_n\, ,
	&&[W^m , Z_n ] = \varepsilon \delta^m{}_n\, ,
	&&[ X^m{}_i , Y^j{}_n ] = \varepsilon \delta^m{}_n \delta^j{}_i\, .
\end{align}
in the B-type Omega background (Section \ref{sec:Wilson-Omega}).  The complex moment maps associated to each gauge node are the normal-ordered combinations
\begin{align}
	\mu_L = \,:\!U^m V_m \!:\,,
	&&\mu_R = \,:\!W^m Z_m  \!:\,,
	&&(\mu_C)^m{}_n = \,:\!U^m V_n + W^m Z_n + X^m{}_i Y^i{}_n  \!: \;.
\end{align}
Then, abstractly, the algebra of bulk local operators --- the quantized Higgs-branch chiral ring --- consists of $G$-invariant polynomials, modulo the complex moment maps:
\begin{align}\label{Ops1}
	\End^\varepsilon_B(\id) = \C_\varepsilon[\CM_H] = \big[\C_\varepsilon[U,V,W,Z,X,Y] /  \left( \mu_L + t_L ,\, \mu_R + t_R,\, \mu_C + t_C \right) \big]^G\,.
\end{align} 
Here $t_L,t_R,t_C$ are the complex FI parameters for the respective gauge nodes. Also, as discussed in Section \ref{sec:Wilson-Omega}, the quotient in \eqref{Ops1} should be interpreted as a quotient by a left ideal; the resulting quotient regains the structure of an algebra after imposing $G$-invariance.

The $G$-invariant subspace of $\C_\varepsilon[U,V,W,Z,X,Y]$ is generated by ``loop monomials,'' \emph{i.e.} products of hypermultiplets along paths in the quiver that are either closed, or begin/end on the $PSU(2)_F$ flavor node (\cf\ \cite{BFHH-Hilbert}). To generate this space as an algebra, it is sufficient to consider a finite set of minimal paths. The closed loops are all fixed by the moment-map constraints, and we are left with 1) the path starting on the $PSU(2)_F$ node, going to $U(2)_C$, and coming back; and 2) the path starting on $PSU(2)_F$, going to $U(2)_C$, looping around either the left or right part of the quiver (the two choices are not independent), and coming back to $PSU(2)_F$. This ultimately leads to six algebra generators, which are conveniently grouped as
\begin{align}\label{generators1}
	\begin{pmatrix}
		\frac{1}{2}J^0 &	J^+	\\
		J^-		& -\frac{1}{2}J^0 
	\end{pmatrix}^i_{\;j}  &:= X^m{}_j Y^i{}_m - \frac 12 X^m{}_k Y^k{}_m \delta^i{}_j, \nonumber \\
	\begin{pmatrix}
		\frac 1 2 M^0	&	M^+	\\
		M^- 	& - \frac 1 2 M^0 
	\end{pmatrix}^i_{\;j} &: = (U^mV_n - W^mZ_n )X^n{}_j Y^i{}_m - \frac 1 2 (U^mV_n - W^mZ_n )X^n{}_k Y^k{}_m \delta^i{}_j.
\end{align}
On the RHS we have removed the traces, which can be solved for in terms of FI parameters and $\varepsilon$ using the moment-map constraints. (See Appendix \ref{app:Wilson}.)

The operators $J^-,J^\pm$ are the components of the complex moment map for $PSU(2)_F$. The operators $M^0,M^\pm$ are in an adjoint representation of $PSU(2)_F$. They are precisely mirror to the monopole operators $V^0,V^\pm,W^0,W^\pm$ described in Section \eqref{sec:SQCD-bulk},
\be V^0,V^\pm;W^0,W^\pm\quad\overset{MS}\longleftrightarrow\quad J^0,J^\pm;M^0,M^\pm\,.\ee
Their charges under the Cartan $U(1)_F\subset PSU(2)_F$ and a Cartan $U(1)_H$ of the $SU(2)_H$ R-symmetry (in units where hypers have R-charge 1/2) are
\be \label{W-charges-bulk}
 \begin{array}{c|cccc}  &J^0&J^\pm & M^0 & M^\pm \\\hline
U(1)_F & 0&\pm 1&0&\pm 1 \\
U(1)_H & 1&1&2&2 \end{array}
\ee
The entire Higgs-branch chiral ring $\End^\varepsilon_B(\id)$ is generated by these six operators modulo two relations of R-charge 3 and 4, which are given in \eqref{relJM}, \eqref{relMM} of Appendix \ref{app:Wilson}. The two relations are identical to the relations among $V$'s and $W$'s from Section \ref{sec:SQCD-bulk} modulo the swap of mass and FI parameters $m_L,m_C,m_R \leftrightarrow t_L,t_C,t_R$.

\subsubsection{Wilson-line algebra}
\label{sec:W2-ops}

Now consider the half-BPS Wilson line $\W_{\mb 2}$, in the fundamental representation of $U(2)_C$. Local operators bound to the fundamental Wilson line correspond to polynomials in the hypermultiplet scalars that transform as $(0,\mathbf 2 \otimes \bar {\mathbf 2},0) \simeq  (0,\End(\mathbf 2),0)$ of the gauge group, subject to a \emph{modified} moment map relation. Following Section \ref{sec:Wilson-Omega} we find, abstractly,
\be
\label{WB-eg}
\begin{array}{c} \text{End}^\varepsilon_B(\W_{\mb 2}) = \big[\C_\varepsilon[X,Y,U,V,W,Z]\otimes \text{End}(\mb{\bar 2})/\CI \big]^G\,,  \\[.2cm]
 \CI := (\mu_L + t_L)\otimes \text{id}_{\mb{\bar 2}},\mu_C \otimes \text{id}_{\mb{\bar 2}} - \varepsilon \rho_{\mb{\bar 2}} + t_C \otimes \text{id}_{\mb{\bar 2}},(\mu_R + t_R)\otimes \text{id}_{\mb{\bar 2}})\,.
\end{array}
\ee

By carefully considering the index contractions required for a polynomial to belong to $\End(\mb 2)$, it is not difficult to see that $\End^\varepsilon_B(\W_{\mathbf 2})$ is generated by two types of operators.

First there are operators of the form $\CO \delta^m{}_n$, where $\CO$ are elements of the bulk algebra $\End^\varepsilon_B(\mathds{1})$. At $\varepsilon=0$ (without Omega-background) these operators should just be interpreted as restrictions of the bulk chiral-ring operators to the Wilson line. In particular, at $\varepsilon=0$, they obey the same two relations as in the bulk (times $\delta^m{}_n$). It is important to note, however, that when $\varepsilon\neq 0$ the bulk relations get deformed on the line, in a highly nontrivial way --- ultimately due to shifting the $\mu_C$ moment map in \eqref{WB-eg}. This is physically consistent, since the Omega background prevents the operation of moving a local operator from the bulk on to the Wilson line.
 The deformed relations are given in \eqref{relJM2}--\eqref{relMM2}. 

The second type of operators come from paths in the quiver that start and end on the $U(2)_C$ node (without taking a trace there). The paths looping to the left and right, and the path going down, lead to operators
\begin{align} 
 (\Psi_L)^m{}_n &:= U^m V_n\,,  \notag \\
 (\Psi_R)^m{}_n &:= W^m Z_n\,, \label{generators2} \\
 \begin{pmatrix}
		\frac 1 2 (N^0)^m{}_n && (N^+)^m{}_n \\
		(N^-)^m{}_n && - \frac 1 2(N^0)^m{}_n
	\end{pmatrix}^i_{\; j} &:= X^m{}_j Y^i{}_n - \frac 1 2 (X^m{}_k Y^k{}_n) \delta^i{}_j\,.  \notag
\end{align}
Just as for bulk local operators, the trace on the RHS be solved for in terms of $\Psi_L, \Psi_R$, FI parameters, and $\varepsilon$ using the moment map constraints. Under the action of $PSU(2)_F$ (generated by $J^\pm, J^0$), the operators $\Psi_L$ and $\Psi_R$ transform trivially, while the operators $N^\pm, N^0$ transform in the adjoint representation. It turns out that the operators $M^\pm, M^0$, which were independent generators of $\End^\varepsilon_B(\mathds{1})$, can now be written in terms of $\Psi$ and $N$ as
\be\begin{array}{rl}
	M^{\pm} &= \{\Psi_L - \Psi_R, N^{\pm}\} + (t_L - t_R) N^{\pm} - J^{\pm} (\Psi_L - \Psi_R) - (t_L - t_R)J^{\pm}\,, \\[.2cm]
	M^0 &= \{\Psi_L - \Psi_R, N^0\} + (t_L - t_R)N^0 - J^0 (\Psi_L - \Psi_R)- (t_L - t_R)J^0\,.
\end{array} \ee
Note that, in these formulas, $M,J$ mean $M\delta^m{}_n$, $J\delta^m{}_n$; and every multiplication implicitly involves a matrix multiplication in $\End(\mb 2)$. We will usually leave this implicit.

Altogether, we find generators of $\End^\varepsilon_B(\W_{\mathbf 2})$ given by $J^0,J^{\pm}, N^\pm, N^0, \Psi_L, \Psi_R$.
Under the Cartans $U(1)_F\times U(1)_H$ of flavor and R-symmetry, they have charges
\be \label{W-charges-line}
\begin{array}{c|cccccc}  &J^0&J^\pm & N^0 & N^\pm & \psi_L & \psi_R \\\hline
U(1)_F & 0&\pm 1&0&\pm 1 &0&0 \\
U(1)_H & 1&1&1&1 & 1 & 1 \end{array}
\ee
These generators satisfy many relations, which are derived and summarized in Appendix \ref{app:Wilson}.
 In particular, at R-charge 2, we find 5 relations transforming as singlets of $PSU(2)_F$, as well as three triplet and one pentuplet of $PSU(2)_F$. Specializing them (for simplicity) to zero values of FI parameters $t_L=t_C=t_R=0$, they become:
\begin{align}
\text{singlets}: &\quad \Psi_L^2=0\,,\quad \Psi_R^2=0\,, \notag \\
  &\quad  2\{N^+,N^-\}+(N^0)^2+\{\Psi_L,\Psi_R\}-4\varepsilon(\Psi_L+\Psi_R)=0 \notag \\
  &\quad 2\{J^+,J^-\}+(J^0)^2 - 4\{\Psi_L,\Psi_R\} + \varepsilon^2 = 0 \notag\\
  &\quad 2J^+ N^- + 2 J^- N^+ +J^0 N^0 + 2\{\Psi_L,\Psi_R\} - 5 \varepsilon (\Psi_L+ \Psi_R) - 6 \varepsilon^2 = 0\\
\text{triplet}: &\quad \{\Psi_L+\Psi_R,N^+\} + [N^0,N^+]-4\varepsilon N^+ = 0 \quad\text{(+$PSU(2)_F$ conjugates)}  \notag \\
  &\quad  [N^+, N^0] - J^+ (\Psi_L + \Psi_R) = 0 \quad\text{(+$PSU(2)_F$ conjugates)} \notag \\
  &\quad  [N^+, \Psi_L + \Psi_R] + \{N^+, N^0\} - 2 J^+ N^0 = 0 \quad\text{(+$PSU(2)_F$ conjugates)} \notag \\
\text{pentuplet}: &\quad (N^+-J^+)N^+ = 0 \quad\text{(+$PSU(2)_F$ conjugates)} \notag
\end{align}
For the triplets and pentuplet, we have only given the highest-weight vector in the multiplet.

The Hilbert series (or graded trace) of the algebra $\text{End}_B^\varepsilon(\W_{\mb 2})$ is fairly easy to compute, and is discussed in Appendix \ref{app:Hser}. It provides a useful way to organize the generators and relations discussed here. The computation of the Hilbert series, at least up to R-charge 2, is consistent with the generators and relations that we find. This does not completely rule out that there may be additional generators/relations entering at R-charge 3 and higher. For the remainder of the paper, we will content ourselves with matching the generators and relations above!

\subsection{Vortex line}\label{sec:FundamentalVortexLine}

We now return to the conifold vortex line $\V_{\rm con}$ in $U(2)$ SQCD, defined by the coupling to $\SQM_A$ quiver quantum mechanics shown in \eqref{fig:quiver24}. It follows from the brane constructions of \cite{AsselGomis} and S-duality that the vortex line $\V_{\rm con}$ should be the 3d mirror of the Wilson line $\W_{\mb 2}$. We would like to compute parts of its endomorphism algebra $\text{End}_A^\varepsilon(\V_{\rm con})$, and match them to the generators and relations of $\text{End}_B^\varepsilon(\W_{\mb 2})$ discussed above.

\subsubsection{Algebraic description}
\label{sec:con-alg}

We begin by translating the quiver quantum mechanics to an algebraic characterization of the vortex line. The quiver we are looking at is
\be \raisebox{-.5in}{\includegraphics[width=4in]{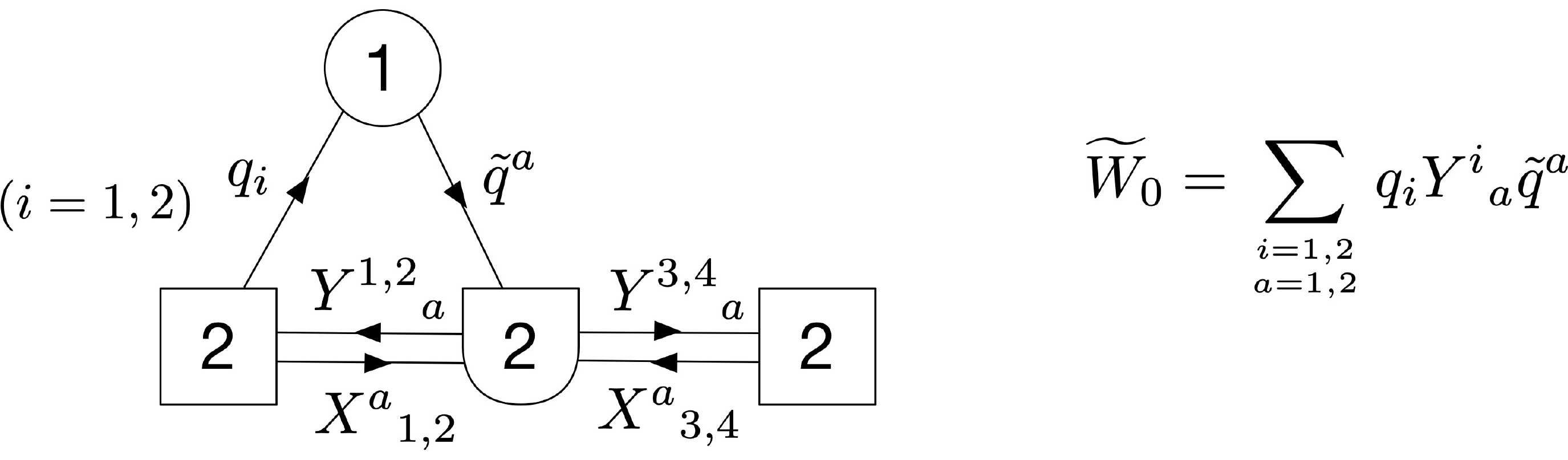}} \label{Q24-2} \ee

The 1d $\SQM_A$ quantum mechanics is a $U(1)$ gauge theory with two chiral multiplets $q_i$ of charge $+1$ and two chiral multiplets of $\tilde q^a$ of charge $-1$. Part of the \emph{definition} of the vortex line is a choice of real 1d FI parameter $t_{1d}$, which is taken to be positive; then the 1d moment-map constraint
\be |q|^2 - |\tilde q|^2 + t_{1d}=0 \ee
ensures that $\tilde q\neq 0$. The entire 1d quantum mechanics is then equivalent in $Q_A$-cohomology to a sigma-model onto the smooth Higgs branch
\begin{align} \CE &= \big\{q,\tilde q\,\big|\,|q|^2 - |\tilde q|^2 + t_{1d}=0 \big\}/U(1) \notag \\
 &\simeq \big\{(q,\tilde q)\in \C^2\oplus \C^2\,\big|\, \tilde q\neq 0\}/\C^*  \\
 &\qquad = \CO(-1)\oplus \CO(-1)\,\to\,\cp^1\,. \notag
\end{align}
This is the standard toric construction of the total space of the bundle $\CO(-1)\oplus\CO(-1)\to \cp^1$, \emph{i.e.} the resolved conifold. The base is parameterized by the projective coordinates $\tilde q=(\tilde q^1,\tilde q^2)$. The ring of global functions on $\CE$ (the ``chiral ring'' of the quantum mechanics) is generated by the four components of the matrix
\be p^a{}_i := \tilde q^aq_i = \bp \tilde q^1q_1 & \tilde q^1  q_2 \\ \tilde  q^2q_1 & \tilde q^2  q_2  \ep\,,  \ee
subject to the obvious relation that  $\det p =0$. Explicitly,  $\C[\CE]=\C[p]/(\det p)$.

When coupling the 1d quantum mechanics to the 3d bulk theory, the bulk hypermultiplets are split into two sets: $(X_{(1,2)},Y^{(1,2)}) = (X^a{}_{i=1,2},Y^{i=1,2}{}_a)$ with flavor indices $i=1,2$; and $(X_{(3,4)},Y^{(3,4)}) = (X^a{}_{i=3,4},Y^{i=3,4}{}_a)$ with flavor indices $i=3,4$. The splitting is preserved by a subgroup $(U(2)_F\times U(2)'_F)/U(1)$ of the bulk flavor symmetry $PSU(4)_F$. The quiver quantum mechanics has flavor symmetry $(U(2)\times \wt{U(2)})/U(1)$, acting by rotations of $q,\tilde q$; and the 1d-3d coupling identifies $U(2)$ with the bulk flavor symmetry $U(2)_F$, and $\wt{U(2)}$ with the bulk gauge symmetry. This is all efficiently encoded in the quiver \eqref{Q24-2}. The 1d-3d superpotential $\wt W_0$ preserves the split flavor symmetry, bulk gauge symmetry, etc.

It is useful to describe the symmetries and superpotential couplings geometrically, in terms of the conifold. The 1d  flavor symmetry $\wt{U(2)}$ acts on the $\cp^1$ base of the conifold, and is gauged in coupling to the bulk. The 1d flavor symmetry $U(2)$ acts on the $\C^2$ fibers of $\CE$. Altogether, $\wt{U(2)}\times U(2)$ act by left and (inverse) right multiplication on the matrix $p$, and leave invariant the superpotential
\be \wt W_0 = \text{Tr}(Y^{(1,2)}p) \label{W-con} \ee
Via the usual 1d-3d F-term constraints, the superpotential sets
\be X_{(1,2)}(z) = \frac 1z p + \text{regular}\,,\qquad Y^{(1,2)}(z) = \lambda(p-(\text{Tr}\,p)\mb 1) + z(\text{regular})\quad (\lambda\in \C)\,, \ee
allowing the $2\times 2$ matrix of hypers $X_{(1,2)}$ to have a pole of rank 1, and, dually restricting the zero-mode of $Y^{(1,2)}$ to be orthogonal to the pole in $X$. This is a gauge-invariant description of the line operator $\V_{\rm con}$, precisely along the lines of Section \ref{sec:VL-genQM}.

Finally, we may gauge-fix. Coupling to the $\cp^1$ base of the conifold breaks the bulk gauge symmetry $G\to T$ to its maximal torus. In algebraic terms, it breaks $G(\CO)\to \CI$ to the Iwahori subgroup familiar from Section \ref{sec:IwahoriLine}. We may use the bulk gauge symmetry to set $\tilde q=(1,0)^T$, or
\be  p = \bp q_1 & q_2 \\  0&0 \ep\,, \ee
which is invariant under left-multiplication by the standard Iwahori. Then we find that $\V_{\rm con}$ is characterized by the algebraic data
\be \label{Vcon-alg} \V_{\rm con}:\; \CG_0=\CI\,,\quad    \CL_0 \,=\,\bigg\{ X\in  \bp z^{-1}\CO & z^{-1}\CO &\CO &\CO \\ \CO&\CO&\CO&\CO \ep\,,\quad 
 Y^T \in \bp z\CO  & z\CO &\CO & \CO \\ \CO&\CO&\CO&\CO \ep \bigg\}\,. \ee

\subsubsection{Vacuum and disc Hilbert space}
\label{sec:fundVortexHilbertSpace}

We work with the same conventions as in Section \ref{sec:IwahoriLine}. Indeed, the only difference between this analysis and that of the Iwahori line is the additional singularity in $\CL_0$.

We choose the same vacuum $\nu$ at infinity, which sets
\be  \nu:\quad  X^a{}_i = \delta^a{}_i\,,\qquad  Y  \equiv 0 \label{con-vac} \ee
Then, in the presence of the line operator $\V_{\rm con}$, we find a disc moduli space with $Y\equiv 0$ on the entire disc, and
\begin{align} \label{MD-con} \CM_D(\CB_\nu;\V_{\rm con}) &\simeq \CI[z]\backslash \{ X(z)\in \CL_0 \cap R[z,z^{-1}] \\
 & \hspace{.6in} \text{s.t.}\; \det X_{(1,2)}\neq 0,\, \text{deg}\,\det X_{(3,4)}< \deg \det X_{(1,2)}\}\,. \notag\end{align}
This breaks up into components $\CM_D^\n(\CB_\nu;\V_{\rm con})$ labeled by vortex numbers
\be \n = \text{deg}\,\det X_{(1,2)} \geq -1\,. \ee
Each component is further covered by affine charts $\CM_D^{\n,k,\sigma}$, labeled by $\sigma\in \{1,w\}$ and cocharacters $k=(k_1,k_2)$ with $k_1\geq -1$, $k_2\geq 0$, such that $k_1+k_2=\n$. In each chart we may fully fix the Iwahori gauge symmetry, finding
\be \label{iwahoriGaugeFixedC}
\begin{array}{rl}
	\CM_D^{\n,k,1}: &X(z) = \begin{pmatrix}
		z^{k_1} + \sum_{d=-1}^{k_1 - 1} x^1{}_{1 , d} z^d &  \sum_{d=-1}^{k_2 - 1} x^1{}_{2 , d} z^d & \sum_{d=0}^{k_1 - 1}  x^1{}_{3 , d}z^d & \sum_{d=0}^{k_1 - 1}  x^1{}_{4 , d}z^d\\
		\sum_{d=0}^{k_1} x^2{}_{1 , d} z^d  & z^{k_2} + \sum_{d=0}^{k_2 - 1} x^2{}_{2 , d} z^d & \sum_{d=0}^{k_2 - 1}  x^2{}_{3 , d}z^d & \sum_{d=0}^{k_2 - 1}  x^2{}_{4 , d} z^d
	\end{pmatrix}\\
	&\\
	\CM_D^{\n,k,w}: & X(z) = \begin{pmatrix}
		\sum_{d=-1}^{k_1 - 1} x^1{}_{1 , d} z^d &  z^{k_2} + \sum_{d=-1}^{k_2 - 1} x^1{}_{2 , d} z^d & \sum_{d=0}^{k_2 - 1}  x^1{}_{3 , d}z^d & \sum_{d=0}^{k_2 - 1}  x^1{}_{4 , d} z^d\\
		z^{k_1} + \sum_{d=0}^{k_1-1} x^2{}_{1 , d} z^d  & \sum_{d=0}^{k_2} x^2{}_{2 , d} z^d & \sum_{d=0}^{k_1 - 1}  x^2{}_{3 , d}z^d & \sum_{d=0}^{k_1 - 1}  x^2{}_{4 , d} z^d
	\end{pmatrix}\\
\end{array}
\ee 
These are almost identical to the charts \eqref{Iwahori-charts} in Section \ref{sec:IwahoriVortexHilbertSpace}; the only difference is the inclusion of extra polar modes $x^1{}_{i,-1}$ for $i=1,2$.

Under combined gauge/flavor/$U(1)_\varepsilon$ action, the coordinates in each chart transform as in \eqref{torusAction}. There is a single fixed point at the origin of each chart and the required compensating gauge transformation is given by (\ref{gaugeFixingIwahori}). The equivariant weight of the fixed point labeled by $(k, \sigma)$ is
\begin{align}
	\begin{array}{rl}
		\omega_{\mathfrak{n},k,\sigma}  = & (-1)^{\sigma}(m_1-m_2+(k_1 - k_2)\varepsilon) \\[.2cm]
		& \times  (-k_{\sigma(1)} - 1)\varepsilon(m_{\sigma(2)}-m_{\sigma(1)}+(k_{\sigma(2)}-k_{\sigma(1)}-1)\varepsilon)\\[.2cm]
		& \times \prod\limits_{i=1}^2\prod\limits_{a = 1}^2\prod\limits_{l=0}^{k_i-1}(m_i - m_a + (l-k_a)\varepsilon) \prod\limits_{i=3}^4\prod\limits_{a = 1}^2\prod\limits_{l=0}^{k_a-1}({m}_{i} - m_a + (l-k_a)\varepsilon)\,.
	\end{array}
\end{align}
Putting everything together, we find
\begin{align}
	\CH(\CB_\nu, \V_{\rm con}) = \bigoplus_{k, \sigma} \C \ket{ \n , k ,\sigma } \hspace{1cm} \ket{ \n , k , \sigma } = \frac{1}{\omega_{\n , k,\sigma}} \delta_{\n , k, \sigma},
\end{align} where $\delta_{\n,k,\sigma}$ is the fundamental class of the fixed point at the origin of $\CM_D^{\n,k,\sigma}$.

\subsubsection{Operators}\label{sec:fundVortexOps}

Finally, we move on to the operator algebra itself. We expect to find $\text{End}_A^\varepsilon(\V_{\rm con})$ represented in the equivariant cohomology of the raviolo space $\CM_\rav(\CB_\nu; \V_{\rm con},\V_{\rm con})$. The raviolo space is, as usual,
\begin{align}
 	\CM_\rav(\CB_\nu; \V_{\rm con},\V_{\rm con}) = \CI[z]' \backslash \{ (X' , g , X)\;|\; X'=gX \} / \CI[z]\,,
\end{align}
subject as well to the disc constraints $X,X'\in \CL_0\cap R[z,z^{-1}]$ and $\det X_{(1,2)},\det X'_{(1,2)}\neq 0$.
This space breaks up into components labeled by vortex numbers $\n',\n$ on the top and bottom disks. As was the case with operators on the Iwahori line, we expect to have operators labeled by $\CI[z]$-equivariant cohomology classes of $\CI'[z] \backslash G[z, z^{-1}]$, which can be organized according to their support. The main difference between this computation and the Iwahori line is that there are more interesting compatibility requirements between $g\in \CO_A$ and various values of $X,X'$.

We give a \emph{brief} summary of the local operators of monopole number 0,$\pm1$ and R-charge 1, which end up matching our generators $J^0,J^\pm,N^0,N^\pm,\Psi_L,\Psi_R$ from Section \ref{sec:W2-ops}. Many more details are in Appendix \ref{app:conifoldops}.

Let's begin with the monopole number zero case $\n' = \n$. There is one simple $\cp^1$ cycle that is identical to the one found for the Iwahori line, namely it is parameterized by affine charts
\begin{align}
	\begin{pmatrix}
		1 & 0 \\
		c & 1 
	\end{pmatrix},
	&&
	\begin{pmatrix}
		0 & 1 \\
		1 & d 
	\end{pmatrix}
\end{align}
with transition function $d = 1 / c$. Just as before, this cycle can be realized as $\CI[z]' \backslash \CO_{(0,0)}$, where $\CO_{(0,0)}$ is the $G[z]' \times G[z]$ orbit of $g = \bsp 1&0\\0&1 \esp$ in $G[z,z^{-1}]$. Unlike the case with the Iwahori line, this is class does not fiber over the entire space of $X$'s, but instead only over the subvariety of $\CM_D$ locally defined by $x^1{}_{1,-1} = x^1{}_{2,-1} = 0$. (Otherwise, these gauge transformations would bring a pole into the bottom row of $X$.) Thus, we consider the subvariety 
\be \CS_{(0,0)} = \CI[z]'\backslash  \big\{\big(X',g\in \CO_{(0,0)},X\big)\big\}/\CI[z] \;\subset\; \bigsqcup_{\n\geq 0} \CM_{\rm rav}^{\n,\n} \;\subset\; \CM_{\rm rav} \ee
which is a $\cp^1$ fibration over the subvariety of $\CM_D$ locally defined by the equation $x^1{}_{1,-1} = x^1{}_{2,-1} = 0$.

There is another $\cp^1$ cycle given by the closure of $\CI[z]' \backslash \CO_{w, (1,-1)}$, where $\CO_{w, (1,-1)}$ is the $\CI[z]' \times \CI[z]$ orbit of $w z^{(1,-1)}$ in $G[z, z^{-1}]$, that we parameterize with charts
\begin{align}
	\begin{pmatrix}
		1 & b z^{-1} \\
		0 & 1 
	\end{pmatrix},
	&&
	\begin{pmatrix}
		a & z^{-1} \\
		z & 0
	\end{pmatrix}
\end{align}
with transition function $b = 1 / a$. Just as before, this collection of gauge transformations does not fit into correspondences above all $X$'s, but instead the subvariety of $\CM_D$ locally defined by submanifold $x^2{}_{3,0} = x^2{}_{4,0} =0$, \ie \, we consider the subvariety 
\be \CS_{w,(1,-1)} = \CI[z]'\backslash  \big\{\big(X',g\in \CO_{w, (1,-1)},X\big)\big\}/\CI[z] \;\subset\; \bigsqcup_{\n\geq 0} \CM_{\rm rav}^{\n,\n} \;\subset\; \CM_{\rm rav} \ee
which is a $\cp^1$ fibration over the subvariety of $\CM_D$ locally defined by the equation $x^2{}_{3,0} = x^2{}_{4,0} =0$.

We can evaluate the action on of the (fundamental classes of the) cycles found above by the formula (\ref{fixedPointActions}), where the normal bundle factors are given by
\be
	e(N_{\CS_{(0,0)}}| \sigma'; \sigma, k) = p(-m_{\sigma(1)} - (k_{\sigma(1)} + 1)\varepsilon)
\ee
for the first $\cp^1$ and
\be
e(N_{\CS_{w,(1,-1)}}|\one; k, \sigma) = \hat p(-m_{\sigma(2)} - k_{\sigma(2)}\varepsilon) = e(N_{\CS_{w,(1,-1)}}|w z^{(1,-1)}; k, \sigma)
\ee
for the second $\cp^1$, as can be seen by examining the equivariant weights of $x^1{}_{1,-1}, x^1{}_{2,-1}$ and $x^2{}_{3,0}, x^2{}_{4,0}$, respectively. We then find operators $ \partial_1,  \partial_2$ arising from the fiberwise fundamental classes with actions given by
\begin{align}
	 \partial_1 \ket{ \n , k , \sigma } &= \frac{(-1)^\sigma}{m_1 - m_2 + (k_1 - k_2 ) \varepsilon }  \left( p ( - m_1 - ( k_1 + 1 ) \varepsilon ) \ket{ \n , k ; 1} + p ( - m_2 - ( k_2 + 1 ) \varepsilon ) \ket{\n , k ; w } \right) \nonumber \\
	 \partial_2 \ket{ \n , k , \sigma } &= \frac{(-1)^\sigma}{m_1 - m_2 + (k_1 - k_2 + (-1)^\sigma) \varepsilon }  ( \hat p ( - m_{\sigma(2)} - k_{\sigma(2)} \varepsilon ) \ket{ \n , k , \sigma } \nonumber \\
		& \qquad \qquad \qquad \qquad + \hat p ( - m_{\sigma(1)} - ( k_{\sigma(1)} + 1 ) \varepsilon ) \ket{\n , k + e_{\sigma(1)} - e_{\sigma(2)} , w \sigma } ) . 
\end{align}
Besides these, at monopole number zero, there are operators $\varphi_a$. They should be interpreted as dressed versions of the identity operator, supported on the orbit $\CI[z]' 1 \CI[z]$, just as on the Iwahori line.%
\footnote{We can also obtain the action of higher classes on these $\cp^1$ cycles. Only the fundamental classes will be necessary for comparing to $\End^\varepsilon_B(\W_{\mathbf 2})$.}

Let's start our comparison between $\End^\varepsilon_B(\W_{\mathbf 2})$ and $\End^\varepsilon_A(\V_{\rm con})$. In our analysis of the fundamental Wilson line in Section \ref{sec:FundamentalWilsonLine}, we found four independent generators of R-charge 1 with zero charge under the $PSU(2)$ flavor symmetry: $J^0, \Psi_L, \Psi_R,$ and $N_0$. These should be mirror to four operators of R-charge 1 and monopole number zero on the fundamental vortex line. Indeed we have found four such operators: $\partial_1, \partial_2,$ and $\varphi_a$, so it only remains to confirm that they satisfy the same relations. The appropriate\footnote{There are some ambiguities in this identification, some of these ambiguities are discussed in Appendix \ref{app:conifold}. An example of such an ambiguity arises from the $\Z_2$ symmetry exchanging $U(1)_L$ and $U(1)_R$.} linear combinations mirror are given by
\begin{align}
	(J^0)^! & = -2(\varphi_1 + \varphi_2) + \varepsilon, \nonumber\\
	(\Psi_L)^! &= \partial_1 + \varphi_1 + m_1 - \tfrac 1 2 \varepsilon , \nonumber \\
	(\Psi_R)^! &=  \partial_2 + \varphi_2 + m_3 + \tfrac 1 2 \varepsilon  , \nonumber \\
	(N^0)^! &=  \partial_1 -  \partial_2 - \varphi_1 - \varphi_2 - m_3 - m_4.
\end{align}

To obtain the positive monopole operators, we examine the correspondence spaces at $\n' = \n + 1$. Just as above, there are two simple $\cp^1$ cycles in $\CI'[z] \backslash G[z, z^{-1}]$ that fit into such correspondences.  The first is the closure of $\CI[z]' \backslash \CO_{(1,0)}$, where $\CO_{(1,0)}$ is the $\CI[z]'\times\CI[z]$ orbit of $z^{(1,0)}$ in $G[z, z^{-1}]$, which is covered by affine charts
\be
	\begin{pmatrix}
		z & 0 \\
		z c & 1
	\end{pmatrix} \hspace{1cm}
	\begin{pmatrix}
		0 & 1 \\
		z & d
	\end{pmatrix},
\ee
with transition function $d = 1/ c$. The second $\cp^1$ class is the closure of $\CI[z]' \backslash \CO_{(0,1)}$, where $\CO_{(0,1)}$ is the $\CI[z]'\times\CI[z]$ orbit of $z^{(0,1)}$ in $G[z, z^{-1}]$, which is covered by affine charts
\be
	\begin{pmatrix}
			1 & b \\
			0 & z
		\end{pmatrix} \hspace{1cm}
		\begin{pmatrix}
			a & 1 \\
			z & 0
		\end{pmatrix},
\ee
with transition function $b = 1/ a$. Both of these cycles fit above the entire space of $X$'s, \ie\, we have two subvarieties of $\CM_{\rav}$ given by
\be
\CS_{(1,0)} = \CI[z]'\backslash  \big\{\big(X',g\in \CO_{(1,0)},X\big)\big\}/\CI[z] \;\subset\; \bigsqcup_{\n\geq 0} \CM_{\rm rav}^{\n+1,\n} \;\subset\; \CM_{\rm rav}
\ee
and 
\be
	\CS_{(0,1)} = \CI[z]'\backslash  \big\{\big(X',g\in \CO_{(0,1)},X\big)\big\}/\CI[z] \;\subset\; \bigsqcup_{\n\geq 0} \CM_{\rm rav}^{\n+1,\n} \;\subset\; \CM_{\rm rav}
\ee
which are both $\cp^1$ fibrations over $\CM_D$. From these subvarieties we get two monopole operators $V^+_1, V^+_2$ associated to the fiberwise fundamental classes and their action on $\CH(\CB_\nu, \V_{\rm con})$ is given by
\begin{align}
	V^+_1 \ket{ \n , k , \sigma} &= 
		(-1)^\sigma \frac{\hat p ( - m_{\sigma (1)} - ( k_{\sigma(1)} + 1 ) \varepsilon) }{m_1 - m_2 + (k_1 - k_2 + ( -1 )^\sigma) \varepsilon } \bigg( p ( - m_{\sigma(1)} - ( k_{\sigma(1)} + 2 ) \varepsilon ) \ket{\n+1, k + e_{\sigma(1)} , \sigma } \nonumber \\
	&\qquad \qquad \qquad \qquad + p ( - m_{\sigma(2)} - ( k_{\sigma(2)} + 1 ) \varepsilon ) \ket{\n+1 , k + e_{\sigma(1)} , w \sigma } \bigg) , \nonumber \\
	V^+_2 \ket{ \n , k , \sigma} &= 
		(-1)^\sigma \frac{p ( - m_{\sigma (2)} - ( k_{\sigma(2)} + 1 ) \varepsilon) }{m_1 - m_2 + (k_1 - k_2 ) \varepsilon } \bigg( \hat p ( - m_{\sigma(2)} - ( k_{\sigma(2)} + 1 ) \varepsilon ) \ket{\n+1, k + e_{\sigma(2)} , \sigma } \nonumber \\
	&\qquad \qquad \qquad \qquad + \hat p ( - m_{\sigma(1)} - ( k_{\sigma(1)} + 1 ) \varepsilon ) \ket{\n+1 , k + e_{\sigma(1)} , w\sigma } \bigg) .
\end{align}

Finally, let's consider negative monopole operators, that is, correspondences with $\n' = \n -1$. Again, there are two simple $\cp^1$ classes that fit into such correspondences. The first corresponds to the closure of $\CI[z]' \backslash \CO_{(-1,0)}$, where $\CO_{(-1,0)}$ is the $\CI[z]'\times\CI[z]$ orbit of $z^{(-1,0)}$ in $G[z,z^{-1}]$, and is covered by affine charts which are $z^{-1}$ times those found in $\CI[z]' \backslash \CO_{(0,1)}$:
\begin{align}
	\begin{pmatrix}
		z^{-1} & b z^{-1} \\
		0 & 1 
	\end{pmatrix}
	&&
	\begin{pmatrix}
		a z^{-1} & z^{-1} \\
		1 & 0 
	\end{pmatrix}.
\end{align}
with the same transition function as before. Similarly, the second $\cp^1$ class comes from the closure of $\CI[z]' \backslash \CO_{(0,-1)}$, where $\CO_{(0,-1)}$ is the $\CI[z]'\times\CI[z]$ orbit of $z^{(0,-1)}$ in $G[z,z^{-1}]$, and is covered by affine charts which are $z^{-1}$ times those found in $\CI[z]' \backslash \CO_{(1,0)}$:
\begin{align}
	\begin{pmatrix}
		1 & 0 \\
		c & z^{-1} 
	\end{pmatrix}
	&&
	\begin{pmatrix}
		0 & z^{-1} \\
		1 & d z^{-1} 
	\end{pmatrix}
\end{align}
with the same transition function as before. As usual, these cycles will give us operators $V^-_1$ and $V^-_2$ arising from the fiberwise fundamental classes.

The subvariety
\be
\CS_{(-1,0)} = \CI[z]'\backslash  \big\{\big(X',g\in \CO_{(-1,0)},X\big)\big\}/\CI[z] \;\subset\; \bigsqcup_{\n\geq 0} \CM_{\rm rav}^{\n-1,\n} \;\subset\; \CM_{\rm rav}
\ee
is a fibration over a subvariety of $\CM_D^\n$ and the normal bundle factors at the fixed points are given by
\be
e(N_{\CS_{(-1,0)}}| z^{(-1,0)}; k, \sigma) = p(-m_{\sigma(1)} - (k_{\sigma(1)} + 1)\varepsilon)\hat p(-m_{\sigma(1)}-k_{\sigma(1)}\varepsilon)
\ee
and
\be
e(N_{\CS_{(-1,0)}}| w z^{(0,-1)}; k, \sigma) = p(-m_{\sigma(1)} - (k_{\sigma(1)} + 1)\varepsilon)\hat p(-m_{\sigma(2)}-k_{\sigma(2)}\varepsilon),
\ee
where $p(x) = (x+m_1)(x+m_2)$ and $\hat p(x) =(x+m_3)(x+m_4)$. Similarly, the subvariety
\be
\CS_{(0,-1)} = \CI[z]'\backslash  \big\{\big(X',g\in \CO_{(0,-1)},X\big)\big\}/\CI[z] \;\subset\; \bigsqcup_{\n\geq 0} \CM_{\rm rav}^{\n-1,\n} \;\subset\; \CM_{\rm rav} \ee
is a fibration over a subvariety of $\CM_D^\n$. The normal bundle factors at the fixed points are given by
\be
e(N_{\CS_{(0,-1)}}| z^{(0,-1)}; k, \sigma) = p(-m_{\sigma(2)} - k_{\sigma(2)}\varepsilon)\hat p(-m_{\sigma(2)}-k_{\sigma(2)}\varepsilon)
\ee
and
\be
e(N_{\CS_{(0,-1)}}| w z^{(0,-1)}; k, \sigma) = p(-m_{\sigma(1)} - (k_{\sigma(1)} + 1)\varepsilon)\hat p(-m_{\sigma(2)}-k_{\sigma(2)}),
\ee
Finally, the action of $V^-_1$ and $V^-_2$ are given by
\begin{align}
	V^-_1 \ket{\mathfrak{n}, k, \sigma} = \frac{(-1)^\sigma}{m_2 - m_1 + (k_2- k_1) \varepsilon} \bigg[\ket{\mathfrak{n}-1, k-e_{\sigma(1)}, \sigma} + \ket{\mathfrak{n}-1, k - e_{\sigma(2)}, w \sigma} \bigg],
\end{align}
and
\begin{align}
	V^-_2 \ket{\mathfrak{n}, k, \sigma} = \frac{(-1)^\sigma}{m_1 - m_2 + (k_1 - k_2 + (-1)^\sigma) \varepsilon}\bigg[\ket{\mathfrak{n}-1, k-e_{\sigma(2)}, \sigma}+ \ket{\mathfrak{n}-1, k - e_{\sigma(2)}, w \sigma} \bigg].
\end{align}

We are now in a position to compare with the remaining generators of the algebra discussed in Section \ref{sec:FundamentalWilsonLine}. As discussed in Appendix \ref{app:conifold}, the operators generating the $PSU(2)_{top}$ topological flavor symmetry are given by the linear combinations
\be
(J^{\pm})^! = V^{\pm}_1 \mp V^{\pm}_2.
\ee
We can then act on $(N^0)^!$ to find
\be
(N^\pm)^! = V^{\pm}_1.
\ee
With these identifications, it is straightforward to check that they satisfy the same relations as the corresponding operators on $\End^\varepsilon_B(\W_{\mathbf 2})$.
\footnote{Just as with the other generators, there are some ambiguities in this identification. Some of these ambiguities are discussed in Appendix \ref{app:conifold}.}

\appendix

\section{SUSY algebra}
\label{app:SUSY}

The $3$d $\CN=4$ algebra is generated by 8 supercharges $Q^{a \dot a}_\alpha$ and is of the form
\begin{align}\label{SUSYAlgebra}
	\{ Q^{a \dot a}_\alpha , Q^{b \dot b}_\beta \} = \epsilon^{ab} \epsilon^{\dot a \dot b } P_{\alpha \beta} -i \epsilon_{\alpha \beta} \left( \epsilon^{ab} m^{\dot a \dot b} + \epsilon^{\dot a \dot b} t^{ab} \right) .
\end{align}
Here $\alpha \in \{ + , - \}$ are spinor indices labeling states transforming under the Euclidean spin group $SU(2)_E$. Upper indices transform in the fundamental $\square$ representation of $SU(2)_E$ with lower indices in the anti-fundamental $\overline \square$. The isomorphism between the fundamental and anti-fundamental representations of $SU(2)$ is implemented by the epsilon tensor and its inverse
\begin{align}
	X^\alpha = \epsilon^{\alpha \beta} X_\beta\,,
	&&X_\alpha = \epsilon_{\alpha \beta} X^\beta\,,
\end{align}
with
\begin{align}
	\epsilon^{+-} = \epsilon_{-+} = 1.
\end{align}

Lower-case Latin indices $a, \dot a$ on the other hand transform under the $SU(2)_H$ and $SU(2)_C$ R-symmetries, respectively, and have the same conventions as Euclidean spinor indices. The mass and FI parameters $m^{\dot a \dot b}$ and $t^{ab}$ are central charges in the symmetric tensor representation $\text{Sym}^2(\square)$ of $SU(2)_C$ and $SU(2)_H$, respectively
\begin{align}
	m^{\dot a \dot b} = m^{(\dot a \dot b)},
	\qquad \qquad t^{ab} = t^{(ab)} .
\end{align}
In the transformation laws for fundamental fields presented below they will be realized by the action of some $m^{\dot a \dot b} \in \mathfrak f, ~t^{ab} \in \mathfrak f_t$ where $\mathfrak f$ is a Cartan subalgebra of the group of global symmetries acting on hypermultiplets and $\mathfrak f_t$ is the algebra of topological symmetries.
(More technically, we might say $m \in \mathfrak f\otimes \text{Sym}^2(\square)_C$, $t \in \mathfrak f_t\otimes \text{Sym}^2(\square)_H$.)

The isomorphism between the symmetric tensor representation of $SU(2)_E$ and the adjoint representation (a spacetime vector) is implemented by the sigma matrices
\begin{align}
	\sigma_\mu^{\alpha \beta} = 
		\left \{ 
			\begin{pmatrix}
				1 & 0 \\
				0 & -1
			\end{pmatrix} ,
			\begin{pmatrix}
				- i & 0 \\
				0 & - i
			\end{pmatrix} ,
			\begin{pmatrix}
				0 & -1 \\
				-1 & 0
			\end{pmatrix}
		\right \}\,.
\end{align}
Here $\mu\in{1,2,3}$ indexes a basis for the adjoint representation. For $SU(2)_C$ and $SU(2)_H$, the isomorphism is implemented by identical sigma matrices that we denote $\sigma^{\dot a\dot b}_{\dot I}$ and $\sigma^{ab}_I$, respectively.
Lowering indices, we also have
\begin{align}
	(\sigma^\mu )^\alpha{}_\beta &= 
		\left \{ 
			\begin{pmatrix}
				0 & 1 \\
				1 & 0
			\end{pmatrix} ,
			\begin{pmatrix}
				0 & -i \\
				i & 0
			\end{pmatrix} ,
			\begin{pmatrix}
				1 & 0 \\
				0 & -1
			\end{pmatrix}
		\right \} ,
	\nonumber \\[.2cm]
	\sigma^\mu _{\alpha \beta} &=
		\left \{ 
			\begin{pmatrix}
				- 1 & 0 \\
				0 & 1
			\end{pmatrix} ,
			\begin{pmatrix}
				- i & 0 \\
				0 & - i
			\end{pmatrix} ,
			\begin{pmatrix}
				0 & 1 \\
				1 & 0
			\end{pmatrix}
		\right \} .
\end{align}
The traceless Hermitian matrices $(\sigma^\mu )^\alpha{}_\beta$ are the usual Pauli matrices. In this form they will often simply be denoted `$\sigma^\mu$' in matrix notation, and they satisfy the algebra
\begin{align}
	\sigma^\mu \sigma^\nu = \delta^{\mu \nu} \id + i \epsilon^{\mu \nu \lambda} \sigma_\lambda \,.
\end{align}
Adjoint $SU(2)$ indices are lowered and raised with the metric $\delta_{\mu \nu}$ (similarly: $\delta_{IJ}$, $\delta_{\dot I\dot J}$); and the totally antisymmetric tensor is denoted by $\epsilon^{\mu \nu \lambda}$ where
\begin{align}
	\epsilon^{123} = \epsilon_{123} = 1 \, .
\end{align}
Some useful identities for manipulating sigma matrices in these conventions are
\begin{gather}
	[ \sigma^\mu , \sigma^\nu ] = 2 i \epsilon^{\mu \nu \lambda} \sigma_\lambda \,, 
	\qquad \qquad \Tr ( \sigma^\mu \sigma^\nu ) = 2 \delta^{\mu \nu} \,, 
	\qquad \qquad  \Tr ( \sigma^\mu \sigma^\nu \sigma^\lambda ) = 2 i \epsilon^{\mu \nu \lambda} \,,
	\nonumber \\
	(\sigma^\mu)^\alpha{}_\beta(\sigma_\mu)^\gamma{}_\delta = 2\delta^\alpha{}_\delta\delta_\beta{}^\gamma - \delta^\alpha{}_\beta\delta^\gamma{}_\delta\,,
	\qquad \qquad (\sigma^\mu)_{\alpha\beta}(\sigma_\mu)_{\gamma\delta} = 2\epsilon_{\beta(\gamma}\epsilon_{\delta)\alpha}\,.
\end{gather}

We will often use the isomorphism $\sigma$ implicitly, writing vectors as bi-spinors and vice versa. Given any (co)vector $v_\mu$ we set
\begin{align}
	v_{\alpha \beta} := \sigma^\mu_{\alpha \beta} v_\mu\,,\;\qquad \text{or}
	\qquad \qquad
	v_\mu := \frac 1 2 \sigma_\mu^{\alpha \beta} v_{\alpha \beta} \,.
\end{align}
For instance, the momentum operator $P_\mu = - i \partial_\mu$ as a bi-spinor is
\begin{align}
	&P_{\alpha \beta} = 
	\begin{pmatrix}
		- 2 P_{\bar z} & P_t \\
		P_t & 2 P_z
	\end{pmatrix},
	&&\partial_{\alpha \beta} =
	\begin{pmatrix}
		- 2 \partial_{\bar z} & \partial_t \\
		\partial_t & 2 \partial_z
	\end{pmatrix}.
\end{align}
Similarly, letting $m^{\dot I}, t^{I}$ denote the mass/FI parameters in the adjoint representations of $SU(2)_C$ and $SU(2)_H$, respectively, and defining real and complex combinations as
\begin{align}
	&m_\C = \tfrac12(m_1 - i m_2)\,, &&m_\R = -m_3\,, \nonumber \\
	&t_\C = \tfrac12(t_1 - i t_2)\,, &&t_\R = -t_3\,,
\end{align}
we find that
\begin{align}
	&m^{\dot a \dot b} = 
	\begin{pmatrix}
		2m_\C & m_\R \\
		m_\R & - 2\bar m_\C
	\end{pmatrix},
	&&t^{ab} =
	\begin{pmatrix}
		2t_\C & t_\R \\
		 t_\R & -2 \bar t_\C
	\end{pmatrix}.
\end{align}
In terms of $P_t,P_z,P_{\bar z}$, the SUSY algebra takes the form
\begin{gather}
	\{ Q^{a \dot a}_+ , Q^{b \dot b}_+ \} = - 2 \epsilon^{ab} \epsilon^{\dot a \dot b} P_{\bar z} \,,
	\qquad \qquad
	\{ Q^{a \dot a}_- , Q^{b \dot b}_- \} = 2 \epsilon^{ab} \epsilon^{\dot a \dot b} P_{z} \,,
	\nonumber \\
	\{ Q^{a \dot a}_+ , Q^{b \dot b}_- \} = \epsilon^{ab} \epsilon^{\dot a \dot b} P_t  -i\epsilon^{ab} m^{\dot a \dot b} -i \epsilon^{\dot a \dot b} t^{ab} \,.	
\end{gather}

\subsection{Vectormultiplets}

An off-shell 3d $\CN=4$ vector multiplet consists of the fields
\be \label{vec-fields}
	A_\mu\,,\qquad 
	\phi^{\dot a \dot b}\,,\qquad 
	\lambda^{a \dot a}_\alpha\,, \qquad 
	D^{ab}\,.
\ee
Here $A_\mu$ is a connection 1-form; $\phi^{(\dot a \dot b)}$ is a scalar field in the adjoint representation of the $SU(2)_C$ R-symmetry, with real component $\sigma$ and complex component $\varphi$
\begin{align}
	\phi^{\dot a \dot b} =
	\begin{pmatrix}
		2\varphi && \sigma \\
		 \sigma && - 2\bar \varphi 
	\end{pmatrix};
\end{align}
$\lambda^{a \dot a}_\alpha $ is a complex gaugino in the bi-fundamental of $SU(2)_H\times U(2)_C$; and $D^{(ab)}$ is an auxiliary field in the adjoint of $SU(2)_H$. For gauge group $G$, all the fields in \eqref{vec-fields} transform additionally in the Lie algebra $\mathfrak g$ (or the complexified lie algebra $\mathfrak g_\C$, in the case of $\varphi$ and the gauginos $\lambda$).

We will work with ``physics conventions,'' in which the real Lie algebra $\mathfrak g$ is generated by \emph{Hermitian} matrices. This has the advantage that ``real'' masses $m_\R$ and FI parameters $t_\R$ will actually take real values. It has a familiar disadvantage that an extra factor of $i$ appears in Lie algebra structure constants: $[T^a,T^b]=if^{ab}{}_c T^c$, and in covariant derivatives.
The $G$-covariant derivative takes the form
\begin{align}
	d_A = d - iA\,,
\end{align}
and the field strength is
\begin{align}
	F = i[ d_A , d_A ]= d A - iA \wedge A\,. 
\end{align}

In three dimensions the field strength may be dualized to a vector $(*F)_\mu=\frac12\epsilon_{\mu\nu\lambda}F^{\nu\lambda}$, or a traceless Hermitian bispinor
\be F^\alpha{}_\beta = 2  (\sigma^\mu)^\alpha{}_ \beta ( \star F)_\mu = -i(\sigma^{\mu \nu})^\alpha{}_\beta F_{\mu \nu}\,, \ee
where, in matrix notation, $\sigma^{\mu \nu} := \frac 1 2 [ \sigma^\mu , \sigma^\nu] = i \epsilon^{\mu\nu\lambda}\sigma_\lambda$\,. Explicitly,
\be F^\alpha{}_\beta = 4i \bp -F_{z\bar z} & F_{zt} \\ -F_{\bar z t} & F_{z\bar z} \ep\,,\qquad \text{or}\qquad
 F_{\alpha\beta} = 4i \bp F_{\bar z t} & -F_{z\bar z} \\ -F_{z\bar z} & F_{zt} \ep\,. \ee
Note that $F_{\alpha\beta}$ is symmetric.
The Bianchi identity $d_A F = 0$ then reads
\be (d_A)_{\gamma [\alpha} F^\gamma{}_{\beta ]} = 0\,,
	\qquad\text{or}\qquad
	d^{\alpha \beta}_A F_{\alpha \beta} = 0\,.
\ee

Finally, we can state the transformation rules for the 3d $\CN=4$ vectormultiplet: 
\begin{align}
	Q^{a \dot a}_\alpha A_{\beta\gamma} &=  \lambda^{a \dot a}_{(\beta} \epsilon_{\gamma ) \alpha} \,,
	\qquad \qquad Q^{a \dot a}_\alpha \phi^{\dot b \dot c} = i\lambda^{a ( \dot b}_\alpha \epsilon^{\dot c ) \dot a } \,, \nonumber \\
	Q^{a \dot a}_\alpha \lambda^{b \dot b}_\beta &=  \frac 1 2  \epsilon^{a b} \epsilon^{\dot a \dot b} F_{\alpha \beta} -  \epsilon^{a b} (d_A )_{\alpha \beta} \phi^{\dot a \dot b} -i \epsilon_{\alpha \beta} \epsilon^{\dot a \dot b} D^{ab} + \frac 1 2 \epsilon_{\alpha \beta} \epsilon^{ab} [ \phi^{\dot a}{}_{\dot c} , \phi^{\dot c \dot b} ] \,, \nonumber \\
	Q^{a \dot a}_\alpha D^{bc} &= - (d_A )_\alpha{}^\beta \epsilon^{a (b} \lambda^{c) \dot a}_\beta - [ \phi^{\dot a}{}_{\dot b} , \epsilon^{a(b} \lambda^{c) \dot b}_\alpha ] \, .
\end{align}
One may check that the algebra of supersymmetries acting on the fields satisfies
\begin{align} \label{close-vec}
	\{ Q^{a \dot a}_\alpha , Q^{b \dot b}_\beta \} = \epsilon^{ab} \epsilon^{\dot a \dot b } P_{\alpha \beta} -i \epsilon_{\alpha \beta} \left( \epsilon^{ab} \phi^{\dot a \dot b} + \frac 1 {g^2} \epsilon^{\dot a \dot b} D^{ab} \right) .
\end{align}
Here $\phi^{\dot a \dot b}$ acts on fields as an infinitesimal $\mathfrak g$-gauge transformation. Similarly, the $D$-term acts as an infinitesimal topological symmetry; explicitly, it ``acts'' as zero on $\phi$ and $\lambda$, but acts as a translation of the dual photon $\gamma$, which satisfies
\begin{align}
	\frac 1 {2 g^2} \Tr (F_{\alpha \beta} ) = \partial_{\alpha \beta} \gamma \,.
\end{align}

Note that upon using the equation of motion 
\be \frac{1}{g^2}D^{ab} = t^{ab} \qquad\text{(in the absence of matter)}\,, \ee
and restricting to gauge-invariant combinations of vectormultiplet fields (on which $\phi$ acts as zero), the algebra 
 \eqref{close-vec} reduces to the general form \ref{SUSYAlgebra}.
Mass parameters could also be introduced, as scalars in background vectormultiplets associated to a flavor symmetry; in \eqref{close-vec} this amounts to replacing  $\phi \leadsto m$.

\subsection{Hypermultiplets}

A hypermultiplet contain an $SU(2)_H$ doublet of complex scalar fields and an $SU(2)_C$ doublet of complex fermions. It's convenient to introduce an additional $SU(2)'$ spinor index $A\in\{1,2\}$, writing the scalars as $X^{aA}$ subject to a reality condition
\begin{align}\label{reality}
	(X^{aA})^* = X_{a A}\,.
\end{align}
This makes manifest the full $SO(4) \simeq SU(2)_H \times SU(2)'$ symmetry of the four real scalars in the hypermultipet. With respect to a 3d $\CN=2$ subalgebra, the fields $X^{+1} = X$ and $X^{+2} = Y$ are chiral, whereas $X^{-2} = \bar X$ and $X^{-1} = - \bar Y$ are anti-chiral. Altogether, we have
\begin{align} \label{XXY}
	X^{aA} =
	\begin{pmatrix}
		X & Y \\
		- \bar Y & \bar X
	\end{pmatrix} .
\end{align}
Similarly, we write the fermions as $\psi^{\dot a A}_\alpha$. In Lorentzian signature they would obey a reality constraint $ (\psi^{\dot A}_\alpha)^\dagger \sim \psi_{\alpha \dot a A}$; but in Euclidean signature the components of $\psi^{\dot a A}_\alpha$ are independent, and $\bar\psi$ does not appear in the action or integration measure.
The supersymmetry transformations for a single free hypermultiplet are simply
\begin{align}
	Q^{a \dot a}_\alpha X^{b A} = i \epsilon^{ab} \psi^{\dot a A}_\alpha\,,
	\qquad \qquad
	Q^{a \dot a}_\alpha \psi^{\dot b A}_\beta =  \epsilon^{\dot a \dot b} \partial_{\alpha \beta} X^{a A} \,.
\end{align}

The $SU(2)'$ indices are raised and lowered by antisymmetric tensors $\Omega^{AB}$ and $\Omega_{AB}$.  We'll use the convention $\Omega^{12} = \Omega_{21} = 1$. The tensor $\Omega^{AB}$ (resp. $\Omega_{AB}$) has a geometric interpretation as the holomorphic Poisson structure (resp. symplectic structure) on the ``target space'' $T^* \C$ of the theory of a free hypermultiplet.

For a collection of $N$ hypermultiplets, the extra symmetry $SU(2)'$ is extended to $USp(N)$, and the index $A$ takes values $A = 1 , \dots , 2N$. It is raised and lowered by the tensors
\be \Omega^{AB} = \bp 0& \id_N \\ -\id_N&0\ep \,,\qquad \Omega_{AB} = \bp 0& -\id_{N}\\ \id_N&0\ep \ee
(where $\id_N$ denotes the $N\times N$ identity matrix),
which now play the role of holomorphic Poisson/symplectic tensors on $T^*\C^N$. The reality constraint on scalars continues to take the form (\ref{reality}). We will typically split the scalars into chiral halves, generalizing \eqref{XXY},
\be \begin{array}{c}X^i = X^{+,i}\,,\qquad Y_i = X^{+,N+i}\,, \\[.1cm]
 \ol X_i = X^{-,N+i}\,,\qquad \ol Y^i = - X^{-,i} \end{array}\qquad i = 1,...,N\,, \ee
with $X^i,\ol Y^i$ transforming in the fundamental representation of $U(N)$, and $Y_i,\ol X_i$ in the dual.

We may couple a collection of $N$ hypermultiplets to a $G$ gauge symmetry by identifying $G$ with a subgroup of $USp(N)$. (Equivalently, we specify how the hypermultiplets transform in a unitary symplectic representation of $G$.) The on-shell SUSY transformations then become
\begin{align}
	Q^{a \dot a}_\alpha X^{b A} = i \epsilon^{ab} \psi^{\dot a A}_\alpha\,,
	\qquad \qquad
	Q^{a \dot a}_\alpha \psi^{\dot b B}_\beta = \big(\epsilon^{\dot a \dot b} (d_A)_{\alpha \beta}  + \epsilon_{\alpha \beta}\phi^{\dot a \dot b} \big)\cdot X^{a B}
\end{align}
where $d_A$ is the $G$-covariant derivative and $\phi\, \cdot X$ denotes an infinitesimal gauge transformation generated by $\phi$ in the appropriate unitary symplectic representation of $G$.

\subsection{Moment maps}
\label{app:moment}

In a gauge theory with hypermultiplet matter, the equations of motion set the auxiliary field $D^{ab}$ in the vectormultiplet to
\be  D^{ab} = \mu^{ab} + t^{ab}\,, \ee
where
\be \mu^{ab} = \begin{pmatrix}
		2\mu 	&  \mu_\R \\
		 \mu_\R 	& - 2\bar \mu
	\end{pmatrix}
\ee
 is the triplet of hyperk\"ahler moment maps. Recall that the moment maps take values in the dual of the Lie algebra, $\mu_\R \in \mathfrak g^*$ and $\mu\in \mathfrak g^*_\C$. We can describe them explicitly as follows.  Let $\{(\tau_k)^A{}_B\}_{k=1}^{\text{rank}\,G}$ denote a basis of generators of $\mathfrak g$, as elements of $\mathfrak{usp}(N)$. Then for each generator $\tau_k$,
 \be \langle \tau_k ,\mu^{ab}\rangle = - X^a{}_A (\tau_k)^A{}_B X^{bB}\,. \label{gen-moment} \ee

In this paper we will always assume that hypermultiplets transform in a representation of the form $R\oplus R^*$, where $R$ is a unitary representation of $G$, and $R^*$ its dual. In this case, $G$ acts as a subgroup of $U(N)$, and the moment maps may similarly be interpreted as elements of $\mathfrak u(N)^*$ or $\mathfrak u(N)^*_\C$. Letting $\{(T_k)^i{}_j\}_{k=1}^{\text{rank}\,G}$ denote the generators of $\mathfrak g$, as elements of $\mathfrak u(N)$, we have
\be \tau_k  = \bp T_k & 0 \\ 0 & - T_k \ep \in \mathfrak{usp}(N)\,. \ee
The general expression \eqref{gen-moment} for the moment maps simplifies to
\be -\langle T_k,\mu \rangle = Y_i (T_k)^i{}_j X^j\,,\qquad -\langle T_k,\mu_\R \rangle = \ol X_i (T_k)^i{}_j X^j - Y_i (T_k)^i{}_j \ol Y^j\,.\ee
For instance, if $G=U(1)$ acts on a single hypermultiplet, with charge generator
\be T = 1 \in \mathfrak u(1)\,,\qquad \tau = \bp 1&0 \\ 0&-1\ep \in \mathfrak{usp}(1)\,,\ee
so that $X,Y$ have charges $\pm 1$, then the moment maps are familiar expressions
\be \mu = XY\,,\qquad \mu_\R = |X|^2-|Y|^2\,.\ee

\section{Half-space setup and representations for the BFN algebra}
\label{app:BFN}

The computation of local operators at junctions of A-type lines that was outlined in Section \ref{sec:comp} is not obviously related to the definition of the Coulomb-branch chiral ring (\emph{a.k.a.} $\text{Hom}_A(\id,\id)$) that was given by Braverman-Finkelberg-Nakajima in the mathematics literature \cite{Nak, BFNII}. 
In this appendix, we discuss two modifications of the setup from Section~\ref{sec:comp} that can be implemented to produce a computational scheme that \emph{directly} recovers and generalizes the BFN construction. 

The reason for not working in the modified scheme from the outset was mentioned in the introduction to Section \ref{sec:comp}\,: doing computations (that match physical expectations) generally requires the mathematics of Borel-Moore homology on infinite-dimensional stacks, which is somewhat sophisticated, and which we did not want to use in this paper. 

We note that many of the proposed constructions in this section have appeared (in slightly different guises) in the mathematics literature. In particular, both \cite{BFN-lines} and \cite{Web2016} use Borel-Moore homology of the sorts of moduli spaces we define here to characterize some special families of A-type line operators in 3d $\CN=4$ gauge theories. Our goal in writing this appendix is two-fold:
\begin{itemize}
\item[1)] To explain one concrete physical setup that (plausibly) leads to the definition of the Coulomb-branch algebra and some line operators seen in the mathematics literature.
\item[2)] To generalize the mathematics constructions to include the entire class of line operators from Section \ref{sec:VL}, given either using $(\CL_0,\CG_0)$ data or quantum-mechanics $(\CE,W_0)$ data.
\end{itemize}
We will \emph{not} do any explicit computations.

\subsection{Modification 1: half-space setup}
\label{sec:halfspace}

Let us fix a 3d $\CN=4$ gauge theory with data $G,R$, and work in the A-twist. Let us also choose an A-type line operator $\CL$ and a $Q_A$-preserving half-BPS boundary condition $\CB$.
By a state-operator correspondence as in Section \ref{sec:cyl}, we expect the cylinder Hilbert space $\CH_D(\CB,\CL)$ to be equivalent to $Q_A$-cohomology of the space of local operators at the intersection of the line $\CL$ and the boundary condition $\CB$. Either space might sensibly be called $\text{Hom}_A(\CB,\CL)$. Nevertheless, the algebraic description of the two spaces is subtly different.

In the case of the cylinder setup, the line operator $\CL$ is supported in an infinitesimal neighborhood of $z=0$, while the boundary condition $\CB$ lies either at some finite radius $|z|=r$ or (in the case of a vacuum boundary condition) asymptotically at $z\to \infty$. Algebraically, this means that we describe the line operator using formal Laurent series in $z$, \emph{e.g.} using a Lagrangian $\CL_0 \in T^*R(\CK)$. However, we must describe an asymptotic boundary condition using Laurent series in $z^{-1}$, denoted $\CK_\infty$ in Section \ref{sec:Mrav}. A finite-distance boundary condition would be described using either polynomials in $z$ or Laurent series that converged in some radius $|z|\leq r$.

In contrast, if we consider a half-space $\C_z\times \R_{t\geq 0}$, with $\CL$ supported in an infinitesimal neighborhood of $z=0$ and $\CB$ supported in an infinitesimal neighborhood of $t=0$,
 we may do the entire analysis in an infinitesimal neighborhood of the point $z=t=0$ at which $\CL$ meets $\CB$.
Algebraically, this means that both lines and boundaries get a (putative) description in terms of formal Laurent series in $z$. This leads to slightly more systematic (proposed) definitions of moduli spaces.

The half-space setup matches the descriptions of moduli spaces in the mathematics literature on Coulomb branches and line operators. Everything is done in terms of formal Laurent series in $z$, and no polynomials or series in $z^{-1}$ are ever present.

We \emph{expect} --- due to the state-operator correspondence --- that both cylinder and half-space setups lead to exactly the same spaces of local operators $\text{Hom}_A(\CB,\CL)$, and exactly the same actions of junctions $\text{Hom}_A(\CL,\CL')$ on these spaces. The equivalence only need to appear after cohomologies of the relevant moduli spaces are computed, since physically the state-operator correspondence only holds after the topological A-twist. It is not yet obvious to us how the equivalence will appear mathematically.

\subsection{Modification 2: finite-distance boundary conditions}
\label{sec:halfB}

When working on a half-space, it is not suitable to use a boundary condition $\CB_\nu$ defined asymptotically by a vacuum $\nu$. We need an honest, half-BPS (in particular, $Q_A$-preserving), finite-distance boundary condition.

One option is to use the ``Lefschetz thimble'' boundary conditions discussed in \cite{BDGH} (generalizing the classical 2d $\CN=(2,2)$ constructions of \cite{HoriIqbalVafa}). For each vacuum $\nu$, there is a $Q_A$-preserving finite-distance boundary condition $\ol\CB_\nu$ that mimics the effect of asymptoting to the vacuum $\nu$. One should be able to do half-space calculations with $\ol\CB_\nu$ that reproduce all the cohomologies $H^*(\CM_D(\CB_\nu,\CL))$, $H^*(\CM_{\rm rav}(\CB_\nu;\CL',\CL))$ from Section \ref{sec:comp}. We leave it to future work, or to the inspired reader, to spell this out.

A different option, which connects to the BFN construction, is the following. There is a finite-distance boundary condition $\CB_R$ that preserves a 2d $\CN=(2,2)$ subalgebra of 3d $\CN=4$ (including $Q_A$), defined by
\begin{itemize}
\item Setting the hypermultiplet scalars $Y$ (which are valued in $R^*$) to zero at the boundary, and extending this to the entire hypermultiplet in a way that preserves 2d $\CN=(2,2)$ supersymmetry. In particular, $X\in R$ will get a Neumann-like boundary condition, so that the values of $X$ are unconstrained at the boundary.

\item Preserving gauge symmetry at the boundary, meaning Neumann boundary conditions for the 3d gauge field, extended to the entire 3d vectormultiplet in a way that preserves 2d $\CN=(2,2)$ SUSY. In particular, the complex scalars $\varphi$ also receive a Neumann-like boundary condition, so their values at the boundary are unconstrained.
\end{itemize}
See \cite{BDGH} for further details. The boundary condition $\CB_R$ is \emph{almost} canonical, though it does depend on a $G$-invariant Lagrangian splitting of the hypermultiplet fields. When writing the hypermultiplet representation as $T^*R$, we have already chosen such a splitting, into $R\oplus R^*$. However, often there are other splittings available, \emph{i.e.} ways to rewrite $T^*R \simeq T^*V$ for some other $G$-representation $V$. We keep track of the choice of splitting in the subscript `$R$' of $\CB_R$.

\subsection{Moduli spaces}
\label{sec:modR}

Now consider a line operator $\CL$ ending on the boundary condition $\CB_R$ described above. Suppose that the line operator is characterized by algebraic data $\CG_0,\CL_0$ as in Section \ref{sec:VL-GR}. Then we expect that local operators at the intersection of $\CL$ and $\CB_R$ can be identified as cohomology classes
\be \text{Hom}_A(\CB_R,\CL) = H^*\big(\CM_{\rm half}(\CB_R,\CL)\big)\,, \label{Ops-app} \ee
where
\be \label{Mhalf} \begin{array}{l}
 \CM_{\rm half}(\CB_R,\CL) =  \left\{ \begin{array}{l}
  \text{solutions to $\SQM_A$ BPS equations on an infinitesimal disc} \\
  \text{\quad around $z=0$, punctured by $\CL$ at $z=0$} \\
  \text{\quad and compatible with $\CB_R$ for $z\neq 0$} \end{array}\right\} \\[.8cm]
  \hspace{.7in}= \left\{ \begin{array}{l} \text{$E,X$ s.t. $E$ is an algebraic $G_\C$ bundle on a formal disc} \\
   \text{\quad and $X(z)\in R(\CK)$ is an algebraic section of an associated} \\
   \text{\quad  $R$-bundle away from $z=0$, with $X\in \CL_0\cap R(K)$} \end{array}\right\}/\CG_0  \\[.8cm]
   \hspace{.7in}= \CG_0 \big\backslash (\CL_0\cap R(\CK))\,.
 \end{array}
\ee
In general, this space is an infinite-dimensional stack, and it seems from the mathematics literature that one correct way to interpret \eqref{Ops-app} is via equivariant Borel-Moore homology.

Local operators at junctions of lines act on spaces $H^*\big(\CM_{\rm half}(\CB_R,\CL)\big)$ by convolution. Just as in Section \ref{sec:cyl}, this furnishes a representation of the category of line operators and its Hom spaces. In the current half-space setup with a boundary condition $\CB_R$, the Hom spaces are represented in cohomologies of raviolo spaces
\be \text{Hom}_A(\CL,\CL') \,\longrightarrow\, H^*\big(\CM_{\rm half}^{\rm rav}(\CB_R;\CL',\CL)\big)\,, \label{junc-app}\ee
where
\be \label{Mhalfrav} \begin{array}{l}
 \CM_{\rm half}^{\rm rav}(\CB_R;\CL',\CL) =  \left\{ \begin{array}{l}
  \text{solutions to $\SQM_A$ BPS equations on two infinitesimal} \\
  \text{\quad  discs punctured by $\CL$ and $\CL'$ at $z=0$} \\
  \text{\quad and equivalent to each other and to $\CB_R$ for $z\neq 0$} \end{array}\right\} \\[.8cm]
  \hspace{.7in}= \left\{ \begin{array}{l} \text{$E',X';g;E,X$ s.t. $E',X'$ and $E,X$ are as in \eqref{Mhalf}} \\
   \text{\quad and $g(z)\in G(\CK)$ is an isomorphism away from $z=0$} \end{array}\right\}/\CG_0'\times \CG_0  \\[.8cm]
   \hspace{.7in}= \begin{array}{l} \CG_0' \big\backslash (\CL_0'\cap R(\CK)) \times G(\CK) \times (\CL_0\cap R(\CK))\big|_{(*)}\big/\CG_0 \\
    \hspace{.7in} X' \hspace{.5in} g \hspace{.8in} X \end{array}
 \end{array}
\ee
subject to a constraint $(*)$ that $X'=gX$. This space is again an infinite-dimensional, singular stack. The mathematical literature indicates that the cohomology \eqref{junc-app} can be interpreted as (renormalized) equivariant Borel-Moore homology \cite{BFNII, Raskin}.

\subsubsection{Monopole number}

The spaces $\CM_D(\CB_\nu;\CL)$ analyzed in the main body of the paper decomposed into (infinitely many) finite-dimensional components  $\CM_D^\n$, labeled by vortex number. This decomposition is ultimately a feature of the vacuum boundary condition $\CB_\nu$. From a physical/analytic perspective, the fact that $\CB_\nu$ breaks gauge symmetry near $|z|\to \infty$ implies that the vortex number $\int_D \text{Tr}\, F$ is well defined.

In contrast, the boundary condition $\CB_R$, which is Neumann on the gauge fields, preserves gauge symmetry. There is no longer a well-defined vortex number. Thus, the spaces $\CM_D(\CB_\nu;\CL)$ do not generally admit further decompositions, and have a single infinite-dimensional component.

Similarly, the raviolo spaces $\CM_{\rm half}^{\rm rav}$ do not decompose according to pairs of vortex numbers $\n,\n'$. They \emph{do} have connected components labeled by a single element $\mathfrak m \in \pi_1(G)$,
which is a winding number for the single algebraic loop group $G(\CK)$ appearing in \eqref{Mhalfrav}. This single $\mathfrak m$ corresponds to the difference $\n'-\n$ of vortex numbers from Section \ref{sec:vxnumber}. Physically, it is the ``monopole number'' of local operators --- their charge under the topological flavor symmetry of the 3d gauge theory. Unlike $\CM_{\rm rav}^{\n',\n}$, the components in the decomposition  $\CM_{\rm half}^{\rm rav} = \bigsqcup_{\mathfrak m} \CM_{\rm half}^{\rm rav,\mathfrak m}$ are generally infinite-dimensional.

\subsubsection{The BFN algebra}

The construction above specializes to reproduce the Braverman-Finkelberg-Nakajima definition of the Coulomb-branch chiral ring. To see this, we take $\CL'=\CL=\id$, and consider the raviolo space
\be \CM_{\rm half}^{\rm rav}(\CB_R;\id,\id) = G(\CO) \big\backslash \underset{X'}{R(\CO)}\times \underset{g}{G(\CK)}\times \underset{X}{R(\CO)\big|_{(*)}} \big/G(\CO)\,, \ee
with a constraint $(*)$ that $X'=gX$.
This is the ``BFN space'' of \cite{Nak,BFNII}, and the Coulomb-branch chiral ring was identified with its Borel-Moore homology
\be \C[\CM_C] \simeq H^*\big(\CM_{\rm half}^{\rm rav}(\CB_R;\id,\id) \big) = H^*_{G(\CO)}\big(R(\CO)\times G(\CK)\times R(\CO)\big|_{(*)}\big/G(\CO)\big)\,. \ee
In this case, the map \eqref{junc-app} representing the space $\C[\CM_C]=\text{Hom}_A(\id,\id)$ seems to be faithful. 

For pure gauge theory, with $R=0$, the boundary condition $\CB_R$ is canonical --- the only choice made is to put Neumann b.c. on the gauge fields. In this case the raviolo space above reduces to
\be  \CM_{\rm half}^{\rm rav}(\CB_R;\id,\id) = G(\CO)\backslash G(K)/G(\CO)  = G(\CO) \backslash \text{Gr}_G\,, \ee
where $\text{Gr}_G$ is the affine Grassmannian; and the corresponding representation of bulk local operators becomes
\be  H^*\big(\CM_{\rm half}^{\rm rav}(\CB_R;\id,\id)\big) = H^*_{G(\CO)}(\text{Gr}_G)\,. \ee
This famous convolution algebra was studied by \cite{BFM} and proposed by Teleman \cite{Teleman-ICM} to be the Coulomb branch chiral ring of pure gauge theory.

\subsubsection{Coupling to quantum mechanics}

If instead a line operator $\CL$ is defined by coupling to quantum-mechanics with target $\CE$ and a $G(\CO)$-invariant superpotential $\wt W_0:\CE\times T^*R(\CO)\to \C$, then the algebraic moduli space $\CM_{\rm half}$ in the presence of a $\CB_R$ boundary condition become
\be \label{Mhalf-QM} \begin{array}{ll}
 \CM_{\rm half}(\CB_R,\CL) &= \left\{ \begin{array}{l} \text{$\alpha, E,X$ s.t. $\alpha\in \CE$, $E$ is a $G_\C$ bundle on a formal disc,} \\
   \text{\quad and $X\in R(\CK)$ is a section of an $R$-bundle away from $z=0$,} \\
   \text{\quad all subject to $\delta W=0$} \end{array}\right\}/G(\CO)  \\[.8cm]
   &= G(\CO) \big\backslash \big(\CE\times R(\CK)\big|_{\delta W=0}\big)\,,
 \end{array}
\ee
where $\delta W=0$ denotes the critical point equations from \eqref{tWcrit}, intersected with $Y=0$ (\emph{i.e.} $y_n= 0$ $\forall n$) due to the $\CB_R$ boundary condition. Similarly, the raviolo space becomes
\be \label{Mhalfrav-QM} \begin{array}{ll}
 \CM_{\rm half}^{\rm rav}(\CB_R;\CL',\CL) &=
  \left\{\begin{array}{l} \text{$\alpha',E',X';g;\alpha,E,X$ s.t. } \\
   \text{\quad each triple $\alpha',E',X$ and $\alpha,E,X$ is as in \eqref{Mhalf-QM}} \\
   \text{\quad and $g:(E,X)\overset{\sim}{\to} (E',X')$ is an isomorphism away from $z=0$}  \end{array} \right\} \\[.8cm]
  &= G(\CO)\big\backslash \underset{\alpha'}{\CE}\times \underset{X'}{R(\CK)}\times \underset{g}{G(\CK)} \times \underset{X}{R(\CK)}\times \underset{\alpha}{\CE} \big|_{(*)} \big/G(\CO)
 \end{array}
 \ee
with a constraint $(*)$ that sets $\delta W'=\delta W=0$ (at $Y'=Y=0$) and $X'=gX$.

For a line operator defined by algebraic data $\CG_0, \CL_0$ that can be engineered by coupling to quantum mechanics with target $\CE_0$ and superpotential $\wt W_0$ --- as in Section \ref{sec:VL-GR} --- the algebraic moduli spaces \eqref{Mhalf-QM} and \eqref{Mhalf} are isomorphic; as are \eqref{Mhalfrav-QM} and \eqref{Mhalfrav}.

\section{Conifold matching}
\label{app:conifold}

In this appendix we address the matching of \be
\text{End}^\varepsilon_{A}(\V_{\rm con}) \simeq \text{End}^\varepsilon_{B}(\W_{\mathbf 2})
\ee in a more complete manner. We start with a discussion of $\End^\varepsilon_{B}(\W_{\mathbf 2})$ and some straightforward relations, followed by simple operators in $\End^\varepsilon_{A}(\V_{\rm con})$ and then explicit matching of operators and parameters across mirror symmetry.

\subsection{Fundamental Wilson line}
\label{app:Wilson}

As in Section \ref{sec:conifold}, it will be useful to identify the bulk algebra, because those operators (in particular, the operators generating the (complexified) $PSU(2)_F$ flavor symmetry) will help organize the algebra of operators on the fundamental Wilson line.

The Higgs-branch chiral ring can be described as the ($U(1)_L \times U(2)_C \times U(1)_R$)-invariant part of the Heisenberg algebra $\C_{\varepsilon}[X^m{}_i, Y^i{}_m, U^m, V_m, W^m,Z_m]$ modulo the complex moment map relations\footnote{In order to impose the moment map constraints in the presence of an $\Omega$-background (where ordering of operators is important), we stick to the conventions described in Section \ref{sec:WL}.
} %
\be
	\mu_L + t_L = 0,  \hspace{1cm} (\mu_C)^m{}_n + t_C \delta^m{}_n = 0, \hspace{1cm} \mu_R + t_R = 0
\ee
where
\be
	\mu_L = :U^m V_m:,  \hspace{1cm} (\mu_C)^m{}_n = :U^m V_n + W^m Z_n + X^m{}_i Y^i{}_n:, \hspace{1cm} \mu_R = :W^m Z_m:.
\ee
As usual, $: ... :$ denotes the normal ordered product.

We denote the generating elements by 
\begin{align}
	X^m{}_j Y^i{}_m &= 
	\begin{pmatrix}
		\frac{1}{2}J^0 &	J^+	\\
		J^-		& -\frac{1}{2}J^0 
	\end{pmatrix}^i_{\; j}
	+ \frac 12 X^m{}_k Y^k{}_m \delta^i{}_j, \nonumber \\
	(U^mV_n - W^mZ_n )X^n{}_j Y^i{}_m &= 
	\begin{pmatrix}
		\frac 1 2 M^0	&	M^+	\\
		M^- 	& - \frac 1 2 M^0 
	\end{pmatrix}^i_{\; j}
	+ \frac 1 2 (U^mV_n - W^mZ_n )X^n{}_k Y^k{}_m \delta^i{}_j.
\end{align}
The pure trace parts are not independent operators, they are simply numbers determined by the FI parameters and $\varepsilon$. For instance, using the moment-map relations, we have (within $\End^\varepsilon_B(\mathds{1})$)
\begin{align}
	X^m{}_k Y^k{}_m = - U^m V_m - W^m Z_m - 2 (t_C-2\varepsilon) = - 2 t_C + t_L + t_R + 2 \varepsilon.
\end{align}
A similar computation yields
\begin{align}
	(U^mV_n - W^mZ_n )X^m{}_k Y^k{}_n = (t_L - t_R) (t_C - t_L - t_R - \varepsilon).
\end{align}

The operators $J^0, J^{\pm}$ generate the (complexified) $PSU(2)_F$ flavor symmetry, while the operators $M^0,M^{\pm}$ transform in the adjoint representation of this $PSU(2)_F$. For instance,
\be
[J^+, J^-] = [X^m{}_2 Y^1{}_m, X^n{}_1 Y^2{}_n] = X^m{}_2[Y^1{}_m, X^n{}_1]Y^2{}_n + X^n{}_1[X^m{}_2,Y^2{}_n]Y^1{}_m = \varepsilon J^0.
\ee
The other commutation relations are computed similarly. The triple $J^0, J^{\pm}$ have a Casimir element we will denote as
\be
J^2 := 2 \{J^+, J^-\} + (J^0)^2.
\ee
In addition to these transformation properties, there are two $PSU(2)_F$ invariant operators $JM$ (of $R$-charge 3) and $M^2$ (of $R$-charge 4) satisfying relations (within $\End^\varepsilon_B(\mathds{1})$)
\be \label{relJM}
JM:= 2 J^+ M^- + 2 J^- M^+ + J^0 M^0 = \tfrac{1}{2}(t_R-t_L)J^2 - \tfrac{1}{2}(t_R^2-t_L^2)(2 t_C - t_L - t_R)
\ee
and
\be \label{relMM}
\begin{array}{rl} 
	M^2:= 2 \{M^+, M^-\} + (M^0)^2 & = \tfrac{1}{4} (J^2)^2-\tfrac{1}{2}(2t_C^2-2t_C(t_L+t_R)+(t_L+t_R)^2 -6 \varepsilon^2)J^2\\
	&  -2 t_C^2(t_L^2 + t_R^2)-t_C(t_L+t_R)(3 t_L^2-2 t_L t_R +3 t_R^2)\\
	& +\tfrac{1}{4}(t_L + t_R)^2(5 t_L^2-6 t_L t_R + 5 t_R^2)\\
	&-\varepsilon^2(4t_C^2-4t(t_L+t_R)+3t_L^2+2 t_L t_R + 3 t_R^2)+4\varepsilon^4.
\end{array}
\ee

Now consider the algebra of operators on a Wilson line transforming in the fundamental representation of $U(2)_C$, \ie\, $\End^\varepsilon_{B}(\W_{\mathbf 2})$. Following Section \ref{sec:WL}, these operators can be realized as elements of $\C_{\varepsilon}[X^m{}_i, Y^i{}_m, U^m, V_m, W^m,Z_m]$ transforming as $\End(\mathbf 2)$, modulo the modified moment map relations given by \eqref{WB-eg}.

In addition to the above operators (times $\delta^m{}_n$), the algebra of operators on this Wilson line is generated by the following operators:
\begin{align}
	U^m V_n = (\Psi_L)^m{}_n,
	&&W^m Z_n = (\Psi_R)^m{}_n,
	&&X^m{}_j Y^i{}_n =
	\begin{pmatrix}
		\frac 1 2 (N^0)^m{}_n && (N^+)^m{}_n \\
		(N^-)^m{}_n && - \frac 1 2(N^0)^m{}_n
	\end{pmatrix}^i_{\; j}+ \frac 1 2 (X^m{}_k Y^k{}_n) \delta^i{}_j.
\end{align}
Just like on the trivial line, the trace $X^m{}_k Y^k{}_n$ can be solved for in terms of $\Psi_L, \Psi_R$, FI parameters, and $\varepsilon$ using the $U(2)_C$ moment map constraint. In the following discussion, we will suppress most indices and juxtaposition of elements should be understood as matrix multiplication. Operators built out of gauge invariant quantities should be understood as being proportional to the identity endomorphism, \ie\, $(\CO^{G-\textrm{invt}})^m{}_n = \CO^{G-\textrm{invt}}  \delta^m{}_n$.

Using the above definitions for our operators, we get the following relations describing their transformation properties under the action of $PSU(2)_F$: \begin{align}
	&[ J^0 , \Psi_L ] = 0 && [ J^\pm , \Psi_L] = 0 \\
	&[ J^0 , \Psi_R] = 0 && [ J^\pm , \Psi_R] = 0 \\
	&[ J^0,N^0]= 0 &&[ J^0 , N^\pm] = \pm 2 \varepsilon N^\pm \nonumber \\ 	
	&[J^\pm, N^0] = \mp \varepsilon N^\pm
	&& [J^\pm , N^\pm] = 0 \nonumber \\
	&[J^\pm, N^\mp] = \pm \varepsilon N^0
\end{align}
In particular, $\Psi_L, \Psi_R$ are scalars, while $N^0, N^{\pm}$ transform in the adjoint representation.

Amongst these new generators there are relations given by
\begin{align}
	&\Psi_L^2 + t_L \Psi_L  = 0 \label{app:psi1rel}\\
	&\Psi_R^2 + t_R \Psi_R  = 0 \label{app:psi2rel}\\
	&2 \{ N^+ , N^- \} + (N^0)^2  + \{ \Psi_L , \Psi_R \} \nonumber \\
	&\qquad - (2 t_C - t_L - 2 t_R + 4 \varepsilon ) (\Psi_L+\Psi_R) - t_C\left (3 t_C - 2 t_L - 2 t_R + 4 \varepsilon \right )= 0, \label{app:NNCasimirrel}\\
	& 2J^+ N^- + 2 J^- N^+ + J^0 N^0 - 2\{\Psi_L, \Psi_R\} \nonumber\\
	& \qquad-(2t_C - t_L + t_R+5\varepsilon) \Psi_L -(2t_C + t_L - t_R+5\varepsilon) \Psi_R-(t_C+2\varepsilon)(t_L + t_R + 3\varepsilon) = 0 \label{app:NJCasimirrel}
\end{align}
which all transform trivially under $PSU(2)_F$. In addition to above scalar relations, there are relations 
\begin{align}
	& \{N^+, \Psi_L + \Psi_R \} + [ N^0 , N^+] - 2(t_C - t_L - t_R + 2 \varepsilon ) N^+ = 0, \label{app:PsiNacommtriplerel}\\
	& [N^+, \Psi_L + \Psi_R] + \{N^+, N^0\}-2J^+ N^0-2\varepsilon N^+ = 0, \label{app:PsiNcommtriplerel}\\
	& [N^+, N^0]+(2t_C -t_L - t_R+5\varepsilon)N^+-(\Psi_L + \Psi_R+t_C + 2\varepsilon)J^+=0, \label{app:NNasymtriplerel}\\
	& N^+(N^+-J^+)=0, \label{app:NNsymmpentrel}
\end{align}
which are highest weight under the $PSU(2)_F$. The first three generate triplets, while the last generates a pentuplet. As an example of how these relations are computed using the framework of Section \ref{sec:WL}, consider the simple relation for $\Psi_L$. We find:
\begin{align}
	&(\Psi_L^2)^m{}_n = U^m V_l U^l V_n = U^m ( \mu_L - \varepsilon) V_n = \mu_L U^m V_n = -t_L (\Psi_L)^m{}_n.
\end{align}

It is interesting to note that, for $2 \times 2$ matrices $A, B$ with mutually commuting matrix elements, there is an identity
\begin{align}
	\Tr[A B] \mathds{1} = \{A,B\} - \Tr[A] B - \Tr[B] A + \Tr[A] \Tr[B] \mathds{1}.
\end{align}
Applying this identity to $A = \Psi_1 - \Psi_2$ and $B = N^{\pm}, N^0$ yields the following relations (within $\End^\varepsilon_B(\W_{\mathbf 2})$):
\begin{align}
	& M^{\pm} = \{\Psi_L - \Psi_R, N^{\pm}\} + (t_L - t_R) N^{\pm} - J^{\pm} (\Psi_L - \Psi_R) - (t_L - t_R)J^{\pm} \nonumber\\
	& M^0 = \{\Psi_L - \Psi_R, N^0\} + (t_L - t_R)N^0 - J^0 (\Psi_L - \Psi_R)- (t_L - t_R)J^0
\end{align}
implying that $M^{\pm}, M^{0}$ are not independent generators within the algebra of local operators bound to this Wilson line.

Even more interestingly, the relations between the gauge invariant operators $J^0, J^{\pm}, M^0, M^{\pm}$ are deformed on the fundamental Wilson line.\footnote{This deformation arises due to the different quotient on the Wilson line. In particular, on the trivial Wilson line we must quotient by \be (\mu_\C)^m{}_n + t \delta^m{}_n = 0\ee whereas on the fundamental Wilson line we must quotient by \be ((\mu_\C)^m{}_n + t \delta^m{}_n)\delta^p{}_q + \varepsilon \delta^m{}_q \delta^p{}_n = 0.\ee These two quotients agree when $\varepsilon = 0$.} For starters, the operator $J^2$ is not independent on the fundamental Wilson line. In particular, a straightforward application of the above $2 \times 2$ matrix identity leads to
\be
J^2 = 4 \{\Psi_L, \Psi_R\}+4 t_R \Psi_L+4 t_L\Psi_R + (t_L + t_R)^2-\varepsilon^2.  \label{J2-rel}
\ee
It is worth mentioning that even though $J^2$ is can be expressed as a product of operators proportional to the identity endomorphism, it is not itself proportional to the identity endomorphism. This unusual feature is purely due to the quotient on the Wilson line for nonzero $\varepsilon$, indeed for $\varepsilon = 0$ the operator $J^2$ is proportional to the identity endomorphism.

Similar computations show that
\be \label{relJM2}
\begin{array}{rl}
	J M & = \tfrac{1}{2}(t_R-t_L)J^2 - \tfrac{1}{2}(t_R-t_L)(t_L + t_R + \varepsilon)(2 t_C - t_L - t_R + \varepsilon)\\
	& + \varepsilon (2t_C-t_L-t_R+ \varepsilon) (\Psi_L-\Psi_R)\\
\end{array}
\ee
and
\be
\begin{array}{rl} \label{relMM2}
	M^2 & = \tfrac{1}{4}(J^2)^2-\tfrac{1}{2}(2t_C^2-2t_C(t_L+t_R)+(t_L+t_R)^2+2\varepsilon(t_L+t_R)+5 \varepsilon^2)J^2\\
	& + 4 \varepsilon(\Psi_L \Psi_R \Psi_L + \Psi_R \Psi_L \Psi_R)-\varepsilon(4t_C-4t_L-4t_R-\varepsilon)\{\Psi_L, \Psi_R\}\\
	& +\varepsilon(4t_C^2-6t_C(t_L+t_R)+4(t_L^2+t_L t_R +t_R^2)+\varepsilon(4t_C-3t_L)-6\varepsilon^2)\Psi_L\\
	& +\varepsilon(4t_C^2-6t_C(t_L+t_R)+4(t_L^2+t_L t_R +t_R^2)+\varepsilon(4t_C-3t_R)-6\varepsilon^2)\Psi_R\\
	&+2t_C^2(t_L^2+t_R^2)-t_C(t_L+t_R)(3t_L^2-2t_L t_R+3t_R^2)+\tfrac{1}{4}(t_L+t_R)^2(5t_L^2-6t_L t_R+5t_R^2)\\
	&+\varepsilon(t_L+t_R)(2t_C^2-2t_C(t_L+t_R)+(t_L+t_R)^2)+\tfrac{1}{2}\varepsilon^2(10t_C^2-14t_C(t_L+t_R)+7t_L^2+6t_L t_R+7 t_L^2)\\
	& - \varepsilon^3(4t_C+t_L+t_R)+\tfrac{9}{4}\varepsilon^4\\
\end{array}
\ee
which differ from those relations found on the trivial line by terms proportional to $\varepsilon$. Again, even though $JM$ and $M^2$ can be expressed as products of operators proportional to the identity endomorphism, they are not themselves proportional to the identity endomorphism.

In the following subsection, we will compare the above analysis to a Hilbert series computation. We find agreement between the above generators and relations and the Hilbert series, up to R-charge 2. The subsequent subsection focuses on identifying a handful of operators in $\End^\varepsilon_A(\V_{\rm con})$ which are putative mirrors to the above. The last subsection of this appendix discusses matching the relations and parameters across mirror symmetry. The operators that we need to match are $J^0, J^{\pm}$, $N^0, N^{\pm}$, $\Psi_L$ and $\Psi_R$; their mirrors should satisfy the various relations listed above.

\subsection{Hilbert series}
\label{app:Hser}

The structure of B-type local operators and relations, both in the bulk and on a Wilson line, can be encoded in a Hilbert series. They are fairly straightforward to compute, \emph{cf.} \cite{BFHH-Hilbert, HM-complete}.

First consider the algebra $\CA=\C[X,Y,U,V,W,Z]/(\mu_L,\mu_C,\mu_R)$ of all polynomials in the hypermultiplet scalars, modulo moment-map relations. Its Hilbert series is the graded trace
\begin{align} \CI[\CA] &:= \text{Tr}_\CA \rho^H f^F s_L^{e_L} s_1^{e_{C,1}} s_2^{e_{C,2}} s_R^{e_R} \notag \\
&= \prod_{\sigma=s_L,s_R}\prod_{m=1,2} \frac{1}{(1-\rho^{\frac12} \frac{s_m}{\sigma})(1-\rho^{\frac12} \frac{\sigma}{s_m})}
\prod_{m=1,2}\prod_{\epsilon,\epsilon'=\pm} \frac{1}{(1-\rho^{\frac12}f^{\frac12\epsilon}s_m^{\epsilon'})} \\
&\quad\times (1-\rho)^4(1-\rho\tfrac{s_1}{s_2})(1-\rho\tfrac{s_2}{s_1})\,, \notag
\end{align} 
where $H,F,e_L,e_{C,m},e_R$ denote the charges under the Cartans $U(1)_H,U(1)_F,U(1)_L,U(1)^2_C,U(1)_R$ of R-symmetry, flavor symmetry, and gauge symmetry; and $\rho,f,s_L,s_m,s_R$ are the corresponding fugacities. The denominator contains the charges of the components of the hypermultiplet fields, while the numerator contains the charges of the moment maps. We could also have introduced FI parameters and an Omega-background; since both of these are flat deformations, they do not alter the Hilbert series.

The Hilbert series of the bulk algebra $\text{End}_B^\varepsilon(\id) = \C_\varepsilon[\CM_H]$ (with or without Omega-deformation) is given by projecting to gauge-singlets:
\begin{align} \CI_\id &:= \text{Tr}_{\text{End}_B(\id)} \rho^H f^F \notag \\
&= \oint \frac{ds_L ds_1 ds_2 ds_R}{(2\pi i)^4} \frac12\Big(1-\frac{s_1}{s_2}\Big)\Big(1-\frac{s_2}{s_1}\Big) \CI[\CA]\,,
\end{align}
where $\tfrac12(1-\tfrac{s_1}{s_2})(1-\tfrac{s_2}{s_1})$ is the Haar measure for $U(2)_C$. This can be computed simply by expanding the integrand as a series in $\rho$, and taking the $s$-independent part of each term. Alternatively, one may use a Jeffrey-Kirwan residue prescription \cite{JeffreyKirwan}. The result of the residue calculation is easily reorganized into
\be \CI_{\mathds{1}} = \frac{(1-\rho^3)(1-\rho^4)}{(1-\rho)(1-\rho f)(1-\rho/f)\times (1-\rho^2)(1-\rho^2f)(1-\rho^2/f)}\,. \ee
This is perfectly consistent with the claim that $\text{End}_B^\varepsilon(\id)$ is generated by gauge-invariant operators $J^0,J^\pm;M^0,M^\pm$ (contributing each factor in the denominator), modulo two relations of R-charge 3 and 4 (contributing the factors in the numerator). As expected, this agrees with the Hilbert series for the Coulomb branch chiral ring of the mirror theory, see \eg\, \cite{CHZ-Hilbert}.

The Hilbert series of the Wilson-line algebra $\text{End}_B(\W_{\mb 2})$ is given by projecting to gauge-adjoints. This is accomplished by inserting a character of the $\text{End}(\mb{\bar 2})$ representation:
\begin{align} \CI_{\mb 2} &:= \text{Tr}_{\text{End}_B(\W_{\mb 2})} \rho^H f^F \notag \\
&= \oint \frac{ds_L ds_1 ds_2 ds_R}{(2\pi i)^4} \frac12\Big(1-\frac{s_1}{s_2}\Big)\Big(1-\frac{s_2}{s_1}\Big)\times  \Big(1+\frac{s_1}{s_2}\Big)\Big(1+\frac{s_2}{s_1}\Big)\CI[\CA] \\
&= \frac{(1-\rho^2)^3\big(1-\rho^3+(1-\rho)(1+f+f^{-1})\rho\big)}{(1-\rho)^2(1-\rho)(1-\rho f)(1-\rho/f)(1-\rho^2)(1-\rho^2 f)(1-\rho^2/f)}\,. \notag
\end{align}
There is no way to simplify this as a product of binomials, as would be the case for the \emph{commutative} ring of functions on a complete intersection. The algebra $\text{End}_B(\W_{\mb 2})$ is intrinsically non-commutative (with or without Omega background), and we should not expect it to have a simple Hilbert series.  We are better off expanding in $\rho$, to find
\be \label{Hilb2} \CI_{\mb 2} = 1+\big[2+2\chi_f(\mb 3)\big]\rho + \big[2+5\chi_{f}(\mb 3)+2\chi_f(\mb 5)\big]\rho^2 + O(\rho^3)\,,  \ee
where $\chi_f(\mb 3) = f+1+f^{-1}$ and $\chi_f(\mb 5)=f^2+f+1+f^{-1}+f^{-2}$ are the characters of the triplet and pentuplet representations of $PSU(2)_F$.

The term in \eqref{Hilb2} of order $\rho$ contains the generators we identified above, all of R-charge~1. There are two $(\Psi_L,\Psi_R)$ in a singlet of $PSU(2)_F$, and 6 $(J^0,J^\pm;N^0,N^\pm)$ forming two triplets. The term of order $\rho^2$ counts the operators of R-charge 2, modulo the relations satisfied by those operators.

The various R-charge 2 operators we have are:\footnote{We distinguish the products $A B$ and $B A$ because the product is composition of endomorphisms; unless some relation specifies commutativity, it should not be assumed.}
\begin{itemize}
	\item the bilinears of the adjoints: $NN, J N, N J, J J \rightarrow 4(\chi_f(\mb 5)+\chi_f(\mb 3)+\chi_f(\mb 1))$
	\item the bilinears of adjoints and scalars: $\Psi N,N \Psi,\Psi J,J \Psi \rightarrow 8 \chi_f(\mb 3)$
	\item the bilinears of scalars: $\Psi \Psi \rightarrow 4 \chi_f(\mb 1)$
\end{itemize} for a total of 
\be \CI_{\mb 2}^{2-\rm ops} = 4 \chi_f(\mb 5)+12 \chi_f(\mb 5)+8 \chi_f(\mb 5). \ee
The various R-charge 2 relations we have are:
\begin{itemize}
	\item the $J$'s generate $PSU(2)_F \rightarrow \chi_f(\mb 3)$
	\item the $N$'s transform as $\mb3$ under $PSU(2)_F \rightarrow \chi_f(\mb 5)+\chi_f(\mb 3)+\chi_f(\mb 1)$
	\item the $\Psi$'s transform as $\mb1$ under $PSU(2)_F \rightarrow 2\chi_f(\mb 3)$
	\item the relations \eqref{app:psi1rel}, \eqref{app:psi2rel}, \eqref{app:NNCasimirrel}, \eqref{app:NJCasimirrel},\eqref{J2-rel} transforming as $\mb1 \rightarrow 5\chi_f(\mb 1)$
	\item the relations \eqref{app:PsiNacommtriplerel}, \eqref{app:NNasymtriplerel}, \eqref{app:PsiNcommtriplerel} transforming as $\mb3 \rightarrow 3\chi_f(\mb 3)$
	\item the relation \eqref{app:NNsymmpentrel} transforming as $\mb5 \rightarrow \chi_f(\mb 5)$
\end{itemize}  for a total of 
\be \CI_{\mb 2}^{2-\rm rels} = 2 \chi_f(\mb 5) + 7 \chi_f(\mb 3) + 6 \chi_f(\mb 1).\ee The difference $\CI_{\mb 2}^{2-\rm ops} - \CI_{\mb 2}^{2-\rm rels}$ exactly matches the $\rho^2$ term  of \eqref{Hilb2}. From this, we conclude that, at least up to R-charge 2, the proposed generators and relations given above are consistent with the Hilbert series computation. This does not completely rule out the appearance of additional generators at higher R-charge, though we find it unlikely.

\subsection{Conifold line operator}
\label{app:conifoldops}

Here we work out in detail the algebra of local operators $\text{End}_A^\varepsilon(\V_{\rm con})$ bound to the ``conifold'' vortex line from Section~\ref{sec:FundamentalVortexLine} in 3d $\CN=4$ SQCD with gauge group $G=U(2)$ and four flavors of fundamental hypermultiplets $(X^a{}_i,Y^i{}_a)$, $a=1,2$, $i=1,2,3,4$. We follow the conventions set out in the main text. The line operator $\V_{\rm con}$ was defined in Section \ref{sec:con-alg} by coupling the bulk theory to 1d quiver quantum mechanics \eqref{Q24-2}, repeated here:
\be \raisebox{-.5in}{\includegraphics[width=5.2in]{quiver24}} \label{fig:app:quiver24} \ee
In $Q_A$-cohomology, this is equivalent to coupling to a 1d $\SQM_A$ sigma-model with target the resolved conifold. We fix a bulk vacuum $\nu$ at $z\to \infty$ given by
\be  \nu:\quad  X^a{}_i = \delta^a{}_i\,,\qquad  Y  \equiv 0\,, \ee
as in \eqref{con-vac}.

\subsubsection{Vortex moduli space}

The disc moduli space
\begin{align} \CM_D(\CB_\nu;\V_{\rm con}) &= \bigsqcup_{\n \geq -1}\CM_D^\n(\CB_\nu;\V_{\rm con}) \notag \\
 &\simeq \CI[z]\backslash \{ X(z)\in \CL_0 \cap R[z,z^{-1}] \\
 & \hspace{.6in} \text{s.t.}\; \det X_{(1,2)}\neq 0,\, \text{deg}\,\det X_{(3,4)}< \deg \det X_{(1,2)}\}\,. \notag\end{align}
described in \eqref{MD-con} is covered by affine charts of the form
\be \label{app:iwahoriGaugeFixedC}
\begin{array}{rl}
	\CM_D^{\n,k,1}: &X(z) = \begin{pmatrix}
		z^{k_1} + \sum_{d=-1}^{k_1 - 1} x^1{}_{1 , d} z^d &  \sum_{d=-1}^{k_2 - 1} x^1{}_{2 , d} z^d & \sum_{d=0}^{k_1 - 1}  x^1{}_{3 , d}z^d & \sum_{d=0}^{k_1 - 1}  x^1{}_{4 , d}z^d\\
		\sum_{d=0}^{k_1} x^2{}_{1 , d} z^d  & z^{k_2} + \sum_{d=0}^{k_2 - 1} x^2{}_{2 , d} z^d & \sum_{d=0}^{k_2 - 1}  x^2{}_{3 , d}z^d & \sum_{d=0}^{k_2 - 1}  x^2{}_{4 , d} z^d
	\end{pmatrix}\\
	&\\
	\CM_D^{\n,k,w}: & X(z) = \begin{pmatrix}
		\sum_{d=-1}^{k_1 - 1} x^1{}_{1 , d} z^d &  z^{k_2} + \sum_{d=-1}^{k_2 - 1} x^1{}_{2 , d} z^d & \sum_{d=0}^{k_2 - 1}  x^1{}_{3 , d}z^d & \sum_{d=0}^{k_2 - 1}  x^1{}_{4 , d} z^d\\
		z^{k_1} + \sum_{d=0}^{k_1-1} x^2{}_{1 , d} z^d  & \sum_{d=0}^{k_2} x^2{}_{2 , d} z^d & \sum_{d=0}^{k_1 - 1}  x^2{}_{3 , d}z^d & \sum_{d=0}^{k_1 - 1}  x^2{}_{4 , d} z^d
	\end{pmatrix}\,,\\
\end{array}
\ee
with  $k_{\sigma(1)} \geq -1$ and $k_{\sigma(2)} \geq 0$. There is a unique  $T_F\times \C^*_\varepsilon$ fixed point at the origin of each chart, and we denote the corresponding classes $\ket{\mathfrak{n}, k, \sigma}$. We normalize the vectors $\ket{\mathfrak{n},k,\sigma}$ as
\begin{align}
\ket{\mathfrak{n},k,\sigma} = \frac{1}{\omega_{\mathfrak{n},k,\sigma}} \delta_{\mathfrak{n}, k, \sigma},
\end{align}
where $\delta_{\mathfrak{n}, k, \sigma}$ is the fundamental class of the fixed point labeled by $(k,\sigma)$ and $\omega_{\mathfrak{n},k,\sigma}$ is the equivariant weight of the (normal bundle to the) fixed point in $\CM_{D}^\n(\CB_\nu,\V_{\rm con})$. It follows that the equivariant cohomology of the vortex moduli space is spanned by vectors $\ket{\mathfrak{n}, k, \sigma}$:
\begin{align}
\CH(\CB_\nu, \V_{\rm con}) = H^*_{T_F \times U(1)_\varepsilon}(\CM_D(\CB_\nu, \V_{\rm con})) = \bigoplus \limits_{k,\sigma} \C \ket{\mathfrak{n},k, \sigma}.
\end{align}

In order for the origins of the above patches to be fixed, there must be a compensating torus gauge transformation satisfying $\varphi_{\sigma(a)} + m_a + (k_a + \frac{1}{2})\varepsilon = 0$. It follows that the Coulomb-branch scalars $\varphi_a$ act as
\begin{align}
	\varphi_a \ket{\mathfrak{n},k,\sigma} = -(m_{\sigma(a)} + (k_{\sigma(a)} + \tfrac{1}{2})\varepsilon)\ket{\mathfrak{n}, k, \sigma}.
\end{align}
Using this compensating gauge transformation, the $T_F \times U(1)_\varepsilon$ equivariant weights can expressed as
\begin{align}
	\begin{array}{rl}
		\omega_{\mathfrak{n},k,\sigma}  = & (-1)^{\sigma}(m_1-m_2+(k_1 - k_2)\varepsilon) \\[.2cm]
		& \times  (-k_{\sigma(1)} - 1)\varepsilon(m_{\sigma(2)}-m_{\sigma(1)}+(k_{\sigma(2)}-k_{\sigma(1)}-1)\varepsilon)\\[.2cm]
		& \times \prod\limits_{i=1}^2\prod\limits_{a = 1}^2\prod\limits_{l=0}^{k_i-1}(m_i - m_a + (l-k_a)\varepsilon) \prod\limits_{i=3}^4\prod\limits_{a = 1}^2\prod\limits_{l=0}^{k_a-1}({m}_{i} - m_a + (l-k_a)\varepsilon)\,.
	\end{array}
\end{align}

\subsubsection{Correspondence spaces}

With the equivariant cohomology of the moduli space of vortices in hand, we move to the spaces of correspondences (raviolo spaces) between them:
\begin{align}
	\CM_\rav(\CB_\nu; \V_{\rm con}, \V_{\rm con}) =   \CI'[z]\big \backslash\{ ( X' ; g ; X) ~ | ~ X' = g X , ~ g \in G[z, z^{-1}] \} \big/ \CI[z]
\end{align}
where $\CI[z]$ is the group of residual (polynomial) gauge transformations on the ``bottom" disk, and similarly for $\CI'[z]$ on the ``top" disk. The group $G[z, z^{-1}]$ is the group of $2 \times 2$ matrices over $\C[z, z^{-1}]$.

As discussed in the main body of the text, $\CM_\rav(\CB_\nu; \V_{\rm con}, \V_{\rm con})$ has disconnected components labeled by the number of vortices on the top and bottom disks
\begin{align}
	\CM_\rav(\CB_\nu; \V_{\rm con}, \V_{\rm con}) = \coprod\limits_{\n',\n} \CM^{\mathfrak{n}', \mathfrak{n}}_\rav(\CB_\nu; \V_{\rm con}, \V_{\rm con})
\end{align}
and is endowed with two projection maps $\pi: \CM_\rav(\CB_\nu; \V_{\rm con}, \V_{\rm con}) \rightarrow \CM_{D}(\CB_\nu,\V_{\rm con})$ (resp. $\pi': \CM_\rav(\CB_\nu; \V_{\rm con}, \V_{\rm con}) \rightarrow \CM_{D}(\CB_\nu,\V_{\rm con})$), corresponding to forgetting $g, X'$ (resp. $g, X$). These projection maps respect the decompositions into spaces of fixed vortex number.

First consider the case where $\mathfrak{n} = \mathfrak{n}'$, \ie \, operators of zero monopole charge. These operators will be matched with the scalar operators $\Psi_L, \Psi_R, J^0, N^0$ found in $\End^\varepsilon_B(\W_{\mathbf 2})$. There is an obvious cohomology class corresponding to ``doing nothing", \ie \, the cohomology class dual to the cycle with $g$ the identity matrix or simply a copy of $\CM^{\mathfrak{n}}_D(\CB_\nu,\V_{\rm con})$ inside $\CM^{\mathfrak{n}, \mathfrak{n}'}_\rav(\CB_\nu; \V_{\rm con}, \V_{\rm con})$. The corresponding operator is the identity operator $1$ in the algebra.

Extracting other cohomology classes is somewhat more delicate. In particular, we will look for $\CI[z]$-equivariant cycles in $\CI'[z] \backslash G[z, z^{-1}]$ and submanifolds $\CS$ of $\CM^{\mathfrak{n}, \mathfrak{n}}_\rav(\CB_\nu; \V_{\rm con}, \V_{\rm con})$ to which they belong. As we consider larger and larger cycles in $\CI'[z] \backslash G[z, z^{-1}]$, the submanifolds of $\CM^{\mathfrak{n}, \mathfrak{n}}_\rav(\CB_\nu; \V_{\rm con}, \V_{\rm con})$ that they belong to will shrink.

As an example of this process consider $\CI[z]' \backslash G[z] \subset \CI[z]' \backslash G[z,z^{-1}]$. This can be parameterized by the $2 \times 2$ matrix
\be
\begin{pmatrix}
		a & b \\
		c & d
	\end{pmatrix},
\ee
with $a d - b c \in \C^*$, modulo the left action of $\CI[z]'$. When $b(0):=b_0 \neq 0$ it is possible to act with
\be
g = \begin{pmatrix}
	\frac{-b}{a d - b c} & \frac{d}{a d - b c}\\
	\frac{1}{a d - b c}\big(a-\tfrac{a_0}{b_0}b \big)& \frac{1}{a d - b c}\big(\tfrac{a_0}{b_0}d- c\big)\\
\end{pmatrix} \in \CI[z],
\ee
where $a(0):= a_0$, to find an affine chart given by
\be
	\begin{pmatrix}
		1 & 0\\
		\tfrac{a_0}{b_0}& 1\\
	\end{pmatrix}.
\ee
Similarly, when $a_0 \neq 0$ we find an affine chart given by
\be
\begin{pmatrix}
	0 & 1\\
	1 & \tfrac{b_0}{a_0}\\
\end{pmatrix}.
\ee
We conclude that this (double) orbit is a copy of $\cp^1$ parametrized by two charts given by
\begin{align}
	\begin{pmatrix}
		1 & 0 \\
		c & 1
	\end{pmatrix}
	\qquad 
	\begin{pmatrix}
		0 & 1 \\
		1 & d
	\end{pmatrix},
\end{align}
where $d = \tfrac1 c$ away from $c = 0$.

This class doesn't fit into correspondences over all $X$ as this could place a pole in the bottom row of $X$. Instead, we find that it only fits into the submanifold  $\CS_{(0,0)} \subset \CM^{\mathfrak{n}, \mathfrak{n}}_\rav(\CB_\nu; \V_{\rm con}, \V_{\rm con})$ locally defined by the equations $x^1{}_{i,-1} = 0$, where $i = 1,2$, An interesting class in the equivariant cohomology of the full correspondence space is the fundamental class of this submanifold. This class can be expressed in terms of torus fixed points of $\CS^{(1)}$, which can be realized as a fixed point in the ``base" times a fixed point in the ``fiber" over that base fixed point. The fiber fixed points are the origins of the above patches, \ie
\begin{align}
	g = \mathds{1} = \begin{pmatrix}
		1 &  0 \\
		0  & 1
	\end{pmatrix} \hspace{1cm} g = w = \begin{pmatrix}
		0 &  1 \\
		1  & 0
	\end{pmatrix}.
\end{align}
We therefore have an operator given by
\begin{align}
	\partial_1 = \frac{\delta^{(1)}_{\mathds{1}, 0} - \delta^{(1)}_{w, 0}}{\varphi_2 - \varphi_1}.
\end{align}
where $\delta^{(1)}_{\sigma', 0}$ is a sum over the base fixed points for the fiber fixed point labeled by $\sigma'$. The denominator $\varphi_2 - \varphi_1$ is the equivariant weight of the $\mathds{1}$ fixed point. In order to determine the action of $\partial_1$ on $\ket{\mathfrak{n},k,\sigma}$, it suffices to understand the action of the fixed point classes which follows from \eqref{fixedPointActions} and knowledge of the equivariant weight $e(N_{\CS_{(0,0)}}|\sigma'; k,\sigma)$ of the (normal bundle to the) submanifold. A straightforward computation shows that
\be
e(N_{\CS_{(0,0)}}|\sigma'; k,\sigma) = p(-m_{\sigma(1)}-(k_{\sigma(1)}+1)\varepsilon)
\ee
which is simply given by the product of the equivariant weights of the coordinates $x^1{}_{i,-1}$ within the appropriate chart. Putting this together, the action of $\partial_1$ on the class $\ket{\n,k,\sigma}$ is
\be
\partial_1 \ket{ \n , k , \sigma }= \frac{(-1)^\sigma}{m_1 - m_2 + (k_1 - k_2 ) \varepsilon }  \left( p ( - m_1 - ( k_1 + 1 ) \varepsilon ) \ket{ \n , k , 1} + p ( - m_2 - ( k_2 + 1 ) \varepsilon ) \ket{\n , k , w } \right)
\ee
It is worth noting that when acting on a state $\ket{\mathfrak{n},k,\sigma}$ with $k_{\sigma(1)} = -1$, the second term should vanish. This is because the $w$ fixed point does not fit into correspondences over those fixed points, in particular there are no fixed points with $k_{\sigma(1)} < -1$.

A similar procedure can be applied to the cycle for the closure of $\CI[z]' \backslash \CO_{w,(1,-1)}$, where $\CO_{w,(1,-1)}$ is the $\CI[z]' \times \CI[z]$ orbit of $w z^{(1,-1)}$, which is also a copy of $\cp^1$. This cycle can parameterized by the affine charts
\begin{align}
	\begin{pmatrix}
		1 & b z^{-1} \\
		0 & 1 
	\end{pmatrix},
	&&
	\begin{pmatrix}
		a & z^{-1} \\
		z & 0
	\end{pmatrix}
\end{align}
where $b = 1 / a$ on the overlap. This cycle fits into a submanifold $\CS_{w,(1,-1)}$ described locally by the equations $x^{2}{}_{i',0} = 0$, where $i' = 3,4$, and so
\be
e(N_{\CS_{w,(1,-1)}} |\one; k, \sigma) = \hat p(-m_{\sigma(2)}-k_{\sigma(2)}\varepsilon) = e(N_{\CS_{w,(1,-1)}} |wz^{(1,-1)}; k, \sigma).
\ee
From this, is straightforward to see that fundamental class of this submanifold, which we denote $\partial_2$, acts on $\ket{\n,k,\sigma}$ as
\be
\begin{array}{rl}
	\partial_2 \ket{\mathfrak{n}, k; \sigma} & \ds = \frac{(-1)^\sigma}{m_2 - m_1 + (k_2 - k_1-(-1)^\sigma) \varepsilon}\bigg[\hat p(-m_{\sigma(2)}-k_{\sigma(2)}\varepsilon)\ket{\mathfrak{n}, k; \sigma} \nonumber\\&
	\quad +\, \hat p(-m_{\sigma(1)}-(k_{\sigma(1)}+1)\varepsilon)\ket{\mathfrak{n}, k+e_{\sigma(1)}-e_{\sigma(2)}; w \sigma} \bigg]
\end{array}
\ee 
where $\hat{p}(x) = (x+m_3)(x+m_4)$ and $e_1, e_2$ are the lattice vectors $(1,0)$ and $(0,1)$. Just as with $\partial_1$, when acting on a state $\ket{\mathfrak{n},k;\sigma}$ with $k_{\sigma(2)} = 0$, the second term should vanish. Together with $\varphi_a$, the operators $\partial_1, \partial_2$ will be all that are needed to match with the $\End^\varepsilon_B(\W_{\mathbf 2})$ operators $\Psi_L, \Psi_R, J^0, N^0$.

Now let's move to operators of monopole charge 1. Operators of higher monopole charge can be found in a similar fashion, but these simple operators are sufficient for matching the operators in $\End^\varepsilon_B(\W_{\mathbf 2})$ discussed above. There are two simple $\cp^1$ cycles in $\CI[z]' \backslash G[z, z^{-1}]$ that fit into correspondences over all $X$.

The first $\cp^1$ arises as the closure of $\CI[z]' \backslash \CO_{(1,0)}$, where $\CO_{(1,0)}$ is the $\CI[z]' \times \CI[z]$ orbit of $z^{(1,0)}$ in $G[z,z^{-1}]$. A straightforward computation shows that this $\cp^1$ cycle admits affine charts
\begin{align}
	g = \begin{pmatrix}
		z &  0 \\
		z c  & 1\\
	\end{pmatrix} \hspace{1cm} \text{and} \hspace{1cm} g = \begin{pmatrix}
		0 &  1 \\
		z  & d\\
	\end{pmatrix},
\end{align}
with transition function $d = \tfrac 1 c$. The second arises as the closure of $\CI[z]' \backslash \CO_{(1,0)}$, where $\CO_{(0,1)}$ is the $\CI[z]' \times \CI[z]$ orbit of $z^{(0,1)}$ in $G[z,z^{-1}]$. It admits affine charts
\begin{align}
	g = \begin{pmatrix}
		1 &  b \\
		0  & z\\
	\end{pmatrix} \hspace{1cm} \text{and} \hspace{1cm} g = \begin{pmatrix}
		a &  1 \\
		z  & 0\\
	\end{pmatrix}
\end{align}
with transition function $a = \tfrac 1 b.$

Using the same procedure described for the operators of zero monopole charge, there are two interesting operators $V^+_1$ and $V^+_2$ given by the fundamental classes of the submanifolds determined by these cycles. Their actions on $\ket{\mathfrak{n},k,\sigma}$ are given by
\begin{align}
	V^+_1 \ket{\mathfrak{n}, k, \sigma} &= \frac{(-1)^\sigma \hat p(-m_{\sigma(1)}-(k_{\sigma(1)}+1)\varepsilon)}{m_1 - m_2 + (k_1 - k_2+(-1)^\sigma) \varepsilon} \\
	&\hspace{.2in}\times  \bigg[p(-m_{\sigma(1)}-(k_{\sigma(1)}+2)\varepsilon)\ket{\mathfrak{n}+1, k+e_{\sigma(1)}, \sigma} \nonumber \\
	&\hspace{.3in} + p(-m_{\sigma(2)}-(k_{\sigma(2)}+1)\varepsilon)\ket{\mathfrak{n}+1, k + e_{\sigma(1)}, w \sigma} \bigg],  \notag
\end{align}
and
\begin{align}
	V^+_2 \ket{\mathfrak{n}, k, \sigma} &= \frac{(-1)^\sigma p(-m_{\sigma(2)}-(k_{\sigma(2)}+1)\varepsilon) }{m_1 - m_2 + (k_1 - k_2) \varepsilon} \\
	&\hspace{.2in} \times  \bigg[\hat p(-m_{\sigma(2)}-(k_{\sigma(2)}+1)\varepsilon)\ket{\mathfrak{n}+1, k+e_{\sigma(2)}, \sigma} \nonumber\\
	&\hspace{.3in} + \hat p(-m_{\sigma(1)}-(k_{\sigma(1)}+1)\varepsilon)\ket{\mathfrak{n}+1, k + e_{\sigma(1)}, w \sigma} \bigg].  \notag
\end{align}

Finally, consider operators of monopole charge -1. Just as with the operators of monopole charge $+1$, there are two simple $\cp^1$ classes in $\CI[z]' \backslash G[z, z^{-1}]$ that fit this description. The first is given by $\CI[z]' \backslash \CO_{(-1,0)}$ and admits affine charts
\begin{align}
	g = \begin{pmatrix}
		z^{-1} & z^{-1} b \\
		0  & 1\\
	\end{pmatrix} \hspace{1cm} \text{and} \hspace{1cm} g = \begin{pmatrix}
		z^{-1} a &  z^{-1} \\
		1  & 0\\
	\end{pmatrix},
\end{align}
with transition function $b = \tfrac 1 a$. Just like the monopole 0 operators, this cycle does not fit into correspondences over all $X$. One finds that the normal bundle factors are given by
\be
e(N_{\CS_{(-1,0)}}|z^{(-1,0); k, \sigma}) = p(-m_{\sigma(1)}-(k_{\sigma(1)}+1)\varepsilon) \hat p(-m_{\sigma(1)}-k_{\sigma(1)}\varepsilon)
\ee
and 
\be
e(N_{\CS_{(-1,0)}}|w z^{(0,-1); k, \sigma}) = p(-m_{\sigma(1)}-(k_{\sigma(1)}+1)\varepsilon) \hat p(-m_{\sigma(2)}-k_{\sigma(2)}\varepsilon).
\ee

The second $\cp^1$ arises as $\CI[z]' \backslash \CO_{(0,-1)}$ and admits affine charts \begin{align}
	g = \begin{pmatrix}
		1 &  0 \\
		c  & z^{-1}\\
\end{pmatrix} \hspace{1cm} \text{and} \hspace{1cm} g = \begin{pmatrix}
	0 &  z^{-1}1 \\
	1  & z^{-1}d\\
\end{pmatrix},
\end{align} with transition function $d = \tfrac 1 c$. Again, this cycle does not fit into correspondences over all $X$. The normal bundle factors are given by
\be
e(N_{\CS_{(0,-1)}}|z^{(0,-1)}; k, \sigma) = p(-m_{\sigma(2)}-k_{\sigma(2)}\varepsilon) \hat p(-m_{\sigma(2)}-k_{\sigma(2)}\varepsilon)
\ee
and 
\be
e(N_{\CS_{(0,-1)}}|w z^{(0,-1)}; k, \sigma) = p(-m_{\sigma(1)}-(k_{\sigma(1)}+1)\varepsilon) \hat p(-m_{\sigma(2)}-k_{\sigma(2)}\varepsilon).
\ee

From the fundamental classes of these submanifolds we get operators $V^-_1$ and $V^-_2$. Their actions on $\ket{\mathfrak{n},k;\sigma}$ are given by \begin{align}
	V^-_1 \ket{\mathfrak{n}, k, \sigma} = \frac{(-1)^\sigma}{m_2 - m_1 + (k_2- k_1) \varepsilon} \bigg[\ket{\mathfrak{n}-1, k-e_{\sigma(1)}, \sigma} + \ket{\mathfrak{n}-1, k - e_{\sigma(2)}, w \sigma} \bigg],
\end{align} and \begin{align}
	V^-_2 \ket{\mathfrak{n}, k, \sigma} = \frac{(-1)^\sigma}{m_1 - m_2 + (k_1 - k_2 + (-1)^\sigma) \varepsilon}\bigg[\ket{\mathfrak{n}-1, k-e_{\sigma(2)}, \sigma}+ \ket{\mathfrak{n}-1, k - e_{\sigma(2)}, w \sigma} \bigg].
\end{align} Just as was the case earlier, it should be understood that any occurrence of a vector $\ket{\mathfrak{n}-1, k';\sigma'}$ with $k'_{\sigma'(1)}< -1$ or $k'_{\sigma'(2)} < 0$ should be set to zero.

All together, we have the 8 operators $\varphi_1$, $\varphi_2$, $\partial_1$, $\partial_2$, $V^{\pm}_1$ and $V^{\pm}_2$. They are all $R$-charge 1 and have the correct flavor charges to match the operators of $R$-charge 1 in $\End^\varepsilon_B(\W_{\mathbf 2})$. With the above expressions, it is straightforward to compute relations between these operators. In the next subsection we address the matching of these operators with those in $\End^\varepsilon_B(\W_{\mathbf 2})$.

\subsection{Matching across mirror symmetry}

We now move to matching various local operators and parameters across mirror symmetry. We use the notation where the mirror of an operator $\CO$ in $\End^\varepsilon_B(\W_{\mathbf 2})$ denoted by the operator $\CO^!$ in $\End^\varepsilon_A(\V_{\rm con})$. There are many sources of ambiguities in the matching across mirror symmetry coming from composing with an automorphism of the algebra of operators. We will mention several of these along the way.

As a first step, let us identify the operators mirror to those generating the (comlexified) $PSU(2)$ flavor symmetry. It is well known that, under mirror symmetry, the Higgs-branch flavor torus is mapped to the topological symmetry rotating the dual photon, although there can be mixing with $U(1)_\varepsilon$. This tells us that the scalar operator $J^0$, which measures charge under the Higgs-branch flavor torus, should be mapped to an operator $(J^0)^! = -2(\varphi_1 + \varphi_2) + c$, for $c$ some linear function of $\varepsilon$. The operators $(J^\pm)^!$ should then have charges $\pm 2$ with respect to $(J^0)^!$, \ie\, they should increase monopole charge by $\pm 1$; therefore they should be linear combinations of the operators $V^{\pm}_1, V^{\pm}_2$.

A convenient way to pin down the forms of $(J^{\pm})^!$ is to consider the operator \begin{align}
	[a_+ V^+_1 + b_+ V^+_2, a_- V^-_1 + b_- V^-_2]
\end{align} and require that $\ket{\mathfrak{n},k,\sigma}$ is an eigenvector thereof. This constraint automatically implies that $a_\pm = \mp b_\pm$, and to get something that matches the form of $(J^0)^!$ above implies that $a_+ a_- = 1$ for $c = \varepsilon.$\footnote{It is important to remember that $m_i$ are equivariant parameters for the $P(U(2)\times U(2)')$ flavor symmetry, chosen so that $m_1 + m_2 + m_3 +  m_4 = 0$.} We shall choose $a_+ = a_- = 1$, namely \begin{align}
	& (J^{\pm})^! = V^\pm_1 \mp V^\pm_2\nonumber\\
	& (J^0)^! = -2(\varphi_1 + \varphi_2) + \varepsilon.
\end{align} We conclude that the operators $(J^0)^!, (J^{\pm})^!$ generate the (complexified) $PSU(2)$ Coulomb-branch flavor symmetry and are mirror to the operators $J^0, J^{\pm}$.\footnote{There is some ambiguity in this identification arising from automorphisms of $PSU(2)$. For example, $(J^0)^! = 2(\varphi_1 + \varphi_2) - \varepsilon$, $(J^{\pm})^! = V^\mp_1 \pm V^\mp_2$ is an equally good identification.}

Now, consider the operators $\Psi_L, \Psi_R$ and their mirrors. They satisfy relations \eqref{app:psi1rel} and \eqref{app:psi2rel}, thus we seek operators with zero monopole charge satisfying similar relations, \ie\ we will look for two independent operators $(\Psi_L)^!$ and $(\Psi_R)^!$ that schematically satisfy $((\Psi_p)^!)^2 + \lambda_p (\Psi_p)^! = 0$ for $p = L,R$ and $\lambda_p$ some linear function of $m_i, \varepsilon$. The precise values of $\lambda_p$ can be determined via a brane construction of these theories and line operators by tracking how positions of branes are exchanged, but we shall ignore this route and find the same values of $\lambda_p$ independent of a brane construction. The most general form of $(\Psi_p)^!$ is given by \begin{align}
	a_p \partial_1 + b_p \partial_2 + c_p \varphi_1 + d_p \varphi_2 + e_p 
\end{align} where $e_p$ is some linear function of $m_i, \varepsilon$. With this parameterization, it is a straightforward computation to solve the equation $((\Psi_p)^!)^2 + \lambda_p (\Psi_p)^! = 0$. We find two solutions\footnote{There is an ambiguity in assigning which solution to label $(\Psi_L)^!$ versus $(\Psi_R)^!$, but this is in agreement with the $\Z_2$ symmetry exchanging the two $U(1)$ factors (and hence $\Psi_L$ and $\Psi_R$). There is a further ambiguity associated to the redefinition $\Psi_p \mapsto \Psi_p':= \lambda_p \one - \Psi_p$ which also satisfies $(\Psi_p')^2 + \lambda_p \Psi_p' = 0$.} given by \begin{align}
	(\Psi_L)^! = \partial_1 + \varphi_1 + m_1 - \tfrac{\varepsilon}{2} \hspace{1cm} (\Psi_R)^! = \partial_2 + \varphi_2 + m_3 + \tfrac{\varepsilon}{2}
\end{align} for $\lambda_L = m_2 - m_1$ and $\lambda_R = m_4 - m_3$. This identification automatically tells us that the mirrors of the FI parameters $t_L$ and $t_R$ should be identified as \begin{align}
	(t_L)^! = m_2 - m_1 \hspace{1cm} (t_R)^! = m_4 - m_3.
\end{align}

Finally, let's move to the operators transforming in the adjoint representation of the flavor symmetry. A straightforward approach is to find highest and lowest weight vectors and then check that they indeed agree ``in the middle," \ie\, that raising the lowest weight operator yields the same operator as lowering the highest weight operator. The only operators with the desired $PSU(2)$ charges of $(N^{\pm})^!$ are linear combinations of $V^{\pm}_1, V^{\pm}_2$, and the choice should be linearly independent of $(J^{\pm})^!$. It turns out that the space of viable linear combinations is reduced to either $V^{\pm}_1$ or $\mp V^{\pm}_2$ when matching the other relations.\footnote{This remaining ambiguity arises from the automorphism $N^0 \mapsto (N^0)' = J^0 - N^0, N^\pm \mapsto (N^\pm)' = J^\pm - N^\pm$.} We choose \begin{align}
	(N^{\pm})^! = V^{\pm}_1.
\end{align} With this choice, we find that we should identify \begin{align}
	(N^0)^! = \partial_1 - \partial_2 - \varphi_1 - \varphi_2 - m_3 - m_4.
\end{align} It is straightforward, although tedious, to check that the relations in $\End^\varepsilon_B(\W_{\mathbf 2})$ are indeed satisfied under the final identification \begin{align}
	(t_C)^! = m_4 - m_1.
\end{align}


\bibliographystyle{JHEP}
\bibliography{lineops}

\end{document}